\pdfoutput=1
\PassOptionsToPackage{dvipsnames,table}{xcolor}
\documentclass[acmsmall,screen,nonacm]{acmart}

\newtoggle{EXTND} %
 \toggletrue{EXTND} %
\iftoggle{EXTND}
{
\settopmatter{printccs=true,printacmref=false,printfolios=true}
\renewcommand\footnotetextcopyrightpermission[1]{}
\pagestyle{plain}
}
{
\settopmatter{}
}

\usepackage{marvosym,listings,etoolbox}
\usepackage{amsmath}
\usepackage{mathtools}
\usepackage{graphicx}
\usepackage{booktabs}
\usepackage{mathpartir}
\usepackage{stmaryrd}
\usepackage[noabbrev]{cleveref}
\usepackage{enumitem}
\usepackage{array}
\usepackage{tabularx}
\usepackage{wrapfig}
\usepackage{subcaption}
\usepackage{xspace}
\usepackage{float}
\usepackage{placeins}
\usepackage{makecell}
\usepackage{adjustbox}
\usepackage{balance}
\usepackage{algorithm}
\usepackage{algpseudocode}

\usepackage{bm}
\usepackage{stmaryrd}

\input{macros}

\begin{abstract}
Array-of-structures (\AoS{}) to structure-of-arrays (\SoA{}) is a classic compiler transformation that improves memory locality and enables data-parallel execution.
Existing \AoS{}-to-\SoA{} transformations target regular, array-based programs in imperative languages, such as \texttt{C}.
However, many applications traverse tree-shaped data such as ASTs in compilers, DOM trees in browsers, and k-d trees in scientific workloads.
Prior work speeds up functional programs over trees by serializing algebraic datatypes (ADTs) into one contiguous buffer.
Such a representation interleaves the fields of every node, similar to an \AoS{} layout.
We introduce a \fullyfactored{} layout of ADTs, which stores its fields in separate buffers, producing a layout that resembles \SoA{}.
In a \fullyfactored{} layout, a traversal can focus on relevant fields and ignore the rest, similar to extracting a column from a \SoA{} representation; in this setting, traversals over scalar fields can be vectorized.
Working with raw byte arrays is error-prone, though: wrong indexing can silently read bytes from a wrong field or write past the end of a value.
We therefore introduce \newcalc{}, a type-safe language for expressing traversals over \fullyfactored{} ADTs.
Our compiler, \colobus{}, transforms a simple functional language with ADTs reminiscent of Haskell to \texttt{C}, using \newcalc{} as the intermediate representation.
\colobus{} generates traversals over byte arrays in the \fullyfactored{} representation and performs optimizations such as field skipping and vectorization.
Across a suite of benchmarks, \colobus{} shows the average speedups of $3.8\times$ and $5.2\times$ over fold and map passes, respectively.
\end{abstract}

\begin{document}

\title{SoCal: A Language for Memory-Layout Factorization of~Recursive Datatypes}

\author{Vidush Singhal}
\orcid{0000-0001-6912-3840}
\affiliation{
  \institution{Purdue University}
  \city{West Lafayette}
  \state{Indiana}
  \country{USA}
}
\email{singhav@purdue.edu}

\author{Mikah Tal Kainen}
\affiliation{
  \institution{Purdue University}
  \city{West Lafayette}
  \state{Indiana}
  \country{USA}
}
\email{mkainen@purdue.edu}

\author{Logan Anderson}
\affiliation{
  \institution{Purdue University}
  \city{West Lafayette}
  \state{Indiana}
  \country{USA}
}
\email{anderslt@purdue.edu}

\author{Artem Pelenitsyn}
\orcid{0000-0001-8334-8106}
\affiliation{
  \institution{Commonwealth University of Pennsylvania}
  \state{Pennsylvania}
  \country{USA}
}
\email{apelenit@commonwealthu.edu}

\author{Michael Vollmer}
\orcid{0000-0002-0496-8268}
\affiliation{
  \institution{University of Kent}
  \city{Canterbury}
  \country{United Kingdom}
}
\email{m.vollmer@kent.ac.uk}

\author{Michael H. Borkowski}
\orcid{0009-0000-8599-2237}
\affiliation{
  \institution{Purdue University}
  \city{West Lafayette}
  \state{Indiana}
  \country{USA}
}
\email{mhborkow@purdue.edu}

\author{Milind Kulkarni}
\orcid{0000-0001-6827-345X}
\affiliation{
  \institution{Purdue University}
  \city{West Lafayette}
  \state{Indiana}
  \country{USA}
}
\email{milind@purdue.edu}

\renewcommand{\shortauthors}{Singhal et al.}
\hypersetup{
  pdftitle={SoCal: A Language for Memory-Layout Factorization of Recursive Datatypes},
  pdfauthor={Vidush Singhal, Mikah Tal Kainen, Logan Anderson, Artem Pelenitsyn, Michael Vollmer, Michael H. Borkowski, Milind Kulkarni}
}

\keywords{Structure of Arrays, Location Calculus, Data-layout Optimization, Compiler Optimization, Tree traversals}

\ccsdesc[500]{Software and its engineering~Compilers}
\ccsdesc[500]{Software and its engineering~Programming languages}
\ccsdesc[300]{Software and its engineering~Program optimization}

\maketitle

\section{Introduction}

Tree traversals arise in a wide range of application domains. For instance, compilers traverse abstract syntax trees (ASTs) to perform optimizations; web browser engines traverse DOM trees to render HTML pages, and scientific-computing applications traverse k-d trees when evaluating numerical kernels.
In such applications, the performance of tree traversals is critical, but achieving high performance has proved challenging with standard tree representations.

The classic representation of trees uses pointers, with each node allocated individually, and pointers connecting parents to children.
Traversing such a tree leads to \emph{pointer chasing}: each node can be accessed only after the pointer to it has been loaded, and because nodes lie in non-contiguous memory locations, each load may miss in the cache~\cite{pchasing, pointer-chasing, Marmoset, ecoop17-gibbon}.
In contrast, if the tree's structure is known in advance, programmers can store the tree as an array to boost performance~\cite{arraybasedmaxnodetree,bantchev2007representing}.
Such array-based representations improve locality but apply only to trees of a known shape.

Trees, in practice, have irregular, input-dependent shapes, so nodes cannot be stored at predictable offsets.
Tackling this problem, the \gibbon{} compiler~\cite{ecoop17-gibbon} does not need to know the shape of the tree in advance.
It compiles traversals to operate over a \emph{packed} representation of the ADT: the ADT is serialized into a single, contiguous memory buffer in preorder.
Each node's constructor tag is followed by its fields. Recursive fields, such as the subtrees of a binary tree, are serialized inline in the same buffer.
Because this buffer contains raw bytes from various parts of the logical tree, accessing it safely is tricky.
Hence, \gibbon{} is based on \oldcalc{}~\cite{LoCal}, a formal language that lets programs build and traverse the serialized representation in a type-safe manner.
The compiler derives offsets from the datatype definition, and a traversal discovers the shape of the tree at run time by pattern matching on its tags.

While the tree representation in \gibbon can produce highly compact layouts with good spatial locality, it introduces new challenges.
For example, the size of a subtree, in general, cannot be known statically.
Hence, accessing the right subtree in a binary tree is non-trivial because the left subtree gets in the way.
As a result, the program needs to either traverse the left subtree first or keep a {\em random-access} pointer to skip it~\cite{LoCal}.
Inserting and deleting elements is also costly: inserting an element requires either copying the entire modified tree to a new buffer or introducing an {\em indirection} (pointer) to the old buffer, which avoids copying but reduces locality.
Most importantly, placing all fields of a tree in one buffer prevents or hurts important optimizations, such as selective traversal of fields, prefetching, loopification, and vectorization.

To illustrate the problem, consider a cons list, defined by the following algebraic datatype\footnote{Following the convention of \gibbon{} papers~\cite{LoCal, Marmoset}, we use Haskell-like syntax for the source throughout this paper.}:

\begin{lstlisting}[style=haskellblock]
data List = Cons Int List | Nil
\end{lstlisting}
\gibbon{} writes the list into a single buffer: a \lstinline|Cons| tag followed by an integer, followed by further \lstinline|Cons| cells, and finally a \lstinline|Nil| tag (Figure~\ref{fig:aos_vs_soa}~(a)).
The buffer thus resembles an array whose elements are heterogeneous structures with several fields, that is, an \emph{array-of-structures} (\AoS{}).

\begin{figure}[t]
\centering
\captionsetup[subfigure]{justification=centering,singlelinecheck=true}
\tikzset{
  aossoacell/.style={
    draw=black!75,
    line width=0.55pt,
    minimum width=0.60669cm,
    minimum height=0.40cm,
    inner sep=0pt,
    font=\ttfamily\fontsize{7}{8}\selectfont,
    text opacity=1
  },
  aossoatag/.style={
    aossoacell,
    fill={rgb, 255:red, 220; green, 220; blue, 220},
    fill opacity=0.6,
    text=black
  },
  aossoaval/.style={
    aossoacell,
    fill=hstype,
    fill opacity=0.4,
    text=black
  },
  aossoanil/.style={
    aossoacell,
    fill={rgb, 255:red, 220; green, 220; blue, 220},
    fill opacity=0.6,
    text=black
  },
  aossoahead/.style={font=\rmfamily\scriptsize}
}
\begin{subfigure}[t]{0.47\columnwidth}
\centering
\begin{adjustbox}{max width=\linewidth,center}
\begin{tikzpicture}
\def\cellstep{0.60669}
\foreach \i in {0,2,4,6,8} {
  \node[aossoatag, anchor=west] at ({\i*\cellstep},0) {Cons};
}
\foreach \i/\txt in {1/1,3/2,5/3,7/4,9/5} {
  \node[aossoaval, anchor=west] at ({\i*\cellstep},0) {\txt};
}
\node[aossoanil, anchor=west] at ({10*\cellstep},0) {Nil};
\end{tikzpicture}
\end{adjustbox}
\caption{\unfactored{} representation}
\label{fig:aos_vs_soa_a}
\end{subfigure}
\hfill
\begin{subfigure}[t]{0.51\columnwidth}
\centering
\begin{adjustbox}{max width=\linewidth,center}
\begin{tikzpicture}
\def\cellstep{0.60669}
\def\buffergap{0.30}
\def\tagx{0.22}
\def\valx{\tagx + 6*\cellstep + \buffergap}
\node[aossoahead] at ({\tagx + 3.0*\cellstep},0.50) {Tags};
\foreach \i in {0,1,2,3,4} {
  \node[aossoatag, anchor=west] at ({\tagx+\i*\cellstep},0) {Cons};
}
\node[aossoanil, anchor=west] at ({\tagx+5*\cellstep},0) {Nil};
\node[aossoahead] at ({\valx + 2.5*\cellstep},0.50) {Values};
\foreach \i/\txt in {0/1,1/2,2/3,3/4,4/5} {
  \node[aossoaval, anchor=west] at ({\valx+\i*\cellstep},0) {\txt};
}
\end{tikzpicture}
\end{adjustbox}
\caption{\fullyfactored{} representation}
\label{fig:aos_vs_soa_b}
\end{subfigure}
\caption{\unfactored{} vs. \fullyfactored{} Representation of a Haskell-Style Cons List in \gibbon{}.}%
\label{fig:aos_vs_soa}
\end{figure}

A classic performance optimization transforms an \AoS{} representation into one array per field of the structure, a \emph{structure-of-arrays} (\SoA{}) representation (Figure~\ref{fig:aos_vs_soa}~(b)), with at least two benefits.

First, a loop over a homogeneous array of primitive values is more efficient when vectorized compared to an \AoS{} array: it fills vector registers with loads of consecutive elements of the same type, without gathering or masking.
In addition, loop-based traversals that support vectorization can be mapped to GPU kernels for SIMT-style execution.

Second, programs often read only some fields of a record\footnote{For example, most database queries use only a fraction of each record~\cite{ailamaki2001weaving}; or in Java programs, many classes have cold fields that lie off the critical traversal path~\cite{Chilimbi1999b}.}. 
A traversal that reads a subset of fields can skip traversing the unused fields' buffers entirely, whereas in an \AoS{} representation it must read past every field. Skipping fields matters even more when a field is itself a recursive structure: consider a list whose cells each hold a nested \lstinline{List}:
\begin{lstlisting}[style=haskellblock]
data NestedList = NCons Int List NestedList | End
\end{lstlisting}
A traversal that reads only the integer of each cell must step over a whole serialized \lstinline{List} to reach the next cell.
Its accesses skip over large, variable-sized regions of memory, which breaks the sequential access pattern that hardware prefetchers rely on.
Worse, the hardware prefetcher may fill the cache with the elements of the nested lists, which the traversal never reads, and these unused elements can evict cache lines that the program still needs.
In a \SoA{}-like representation, the \lstinline{Int} and \lstinline{List} fields reside in separate buffers.
The integers are contiguous, so the prefetcher stays ahead of the traversal, and the traversal issues no loads to the region containing the \lstinline{List}.
In our evaluation (Section~\ref{sec:eval}), a reduction over such a list is $6.1\times$ faster in the \SoA{}-like representation.

While \AoS{}-to-\SoA{} transformations live in folklore and are often implemented for arrays in imperative languages, there has not been a formal mechanism for automatically producing \SoA{} representations for general recursive ADTs.
The challenge lies in code transformations that need to happen to adapt to the new representation.
In an \AoS{} list, the code maintains one {\em cursor}---an abstract pointer that accesses elements in the buffer---to manage the traversal. 
In a \SoA{} list, the traversal must keep one cursor {\em per buffer}, and each read step must advance each cursor past exactly the data it reads.
Reading a \lstinline|Cons| cell, for example, advances the cursor in the data constructor region past one tag and the integer cursor past one integer, after which both cursors point to the start of the tail.
The problem worsens in hand-written \texttt{C} where nothing ensures these steps are performed correctly: a program whose cursor is advanced in the wrong buffer, or not advanced at all, may silently read the data of another field or write past the end of a value.
Lastly, every new datatype and every traversal must repeat this bookkeeping.

This paper introduces the \textit{\fullyfactored{} layout}, which places different fields of an ADT in different buffers, creating a \SoA{}-like representation for ADTs.
To make traversals over this layout safe, we introduce the language \ourcalc{}.
Its types record where each value is stored and how each location was reached, so the type checker rejects ill-formed programs, including those with the cursor errors described above.
Finally, we introduce \colobus{}, a compiler that applies the \AoS{}-to-\SoA{} transformation to general recursive ADTs.
Programmers write ordinary functional programs, which \colobus{} compiles to \ourcalc{} and then lowers to \texttt{C} for high performance.

\subsection*{Contributions}

\begin{itemize}
  \item
    The \fullyfactored{} layout for serialized ADTs (Section~\ref{sec:background:c}).
  \item
    \ourcalc{}, a typed language for writing traversals over general recursive ADTs serialized in the \fullyfactored{} layout.
    We prove progress, preservation and type-safety for \ourcalc{}, and mechanize every result in Lean~4 (Section~\ref{sec:lang}).
  \item
    \colobus{}, a compiler built on \gibbon{} that compiles functional programs to \texttt{C}, using \ourcalc{} as a typed intermediate language, and chooses the \unfactored{}, \fullyfactored{} or \hybridfactored{} layout per datatype (Section~\ref{sec:compiler-implementation}).
  \item
    Mutable cursors, which replace \gibbon's functional-style cursor passing, eliminate cursor copying and enable tail-call optimization (Section~\ref{sec:mutable-cursors}).
  \item
    Optimizations that the \fullyfactored{} layout enables: dependence-free map traversals are loopified per buffer and vectorized, and buffers that a map leaves unchanged are shared with its input instead of copied (Section~\ref{sec:factored-opts}).
\end{itemize}

Section~\ref{sec:eval} evaluates \colobus{} against \gibbon{}~\cite{LoCal} on benchmarks from compilers, databases, document layout and spatial trees, where its \fullyfactored{} code is faster on 69 of 71 fold passes ($3.8\times$ faster, on average) and 28 of 29 map passes ($5.2\times$ faster, on average).

\section{Compiling to Serialized Data Representations}
\newcommand{\SubFigRef}[2]{\hyperlink{#1}{Figure~\ref*{#2}}}

This section follows \lstinline|sumTree| from source to \texttt{C} in both the flat and factored layouts (\secref{sec:background:c}), then introduces the typed intermediate language that connects them~(\secref{sec:background:middle}).

\subsection{Serialized Data Representations}%
\label{sec:background:c}

Our running example is a binary tree with integer leaves: \lstinline[style=haskell,columns=fullflexible,keepspaces=true]!data Tree = Node Tree Tree | Leaf Int!
\par\noindent
The function in \figref{lst:sumtree-source} pattern matches on the constructors \lstinline|Node| and \lstinline|Leaf| to sum the leaves of a tree. In a pointer-based representation of this datatype, each internal node stores pointers to its two children.
Mutations, insertions, and deletions are then cheap, because they rewrite only a few pointers, but every traversal suffers from pointer chasing~\cite{pchasing, ecoop17-gibbon, Marmoset}. 
The \gibbon compiler~\cite{ecoop17-gibbon, LoCal} avoids these costs by operating directly on a serialized, {\em \unfactored{}} representation of the tree.
\begin{figure*}[t]
\centering
\captionsetup[subfigure]{justification=centering,singlelinecheck=true}
\hfill
\begin{subfigure}[b]{0.31\linewidth}
\centering
\begin{lstlisting}[style=haskellblock]
sumTree :: Tree -> Int
sumTree tree =
  case tree of
    Leaf val -> val
    Node l r ->
      sumTree l + sumTree r
\end{lstlisting}
\caption{Source}%
\label{lst:sumtree-source}
\end{subfigure}
\hfill
\begin{subfigure}[b]{0.32\linewidth}
\centering
\hypertarget{fig:sumtree_impl_a_anchor}{}
\begin{lstlisting}[language=C,style=cblock]
RetProd sumTree(Cursor cur) {
  Cursor ct = cur + 1;
  if (*cur == LEAF_TAG) {
    int n = *(int*)ct;
    Cursor end = ct + 8;
    return {n, cur, end}; }
  RetProd l = sumTree(ct);
  RetProd r = sumTree(l.field2);
  return {l.field0 + r.field0,
          cur, r.field2}; }
\end{lstlisting}
\caption{C, \unfactored{}}%
\label{lst:aos-sumtreec}
\end{subfigure}
\hfill
\begin{subfigure}[b]{0.35\linewidth}
\centering
\hypertarget{fig:sumtree_impl_c_anchor}{}
\begin{lstlisting}[language=C,style=cblock]
RetProd sumTree(Cursor ca[2]) {
  Cursor ct = ca[0], ci = ca[1];
  if (*ct == LEAF_TAG) {
    int n = *(int*)ci;
    Cursor out[2] = {ct+1, ci+8};
    return {n, ca, out}; }
  Cursor in[2] = {ct+1, ci};
  RetProd l = sumTree(in);
  RetProd r = sumTree(l.field2);
  int s = l.field0 + r.field0;
  return {s, ca, r.field2}; }
\end{lstlisting}
\caption{C, \fullyfactored{}}%
\label{lst:soa-sumtreec}
\end{subfigure}
\hfill
\caption{\lstinline|sumTree| source and packed C versions using \unfactored{} and \fullyfactored{} ADT layouts.}%
\label{fig:sumtree_impl}
\end{figure*}

\SubFigRef{fig:sumtree_impl_a_anchor}{lst:aos-sumtreec} shows \gibbon-style compilation of \lstinline|sumTree| to \texttt{C}.
The generated code takes a byte array as input (\cil{Cursor} is an alias for \cil{char*}) and inspects the tag of the current node.
After reading a tag or an integer, it bumps the \cil{Cursor} past it, so the cursor always points to the next unread byte of the tree.  %
In this layout, all fields of the tree share one buffer, which keeps the data compact.

In \colobus{}'s \textit{\fullyfactored{}} layout, the integers of the leaves lie in a buffer separate from the constructor tags, so the generated code tracks two cursors, one per buffer (\SubFigRef{fig:sumtree_impl_c_anchor}{lst:soa-sumtreec}).
Traversals over the factored tree can therefore receive both benefits from the \SoA{} transformation.
A traversal that reads only the tags, such as one that counts the leaves, never touches the buffer of integers.
A map that adds a constant to every leaf takes advantage of the uniform offsets implicit in a homogeneous buffer to read the integers from the \lstinline|Int| buffer, enabling optimizations such as efficient vectorization (Section~\ref{sec:factored-opts}).

These optimizations over the \fullyfactored{} layout can dramatically improve performance.
Consider three traversals of \lstinline|Tree|: \texttt{buildTree} builds a complete tree of a given depth, \texttt{add1Tree} is a map that adds a constant to every leaf, and \texttt{sumTree} is a fold that returns the sum of the leaves.
The best \colobus{} configuration is faster than \gibbon{} on all three, and in particular by $3.2$--$5.0\times$ on \texttt{add1Tree} and \texttt{sumTree}.

\subsection{\oldcalc / \ourcalc middle-end}%
\label{sec:background:middle}

\begin{figure}
\begin{subfigure}[b]{0.42\linewidth}
\begin{lstlisting}[style=localblock,aboveskip=0pt,belowskip=0pt,linewidth=\linewidth]
sumTree : (*@$\forall l^r$@*) . Tree @ (*@$l^r$@*) (*@$\rightarrow$@*) Int
sumTree [(*@$l^r$@*)] t = case t of
  Leaf (n : Int @ (*@$l_n^r$@*)) (*@$\rightarrow$@*) n
  Node (a : Tree @ (*@$l_a^r$@*))
       (b : Tree @ (*@$l_b^r$@*))
       (*@$\rightarrow$@*) (sumTree [(*@$l_a^r$@*)] a)
          + (sumTree [(*@$l_b^r$@*)] b)
\end{lstlisting}
\subcaption{\oldcalc}%
\label{lst:sumtree-local}
\end{subfigure}
\hfill
\begin{subfigure}[b]{0.56\linewidth}
\begin{lstlisting}[style=localblock,frame=none,aboveskip=0pt,belowskip=0pt,linewidth=\linewidth]
sumTree : (*@$\forall_{\ngLoc^{\ngReg}}$@*) . Tree @ (*@$\ngLoc^{\ngReg}$@*) (*@$\rightarrow$@*) Int
sumTree t = case t of
  Leaf (n : Int @ (*@$\ngLoc_n^{\ngReg.\ngFieldKey{\mathsf{Leaf}}{1}}$@*)) (*@$\rightarrow$@*) n
  Node (a : Tree @ (*@$\ngLoc_a^{\ngReg}$@*))
       (b : Tree @ (*@$\ngLoc_b^{\ngReg}$@*))
       (*@$\rightarrow$@*) (sumTree [(*@$\ngLoc_a^{\ngReg}$@*)] a)
          + (sumTree [(*@$\ngLoc_b^{\ngReg}$@*)] b)
\end{lstlisting}
\subcaption{\ourcalc}%
\label{lst:sumtree-socal}%
\end{subfigure}
\caption{\lstinline|sumtree| implemented in \oldcalc and \ourcalc}%
\label{lst:sumtree-calc-compatrion}
\end{figure}

Hand-written \texttt{C} over serialized data is prone to the cursor errors described in the introduction.
To avoid this problem, \gibbon uses \oldcalc~\cite{LoCal}, an intermediate language whose operations on serialized data are type-safe.
\oldcalc abstracts \texttt{C}-style pointers into \emph{locations}, which are more limited than pointers but expressive enough for typical tree traversals. \ourcalc{} plays the same role for the \fullyfactored{} layout in \colobus{}. 

Figure~\ref{lst:sumtree-calc-compatrion} contrasts the locations of the two languages on \lstinline|sumTree|.
In \oldcalc{} (Figure~\ref{lst:sumtree-local}), \lstinline{sumTree} takes location variables as arguments to reach the fields and subtrees of the \unfactored{} tree.
Here $l^r$ is a polymorphic location variable in region $r$.
The typing discipline regulates its use: the program must pattern match on the tree, determining whether the current node is a \lstinline{Leaf} or a \lstinline{Node}, before it can access the locations of the node's fields, each of which is placed at a particular location.

The \ourcalc{} version (Figure~\ref{lst:sumtree-socal}) addresses two buffers, so its locations capture more structure.
A \oldcalc{} region holds a value's tags and fields in one sequence, whereas a \ourcalc{} \emph{region}, written $\ngReg$, consists of several buffers.
Within a region, a \emph{path} $\ngPath$ selects one buffer: a path is a sequence of \emph{field keys} $\ngFieldKey{\DC}{j}$, each identifying the $j$-th field of constructor $\DC$.
The \emph{empty path} $\ngEmptyPath$ selects the buffer of the datatype's constructor tags, where every value of the datatype starts.
A \emph{position} $\ngReg.\ngPath$ pairs a region with a path.
A \emph{location atom} $\ngLoc$ is a bare name with no structure of its own; it acquires structure only when paired with a position in a region.
That pairing, written $\ngLoc^{\ngReg.\ngPath}$, is a \emph{location}: a place in a particular buffer of the region.
One atom can thus reference several positions at once.
We abbreviate $\ngLoc^{\ngReg.\ngEmptyPath}$ as $\ngLoc^{\ngReg}$.
Section~\ref{sec:lang} gives the formal definitions.

For type \lstinline{Tree}, the only field key is $\ngFieldKey{\mathsf{Leaf}}{1}$, so a tree has two paths: $\ngEmptyPath$, whose buffer holds the constructor tags, and $\ngFieldKey{\mathsf{Leaf}}{1}$, whose buffer holds the integer of every \lstinline{Leaf} in the tree, regardless of depth, in traversal order.
In Figure~\ref{lst:sumtree-socal}, the argument \lstinline{t} is located at $\ngLoc^{\ngReg}$, the empty path of region $\ngReg$.
The \lstinline{Leaf} pattern locates the corresponding integer at $\ngLoc_n^{\ngReg.\ngFieldKey{\mathsf{Leaf}}{1}}$, in the buffer of leaf integers, while the two subtrees of a \lstinline{Node} are located at $\ngLoc_a^{\ngReg}$ and $\ngLoc_b^{\ngReg}$, in the tag buffer.
In general, a field whose type differs from the datatype of its constructor is \emph{factored}: it gets a field key, and hence a separate buffer.
A self-recursive field, such as a subtree of a \lstinline{Node}, is not factored and is serialized inline in the tag buffer.
The tags of different data constructors cannot be split into separate buffers because the traversal order over the datatype (in this case, preorder) must be preserved; all tags therefore reside in the tag buffer.
A factored field of \lstinline|Int| or another primitive type occupies a single buffer.
A factored field of another datatype is factored recursively: its position holds that datatype's constructor tags, and each of that datatype's own factored fields gets a position that extends the field's path.
Intuitively, then, a path traces through the fields of a complex ADT until it reaches a primitive field, uniquely identifying the buffer holding that primitive field’s data in the region.

\begin{figure}[t]
\centering
\begin{subfigure}[b]{0.70\textwidth}
\begin{lstlisting}[style=localblock,linewidth=\linewidth]
buildList : (*@$\forall_{\ngLoc^{\ngReg}}$@*) . Int -> List @ (*@$\ngLoc^{\ngReg}$@*)
buildList len =
  if len <= 0 then (Nil (*@$\ngLoc^{\ngReg}$@*))
  else -- a Cons carries one factored field: write it, then step past it
       let x : Int @ (*@$\ngLoc^{\ngReg.\ngFieldKey{\mathsf{Cons}}{1}}$@*) = 1 in
       letloc (*@$\ngLoc_1^{\ngReg}$@*) = advanceField (*@$\ngLoc^{\ngReg}$@*) (Cons,1) beyond (Int @ (*@$\ngLoc^{\ngReg.\ngFieldKey{\mathsf{Cons}}{1}}$@*)) in
       -- reserve the Cons tag; the tail starts just past it
       letloc (*@$\ngLoc_r^{\ngReg}$@*) = advanceTags (*@$\ngLoc_1^{\ngReg}$@*) for Cons in
       let rst : List @ (*@$\ngLoc_r^{\ngReg}$@*) = buildList [(*@$\ngLoc_r^{\ngReg}$@*)] (len - 1) in
       -- write the Cons tag, claiming the payload and the tail
       (Cons (*@$\ngLoc^{\ngReg}$@*) x rst)
\end{lstlisting}
\caption{\lstinline|buildList| in \ourcalc{}}%
\label{fig:buildlist-socal}
\end{subfigure}\hfill
\begin{subfigure}[b]{0.28\textwidth}
\centering
\tikzset{
  lcell/.style={draw=black!75, line width=0.55pt, minimum width=0.95cm,
    minimum height=0.42cm, inner sep=0pt, font=\ttfamily\fontsize{7}{8}\selectfont},
  ltag/.style={lcell, fill={rgb, 255:red, 220; green, 220; blue, 220},
    fill opacity=0.6, text opacity=1},
  lval/.style={lcell, fill=hstype, fill opacity=0.4, text opacity=1},
  lidx/.style={font=\fontsize{6}{7}\selectfont, text=black!60},
  lloc/.style={font=\fontsize{7.5}{8.5}\selectfont},
  lhead/.style={font=\fontsize{7.5}{8.5}\selectfont}
}
\begin{tikzpicture}
  \def\w{0.95}
  \def\py{-1.55}
  \node[lhead, anchor=west] at (0,0.62) {Tags at $\ngReg$};
  \node[ltag, anchor=west] at (0*\w,0) {Cons};
  \node[ltag, anchor=west] at (1*\w,0) {Cons};
  \node[ltag, anchor=west] at (2*\w,0) {Nil};
  \node[lhead, anchor=west] at (0,{\py+0.62}) {Payloads at $\ngReg.\ngFieldKey{\mathsf{Cons}}{1}$};
  \node[lval, anchor=west] at (0*\w,\py) {1};
  \node[lval, anchor=west] at (1*\w,\py) {1};
  \foreach \i in {0,1,2} { \node[lidx] at ({\i*\w+\w/2},0.32) {\i}; }
  \foreach \i in {0,1} { \node[lidx] at ({\i*\w+\w/2},{\py+0.32}) {\i}; }
  \node[lloc] at ({0*\w+\w/2},-0.40) {$\ngLoc,\ \ngLoc_1$};
  \node[lloc] at ({1*\w+\w/2},-0.40) {$\ngLoc_r$};
  \node[lloc] at ({0*\w+\w/2},{\py-0.40}) {$\ngLoc$};
  \node[lloc] at ({1*\w+\w/2},{\py-0.40}) {$\ngLoc_1,\ \ngLoc_r$};
\end{tikzpicture}
\caption{The list it builds}%
\label{fig:buildlist-layout}
\end{subfigure}
\caption{\lstinline|buildList| and the layout it builds. (b)~A two-element list in
  region $\ngReg$. Each location's layout cursor holds one index for the root position
  and a layout cursor for each factored field's position.}%
\label{fig:buildlist}
\end{figure}

Reading a factored value advances cursors over buffers but building one also writes to those cursors. The writes must happen in a fixed order.
Consider \lstinline|buildList|, which builds a list of a given length, using the \lstinline|List| datatype of the introduction:
\begin{code}
buildList :: Int -> List
buildList len = if len <= 0 then Nil else let rst = buildList (len - 1) in Cons 1 rst
\end{code}
\figref{fig:buildlist-socal} shows the same function in \ourcalc{}, compiled for the \fullyfactored{} layout.
The output location $\ngLoc^{\ngReg}$, at the empty path of region $\ngReg$, is quantified in the type of \lstinline{buildList}, and each call passes it explicitly, as in \lstinline{buildList [}$\ngLoc_r^{\ngReg}$\lstinline{] (len - 1)}.
The list occupies two positions in $\ngReg$, so $\ngLoc^{\ngReg}$ addresses the tag buffer and $\ngLoc^{\ngReg.\ngFieldKey{\mathsf{Cons}}{1}}$ the buffer of the \lstinline|Cons| integers.

The \lstinline|Nil| branch simply writes the \lstinline|Nil| tag at $\ngLoc^{\ngReg}$.
In the \lstinline|Cons| branch, the program first writes the integer at the factored position and uses \lstinline{advanceField} to move past it.
\lstinline{advanceTags} then reserves the cell for the \lstinline|Cons| tag and binds the location just past it, $\ngLoc_r^{\ngReg}$, where the recursive call builds the tail of the list in place.
Finally, the program writes the \lstinline|Cons| tag at $\ngLoc^{\ngReg}$.
\ourcalc{}'s type system checks this order: a constructor's tag is written only after every one of its fields (Section~\ref{sec:lang}).

\section{\ourcalc{}: A Language for Formalizing the Factored Layout}
\label{sec:lang}

Section~\ref{sec:background:middle} introduced \ourcalc{} through examples; this section defines it.
\ourcalc{}'s types record where each value is stored, and its location
expressions record how each location was computed from earlier ones. Together
they fix both a value's layout and the order in which it is written.

\subsection{Grammar of \ourcalc}\label{sec:terms-values-programs}

\begin{figure*}[!t]
  \centering
  \small
  \input{formal_grammar_meta}

  \socalgrammarcolumns

  \normalsize
  \caption{Grammar of \ourcalc{} \captionscrunch}
  \phantomsection
  \label{fig:grammar}
  \label{fig:typegrammar}
  \label{fig:opergram}
\end{figure*}

Figure~\ref{fig:grammar} gives the grammar. \ourcalc{} is a first-order,
call-by-value, monomorphic functional language with \lstinline{let} expressions,
algebraic datatypes and pattern matching. Every value it builds lives at a
location, and programs construct locations explicitly: \lstinline{letregion}
introduces a region and \lstinline{letloc} binds a location. Regions, field
paths, positions, atoms and locations are as in
\secref{sec:background:middle}, and we abbreviate the root position
$\ngReg.\ngEmptyPath$ to $\ngReg$. We write $\overharpoon{x}$ for a finite
sequence of $x$s. The core calculus has no primitive operations; our examples
use integer literals, arithmetic and \lstinline{if} for readability, and a
literal is typed and written like a nullary constructor.

The location expression $\gramkw{start}$ binds the first location of a region.
A region on which $\gramkw{start}$ has not yet been used is called \emph{unstarted};
$\gramkw{start}$ can be used at most once on a region.
The other three location expressions compute a location from an earlier one, its
\emph{predecessor}. $\gramkw{advanceTags}$ reserves space for a constructor tag.
$\gramkw{after}$ steps past a complete serialized value at a 
valid location, and $\gramkw{advanceField}$ steps past the value of only a particular factored field. 
Each binds the location just past the value they step over.

\ourcalc{} lays out every datatype in factored form (\secref{sec:background:middle}). We write $\ctors{\TYP}$ for the
constructors of $\TYP$ with their field types and $\typeoffield(\DC,j)$ for the
type of the $j$-th field of $\DC$. The \emph{factored fields} of $\TYP$ are
those whose type differs from $\TYP$, each identified by its key
$\ngFieldKey{\DC}{j}$; the remaining fields, of type $\TYP$ itself, are
\emph{self-recursive}. The paths of $\TYP$,
$\ngPathsOf{\TYP}$, are the empty path and every path through its factored
fields, and $\ngTypeAt{\TYP}{\ngPath}$ is the type found at path $\ngPath$.
For \lstinline{List} there are two paths: $\ngEmptyPath$, at type
\lstinline{List}, and $\ngFieldKey{\mathsf{Cons}}{1}$, at type
\lstinline{Int}. \figref{fig:buildlist-layout} shows a two-element list
built by \lstinline|buildList|, with its tags in the buffer at $\ngReg$ and
its payloads in the buffer at $\ngReg.\ngFieldKey{\mathsf{Cons}}{1}$.

At run time, the store $\STOR$ is a log of write events. An event
$(\ngCell{\ngPos}{i},\DC)$ writes the tag $\DC$ in a cell. A cell is a single memory address in the store where 
a constructor may be written, and denoted by indexing into a position as $\rho[i]$. A location map $\MENV$ sends locations to
\emph{layout cursors}. A cursor for $\TYP$ holds an index into the root buffer
and, recursively, one cursor for each factored field: we write
$\ngLayoutCursorT{\TYP}$ for the cursors with exactly one component
$(\DC,\,j,\,c_{\DC,j})$ for each factored field $\ngFieldKey{\DC}{j}$ of $\TYP$,
where $c_{\DC,j} \in \ngLayoutCursorT{\typeoffield(\DC,j)}$. $\MENV$ has at most
one key per atom, sending $\ngLoc^{\ngReg.\ngPath}$ to a cursor for the type
at that position, $\ngTypeAt{\TYP_{\ngReg}}{\ngPath}$, where $\TYP_{\ngReg}$ is
the type of region $\ngReg$. $\MENV(\ngLoc^{\ngReg.\ngPath})$ is defined when
some $\ngLoc^{\ngReg.\ngPath_0} \in \ngDom\MENV$ has
$\ngPath = \ngPath_0.\ngPath_1$ for some
$\ngPath_1 \in \ngPathsOf{\ngTypeAt{\TYP_{\ngReg}}{\ngPath_0}}$
(Definition~\ref{def:mresolve}).
Following the keys of a path $\ngPath$ from a cursor $c$ gives its \emph{sub-cursor} at $\ngPath$, and $c(\ngPath)$ is that sub-cursor's root index.
In \figref{fig:buildlist-layout}, $\ngLoc^{\ngReg}$'s cursor holds index 0 in both
buffers. Stepping past the payload gives $\ngLoc_1^{\ngReg}$, whose cursor has
payload index 1 and tag index 0; stepping past the tag then gives
$\ngLoc_r^{\ngReg}$, at index 1 in both buffers, where the tail begins.

\subsection{Static Semantics of \ourcalc{}}\label{sec:static-semantics}

As \lstinline|buildList| showed, the writes that build a factored value must
happen in a fixed order. The type system enforces this order by tracking each
location through three stages: it is allocated, its value is written, and that
value is used as a field by the constructor that contains it. A constructor
may be written only when each of its fields has been written and not yet used 
by another constructor as its field. 
The main judgment is
\[ \socalbindctx
   \ngJT{\Theta}{\ngTyCtx}{\ngAlCtx}{\EXPR : \TYP@\ngLoc^{\ngReg.\ngPath}}{\ngAlCtx'} . \]
Under the region typing environment $\Theta$ and the typing context
$\ngTyCtx$, and starting from the allocation context $\ngAlCtx$, the
expression $\EXPR$ has type $\TYP$ at location $\ngLoc^{\ngReg.\ngPath}$ and
leaves the allocation context $\ngAlCtx'$. Following
\citet{dunfield2013bidir}, $\dashv$ marks the output context.

$\Theta$ gives each region in scope its type. Only \textsc{T-LetRegion}
changes it, extending it for its body. The typing context $\ngTyCtx = (\TENV,
\SENV, \CENV)$ records what is in scope: $\TENV$ maps variables to located
types, the \emph{complete-location set} $\SENV$ holds the locations whose
values have been written in full (i.e., they are complete), and $\CENV$ maps each location bound by a
\lstinline{letloc} to the location expression that bound it. Subexpressions
see an extended $\ngTyCtx$, but a derivation never returns one.

The allocation context $\ngAlCtx = (\AENV, \NENV, \UENV)$, in contrast, is
threaded through the derivation and tracks the progress of construction.
$\AENV$ gives each region its allocation state: $\ngUnstarted$ if the region is
unstarted, and otherwise a map $\ngFam$ from each path of the region's type to
the atom that the next \lstinline{letloc} at that path takes as its predecessor. 
We write $\AENV(\ngReg.\ngPath)$ for the location $\ngFam(\ngPath)^{\ngReg.\ngPath}$, and $\ngConstFam{\ngLoc}$ for the map that
maps every path to $\ngLoc$. $\AENV$ needs an atom per path because $\gramkw{advanceField}$
moves only one of them (rows~2 and~3 of \figref{fig:buildlist-typing}). The
\emph{pending-location set} $\NENV$ holds the locations allocated but not yet
written. A constructor write requires its location to be in $\NENV$ (pending). 
It is removed from $\NENV$ and moved to the~\emph{consumable-location set} $\UENV$ after the write.
Thus, no location is written twice. 
The fields are required to be in $\UENV$ and are removed after the write, so no field is used twice.

\newcommand{\rtvarloc}{\[
  \ngRuleCtx
  \ngRuleWhere{
  \inferrule*[lab=T-Var]{
    \TENV(\var) = \TYP@\ngLoc^{\ngPos} \\
    \ngLoc^{\ngPos} \in \SENV \\
    \TYP = \ngTypeAt{\TYP_{\ngReg}}{\ngPath}
  }{
    \ngJT{\Theta}{\ngTyCtx}{\ngAlCtx}{\var : \TYP@\ngLoc^{\ngPos}}{\ngAlCtx}
  }
  }{
    \ngPos &= \ngReg.\ngPath
  }
  \socalpairsep
  \ngRuleWhere{
  \inferrule*[lab=T-Concrete-Loc]{
    \ngLoc^{\ngPos} \in \SENV \\
    \TYP = \ngTypeAt{\TYP_{\ngReg}}{\ngPath}
  }{
    \ngJT{\Theta}{\ngTyCtx}{\ngAlCtx}
      {\ngLocVal{\ngLoc^{\ngPos}} : \TYP@\ngLoc^{\ngPos}}{\ngAlCtx}
  }
  }{
    \ngPos &= \ngReg.\ngPath
  }
\]
}
\newcommand{\rtlet}{\[
  \ngRuleCtx
  \ngBind{\ngAlCtx'}{\AENV'}{\NENV'}{\UENV'}
  \ngBind{\ngAlCtx''}{\AENV''}{\NENV''}{\UENV''}
  \ngRuleWhere{
  \inferrule*[lab=T-Let]{
    \ngJT{\Theta}{\ngTyCtx}{\ngAlCtx}{\EXPR_1 : \TYP_1@\ngLoc^{\ngPos}}{\ngAlCtx'} \\
    \ngJT{\Theta}{\ngTup{\TENV'}{\SENV'}{\CENV}}{\ngAlCtx'}{\EXPR_2 : \ngLocatedTySub{2}}{\ngAlCtx''}
  }{
    \ngJT{\Theta}{\ngTyCtx}{\ngAlCtx}
      {\letpack{\var:\TYP_1@\ngLoc^{\ngPos}}{\EXPR_1}{\EXPR_2} : \ngLocatedTySub{2}}
      {\ngAlCtx''}
  }
  }{
    \ngPos &= \ngReg.\ngPath \\
    \TENV' &= \TENV[\var \mapsto \TYP_1@\ngLoc^{\ngPos}] \\
    \SENV' &= \SENV \cup \set{\ngLoc^{\ngPos}}
  }
\]
}
\newcommand{\rtlregion}{\[
  \ngRuleCtx
  \ngRuleWhere{
  \inferrule*[lab=T-LetRegion]{
    \ngReg \notin \ngDom\Theta \\
    \ngFreshCtx{\ngReg}{\ngTyCtx}{\ngAlCtx} \\\\
    \ngJT{\Theta[\ngReg \mapsto \ngAnn]}{\ngTyCtx}{\ngTup{\AENV'}{\NENV}{\UENV}}
      {\EXPR : \TYP@\ngLoc'^{{\ngReg}'.\ngPath'}}
      {\ngTup{\AENV''}{\NENV'}{\UENV'}} \\\\
    \ngReg \neq {\ngReg}' \\
    \forall \ngLoc_x^{\ngReg_x.\ngPath_x} \in \NENV'.\; \ngReg_x \neq \ngReg \\
    \forall \ngLoc_x^{\ngReg.\ngPath_x} \in \UENV'.\; \ngPath_x = \ngEmptyPath
  }{
    \ngJT{\Theta}{\ngTyCtx}{\ngAlCtx}
      {\letregty{\ngReg}{\ngAnn}{\EXPR} : \TYP@\ngLoc'^{{\ngReg}'.\ngPath'}}
      {\ngTup{\AENV'' \setminus \ngReg}{\NENV'}{\UENV' \setminus \ngReg}}
  }
  }{
    \AENV' &= \AENV[\ngReg \mapsto \ngUnstarted] \\
    \UENV' \setminus \ngReg &= \set{\ngLoc_x^{\ngReg_x.\ngPath_x} \in \UENV' \mid \ngReg_x \neq \ngReg}
  }
\]
}
\newcommand{\rtllstart}{\[
  \ngRuleCtx
  \ngRuleWhere{
  \inferrule*[lab=T-LetLoc-Start]{
    \socalgrouprows{
      \socalgrouplabel{region} & \AENV(\ngReg.\ngEmptyPath) = \phi \socalpremsep \TYP = \TYP_{\ngReg}
      \ngPremGap
      \socalgrouplabel{fresh} & \ngFreshCtx{\ngLoc}{\ngTyCtx}{\ngAlCtx} \socalpremsep \ngLoc^{\ngReg} \notin \NENV'' \socalpremsep \ngLoc'^{{\ngReg}'.\ngPath'} \neq \ngLoc^{\ngReg}
      \ngPremGap
      \socalgrouplabel{body} & \ngJT{\Theta}{\ngTup{\TENV}{\SENV}{\CENV'}}{\ngTup{\AENV'}{\NENV'}{\UENV}}
      {\EXPR : \TYP'@\ngLoc'^{{\ngReg}'.\ngPath'}}
      {\ngTup{\AENV''}{\NENV''}{\UENV'}}
    }
  }{
    \ngJT{\Theta}{\ngTyCtx}{\ngAlCtx}
      {\letloc{\ngLoc^{\ngReg}}{\ngStartRegTy{\ngReg}{\TYP}}{\EXPR} :
        \TYP'@\ngLoc'^{{\ngReg}'.\ngPath'}}
      {\ngTup{\AENV''}{\NENV''}{\UENV'}}
  }
  }{
    \CENV' &= \CENV \cup \set{\ngLoc^{\ngReg} \mapsto \ngStartRegTy{\ngReg}{\TYP}} \\
    \AENV' &= \AENV[\ngReg \mapsto \ngStarted{\ngConstFam\ngLoc}] \\
    \NENV' &= \NENV \cup \set{\ngLoc^{\ngReg.\ngPath}
      \;\mid\; \ngPath \in \ngPathsOf{\ngReg}}
  }
\]
}
\newcommand{\rtllafter}{\[
  \ngRuleCtx
  \ngRuleWhere{
  \inferrule*[lab=T-LetLoc-After]{
    \socalgrouprows{
      \socalgrouplabel{predecessor} & \AENV(\ngPos) = \ngLoc'^{\ngPos} \socalpremsep \TYP' = \ngTypeAt{\TYP_{\ngReg}}{\ngPath} \socalpremsep \ngLoc'^{\ngPos} \in \SENV \socalpremsep \ngLoc'^{\ngPos} \in \UENV \socalpremsep \ngLoc'^{\ngPos} \notin \NENV
      \ngPremGap
      \socalgrouplabel{pending} & \ngPath \neq \ngEmptyPath \Rightarrow
      \exists\, \ngLoc_x^{\ngPos} \in \NENV \socalpremsep \forall\, \ngPath_c \in \ngPathsOf{\ngTypeAt{\TYP_{\ngReg}}{\ngPath}},\;
      \ngPath_c \neq \ngEmptyPath.\;\;
      \text{no member of}\;\NENV\;\text{stands at}\;
      \ngPos.\ngPath_c
      \ngPremGap
      \socalgrouplabel{fresh} & \ngFreshCtx{\ngLoc}{\ngTyCtx}{\ngAlCtx} \socalpremsep \ngLoc^{\ngPos} \notin \NENV'' \socalpremsep \ngLoc^{\ngPos} \neq \ngLoc''^{{\ngReg}'.\ngPath''}
      \ngPremGap
      \socalgrouplabel{body} & \ngJT{\Theta}{\ngTup{\TENV}{\SENV}{\CENV'}}{\ngTup{\AENV'}{\NENV'}{\UENV}}
      {\EXPR : \TYP''@\ngLoc''^{{\ngReg}'.\ngPath''}}
      {\ngTup{\AENV''}{\NENV''}{\UENV'}}
    }
  }{
    \ngJT{\Theta}{\ngTyCtx}{\ngAlCtx}
      {\letloc{\ngLoc^{\ngPos}}{\ngAfterTy{\TYP'@\ngLoc'^{\ngPos}}}{\EXPR} :
        \TYP''@\ngLoc''^{{\ngReg}'.\ngPath''}}
      {\ngTup{\AENV''}{\NENV''}{\UENV'}}
  }
  }{
    \ngPos &= \ngReg.\ngPath \\
    \CENV' &= \CENV \cup \set{\ngLoc^{\ngPos} \mapsto \ngAfterTy{\TYP'@\ngLoc'^{\ngPos}}} \\
    \AENV' &= \AENV\big[\ngReg \mapsto
        \ngDescOverride{\AENV(\ngReg)}{\ngReg}{\ngPath}{\ngLoc}\big] \\
    \NENV' &= \NENV \cup \set{\ngLoc^{\ngReg.q}
        \;\mid\; q \in \ngDesc{\ngReg}{\ngPath}}
  }
\]
}
\newcommand{\rtlladvtags}{\[
  \ngRuleCtx
  \ngRuleWhere{
  \inferrule*[lab=T-LetLoc-Advance-Tags]{
    \socalgrouprows{
      \socalgrouplabel{predecessor} & \AENV(\ngPos) = \ngLoc'^{\ngPos} \socalpremsep \ngLoc'^{\ngPos} \in \NENV
      \ngPremGap
      \socalgrouplabel{tag} & (\DC\,\overharpoon{\TYP'}) \in
      \ctors{\ngTypeAt{\TYP_{\ngReg}}{\ngPath}} \socalpremsep m = |\set{j \mid \TYP'_j \neq \ngTypeAt{\TYP_{\ngReg}}{\ngPath}}| \socalpremsep |\overharpoon{\TYP'}| > m
      \ngPremGap
      \socalgrouplabel{chain} & \ngLoc_0^{\ngPos} = \ngChainRoot{\CENV}{\ngLoc'^{\ngPos}} \socalpremsep \ngFieldChain{\CENV}{\ngLoc_0^{\ngPos}}{\DC}{m{+}1}{\ngLoc'^{\ngPos}}
      \ngPremGap
      \socalgrouplabel{pending} & \forall\, \ngFieldKey{\DC'}{j'}.\ngPath' \in
      \ngPathsOf{\ngTypeAt{\TYP_{\ngReg}}{\ngPath}}.\;\;
      \forall \ngLoc_x^{\ngPos.\ngFieldKey{\DC'}{j'}.\ngPath'} \in \NENV.\;\;
      \ngLoc_x = \ngLoc_0 \wedge \DC' \neq \DC
      \ngPremGap
      \socalgrouplabel{fresh} & \ngFreshCtx{\ngLoc}{\ngTyCtx}{\ngAlCtx} \socalpremsep \ngLoc^{\ngPos} \notin \NENV'' \socalpremsep \ngLoc^{\ngPos} \neq \ngLoc''^{{\ngReg}'.\ngPath''}
      \ngPremGap
      \socalgrouplabel{body} & \ngJT{\Theta}{\ngTup{\TENV}{\SENV}{\CENV'}}{\ngTup{\AENV'}{\NENV'}{\UENV}}
      {\EXPR : \TYP''@\ngLoc''^{{\ngReg}'.\ngPath''}}
      {\ngTup{\AENV''}{\NENV''}{\UENV'}}
    }
  }{
    \ngJT{\Theta}{\ngTyCtx}{\ngAlCtx}
      {\letloc{\ngLoc^{\ngPos}}{\ngAdvTags{\ngLoc'^{\ngPos}}{\DC}}{\EXPR} :
        \TYP''@\ngLoc''^{{\ngReg}'.\ngPath''}}
      {\ngTup{\AENV''}{\NENV''}{\UENV'}}
  }
  }{
    \ngPos &= \ngReg.\ngPath &\qquad
    \CENV' &= \CENV \cup \set{\ngLoc^{\ngPos} \mapsto
      \ngAdvTags{\ngLoc'^{\ngPos}}{\DC}} \\
    \AENV' &= \AENV\big[\ngReg \mapsto
        \ngDescOverride{\AENV(\ngReg)}{\ngReg}{\ngPath}{\ngLoc}\big] &\qquad
    \NENV' &= \big(\NENV \cup \set{\ngLoc^{\ngReg.q}
        \;\mid\; q \in \ngDesc{\ngReg}{\ngPath}}\big) - \NENV_r \\
    \NENV_r &= \mathrlap{\set{\ngLoc_0^{\ngPos.\ngFieldKey{\DC'}{j'}.\ngPath'}
        \;\mid\; \ngFieldKey{\DC'}{j'}.\ngPath' \in
          \ngPathsOf{\ngTypeAt{\TYP_{\ngReg}}{\ngPath}},\;
        \DC' \neq \DC}}
  }
\]
}
\newcommand{\rtlladvfield}{\[
  \ngRuleCtx
  \ngRuleWhere{
  \inferrule*[lab=T-LetLoc-Advance-Field]{
    \socalgrouprows{
      \socalgrouplabel{predecessor} & \AENV(\ngPos) = \ngLoc'^{\ngPos} \socalpremsep \ngLoc'^{\ngPos} \in \NENV
      \ngPremGap
      \socalgrouplabel{field} & \ngLoc_0^{\ngPos.\ngFieldKey{\DC}{j}} \in \SENV \socalpremsep \ngLoc_0^{\ngPos.\ngFieldKey{\DC}{j}} \in \UENV \socalpremsep \typeoffield(\DC,j) = \TYP'
      \ngPremGap
      \socalgrouplabel{chain} & \ngFieldChain{\CENV}{\ngLoc_0^{\ngPos}}{\DC}{j}{\ngLoc'^{\ngPos}}
      \ngPremGap
      \socalgrouplabel{fresh} & \ngFreshCtx{\ngLoc}{\ngTyCtx}{\ngAlCtx} \socalpremsep \ngLoc^{\ngPos} \notin \NENV'' \socalpremsep \ngLoc^{\ngPos} \neq \ngLoc''^{{\ngReg}'.\ngPath''}
      \ngPremGap
      \socalgrouplabel{body} & \ngJT{\Theta}{\ngTup{\TENV}{\SENV}{\CENV'}}{\ngTup{\AENV'}{\NENV'}{\UENV}}
      {\EXPR : \TYP''@\ngLoc''^{{\ngReg}'.\ngPath''}}
      {\ngTup{\AENV''}{\NENV''}{\UENV'}}
    }
  }{
    \ngJT{\Theta}{\ngTyCtx}{\ngAlCtx}
      {\letloc{\ngLoc^{\ngPos}}{\ngAdvField{\ngLoc'^{\ngPos}}{\DC}{j}{\TYP'@\ngLoc_0^{\ngPos.\ngFieldKey{\DC}{j}}}}{\EXPR}
        : \TYP''@\ngLoc''^{{\ngReg}'.\ngPath''}}
      {\ngTup{\AENV''}{\NENV''}{\UENV'}}
  }
  }{
    \ngPos &= \ngReg.\ngPath &\qquad
    \CENV' &= \CENV \cup \set{\ngLoc^{\ngPos} \mapsto
      \ngAdvField{\ngLoc'^{\ngPos}}{\DC}{j}{\TYP'@\ngLoc_0^{\ngPos.\ngFieldKey{\DC}{j}}}} \\
    \AENV' &= \AENV[\ngReg \mapsto
        \AENV(\ngReg)[\ngPath \mapsto \ngLoc]] &\qquad
    \NENV' &= \NENV \cup \set{\ngLoc^{\ngPos}}
  }
\]
}
\newcommand{\rtdatacon}{\[
  \ngRuleCtx
  \ngRuleWhere{
  \inferrule*[lab=T-DataConstructor]{
    \ngPremBlock{
      \begin{array}{@{}r@{\quad}l@{}}
      \socalgrouplabel{constructor} & \begin{array}[t]{@{}l@{}}
      \ngVecClear(\DC\,\overharpoon{\TYP'}) \in \ctors{\TYP} \qquad
      \TYP = \ngTypeAt{\TYP_{\ngReg}}{\ngPath} \qquad
      \litnum = |\overharpoon{\VAL}| = |\overharpoon{\TYP'}| \\
      m = |\set{j \mid \TYP'_j \neq \TYP}| \qquad
      p = \litnum - m \qquad
      \ngLoc_0^{\ngPos} = \ngLoc^{\ngPos}
      \end{array}
      \ngPremGap
      \socalgrouplabel{chain} & \begin{array}[t]{@{}l@{\,}l@{}}
        \forall\, 1 \le j \le m.\;\; & \CENV(\ngLoc_j^{\ngPos}) =
          \ngAdvField{\ngLoc_{j-1}^{\ngPos}}{\DC}{j}
            {\TYP'_j@\ngLoc^{\ngPos.\ngFieldKey{\DC}{j}}} \\
        p > 0 \Rightarrow & \CENV(\ngLoc'^{\ngPos}_0) =
        \ngAdvTags{\ngLoc_m^{\ngPos}}{\DC} \\
        \forall\, 1 \le k \le p-1.\;\; & \CENV(\ngLoc'^{\ngPos}_k) = \ngAfterTy{\TYP@\ngLoc'^{\ngPos}_{k-1}}
      \end{array}
      \ngPremGap
      \socalgrouplabel{pending} & \begin{array}[t]{@{}l@{}}
      \ngLoc^{\ngPos} \in \NENV \qquad
      \ngLoc^{\ngPos} \notin \ngFieldLinks{\CENV} \qquad
      \forall\, 1 \le j \le m.\;\; \ngLoc_j^{\ngPos} \in \NENV \\
      \ngQuantBody{\forall\, \ngPath' \in \ngPathsOf{\TYP},\;
          \ngPath' \neq \ngEmptyPath.\;\;
        \forall \ngLoc_x^{\ngPos.\ngPath'} \in \NENV.\;\;}{
        \ngLoc_x = \ngLoc
        \;\wedge\; \exists \DC', j', \ngPath''.\;
          \big(\ngPath' = \ngFieldKey{\DC'}{j'}.\ngPath'' \wedge \DC' \neq \DC\big)}
      \end{array}
      \ngPremGap
      \socalgrouplabel{allocation} & \begin{array}[t]{@{}l@{\,}l@{}}
        p > 0 \Rightarrow \AENV(\ngPos) = \ngLoc'^{\ngPos}_{p-1} \qquad
        p = 0 \Rightarrow \big(\AENV(\ngPos) = \ngLoc_m^{\ngPos}
        \wedge \ngLoc_m^{\ngPos} \in \NENV\big)
      \end{array}
      \ngPremGap
      \socalgrouplabel{consumable} & \forall\, 1 \le j \le m.\;\;
        \ngLoc^{\ngPos.\ngFieldKey{\DC}{j}} \in \UENV \qquad
      \forall\, 0 \le t \le p-1.\;\;
        \ngLoc'^{\ngPos}_t \in \UENV
      \ngPremGap
      \socalgrouplabel{fields} & \begin{array}[t]{@{}l@{\,}l@{}}
        \forall\, 1 \le j \le m.\;\; & \ngJT{\Theta}{\ngTyCtx}{\ngAlCtx}{\VAL_j :
          \TYP'_j@\ngLoc^{\ngPos.\ngFieldKey{\DC}{j}}}{\ngAlCtx} \\
        \forall\, 0 \le t \le p-1.\;\; & \ngJT{\Theta}{\ngTyCtx}{\ngAlCtx}{\VAL_{m+1+t} : \TYP@\ngLoc'^{\ngPos}_t}{\ngAlCtx}
      \end{array}
      \end{array}
    }
  }{
    \ngVecClear
    \ngJT{\Theta}{\ngTyCtx}{\ngAlCtx}
      {\datacon{\DC}{\ngLoc^{\ngPos}}{\overharpoon{\VAL}} : \TYP@\ngLoc^{\ngPos}}
      {\ngTup{\AENV'}{\NENV'}{\UENV'}}
  }
  }{
    \ngPos &= \ngReg.\ngPath &\qquad
    \AENV' &= \AENV\big[\ngReg \mapsto
        \ngDescOverride{\AENV(\ngReg)}{\ngReg}{\ngPath}{\ngLoc}\big] \\
    \NENV' &= \NENV - \set{\ngLoc^{\ngPos}}
        - \set{\ngLoc_j^{\ngPos} \mid 1 \le j \le m} - \NENV_r &\qquad
    \NENV_r &= \set{\ngLoc^{\ngPos.\ngFieldKey{\DC'}{j'}.\ngPath'}
        \;\mid\; \ngFieldKey{\DC'}{j'}.\ngPath' \in \ngPathsOf{\TYP},\;
        \DC' \neq \DC} \\
    \UENV' &= \mathrlap{\big(\UENV
        - \set{\ngLoc^{\ngPos.\ngFieldKey{\DC}{j}} \mid 1 \le j \le m}
        - \set{\ngLoc'^{\ngPos}_t \mid 0 \le t \le p-1}\big)
        \cup \set{\ngLoc^{\ngPos}}}
  }
\]
}
\newcommand{\rtcase}{\[
  \ngRuleCtx
  \ngBind{\ngAlCtx'}{\AENV'}{\NENV'}{\UENV'}
  \ngRuleWhereList{
  \inferrule*[lab=T-Case]{
    \ngPremBlock{
      \begin{array}{@{}r@{\quad}l@{}}
      \socalgrouplabel{scrutinee} & \ngJT{\Theta}{\ngTyCtx}{\ngAlCtx}{\VAL : \TYP''@\ngLoc_s^{\ngPos_s}}{\ngAlCtx}
      \ngPremGap
      \socalgrouplabel{coverage} & \begin{array}[t]{@{}l@{}}
      \litnum = |\overharpoon{\pat}| \qquad
      \forall \ind \neq \ind'.\; \DC_\ind \neq \DC_{\ind'} \\
      \forall (\DC'\,\overharpoon{\TYP'}) \in \ctors{\TYP''}.\;
        \exists \ind \le \litnum.\; \DC_\ind = \DC'
      \end{array}
      \ngPremGap
      \socalgrouplabel{clauses} & \forall \ind \le \litnum.\;\;
        \ngJPT{\Theta}{\TYP''@\ngLoc_s^{\ngPos_s}}{\ngTyCtx}{\ngAlCtx}
          {\pat_\ind : \ngLocatedTy}{\ngAlCtx'}
      \end{array}
    }
  }{
    \ngJT{\Theta}{\ngTyCtx}{\ngAlCtx}{\case{\VAL}{\overharpoon{\pat}} : \ngLocatedTy}{\ngAlCtx'}
  }
  }{
    \ngPos_s = \ngReg_s.\ngPath_s \\
    \ind \in \set{1,\ldots,\litnum} \\
    \DC_\ind\;\text{is clause}\;\pat_\ind\text{'s own constructor}
  }
\]
}
\newcommand{\rtpat}{\[
  \ngRuleCtx
  \ngBind{\ngAlCtx'}{\AENV'}{\NENV'}{\UENV'}
  \ngRuleWhere{
  \inferrule*[lab=T-Pattern]{
    \ngPremBlock{
      \begin{array}{@{}r@{\quad}l@{}}
      \socalgrouplabel{constructor} & \ngVecClear(\DC\,\overharpoon{\TYP'}) \in \ctors{\TYP''} \qquad
      \litnum = |\overharpoon{\TYP'}|
        = |\overharpoon{\var:\TYP'@\ngLoc'^{\ngPos}}| \qquad
      m = |\set{j \mid \TYP'_j \neq \TYP''}|
      \ngPremGap
      \socalgrouplabel{binders} & \forall \ind.\; \ngFreshCtx{\ngLoc'_\ind}{\ngTyCtx}{\ngAlCtx} \qquad
      \forall \ind \neq \ind'.\; \ngLoc'_\ind \neq \ngLoc'_{\ind'} \qquad
      \forall \ind \neq \ind'.\; \var_\ind \neq \var_{\ind'} \qquad
      \forall \ind.\; \ngLoc^{\ngPos_e} \neq \ngLoc'_\ind{}^{\ngPos_\ind}
      \ngPremGap
      \socalgrouplabel{body} & \ngJT{\Theta}{\ngTup{\TENV'}{\SENV'}{\CENV}}{\ngAlCtx}
        {\EXPR : \TYP@\ngLoc^{\ngPos_e}}{\ngAlCtx'}
      \end{array}
    }
  }{
    \ngVecClear
    \ngJPT{\Theta}{\TYP''@\ngLoc_s^{\ngPos_s}}{\ngTyCtx}{\ngAlCtx}
      {\caseclause{\datacon{\DC}{\!\!}{(\overharpoon{\var:\TYP'@\ngLoc'^{\ngPos}})}}{\EXPR}
        : \TYP@\ngLoc^{\ngPos_e}}
      {\ngAlCtx'}
  }
  }{
    \ngPos_s &= \ngReg_s.\ngPath_s, \qquad
      \ngPos_e = \ngReg.\ngPath \\
    \ngPos_\ind &= \ngPos_s.\ngFieldKey{\DC}{\ind}
      \;\;\text{for}\; \ind \le m, \qquad
      \ngPos_\ind = \ngPos_s \;\;\text{for}\; m < \ind \le \litnum \\
    \TENV' &= \TENV\big[\var_1 \mapsto \TYP'_1@\ngLoc'_1{}^{\ngPos_1}, \ldots,
      \var_\litnum \mapsto \TYP'_\litnum@\ngLoc'_\litnum{}^{\ngPos_\litnum}\big] \\
    \SENV' &= \SENV \cup \set{\ngLoc'_\ind{}^{\ngPos_\ind}
      \;\mid\; \ind \le \litnum} \\
    \ind &\in \set{1,\ldots,\litnum}
  }
\]
}
\newcommand{\rtapp}{\[
  \ngRuleCtx
  \ngRuleWhereList{
  \inferrule*[lab=T-App]{
    \ngPremBlock{
      \begin{array}{@{}r@{\quad}l@{}}
      \socalgrouplabel{callee} & \mathrm{Function}(\fvar) =
        \big(\forall_{\overharpoon{\ngLoc_\ind^{\ngReg_\ind}},\,\ngLoc^{\ngReg}}.\;
          \overharpoon{\TYP_\ind@\ngLoc_\ind^{\ngReg_\ind}} \ARROW \TYP@\ngLoc^{\ngReg},\;
          \overharpoon{\var_f},\; \EXPR_f\big)
      \ngPremGap
      \socalgrouplabel{arity} & \ngVecClear\litnum = |\overharpoon{\VAL}|
        = |\overharpoon{\TYP_\ind@\ngLoc_\ind^{\ngReg_\ind}}| \qquad
      |\overharpoon{\ngLoc'^{\ngPos'}}| = \litnum + 1 \qquad
      \forall\, \ind,\kappa \le \litnum{+}1.\;\;
        \ngReg_\ind = \ngReg_\kappa \Rightarrow \ngPos'_\ind = \ngPos'_\kappa
      \ngPremGap
      \socalgrouplabel{output} & \begin{array}[t]{@{}l@{}}
      \TYP = \ngTypeAt{\TYP_{{\ngReg}_{out}'}}{\ngPath} \qquad
      \ngLoc_{out}'{}^{\ngPos_{out}} \in \NENV \qquad
      \ngLoc_{out}'{}^{\ngPos_{out}} \notin \ngFieldLinks{\CENV} \qquad
      \AENV(\ngPos_{out}) = \ngLoc_{out}'{}^{\ngPos_{out}} \\
      \ngQuantBody{\forall\, \ngPath_c \in \ngPathsOf{\TYP},\;
          \ngPath_c \neq \ngEmptyPath.\;\;}{
        \ngLoc_{out}'{}^{\ngPos_{out}.\ngPath_c} \notin \ngFieldLinks{\CENV} \\
        \wedge\; \ngLoc_{out}'{}^{\ngPos_{out}.\ngPath_c} \notin \ngSources{\CENV} \\
        \wedge\; \forall \ngLoc_x^{\ngPos_{out}.\ngPath_c} \in \NENV.\;
          \ngLoc_x = \ngLoc_{out}'}
      \end{array}
      \ngPremGap
      \socalgrouplabel{arguments} & \forall\, \ind \le \litnum.\;\;
        \ngJT{\Theta}{\ngTyCtx}{\ngAlCtx}
          {\VAL_\ind : \TYP_\ind@\ngLoc_\ind'{}^{\ngPos'_\ind}}{\ngAlCtx}
      \end{array}
    }
  }{
    \ngVecClear
    \ngJT{\Theta}{\ngTyCtx}{\ngAlCtx}
      {\fapp{\overharpoon{\ngLoc'^{\ngPos'}}}{\overharpoon{\VAL}}
        : \TYP@\ngLoc_{out}'{}^{\ngPos_{out}}}
      {\ngTup{\AENV}{\NENV'}{\UENV \cup \set{\ngLoc_{out}'{}^{\ngPos_{out}}}}}
  }
  }{
    \overharpoon{\var_f},\,\EXPR_f\;\text{are}\;\fvar\text{'s declared parameters
      and body} \\
    \ngPos'_\ind = {\ngReg}'_\ind.\ngPath'_\ind \\
    \ngLoc_{\litnum+1}^{\ngReg_{\litnum+1}}\;\text{is}\; \ngLoc^{\ngReg},
      \;\text{the scheme's own result, so the premise above ranges} \\
    \ngWhereCont\text{over the whole}\;(\litnum{+}1)\text{-vector} \\
    \ngLoc_{out}'{}^{\ngPos_{out}}\;\text{is}\;
      \overharpoon{\ngLoc'^{\ngPos'}}\text{'s}\;(\litnum{+}1)\text{-th
      entry} \\
    \ngPos_{out} = {\ngReg}_{out}'.\ngPath \\
    \NENV' = \NENV - \set{\ngLoc_{out}'{}^{\ngPos_{out}.\ngPath''}
        \;\mid\; \ngPath'' \in \ngPathsOf{\TYP}}
  }
\]
}
\newcommand{\rtfunctiondef}{\[
  \ngBind{\ngTyCtx_0}{\TENV}{\SENV}{\CENV}
  \ngBind{\ngAlCtx_0}{\AENV}{\NENV}{\UENV}
  \ngRuleWhere{
  \inferrule*
   [lab=T-FunctionDef]
   {
    \socalgrouprows{
      \socalgrouplabel{arity} & \ngVecClear\litnum = |\overharpoon{\var}|
      = |\overharpoon{\TYP_\ind@\ngLoc_\ind^{\ngReg_\ind}}| \socalpremsep |\overharpoon{\ngLoc_\ind^{\ngReg_\ind}},\,\ngLoc^{\ngReg}| = \litnum + 1
      \ngPremGap
      \socalgrouplabel{distinct} & \ngVecClear\text{the location atoms of}\;
      \overharpoon{\ngLoc_\ind^{\ngReg_\ind}},\,\ngLoc^{\ngReg}\;\text{are pairwise distinct}
      \ngPremGap
      \socalgrouplabel{} & \text{the variables}\;\overharpoon{\var}\;\text{are pairwise distinct}
      \ngPremGap
      \socalgrouplabel{body} & \ngJT{\Theta}{\ngTyCtx_0}{\ngAlCtx_0}{\EXPR : \TYP@\ngLoc^{\ngReg}}
      {\ngTup{\AENV}{\NENV'}{\UENV'}}
      \ngPremGap
      \socalgrouplabel{outputs} & \NENV \cap \NENV' = \emptyset \socalpremsep \UENV' = \set{\ngLoc^{\ngReg}}
    }
  }
   {
    \ngJFunT{\Theta}{
    \fvar : \forall_{\overharpoon{\ngLoc_\ind^{\ngReg_\ind}},\,\ngLoc^{\ngReg}}.
    \overharpoon{\TYP_\ind@\ngLoc_\ind^{\ngReg_\ind}} \ARROW \TYP@\ngLoc^{\ngReg}; \;
    \fvar\,\overharpoon{\var} = \EXPR}
   }
  }{
    \ngShowAt{\ngTyCtx_0}\;\text{and}\;\ngShow{\ngAlCtx_0},
      \;\text{with} \\
    \TENV &=
      \set{\var_\ind \mapsto \TYP_\ind@\ngLoc_\ind^{\ngReg_\ind}
        \;\mid\; \ind \le \litnum} \\
    \SENV &=
      \set{\ngLoc_\ind^{\ngReg_\ind} \;\mid\; \ind \le \litnum}, \qquad
      \CENV = \emptyset \\
    \AENV &= \set{\ngReg \mapsto \ngStarted{\ngConstFam\ngLoc}} \\
    \NENV &= \set{\ngLoc^{\ngReg.\ngPath}
        \;\mid\; \ngPath \in \ngPathsOf{\ngReg}}, \qquad
      \UENV = \emptyset
  }
\]
}
\newcommand{\rtprogram}{\[
  \ngBind{\ngTyCtx_0}{\TENV}{\SENV}{\CENV}
  \ngBind{\ngAlCtx_0}{\AENV}{\NENV}{\UENV}
  \ngBind{\ngAlCtx'}{\AENV'}{\NENV'}{\UENV'}
  \ngRuleWhereList{
  \inferrule*
   [lab=T-Program]
   {
    \wfdecl\;\overharpoon{\DD} \\
    \litnum = |\overharpoon{\FD}| \\
    \forall \ind \le \litnum.\; \ngJFunT{\Theta}{\FD_\ind} \\
    \forall \ind \neq \ind'.\; \fvar_\ind \neq \fvar_{\ind'} \\
    \ngVecClear\text{no}\;\ngLocVal{\cdot}\;\text{occurs in}\;\overharpoon{\FD}\;\text{or}\;\EXPR \\
    \ngJT{\Theta}{\ngTyCtx_0}{\ngAlCtx_0}{\EXPR : \ngLocatedTy}{\ngAlCtx'} \\
    \NENV \cap \NENV' = \emptyset \\
    \forall\, \ngLoc''^{{\ngReg}''.\ngPath''} \in \UENV'.\;
      \ngPath'' = \ngEmptyPath
   }{
    \ngVecClear
    \ngJProgT{\Theta}{\overharpoon{\DD} \;; \overharpoon{\FD} \;; \EXPR : \ngLocatedTy}{\ngAlCtx'}
   }
  }{
    \ngReg,\ngLoc^{\ngReg},\TYP_{\ngReg}\;\text{are the
      program's bootstrap region, location and type} \\
    \FD_\ind\;\text{is}\;\overharpoon{\FD}\text{'s}\;\ind\text{-th
      declaration and}\;\fvar_\ind\;\text{its own function name, for}\;
      \ind \in \set{1,\ldots,\litnum} \\
    \ngShow{\ngTyCtx_0}\;\text{and}\;\ngShow{\ngAlCtx_0},\;\text{with} \\
    \ngWhereCont\TENV = \emptyset, \qquad
      \SENV = \emptyset, \qquad \CENV = \set{\ngLoc^{\ngReg} \mapsto \ngStartRegTy{\ngReg}{\TYP_{\ngReg}}} \\
    \ngWhereCont\AENV = \set{\ngReg \mapsto \ngStarted{\ngConstFam\ngLoc}} \\
    \ngWhereCont\NENV = \set{\ngLoc^{\ngReg.\ngPath}
        \;\mid\; \ngPath \in \ngPathsOf{\ngReg}}, \qquad
      \UENV = \emptyset
  }
\]
}
\newcommand{\socalfullrules}{}

\begin{figure*}[tp]
  \centering
  \begin{subfigure}{\textwidth}
    \socalrulestack[\textwidth]{%
      \socaljudgment{Typing}{\socalJTyping}%
      \input{paper-formalism/paper-fragments/rule-tletloc-advfield}\socalrulesep
      \input{paper-formalism/paper-fragments/rule-tletloc-advtags}
    }
    \subcaption{Stepping past a factored field and past a tag.}
    \label{fig:types-body-advance}
  \end{subfigure}
  \socalgroupsep
  \begin{subfigure}{\textwidth}
    \socalrulestack[\textwidth]{%
      \input{paper-formalism/paper-fragments/rule-tdataconstructor}
    }
    \subcaption{Writing a constructor.}
    \label{fig:types-body-datacon}
  \end{subfigure}
  \caption{Typing rules that enforce the write order.
    Appendix~\ref{sec:appendix-formal} gives all typing rules. \captionscrunch}
  \phantomsection
  \label{fig:types-body}
\end{figure*}

Figure~\ref{fig:types-body} gives the three
rules that enforce the write order ($\ngPos$ is short for $\ngReg.\ngPath$). 
The contexts used by the body and the conclusion are defined in the \emph{where} block. 
A constructor's \emph{destination} is the location where its tag is written. 
We write $\ngFieldChain{\CENV}{\ngLoc_0^{\ngPos}}{\DC}{j}{\ngLoc'^{\ngPos}}$ as the chain
of constraints in $\CENV$ that starts from the destination $\ngLoc_0^{\ngPos}$ and ends at $\ngLoc'^{\ngPos}$.
The chain captures $\gramkw{advanceField}$ links for the first $j-1$ fields of $\DC$. 

\textsc{T-LetLoc-Advance-Field} steps over the $j$-th field of $\DC$.
This allows the program to move the $j$-th field's current start location to its frontier: the 
position where the $j$-th field of the next $\DC$ will be written. Its
\emph{predecessor} premises require that $\AENV$ stores the predecessor at
$\ngPos$ and that the predecessor is in $\NENV$. The \emph{field} premises
require the field's value, at the destination $\ngLoc_0^{\ngPos}$ extended by
the field's key, to be complete and in $\UENV$. The \emph{chain} premise requires the fields to be
stepped over in order. The new atom is \emph{fresh}: it occurs in neither
$\ngTyCtx$ nor $\ngAlCtx$. The rule records the location expression in $\CENV$,
adds the new location to $\NENV$, and moves $\AENV$ to the new atom at $\ngPath$
only. At the field's own path, $\AENV$ keeps the atom whose location holds the start of the field's value.

\textsc{T-LetLoc-Advance-Tags} steps past the tag of $\DC$ to where its first
self-recursive field will be written. The \emph{chain} premises lookup the destination
as $\ngChainRoot{\CENV}{\ngLoc'^{\ngPos}}$, the start of the chain of
$\gramkw{advanceField}$ links that ends at the predecessor, and require that
the chain has stepped over all $m$ factored fields of $\DC$. 
The self-recursive field is a value of the region's type at $\ngPath$, with its own factored fields, so
the new atom is allocated at every path in the \emph{descendant set}
$\ngDesc{\ngReg}{\ngPath}$, the paths at and below $\ngPath$. All paths in $\AENV$ move to the new atom. 
The \emph{pending} premise allows the destination to have
pending locations only at the fields of other constructors, which $\DC$ will
never write; the rule removes them from $\NENV$.

\textsc{T-DataConstructor} ensures that $DC$'s value is serialized correctly in 
the factored layout. It writes a constructor's tag after all of its
fields, so that completing a value is a single event. Of the $\litnum$ fields,
$m$ are factored and $p$ are self-recursive. The \emph{chain} premises require
$\CENV$ to record the expected links from the destination $\ngLoc^{\ngPos}$:
one $\gramkw{advanceField}$ link past each factored field, then an
$\gramkw{advanceTags}$ link to the first self-recursive field, then an $\gramkw{after}$ link
from each self-recursive field to the next. The \emph{pending} premises require that the
destination, and its $\gramkw{advanceField}$ links are in $\NENV$ and that the
destination which starts the chain is not itself such a link. The
\emph{allocation} premises require that $\AENV$ stores the last location
of the chain at $\ngPos$, so nothing has been allocated past it. 
The \emph{consumable} and \emph{fields} premises require
every field value to be in $\UENV$ with the required type. 
The rule removes the destination, its field links, and the pending locations of fields of other constructors ($\NENV_r$) from $\NENV$.
It uses $K$'s fields from $\UENV$ to build the value, removes them from $\UENV$, and adds the destination to $\UENV$. 
$\AENV$ is reset over the descendant set to the destination location's atom.

\begin{figure*}[!t]
  \begingroup
  \centering
  \formalgrammarkeywords
  \setlength{\tabcolsep}{0pt}
  \vspace{0.3em}
  \fontsize{7.75}{9.3}\selectfont
  \newcommand{\tCodeKw}{%
    \renewcommand{\gramkw}[1]{\mathord{{\socalkeepcolor{FormalCodeKw}\mathrm{##1}}}}%
    \renewcommand{\gramwd}[1]{\gramkw{##1}}}
  \renewcommand{\arraystretch}{1.2}
  \newcommand{\tq}{(\mathsf{Cons},1)}
  \newcommand{\tL}{\ngLoc^{\ngReg}}
  \newcommand{\tLq}{\ngLoc^{\ngReg.q}}
  \newcommand{\tLone}{\ngLoc_1^{\ngReg}}
  \newcommand{\tLr}{\ngLoc_r^{\ngReg}}
  \newcommand{\tLrq}{\ngLoc_r^{\ngReg.q}}
  \newcommand{\aL}{$\ngLoc$}
  \newcommand{\aLone}{$\ngLoc_1$}
  \newcommand{\aLr}{$\ngLoc_r$}
  \newcommand{\tList}{\mathsf{List}}
  \newcommand{\tCons}{\mathsf{Cons}}
  \newcommand{\tNil}{\mathsf{Nil}}
  \newcommand{\tRule}[1]{\textsc{#1}}
  \newcommand{\tGap}{\hspace{0.35em}}
  \newcommand{\tSet}[1]{$\{#1\}$}
  \newcommand{\tNone}{$\emptyset$}
    \begin{tabular}{@{}r@{\tGap}>{\tCodeKw}l@{\tGap}l@{\tGap}c@{\tGap}c@{\tGap}l@{\tGap}l@{}}
      & Code & Rule & $\AENV(\ngReg.\ngEmptyPath)$ & $\AENV(\ngReg.q)$
        & $\NENV$ & $\UENV$ \\ \hline
      $0$ & (function entry) & \tRule{FunctionDef} &
        \aL & \aL & \tSet{\tL,\, \tLq} & \tNone \\ \hline
            Nil & $\datacon{\tNil}{\tL}{}$ & \tRule{DataCon} &
        \aL & \aL & \tNone & \tSet{\tL} \\ \hline
      \multicolumn{7}{@{}l}{\emph{Cons branch}} \\
      $1$ & $\letpack{x : \Int@\tLq}{1}{}$ & \tRule{Let}, \tRule{DataCon} &
        \aL & \aL & \tSet{\tL} & \tSet{\tLq} \\
      $2$ & $\letloc{\tLone}{\ngAdvField{\tL}{\tCons}{1}{\Int@\tLq}}{}$ & \tRule{AdvField} &
        \aLone & \aL & \tSet{\tL,\, \tLone} & \tSet{\tLq} \\
      $3$ & $\letloc{\tLr}{\ngAdvTags{\tLone}{\tCons}}{}$ & \tRule{AdvTags} &
        \aLr & \aLr & \tSet{\tL,\, \tLone,\, \tLr,\, \tLrq} & \tSet{\tLq} \\
      $4$ & $\letpack{\mathit{rst} : \tList@\tLr}{\mathit{buildList}\;[\tLr]\;(\mathit{len}-1)}{}$ & \tRule{Let}, \tRule{App} &
        \aLr & \aLr & \tSet{\tL,\, \tLone} & \tSet{\tLq,\, \tLr} \\
      $5$ & $\datacon{\tCons}{\tL}{x\;\mathit{rst}}$ & \tRule{DataCon} &
        \aL & \aL & \tNone & \tSet{\tL} \\ \hline
      \multicolumn{7}{@{}l@{}}{\rule{0pt}{2.6ex}$\CENV$ when row~5 is checked:\;
        $\{\,\tLone \mapsto \ngAdvField{\tL}{\tCons}{1}{\Int@\tLq},\;\;
        \tLr \mapsto \ngAdvTags{\tLone}{\tCons}\,\}$} \\
    \end{tabular}
  \endgroup

  \caption{Type checking \lstinline|buildList| (\figref{fig:buildlist-socal}). Each row shows $\AENV$, $\NENV$ and $\UENV$ after the step. Here, $\AENV$ has two keys. One at the tag buffer $\ngReg.\ngEmptyPath$ and the other at the payload
    buffer $\ngReg.q$ ($q \equiv \ngFieldKey{\mathsf{Cons}}{1}$). 
    Row~1 writes the literal like a nullary constructor. Row~0 is the state \textsc{T-FunctionDef} starts the body in; both
    branches continue from it.}
  \label{fig:buildlist-example}
  \label{fig:buildlist-typing}
\end{figure*}

\par\noindent\textbf{An example.}
\figref{fig:buildlist-typing} type checks \lstinline|buildList|
(\figref{fig:buildlist-socal}) one step at a time, writing $q$ for the key
$\ngFieldKey{\mathsf{Cons}}{1}$. On entry (row~0), $\AENV$ maps both paths of
$\ngReg$ to $\ngLoc$. $\ngLoc^{\ngReg}$ and $\ngLoc^{\ngReg.q}$ are in $\NENV$. 
\textsc{T-FunctionDef} requires the body to end with nothing in $\NENV$.
Only $\ngLoc^{\ngReg}$ should be in $\UENV$. $\AENV$ should be reset to $\ngLoc$.
In the \lstinline|Nil| branch, \textsc{T-DataConstructor} writes the tag at
$\ngLoc^{\ngReg}$ and, because only \lstinline|Cons| has a field at $q$, drops $\ngLoc^{\ngReg.q}$ from $\NENV$. 
In the \lstinline|Cons| branch, row~1 writes the payload at $\ngLoc^{\ngReg.q}$. Row~2's
$\gramkw{advanceField}$ needs that payload to be complete and in $\UENV$, and
moves $\AENV$ only at the root path. Row~3's $\gramkw{advanceTags}$ checks that
the chain from $\ngLoc^{\ngReg}$ has stepped over the one factored field of
\lstinline|Cons|, and allocates $\ngLoc_r$ at both paths, because the tail has its own \lstinline|Int| field. 
Row~4's recursive call writes the tail at $\ngLoc_r^{\ngReg}$. 
Row~5 checks the constraints recorded in $\CENV$, uses the payload and the tail location from $\UENV$ to construct the \lstinline|Cons| value, 
and restores the state that \textsc{T-FunctionDef} requires.

\subsection{Dynamic Semantics of \ourcalc{}}
\label{sec:dynamic-semantics}
The dynamic semantics shows how a program computes the cursors of
\figref{fig:buildlist-layout} and writes the store. Reduction relates
\emph{configurations} $\ngCfg{\Theta}{\STOR}{\MENV}{\EXPR}$, which consist of
the region typing environment, the store, the location map and the expression
under evaluation. One step is written
\begin{displaymath}
\ngStepCfg{\Theta}{\STOR}{\MENV}{\EXPR}{\Theta'}{\STOR'}{\MENV'}{\EXPR'} .
\end{displaymath}
Every rule except \textsc{D-Ctx} is a \emph{basic contraction}
$\ngStepCfgB{\Theta}{\STOR}{\MENV}{r}{\Theta'}{\STOR'}{\MENV'}{r'}$ of a redex
$r$, and \textsc{D-Ctx} lifts a contraction into an evaluation context
$\ngCtxFill{\ngCtx}{r}$. Only \textsc{D-LetRegion} changes $\Theta$, adding
the region it allocates.

\let\colobusdynamicsend\relax

\newcommand{\rdevalctx}{\[
  \ngCtx \;\gramdef\; [\,] \;\gramor\;
    \letpack{\var:\ngLocatedTy}{\ngCtx}{\EXPR_2} .
\]
}
\newcommand{\rdctx}{\[
  \inferrule*[lab=D-Ctx]{
    \ngStepCfgB{\Theta}{\STOR}{\MENV}{\EXPR}
      {\Theta'}{\STOR'}{\MENV'}{\EXPR'}
  }{
    \ngStepCfg{\Theta}{\STOR}{\MENV}{\ngCtxFill{\ngCtx}{\EXPR}}
      {\Theta'}{\STOR'}{\MENV'}{\ngCtxFill{\ngCtx}{\EXPR'}}
  }
\]
}
\newcommand{\rdletvalregion}{\[
  \inferrule*[lab=D-Let-Val]{}{
    \ngStepCfgB{\Theta}{\STOR}{\MENV}{\letpack{\var:\ngLocatedTy}{\VAL_1}{\EXPR_2}}
      {\Theta}{\STOR}{\MENV}{\subst{\EXPR_2}{\var}{\VAL_1}}
  }
\]
\[
  \ngRuleWhere{
  \inferrule*[lab=D-LetRegion]{
    \ngVecClear\ngReg_f \;\fresh\;\text{for}\;
      \STOR,\MENV,\;\ngCtxFill{\ngCtx}{r},\;
      \overharpoon{\FD}\;\text{and}\;\ngDom\Theta
  }{
    \ngStepCfgB{\Theta}{\STOR}{\MENV}{\letregty{\ngReg}{\TYP}{\EXPR}}
      {\Theta[\ngReg_f \mapsto \TYP]}{\STOR}{\MENV}{\EXPR'}
  }
  }{
    \EXPR' &= \subst{\EXPR}{\ngReg}{\ngReg_f}
  }
\]
}
\newcommand{\rdllstart}{\[
  \ngRuleWhere{
  \inferrule*[lab=D-LetLoc-Start]{
    \TYP = \TYP_{\ngReg} \\
    \ngLoc_f \;\fresh
  }{
    \ngStepCfgB{\Theta}{\STOR}{\MENV}{\letloc{\ngLoc^{\ngReg}}{\ngStartRegTy{\ngReg}{\TYP}}{\EXPR}}
      {\Theta}{\STOR}{\MENV'}{\EXPR'}
  }
  }{
    c &= \ngZeroCursor{\TYP} \in \ngLayoutCursorT{\TYP_{\ngReg}} \\
    \MENV' &= \MENV \cup \set{\ngLoc_f^{\ngReg} \mapsto c} \\
    \EXPR' &= \subst{\EXPR}{\ngLoc}{\ngLoc_f}
  }
\]
}
\newcommand{\rdllafter}{\[
  \ngRuleWhere{
  \inferrule*[lab=D-LetLoc-After]{
    \vwf{\STOR}{\ngReg.\ngPath}{\MENV(\ngLoc'^{\ngReg.\ngPath})}{\TYP'}{c'} \\
    \ngLoc_f \;\fresh
  }{
    \ngStepCfgB{\Theta}{\STOR}{\MENV}{\letloc{\ngLoc^{\ngReg.\ngPath}}{\ngAfterTy{\TYP'@\ngLoc'^{\ngReg.\ngPath}}}{\EXPR}}
      {\Theta}{\STOR}{\MENV'}{\EXPR'}
  }
  }{
    \MENV' &= \MENV \cup \set{\ngLoc_f^{\ngReg.\ngPath} \mapsto c'} \\
    \EXPR' &= \subst{\EXPR}{\ngLoc}{\ngLoc_f}
  }
\]
}
\newcommand{\rdlladvtags}{\[
  \ngRuleWhere{
  \inferrule*[lab=D-LetLoc-Advance-Tags]{
    \MENV(\ngLoc'^{\ngReg.\ngPath}) = c \\
    \ngLoc_f \;\fresh
  }{
    \ngStepCfgB{\Theta}{\STOR}{\MENV}
      {\letloc{\ngLoc^{\ngReg.\ngPath}}{\ngAdvTags{\ngLoc'^{\ngReg.\ngPath}}{\DC}}{\EXPR}}
      {\Theta}{\STOR}{\MENV'}{\EXPR'}
  }
  }{
    \MENV' &= \MENV \cup
      \set{\ngLoc_f^{\ngReg.\ngPath} \mapsto \ngTagNext{c}} \\
    \EXPR' &= \subst{\EXPR}{\ngLoc}{\ngLoc_f}
  }
\]
}
\newcommand{\rdlladvfield}{\[
  \ngRuleWhere{
  \inferrule*[lab=D-LetLoc-Advance-Field]{
    c = \MENV(\ngLoc'^{\ngReg.\ngPath}) \\
    \vwf{\STOR}{\ngReg.\ngPath.\ngFieldKey{\DC}{j}}{c(\DC,j)}{\TYP'}{g} \\
    \ngLoc_f \;\fresh
  }{
    \ngStepCfgB{\Theta}{\STOR}{\MENV}
      {\letloc{\ngLoc^{\ngReg.\ngPath}}{\ngAdvField{\ngLoc'^{\ngReg.\ngPath}}{\DC}{j}{\TYP'@\ngLoc_0^{\ngReg.\ngPath.\ngFieldKey{\DC}{j}}}}{\EXPR}}
      {\Theta}{\STOR}{\MENV'}{\EXPR'}
  }
  }{
    \MENV' &= \MENV \cup \set{\ngLoc_f^{\ngReg.\ngPath} \mapsto
      \ngReplace{c}{\ngFieldKey{\DC}{j}}{g}} \\
    \EXPR' &= \subst{\EXPR}{\ngLoc}{\ngLoc_f}
  }
\]
}
\newcommand{\rddatacon}{\[
  \ngRuleWhere{
  \inferrule*[lab=D-DataConstructor]{
    \ngVecClear\MENV(\ngLoc^{\ngReg.\ngPath}) = \ngCurs{i}\,\overharpoon{CF}
  }{
    \ngVecClear
    \ngStepCfgB{\Theta}{\STOR}{\MENV}{\datacon{\DC}{\ngLoc^{\ngReg.\ngPath}}{\overharpoon{\VAL}}}
      {\Theta}{\STOR'}{\MENV}{\ngLocVal{\ngLoc^{\ngReg.\ngPath}}}
  }
  }{
    \STOR' &= \ngSnoc{\STOR}{\ngCell{\ngReg.\ngPath}{i}}{\DC}
  }
\]
}
\newcommand{\rdcase}{\[
  \ngRuleWhere{
  \inferrule*[lab=D-Case]{
    \ngPremBlock{
      \begin{array}{@{}r@{\quad}l@{}}
      \socalgrouplabel{scrutinee} & \VAL = \ngLocVal{\ngLoc_s^{\ngReg_s.\ngPath_s}} \qquad
      \MENV(\ngLoc_s^{\ngReg_s.\ngPath_s}) = c \qquad
      \ngVecClear c = \ngCurs{i}\,\overharpoon{CF} \qquad
      \STOR(\ngReg_s.\ngPath_s)(i) = \DC
      \ngPremGap
      \socalgrouplabel{clause} & \ngVecClear(\DC\,\overharpoon{\TYP'}) \in \ctors{\TYP} \qquad
      \pat_\kappa =
        \caseclause{\datacon{\DC}{\!\!}{(\overharpoon{\var:\TYP'@\ngLoc'^{\ngPos}})}}{\EXPR_\kappa}
        \;\text{is the clause for}\; \DC
      \ngPremGap
      \socalgrouplabel{arity} & \ngVecClear\litnum = |\overharpoon{\TYP'}|
        = |\overharpoon{\var:\TYP'@\ngLoc'^{\ngPos}}| \qquad
      m = |\set{j \mid \TYP'_j \neq \TYP}| \qquad
      p = \litnum - m
      \ngPremGap
      \socalgrouplabel{fields} & \begin{array}[t]{@{}l@{}}
      d_0 = c \\
      \forall\, 1 \le j \le m.\;\;
        \vwf{\STOR}{\ngReg_s.\ngPath_s.\ngFieldKey{\DC}{j}}
          {d_{j-1}(\DC,j)}{\TYP'_j}{g_j}
        \;\;\text{and}\;\;
        d_j = \ngReplace{d_{j-1}}{\ngFieldKey{\DC}{j}}{g_j}
      \end{array}
      \ngPremGap
      \socalgrouplabel{children} & \begin{array}[t]{@{}l@{}}
      f_0 = \ngTagNext{d_m} \\
      \forall\, 1 \le k \le p.\;\;
        \vwf{\STOR}{\ngReg_s.\ngPath_s}{f_{k-1}}{\TYP}{f_k}
      \end{array}
      \ngPremGap
      \socalgrouplabel{binders} & \ngVecClear\overharpoon{\ngLoc_f} \;\fresh
        \;(\litnum\;\text{of them, pairwise distinct})
      \end{array}
    }
  }{
    \ngVecClear
    \ngStepCfgB{\Theta}{\STOR}{\MENV}{\case{\VAL}{\overharpoon{\pat}}}
      {\Theta}{\STOR}{\MENV'}{\EXPR'}
  }
  }{
    \ngPos_h &= \ngReg_s.\ngPath_s.\ngFieldKey{\DC}{h}
      \;\;\text{for}\; h \le m, \qquad
      \ngPos_h = \ngReg_s.\ngPath_s \;\;\text{for}\; m < h \le \litnum \\
    c_j &= d_{j-1}(\DC,j) \;\;\text{for}\; 1 \le j \le m, \qquad
      c_{m+k} = f_{k-1} \;\;\text{for}\; 1 \le k \le p \\
    \MENV' &= \MENV \cup
      \set{\ngLoc_{f,h}^{\ngPos_h} \mapsto c_h \mid h \le \litnum} \\
    \EXPR' &= \EXPR_\kappa
      \subst{}{\overharpoon{\ngLoc'}}{\overharpoon{\ngLoc_f}}
      \subst{}{\overharpoon{\var}}
        {\overharpoon{\ngLocVal{\ngLoc_{f,h}^{\ngPos_h}}}}
  }
\]
}
\newcommand{\rdapp}{\[
  \ngRuleWhere{
  \inferrule*[lab=D-App]{
    \ngPremBlock{
      \begin{array}{@{}r@{\quad}l@{}}
      \socalgrouplabel{callee} & \mathrm{Function}(\fvar) =
        \big(\forall_{\overharpoon{\ngLoc_\ind^{\ngReg_\ind}},\,\ngLoc^{\ngReg}}.\;
        \overharpoon{\TYP_\ind@\ngLoc_\ind^{\ngReg_\ind}} \ARROW \TYP@\ngLoc^{\ngReg},\;
          \overharpoon{\var},\; \EXPR\big)
      \ngPremGap
      \socalgrouplabel{arity} & \begin{array}[t]{@{}l@{}}
      \ngVecClear|\overharpoon{\VAL}| = |\overharpoon{\var}| = \litnum =
        |\overharpoon{\TYP_\ind@\ngLoc_\ind^{\ngReg_\ind}}| \qquad
      |\overharpoon{\ngLoc'^{{\ngReg}'.\ngPath'}}| = \litnum + 1 \\
      \forall\, \ind, \kappa \le \litnum+1.\;
        \ngReg_\ind = \ngReg_\kappa \implies
        {\ngReg}'_\ind.\ngPath'_\ind = {\ngReg}'_\kappa.\ngPath'_\kappa
      \end{array}
      \end{array}
    }
  }{
    \ngVecClear
    \ngStepCfgB{\Theta}{\STOR}{\MENV}{\fapp{\overharpoon{\ngLoc'^{{\ngReg}'.\ngPath'}}}{\overharpoon{\VAL}}}
      {\Theta}{\STOR}{\MENV}{\EXPR'}
  }
  }{
    \sigma &= \subst{}{\overharpoon{\ngLoc_\ind^{\ngReg_\ind}},\ngLoc^{\ngReg}}
        {\overharpoon{\ngLoc'^{{\ngReg}'.\ngPath'}}} \\
    \EXPR' &= (\mathrm{Freshen}\,\EXPR)\,\sigma
      \subst{}{\overharpoon{\var}}{\overharpoon{\VAL}}
  }
\]
}
\newcommand{\rdvwfdata}{\[
  \inferrule*[lab=VWF-Data]{
    \ngPremBlock{
      \begin{array}{@{}r@{\quad}l@{}}
      \socalgrouplabel{tag} & \STOR(\ngPos)(i) = \DC \qquad
      \ngVecClear(\DC\,\overharpoon{\TYP'}) \in \ctors{\TYP}
      \ngPremGap
      \socalgrouplabel{arity} & \ngVecClear\litnum = |\overharpoon{\TYP'}| \qquad
      m = |\set{j \mid \TYP'_j \neq \TYP}| \qquad
      p = \litnum - m
      \ngPremGap
      \socalgrouplabel{fields} & \begin{array}[t]{@{}l@{}}
      \ngVecClear d_0 = \ngCurs{i}\,\overharpoon{CF} \\
      \forall\, 1 \le j \le m.\;\;
        \vwf{\STOR}{\ngPos.\ngFieldKey{\DC}{j}}
          {\ngProj{d_{j-1}}{\ngFieldKey{\DC}{j}}}{\TYP'_j}{g_j}
        \;\;\text{and}\;\;
        d_j = \ngReplace{d_{j-1}}{\ngFieldKey{\DC}{j}}{g_j}
      \end{array}
      \ngPremGap
      \socalgrouplabel{children} & \begin{array}[t]{@{}l@{}}
      f_0 = \ngTagNext{d_m} \\
      \forall\, 1 \le k \le p.\;\;
        \vwf{\STOR}{\ngPos}{f_{k-1}}{\TYP}{f_k}
      \end{array}
      \end{array}
    }
  }{
    \ngVecClear
    \vwf{\STOR}{\ngPos}{\ngCurs{i}\,\overharpoon{CF}}{\TYP}{f_p}
  }
\]
}

\colobusdynamicsend

\begin{figure*}[!t]
  \centering
  \socalrulestack[\textwidth]{%
    \socaljudgment{Reduction}{\socalJStep}%
    \socalwhereright{\input{paper-formalism/paper-fragments/rule-dletloc-after}}\socalrulesep
    \input{paper-formalism/paper-fragments/rule-dletloc-advfield}\socalrulesep
    \socalwhereright{\input{paper-formalism/paper-fragments/rule-ddataconstructor}}
  }
  \caption{Reduction rules for stepping past a value or field, and for
    constructors. Appendix~\ref{sec:appendix-formal} gives all rules.
    \captionscrunch}
  \phantomsection
  \label{fig:dynamic-body}
\end{figure*}

Each location rule computes a cursor from its predecessor's and binds it to a
fresh atom $\ngLoc_f$. \textsc{D-LetLoc-Start} binds $\ngZeroCursor{\TYP}$, the
cursor with index $0$ at every path, and \textsc{D-LetLoc-Advance-Tags} raises the root index by
one (Appendix~\ref{sec:appendix-formal}). \textsc{D-LetLoc-After} and \textsc{D-LetLoc-Advance-Field} (\figref{fig:dynamic-body}) must find where a
stored value ends, so they read it back with the value well-formedness
judgment $\vwf{\STOR}{\ngPos}{c}{\TYP}{c'}$, whose single rule follows the
layout (Appendix~\ref{sec:appendix-formal}). We call a derivation of this
judgment a \emph{traversal}, and its output $c'$, which lies just past the
value in every buffer the value occupies, its \emph{end cursor}.
\textsc{D-LetLoc-After} binds the end cursor of the value at its predecessor.
\textsc{D-LetLoc-Advance-Field} binds $\ngReplace{c}{\ngFieldKey{\DC}{j}}{g}$,
the predecessor's cursor $c$ with its sub-cursor for $\ngFieldKey{\DC}{j}$
replaced by the field value's end cursor $g$.

\textsc{D-DataConstructor} is the only rule that writes to the store: it
appends the constructor's tag at the destination's root index. 
\textsc{D-Case} reads the scrutinee with a traversal and binds each pattern variable to a fresh location at the cursor the traversal outputs for it.

\subsection{Type-Safety}\label{sec:type-safety}
\label{sec:preservation}

A configuration is \emph{well typed} at result type $\TYP$, written
$\vdash \ngCfg{\Theta}{\STOR}{\MENV}{\EXPR} : \TYP$
(Definition~\ref{def:cfgwf}), when its datatype declarations are well formed,
its functions are well typed, and $\EXPR$ has a typing derivation under
$\Theta$ and an empty variable environment whose contexts satisfy
\emph{store well-formedness},
$\storewfj{\SENV}{\CENV}{\AENV}{\NENV}{\UENV}{\MENV}{\STOR}$
(Definition~\ref{def:storewf}), along with conditions on names and on $\Theta$.

Store well-formedness says that $\MENV$ and $\STOR$ realize the layout the
derivation prescribes. Its first four clauses are:
\begin{itemize}[leftmargin=1.6em,labelwidth=1.2em,labelsep=0.4em,align=left]
\item[(\ref{swf:complete})] \emph{Complete-location consistency.} For every
  $\ngLoc^{\ngReg.\ngPath} \in \SENV$, $\MENV(\ngLoc^{\ngReg.\ngPath})$ is defined
  and a traversal from it exists at type $\ngTypeAt{\TYP_{\ngReg}}{\ngPath}$.
\item[(\ref{swf:constraints})] \emph{Location constraints realized.} $\CENV$ agrees
  with $\MENV$ and $\STOR$ (Definition~\ref{def:creal}).
\item[(\ref{swf:allocation})] \emph{Allocation state consistent.} $\AENV$ agrees
  with $\MENV$ and $\STOR$ (Definition~\ref{def:allocwf}).
\item[(\ref{swf:nursery})] \emph{Pending locations name unwritten cells.} For every
  $\ngLoc^{\ngReg.\ngPath} \in \NENV$, $\MENV(\ngLoc^{\ngReg.\ngPath})$ is defined
  and the cell at its root index in the buffer at $\ngReg.\ngPath$ is unwritten.
\end{itemize}
Progress reads clauses~(\ref{swf:complete}), (\ref{swf:constraints})
and~(\ref{swf:nursery}). Clauses~(\ref{swf:claims})--(\ref{swf:nursery-cross}) are
in Definition~\ref{def:storewf}, and Preservation uses them.
Since no cell is written twice, a cursor begins at most one traversal, so the
end cursor that \textsc{D-LetLoc-After} and \textsc{D-LetLoc-Advance-Field}
compute is well defined.

\begin{theorem}[Progress]
\label{thm:progress}
Let $\vdash \ngCfg{\Theta}{\STOR}{\MENV}{\EXPR} : \TYP$ hold
(Definition~\ref{def:cfgwf}). Then either $\EXPR$ is a value, or there
are $\STOR',\MENV',\EXPR'$ and $\Theta'$ with
$\Theta \sqsubseteq \Theta'$ and
$\ngStepCfg{\Theta}{\STOR}{\MENV}{\EXPR}{\Theta'}{\STOR'}{\MENV'}{\EXPR'}$.
\end{theorem}

\begin{theorem}[Preservation]
\label{thm:preservation}
Let $\vdash \ngCfg{\Theta}{\STOR}{\MENV}{\EXPR} : \TYP$ hold
(Definition~\ref{def:cfgwf}) and let
$\ngStepCfg{\Theta}{\STOR}{\MENV}{\EXPR}{\Theta'}{\STOR'}{\MENV'}{\EXPR'}$.
Then
\[
  \Theta \sqsubseteq \Theta'
  \qquad\text{and}\qquad
  \vdash \ngCfg{\Theta'}{\STOR'}{\MENV'}{\EXPR'} : \TYP .
\]
\end{theorem}

Here $\Theta \sqsubseteq \Theta'$ means that $\Theta'$ extends $\Theta$.
Type-safety is stated for a whole program
$P = \overharpoon{\DD};\overharpoon{\FD};\EXPR_0$, whose main expression writes
its result into an output buffer that the runtime provides. We model that
buffer as \emph{bootstrap data} $b = (\ngReg, \ngLoc, \TYP_b)$: a region, an
atom and the region's datatype. $P$ is \emph{admitted} at $b$
(Definition~\ref{def:admitted}) when \textsc{T-Program} types it under
$\Theta_{P;b}$, the region typing environment determined by $b$ and the
function signatures, and $P$ meets three syntactic conditions: its binders are
distinct, each region name has one type, and every datatype it names is
declared (Definitions~\ref{def:singlybound},
\ref{def:program-region-typing-environment}
and~\ref{def:declared-annotations}). The initial configuration has the empty
store $\STOR_0 = \varepsilon$ and
$\MENV_0 = \{\ngLoc^{\ngReg} \mapsto \ngZeroCursor{\TYP_b}\}$, and
Lemma~\ref{lem:initial-cfgwf} shows that it is well typed.

\begin{theorem}[Type-safety]
\label{thm:type-safety}
Let $P$ be admitted at $b$, with \textsc{T-Program} giving it result type
$\TYP$. If
$\ngCfg{\Theta_{P;b}}{\STOR_0}{\MENV_0}{\EXPR_0} \longrightarrow^{*}
 \ngCfg{\Theta_k}{\STOR_k}{\MENV_k}{\EXPR_k}$,
then $\Theta_{P;b} \sqsubseteq \Theta_k$,
$\vdash \ngCfg{\Theta_k}{\STOR_k}{\MENV_k}{\EXPR_k} : \TYP$, and either
$\EXPR_k$ is a value or $\ngCfg{\Theta_k}{\STOR_k}{\MENV_k}{\EXPR_k}$ steps.
\end{theorem}

Progress is proved by induction on the typing derivation: a non-value is a
redex under a nest of \lstinline{let} binders, the only inductive case, and the
redex steps because clauses~(\ref{swf:complete}), (\ref{swf:constraints})
and~(\ref{swf:nursery}) ensure that $\MENV$ is defined at every location its rule reads and supply a
traversal of every value it reads. Preservation is proved by induction on the
evaluation context; each basic contraction is one case, closed by inverting its
typing rule and re-establishing store well-formedness for the new store and
location map. Type-safety follows by induction on the length of the run, from
Lemma~\ref{lem:initial-cfgwf}. Appendix~\ref{sec:appendix-proof-structure}
outlines the proofs; the full proofs and a Lean~4~\cite{demoura2021lean4}
mechanization of every result in this section, built on
Mathlib~\cite{mathlib2020}, are in the supplementary material.

\section{\colobus{}: A Compiler for \ourcalc{}}
\label{sec:compiler-implementation}

\colobus{} compiles \lamadt{}, a small subset of (a strict version of) Haskell, to efficient C, using a lower-level variant of \ourcalc{} as its intermediate language.
We implement \colobus{} by extending \gibbon{}, which implements \oldcalc{}.
This section describes the input language (Section~\ref{sec:colobus-input}), the pipeline (Section~\ref{sec:colobus-pipeline}), mutable cursors (Section~\ref{sec:mutable-cursors}), and the optimizations that the factored layout enables (Section~\ref{sec:factored-opts}).

\subsection{Input Language and Interface}%
\label{sec:colobus-input}

\lamadt{} is a higher-order, call-by-value language with rank-1 polymorphism, algebraic datatypes, top-level function definitions, pattern matching, and recursion. The programmer chooses a storage layout for each datatype with a pragma.
The example below gives \lstinline|List| the \unfactored{} layout, which is also the default, and \lstinline|Tree| the \fullyfactored{} layout:
\par\vspace{-\baselineskip}\noindent
\begin{lrbox}{\vscodepane}%
\begin{minipage}[t]{0.42\linewidth}
\begin{lstlisting}[style=vscodeblock]
data List = Cons Int List | Nil
{-# ANN type List "Linear" #-}
\end{lstlisting}
\end{minipage}%
\begin{minipage}[t]{0.58\linewidth}
\begin{lstlisting}[style=vscodeblock,firstnumber=3]
data Tree = Node List Tree Tree | Leaf Int
{-# ANN type Tree "Factored" #-}
\end{lstlisting}
\end{minipage}%
\end{lrbox}%
\usebox{\vscodepane}
\par\noindent
A field keeps the layout assigned to its datatype definition, so a layout pragma is global. Here \lstinline|Tree| is factored, but its \lstinline|List| field keeps the \unfactored{} layout. This mix is a \hybridfactored{} layout. 

\colobus{} has compile-time flags that enable mutable cursors (Section~\ref{sec:mutable-cursors}) and each optimization of Section~\ref{sec:factored-opts}.
\gibbon's random-access and indirection pointers are on by default and can be turned off (Appendix~\ref{sec:pointers}). Users combine these flags to tune the generated code to a workload.

\subsection{Pipeline}%
\label{sec:colobus-pipeline}

Compilation lowers \lamadt{} code to C through a sequence of intermediate representations (IRs)~\cite{nanopass}.
The front end normalizes the source program, the middle end makes locations and layout explicit, and the back end emits cursor code and then C.

The front end type checks the program, then makes it first-order and monomorphic: it binds anonymous lambdas, monomorphizes, and specializes functions that take functions as arguments.
It then desugars and floats \lstinline|case| expressions out into dedicated functions.
The result is lowered to \marmosetlang{}, a first-order, monomorphic functional IR with algebraic datatypes and pattern matching~\cite{Marmoset}.
Passes over \marmosetlang{} put terms into A-normal form and simplify them, so that later analyses see a small, uniform language.

A type-directed \lstinline|inferLocations| pass maps \marmosetlang{} to \ourcalc{}.
It gives every packed value a location and generates constraints that relate locations to one another.
For instance, a constraint places a location at the start of a region, after a tag, or after another fully serialized value. The pass discharges each constraint as soon as it can, unifying locations as it goes.
Where two fixed locations cannot be unified, the pass repairs the program by inserting a copy.
Every subsequent middle-end pass is followed by a type check of the location-annotated program.

Like \gibbon{}, \colobus{} supports only pre-order layouts for recursive datatypes. For \AoS{} datatypes, it places a constructor's scalar fields before its packed fields. In \SoA{} data types, each field instead resides in its own buffer. \colobus{} factors a datatype as discussed in Section~\ref{sec:lang} (Algorithm~\ref{alg:build-soa-loc} in Appendix~\ref{sec:appendix-compiler} formalizes the procedure).

The rest of the middle end adds random-access nodes, extra traversals that compute end witnesses~\cite{LoCal}, redirection pointers, region parameters, and bounds checks against the end of a region. It also decides which functions use mutable cursors (as discussed in the next section).
\ourcalc{} is then lowered to \lamcur{}, which replaces symbolic locations by explicit cursor operations: a single cursor for a \unfactored{} value and an array of cursors, one per buffer, for a \fullyfactored{} one (See Appendix~\ref{sec:socal-to-nocal} for more details).
Next, the optimizations of Section~\ref{sec:factored-opts} are run on \lamcur{}.
Finally, the compiler lowers \lamcur{} to a C-like IR and generates C.

\gibbon{} uses three kinds of pointers in serialized data.
A region grows in \textit{chunks} of contiguous memory: when a traversal reaches the end of a chunk, a new chunk of twice the size is allocated, and a \textit{redirection} pointer links it to the previous one.
\textit{Indirection} pointers share data instead of copying it, and \textit{random-access} pointers let a traversal access fields out of order in $\mathrm{O}(1)$ time complexity~\cite{LoCal}.
\colobus{} adapts each of them to the \fullyfactored{} layout, in which a value spans several buffers.
Redirections across all buffers happen at the same time (i.e., if one buffer fills up, all associated buffers will allocate a new chunk). 
A traversal thus need only check for a redirection in the tag buffer; upon encountering one, all cursors in the cursor array will move to the next chunk.
An indirection to a factored value, and a random-access pointer to a factored field, hold one pointer per buffer that the value or field occupies, and all relevant cursors will follow the pointers as necessary (see Appendix~\ref{sec:pointers} for more details).

\subsection{Mutable Cursors}%
\label{sec:mutable-cursors}

\begin{figure*}[t]
\centering
\captionsetup[subfigure]{justification=centering,singlelinecheck=true}
\begin{subfigure}[t]{0.49\textwidth}
\centering
\hypertarget{fig:add1tree_impl_a_anchor}{}
\begin{minipage}[t]{\linewidth}
\begin{lstlisting}[language=C,style=cblock]
RetProd add1Tree(Cursor curin, Cursor curout){
  Cursor ctin = curin + 1, ctout = curout + 1;
  if (*curin == LEAF_TAG){
    int n = *(int*)ctin;
    *curout = LEAF_TAG;
    *(int*)ctout = n + 1;
    return {curin, ctin + 8, curout, ctout + 8};
  }
  *curout = NODE_TAG;
  RetProd l = add1Tree(ctin, ctout);
  RetProd r = add1Tree(l.field1, l.field3);
  return {curin, r.field1, curout, r.field3};
}
\end{lstlisting}
\end{minipage}
\caption{\unfactored{} (immutable Cursors)}
\label{lst:aos-add1treec}
\end{subfigure}
\hfill
\begin{subfigure}[t]{0.49\textwidth}
\centering
\hypertarget{fig:add1tree_impl_b_anchor}{}
\begin{minipage}[t]{\linewidth}
\begin{lstlisting}[language=C,style=cblock]
void add1Tree(Cursor *curin, Cursor *curout){
  if (**curin == LEAF_TAG){
    **curout = LEAF_TAG;
    (*curin)++; (*curout)++;
    int n = *(int*)(*curin);
    *(int*)(*curout) = n + 1;
    (*curin) += 8; (*curout) += 8;
    return;
  }
  **curout = NODE_TAG;
  (*curin)++; (*curout)++;
  add1Tree(curin, curout);
  add1Tree(curin, curout); // tail call
}
\end{lstlisting}
\end{minipage}
\caption{\unfactored{} (mutable cursors)}
\label{lst:aos-add1treec-2}
\end{subfigure}
\caption{Comparison of Packed \unfactored{} Mutable and Immutable Add1 Tree Implementations}
\phantomsection
\label{fig:add1tree_impl}
\end{figure*}

\gibbon{} generates code that treats a \cil{Cursor} as an immutable value, as in \SubFigRef{fig:sumtree_impl_a_anchor}{lst:aos-sumtreec}.
The analogous code for the \fullyfactored{} layout is shown in \SubFigRef{fig:sumtree_impl_c_anchor}{lst:soa-sumtreec}.
Immutable cursors suit functional programming, but they require extra copying. Because \colobus{} uses multiple cursors per data structure, this copying can introduce substantial overhead.
\colobus{} thus introduces {\em mutable cursors}---cursors that are updated in place---as \SubFigRef{fig:add1tree_impl_b_anchor}{lst:aos-add1treec-2} shows for \lstinline{add1tree} in the \unfactored{} layout.

Updating cursors in place lets \colobus{} realize the performance benefits of destination-passing style (DPS), in which functions write the values they compute to destinations passed as arguments instead of returning them.
In other words, in \ourcalc{}, a function receives the output location of the value it builds and writes the value there, in a region that its caller has already allocated.
In addition to saving on cursor copying overhead, this DPS approach enables straightforward tail-call optimization.

Figure~\ref{fig:add1tree_impl} shows the difference for \cil{add1} over \lstinline{Tree} (for simplicity, in the \unfactored{} layout).
With mutable cursors, a function receives a \cil{Cursor*} and updates it in place, so after a recursive call the caller finds the end witness by dereferencing that cursor and passes the same cursor to the next call.
The second recursive call of \cil{add1} is then in tail position, and GCC and LLVM apply tail-call optimization to it\footnote{Mutable cursors with tail-call optimization also keep stack use bounded: on large inputs, some runs with immutable cursors crash or exceed the time limit (Section~\ref{sec:eval}), while every run with mutable cursors and tail-call optimization completes.}.

Traversals over the \fullyfactored{} layout gain even more from mutable cursors (Section~\ref{sec:eval}): because \SoA{} layout requires {\em arrays} of cursors, copying overheads can often dominate performance. 
Mutable cursors allow the cursor array to be passed by reference, and updated in place.

\subsection{Optimizations Enabled by the Factored Layout}%
\label{sec:factored-opts}

The evaluation (Section~\ref{sec:eval}) distinguishes two kinds of traversals.
A \emph{fold} consumes a value and returns a result of another type, such as a sum.
A \emph{map} returns a new value of the same datatype.

For folds, the benefit of the factored layout is direct: if a fold does not need to visit a field, the \SoA{} layout allows it to skip reading the field's buffer entirely, while an \AoS{} layout requires reading past the unused fields (Section~\ref{sec:background:c}).
Section~\ref{sec:eval} measures this benefit across the suite. 

Maps are harder. \colobus{} implements a number of optimizations that can improve the performance of maps over \SoA{} layouts. We are particularly interested in \emph{dependence-free} maps, in which no output depends on the result of a recursive call (this is analogous to an absence of loop-carried dependences in a for loop):

\par\noindent\textbf{Loopification.}
The first step is to turn dependence-free maps into loops. 
Note that in a \fullyfactored{} layout, each buffer (other than the tag buffer) is essentially a homogeneous array. 
Consider \cil{add1Tree}, which adds $1$ to every integer in a tree. 
In factored form, this looks like a function that uniformly adds $1$ to each element in an integer array---something that can be readily implemented with a loop!

A programmer marks a map that is safe to turn into iterative code with the annotation:\newline
\lstinline|{-# ANN f "OPT:MayVectorize" #-}|, a promise that its recursive calls are independent. The compiler checks this annotation, but does not attempt to infer it.
It rejects the function if the result of a recursive call flows into a write, a tag, a branch condition, an argument of another call, or the function's result, and it then leaves the function unchanged.

To enable downstream optimizations such as vectorization (discussed below), \colobus{} attempts to turn dependence-free maps into for-loops. 
Creating such loops appears to be challenging, as a map over a serialized tree has no readily computable bound.
Here, \colobus{} leverages its whole-program nature---specifically the fact that it also controls the functions that {\em generate} the trees:
producers of a factored value record how many elements they wrote to each chunk of each buffer.
The programmer requests these counts by annotating the producer with \lstinline|{-# ANN mkTree "OPT:StoreScalarCounts" #-}|. 
Each count is stored in a chunk's footer, the metadata at the end of the chunk.

Maintaining this data efficiently is subtle.
A function receives only the end of an input region, so it can reach only the last footer directly.
Counts are therefore stored cyclically: the last footer holds the count of the first chunk, and every other footer holds the count of the next chunk.
A loop reads the first count at the region's end, and each later count in the footer of the chunk it has just finished.
Updating a count per element would require locating another chunk's footer each time.
Instead, a producer keeps a running count and writes it once per chunk: when the chunk fills and writing moves to a new chunk (Appendix~\ref{sec:pointers}), and, for the last chunk, when the producer returns.
With this deferred counting, storing counts only increases build time by roughly 3\% (in geometric mean) over our benchmarks and by at most 14\% ($S_{\ell bv}^{\scriptscriptstyle +av}$ against $S_{rm}^{\scriptscriptstyle +av}$, the \fullyfactored{} layout with and without counts).

Armed with the element counts in the chunk footers, \colobus{} can generate for-loops that traverse the chunks to perform map operations.
A loopified map has two levels of loops.
An outer loop visits the chunks of the input, following redirection pointers; inside it, a counted loop processes the elements of one chunk, using the stored count as its bound.
In the \fullyfactored{} layout there is one such loop per buffer, including the tag buffer, and each loop reads and writes homogeneous data.
The loopified map also records the counts of its output, so its result can itself be loopified.

Loopification also applies to the \unfactored{} layout, which we implement for comparison.
There, a single loop walks the whole buffer, up to the end of the input value, with a switch on the tag in its body.
This loop needs no counts, and counts would not help: the loop still walks a heterogeneous buffer in which tags and fields are interleaved, so it cannot be vectorized efficiently.

\par\noindent\textbf{Selective buffer sharing.}
Selective buffer sharing removes the loops that do nothing but copy. 
\colobus{}, like \gibbon{}, never updates written data in place: it keeps the write-once discipline that the formalism requires for the store to be well formed (Section~\ref{sec:lang}).
A map must therefore produce a new value, and even a field it does not change is copied to the output.
The \SoA{} layout cleanly avoids this copying by \textit{sharing} the buffers of the unmodified fields: the output refers to the input's buffer instead of copying it.

Sharing a buffer is simpler once the traversal is a loop rather than a recursion because the latter must decide which cursors to selectively thread through each recursive call. 
We therefore first turn dependence-free maps into loops, then share the buffers whose loops do nothing but copy.

In place of the copied data, it writes a selective-indirection tag into such output buffers along with pointers pointing to the data in the input buffer. Sharing always includes the tag buffer, because a tag in a scalar buffer cannot be distinguished from scalar data.
The tag buffer therefore also carries a mask recording which buffers are shared, and a consumer follows the pointers of exactly those buffers
thus reusing the data from the input.

\par\noindent\textbf{Loop fusion.}
An optional pass fuses adjacent loops that iterate over the buffers of the same constructor, since those buffers hold the same number of elements in every chunk.

\par\noindent\textbf{Vectorization.}
Finally, a vectorizer rewrites the per-buffer loops to use SIMD instructions.
For each store in a loop body, it builds a small expression DAG that describes how the stored value is computed from the let-bound expressions in A-normal form.
The root of the DAG is the store; its leaves are loads of the current element and loop-invariant values; its interior nodes are arithmetic operations and conditionals.
A loop is vectorized only if every operation in every store's DAG has a SIMD counterpart for its element type.
Loop-invariant values are hoisted out of the loop and broadcast across the vector lanes.
Because each buffer holds homogeneous elements, the stride is constant and known at compile time.
The vectorized loop processes a fixed number of elements per iteration, determined by the element types, and a scalar loop handles
any residual elements at the tail of the loop.
The generated code targets 256-bit AVX2 registers by default.

We also extended \gibbon{} with integers of 8, 16, 32, and 64 bits, written \lstinline|Int8| to \lstinline|Int64|.
Narrow integers fit more elements in a vector register, and Section~\ref{sec:eval-width} uses them to show the trade-off between computation and memory traffic.

\colobus{} runs the optimizations for dependence-free maps in this order: loopification of \unfactored{} traversals, loopification of \fullyfactored{} traversals, propagation of scalar counts, selective buffer sharing, loop fusion, and vectorization. 
In Section~\ref{sec:eval}, Figure~\ref{fig:pldi-stages} measures their performance effects.

\section{Evaluation}%
\label{sec:eval}

\newcommand{\TableJumpRef}[1]{\hyperref[#1]{Table~\ref*{#1}}}
\newcommand{\PassName}[1]{\textit{#1}}
\newcommand{\PackedInput}[1]{%
  \begingroup
    \setlength{\textfloatsep}{4pt plus 1pt minus 1pt}%
    \setlength{\floatsep}{3pt plus 1pt minus 1pt}%
    \setlength{\intextsep}{3pt plus 1pt minus 1pt}%
    \setlength{\abovecaptionskip}{3pt}%
    \setlength{\belowcaptionskip}{2pt}%
    \input{#1}%
  \endgroup
}
\setcounter{topnumber}{4}
\setcounter{bottomnumber}{2}
\setcounter{totalnumber}{6}
\renewcommand{\topfraction}{0.95}
\renewcommand{\bottomfraction}{0.90}
\renewcommand{\textfraction}{0.05}
\renewcommand{\floatpagefraction}{0.85}

We evaluate~\colobus{} on a suite of programs that traverse list- and tree-shaped
data with maps and folds, following the methodology of earlier \gibbon{}
evaluations. The suite has two parts. 
Real-world programs measure the performance of the factored layout on a mix of realistic traversals; synthetic programs
each vary a single property, so that its effect can be measured in isolation.

The eleven real-world programs come from domains that process tree-shaped data.
\PassName{Compiler} analyzes and rewrites a linear instruction-stream IR in the style
of LLVM, and \PassName{DBQuery} analyzes database query plans. Four programs are adapted 
from spatial and scientific computing: \PassName{KDTree} executes 3D search queries,
such as nearest-neighbor search, over a k-d tree; \PassName{OctTree} partitions
space hierarchically for N-body and FMM-style computations; \PassName{ColorOctree}
uses an octree for color quantization; and \PassName{PiecewiseFunctions} represents
functions as adaptive trees, similar to scientific simulations. \PassName{DecisionTree} and
\PassName{DecisionTreeClassify} perform inference, model inspection and
classification over a binary decision tree. The remaining three come from systems
software: \PassName{DomTree} traverses browser-style document and layout trees,
\PassName{Trie} analyzes and updates a prefix index, and \PassName{ObjectGraph}
profiles a heap graph with GC-like passes.

The thirteen synthetic programs fall into three groups. \PassName{List},
\PassName{MonoTree} and \PassName{TernaryTree} are small baselines, each with one map
and between one and three folds over a simple datatype. 
The cells of \PassName{LinearListReduction}'s list carry eleven fields, of which its folds read
two. Each cell of \PassName{ReduceNestedList}'s list holds a nested list that its
reduction never reads. Both programs measure the benefit of skipping dead data.
\PassName{Add1TreeInt}$N$ and \PassName{ArithmeticIntensityInt}$N$ apply a map to the
leaves of a tree at four integer widths, $N = 8$, 16, 32 and 64. The first adds one to
each leaf, and the second applies a chain of arithmetic, so together they show how
element width and computation per element affect vectorization.

The evaluation tests four claims made in the preceding sections. First, the
\fullyfactored{} layout lets a fold skip the fields it does not read
(Section~\ref{sec:background:c}). Second, mutable cursors avoid the cursor copying of
\gibbon{}'s immutable cursors and restore tail calls (Section~\ref{sec:mutable-cursors}).
Third, loopification and selective buffer sharing make maps over the \fullyfactored{}
layout faster than over the \unfactored{} one (Section~\ref{sec:factored-opts}). Fourth,
the \fullyfactored{} layout lets maps be vectorized, with a gain that depends on the
width of the elements. Section~\ref{sec:eval-suite} tests the first three claims
across the suite, and \PassName{KDTree} and \PassName{Compiler} examine them in detail;
Appendix~\ref{sec:appendix-eval} gives the per-pass tables for every program.

\subsection{Experimental platform}

All experiments ran on an Intel Core i7-12700K with 8 performance cores and 4 efficiency
cores (20 hardware threads), under Ubuntu 26.04.1 LTS with Linux 7.0.0-42 and glibc 2.43.
Every benchmark process was pinned to the same performance CPU core, which runs at a base
frequency of 3.6\,GHz and up to 5.1\,GHz with turbo enabled, under the \texttt{performance}
frequency governor. That core has a 48\,KiB L1 data cache, a 32\,KiB L1 instruction cache
and a 1.25\,MiB L2 cache, and the 25\,MiB L3 cache is shared by all cores. We compiled the
generated C with GCC 16.2.0 for AVX2 in every configuration. The processor does not support
AVX-512, so AVX2's 256-bit registers give the vectorized loops the most lanes
(Section~\ref{sec:eval-width}).

\par\noindent\textbf{Configurations.} We name a configuration by its layout, $A$ for the
\unfactored{} layout and $S$ for the \fullyfactored{} one. Subscripts describe the code
that \colobus{} generates: $r$ for recursive traversals, $i$ or $m$ for immutable or
mutable cursors, $\ell$ for loopified traversals, $b$ for selective buffer sharing and
$v$ for the SIMD pass. A superscript $+av$ or $-av$ says whether the C
auto-vectorizer is on or off. \gibbon{}~\cite{ecoop17-gibbon, LoCal}, the vanilla compiler, is $A_{ri}^{\scriptscriptstyle +av}$: the
\unfactored{} layout with recursive traversals and immutable cursors.
Table~\ref{tab:pldi-legend} in the appendix lists every configuration.

\subsection{End-to-end times}
\label{sec:eval-e2e}

\begin{figure}[tp]
  \centering
  \includegraphics{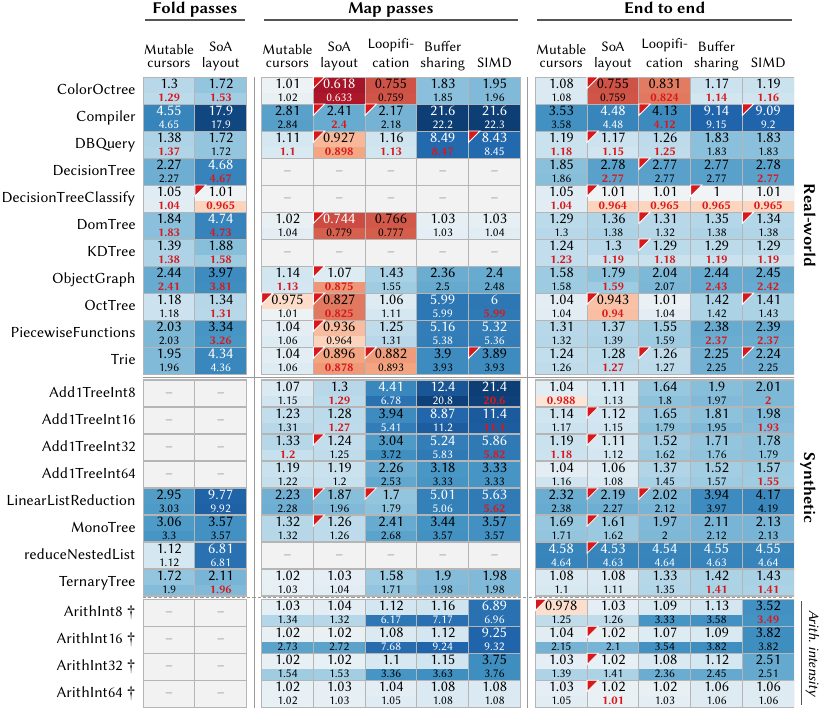}
  \caption{Speedup over \gibbon{} ($A_{ri}^{\scriptscriptstyle +av}$).
    Each column includes every stage to its left.
    The fold and map panels sum a program's passes of that kind, and the end-to-end
    panel adds the build time. In each cell the upper number has the C auto-vectorizer
    off and the lower one has it on. Blue is faster than \gibbon{}, red
    is slower; a dash marks an absence of that kind of pass.
    \captionscrunch}
  \label{fig:pldi-stages}
\end{figure}

We first measure the combined effect of the layout and the optimizations.
Figure~\ref{fig:pldi-stages} shows how each stage of the pipeline changes the
running time relative to \gibbon{}. With every stage enabled
(SIMD), \colobus{} is faster end to end than \gibbon{} on every program except \PassName{DecisionTreeClassify} with the auto-vectorizer on ($0.97\times$). A
classification query follows a single root-to-leaf path, so it reads only a small
part of the tree, and the layout matters little. \PassName{List} is
not shown, because its \gibbon{} run crashed with a segmentation fault.
In the figure, a red corner marks a stage that made the program slower
than the column to its left, and a bold red lower number marks a configuration that the C
auto-vectorizer made slower. In rows marked \dag, \gibbon{} is itself built
with the auto-vectorizer off. ArithInt$N$ abbreviates
\PassName{ArithmeticIntensityInt}$N$.

The C auto-vectorizer is not always a gain. The \fullyfactored{} layout gives GCC more
code to vectorize, including SLP vectorization of updates to a factored
value's array of cursors. GCC's cost model judges these rewrites profitable, but the
generated code costs more than it saves. In \PassName{KDTree}'s recursive passes it adds
instructions to every call and, in two passes, moves values between scalar registers and
vector lanes on the hot path. With the auto-vectorizer on, every \fullyfactored{} pass of
\PassName{KDTree} is therefore 8--49\% slower, while the \unfactored{} passes, which GCC
barely vectorizes, stay within 2\% of their time without it. Disabling only the SLP
vectorizer removes the slowdown, whereas the gains of loopified maps come from GCC's loop
vectorizer (Appendix~\ref{sec:appendix-autovec}).
The rest of this section separates the contributions of the individual stages.

\subsection{Suite-wide results}
\label{sec:eval-suite}

\par\noindent\textbf{Skipping dead data.}
The first claim predicts that folds gain most when they leave fields unread. 
Across the suite, with mutable cursors, the \fullyfactored{} layout speeds up the 68 fold
passes that leave a field unbound by $1.89\times$ in geometric mean (63 run faster), but the
6 that bind every field by only $1.29\times$ (5 run faster).
As \PassName{verifyIR} shows, a pass can bind a field without reading it.
\PassName{LinearListReduction}, whose \PassName{reduction} pass reads 2 of its 11 fields,
shows the effect most directly: over the \fullyfactored{} layout the pass is $5.9\times$ faster
(Table~\ref{tab:pldi-fold-LinearListReduction-soa}) and incurs $182.1\times$ fewer L1
data-cache misses and $600.2\times$ fewer L2 misses
(Tables~\ref{tab:LinearListReduction_pldi_counters_by_config} and~\ref{tab:LinearListReduction_pldi_counters_by_config-soa}).
Likewise, \PassName{ReduceNestedList}, the nested-list reduction of the introduction, is $6.1\times$ faster. In maps, buffer sharing
gives the same benefit for fields the map leaves unchanged.

\par\noindent\textbf{Comparison against \gibbon{}.}
\gibbon{} ($A_{ri}^{\scriptscriptstyle +av}$) differs from
$A_{rm}^{\scriptscriptstyle +av}$ only in its cursors, so comparing the two tests the second claim. Immutable cursors are
slower than mutable ones on all but 3 of 100 passes; 9 immutable-cursor runs crash
with a segmentation fault and 18 exceed the time limit. With mutable cursors, the \fullyfactored{} layout is faster
than \gibbon{} on 69 of 71 fold passes. On map passes, it is faster on only 15 of 29 before the optimizations of
Section~\ref{sec:factored-opts}, and on 28 of 29 with them, by $5.2\times$ in geometric mean.
Two case studies follow.

\subsection{KDTree}
\label{sec:eval-kdtree}

\PassName{KDTree} tests dead-field skipping on a real-world program. A k-d tree~\cite{bentley1975multidimensional} is a standard data structure for nearest-neighbor search, 
range search and correlation queries over points in a $k$-dimensional space. Our
\PassName{KDTree} benchmark is a 3D spatial index: internal nodes hold split and
bounding-box metadata, and leaves hold points and per-point attributes. Its six
passes are queries: nearest-neighbor search (\PassName{nearestDist}), box range
queries (\PassName{countInRange}, \PassName{sumMassInRange}), correlation and
neighborhood queries (\PassName{twoPointCorrelation},
\PassName{pointCloudNeighborhood}), and a multi-phase traversal
(\PassName{photonMappingBenchmark}).

All six passes are folds, and each reads only some of the tree's fields: between three
and six fields are dead in every pass (Table~\ref{tab:KDTree} in
Appendix~\ref{sec:appendix-eval}). Every pass is faster over
the \fullyfactored{} layout, by $1.08$--$1.63\times$ in its best configuration.
These speedups are what the first claim predicts. 
The C auto-vectorizer, however, makes these \fullyfactored{} passes slower (Section~\ref{sec:eval-e2e}).

\subsection{Compiler}
\label{sec:eval-compiler}

\PassName{Compiler} tests the first three claims on one program: its folds skip dead
data, its passes test tail calls with mutable cursors, and its maps
test loopification and buffer sharing.

This benchmark models a compiler pipeline over a linear instruction-stream IR,
inspired by LLVM IR, with regular instructions, basic-block boundaries and
termination nodes. Its passes check structure (\PassName{verifyIR},
\PassName{instCountPass}, \PassName{blockCountPass}, \PassName{castInstCountPass}),
compute microarchitectural proxies (\PassName{memoryOpStatsPass},
\PassName{branchStatsPass}) and latency and throughput models, detect cycles in the
control flow (\PassName{goHasCycle}), and rewrite the IR (the maps
\PassName{targetReturnPass} and \PassName{stripSideEffectsPass}).

Every fold pass is faster over the \fullyfactored{} layout, by $3.2$--$6.7\times$ in
its best configuration (Table~\ref{tab:Compiler} in Appendix~\ref{sec:appendix-eval}).
Every fold skips dead data, including \PassName{verifyIR}, which binds all nine fields of
the IR datatype but reads only the opcode and the tail to check that $\phi$-instructions
come first in each block. Over the \unfactored{} layout,
reading an opcode still brings the instruction's other fields into the cache.
Over the \fullyfactored{} layout, the pass touches only the
opcode buffer, and \PassName{verifyIR} is $3.9\times$ faster and incurs $224.7\times$ fewer L1 data-cache
misses (Tables~\ref{tab:Compiler_pldi_counters_by_config} and~\ref{tab:Compiler_pldi_counters_by_config-soa}).

Tail-call optimization matters only with mutable cursors. Disabling it in the C
compiler makes the mutable-cursor passes $3.8\times$ slower on the \unfactored{} layout
and $9.1\times$ slower on the \fullyfactored{} layout. It leaves the immutable-cursor
passes unchanged (a geometric mean of $0.99\times$), because immutable cursors keep
recursive calls out of tail position (Section~\ref{sec:mutable-cursors}).

With recursive traversals and mutable cursors, both map passes are slower over the
\fullyfactored{} layout ($0.86\times$ and $0.83\times$ against
$A_{rm}^{\scriptscriptstyle +av}$). Without buffer sharing a map copies every field
(Section~\ref{sec:factored-opts}), and over the \fullyfactored{} layout it writes each
field to its own buffer through its own cursor. The optimizations of
Section~\ref{sec:factored-opts} reverse this: with loopification, selective buffer
sharing and SIMD, the two map passes are $7.6\times$ and $7.5\times$ faster over the
\fullyfactored{} layout than the loopified \unfactored{} configuration
($S_{\ell bv}^{\scriptscriptstyle +av}$ against $A_{\ell}^{\scriptscriptstyle +av}$,
Table~\ref{tab:pldi-map-Compiler}). Without buffer sharing, the
loopified \fullyfactored{} passes are slower than the loopified \unfactored{} ones,
because each pass still runs one loop per buffer, eight in all.
Yet each map changes only one field of an instruction: \PassName{targetReturnPass} scales
its latency and \PassName{stripSideEffectsPass} clears its flags. Buffer sharing removes
the loops that only copy the tags and the other six fields, leaving one loop per pass. The SIMD pass adds nothing
measurable. It leaves \PassName{targetReturnPass}'s loop scalar, because that loop
multiplies 64-bit integers and AVX2 has no packed 64-bit multiplication
(Section~\ref{sec:eval-width}); it vectorizes \PassName{stripSideEffectsPass}'s loop,
which stores a constant zero, without making it faster.
In \PassName{Compiler}, the speedup of the maps therefore comes from buffer sharing, while
vectorization, which the next subsection examines, contributes little.

\subsection{Integer width and vectorization}
\label{sec:eval-width}

\input{tables/pldi/_preamble}
\begin{table}[t]
\centering
\caption{The map passes \textit{add1Tree} of \texttt{Add1TreeInt}$N$ and \textit{arithKernel} of \texttt{ArithmeticIntensityInt}$N$ at $N = 8$, 16, 32 and 64: median time per iteration (ms) under each configuration (Table~\ref{tab:pldi-legend}); $^{-av}$ and $^{+av}$ build the same code with the C auto-vectorizer off and on. \emph{SIMD gain}: speedup from the SIMD pass ($S_{\ell b}^{\scriptscriptstyle -av}/S_{\ell bv}^{\scriptscriptstyle -av}$). \emph{vs.\ \gibbon{}}: speedup of the fastest \fullyfactored{} configuration over \gibbon{} ($A_{ri}^{\scriptscriptstyle +av}$).}
\label{tab:width}
\footnotesize\gibbonnumfont
\setlength{\tabcolsep}{2.5pt}
\gibbonfit{%
\begin{tabular}{l rrrrrrrrrr cc}
\toprule
Width & \gibbon{} & Mutable & \shortstack{AoS\\loop} & SoA & \multicolumn{2}{c}{Loopified} & \multicolumn{2}{c}{\shortstack{Buffer\\sharing}} & \multicolumn{2}{c}{SIMD} & \multicolumn{2}{c}{Speedup} \\
\cmidrule(lr){2-2}\cmidrule(lr){3-3}\cmidrule(lr){4-4}\cmidrule(lr){5-5}\cmidrule(lr){6-7}\cmidrule(lr){8-9}\cmidrule(lr){10-11}\cmidrule(lr){12-13}
 & $A_{ri}^{\scriptscriptstyle +av}$ & $A_{rm}^{\scriptscriptstyle +av}$ & $A_{\ell}^{\scriptscriptstyle +av}$ & $S_{rm}^{\scriptscriptstyle +av}$ & $S_{\ell}^{\scriptscriptstyle -av}$ & $S_{\ell}^{\scriptscriptstyle +av}$ & $S_{\ell b}^{\scriptscriptstyle -av}$ & $S_{\ell b}^{\scriptscriptstyle +av}$ & $S_{\ell bv}^{\scriptscriptstyle -av}$ & $S_{\ell bv}^{\scriptscriptstyle +av}$ & \shortstack{SIMD\\gain} & \shortstack{vs.\\\gibbon{}} \\
\midrule
\multicolumn{13}{@{}l}{\textit{add1Tree}} \\
Int8 & 68.83 & 59.83 & 29.87 & 53.48 & 15.61 & 10.15 & 5.56 & 3.31 & 3.21 & 3.34 & 1.73$\times$ & 21.44$\times$ \\
Int16 & 71.88 & 54.85 & 33.00 & 56.59 & 18.26 & 13.29 & 8.11 & 6.40 & 6.31 & 6.47 & 1.28$\times$ & 11.39$\times$ \\
Int32 & 77.91 & 65.04 & 38.12 & 62.27 & 25.64 & 20.93 & 14.85 & 13.36 & 13.28 & 13.38 & 1.12$\times$ & 5.86$\times$ \\
Int64 & 89.95 & 73.44 & 50.23 & 75.10 & 39.89 & 35.55 & 28.24 & 27.05 & 27.04 & 27.02 & 1.04$\times$ & 3.33$\times$ \\
\midrule
\multicolumn{13}{@{}l}{\textit{arithKernel}} \\
Int8 & 393.00 & 369.60 & 356.90 & 376.50 & 442.10 & 80.29 & 425.70 & 69.09 & 71.88 & 71.18 & 5.92$\times$ & 5.69$\times$ \\
Int16 & 199.80 & 177.50 & 159.00 & 178.60 & 449.10 & 63.17 & 433.30 & 52.49 & 52.45 & 52.07 & 8.26$\times$ & 3.84$\times$ \\
Int32 & 355.90 & 328.70 & 311.80 & 330.40 & 459.80 & 150.70 & 442.30 & 139.60 & 135.20 & 134.90 & 3.27$\times$ & 2.64$\times$ \\
Int64 & 521.80 & 488.20 & 484.60 & 486.10 & 479.70 & 475.00 & 460.70 & 460.70 & 460.60 & 460.60 & 1.00$\times$ & 1.13$\times$ \\
\bottomrule
\end{tabular}}
\end{table}

This subsection tests the fourth claim with the narrow integers of
Section~\ref{sec:factored-opts}. A 256-bit AVX2 register holds 32
8-bit lanes but only four 64-bit ones, and each halving of the width also halves the
bytes a loop moves per element. The upper half of Table~\ref{tab:width} runs
\PassName{add1Tree} at the four widths. The gain from the SIMD pass, measured
with the C auto-vectorizer off, is largest at 8 bits ($1.73\times$) and shrinks at
each wider type, to $1.04\times$ at 64 bits. Against \gibbon{}, the best
\fullyfactored{} configuration is $21.4\times$ faster at 8 bits and $3.3\times$ faster
at 64 bits.

\PassName{add1Tree} does one addition per element, so its running time is dominated by memory traffic
rather than computation. \PassName{ArithmeticIntensityInt}$N$ instead applies a chain of
multiplications and additions to each leaf (lower half of Table~\ref{tab:width}).
With more computation per element, the SIMD pass gives a larger speedup: $5.92\times$ at 8 bits,
$8.26\times$ at 16 bits and $3.27\times$ at 32 bits. The speedup is larger at 16 bits than
at 8 because AVX2 has no packed 8-bit multiplication. GCC therefore widens each vector of
8-bit values to 16 bits (\texttt{vpunpcklbw}, \texttt{vpunpckhbw}), multiplies the two
halves with two 16-bit multiplications (\texttt{vpmullw}), and narrows the result
again (\texttt{vpshufb}); at 16 bits, each multiplication is a single
\texttt{vpmullw}. At 64 bits the pass gives no speedup
($1.00\times$): AVX2 has no packed 64-bit multiplication, so the kernel's
multiplications stay scalar.

Further experiments (Appendix~\ref{sec:appendix-autovec}) show that in loopified code the C
auto-vectorizer's full gain is delivered by GCC's loop vectorizer alone, and that the SIMD
pass gains from AVX2's wider registers on the arithmetic kernel but not on \PassName{add1Tree}.
Vectorization therefore speeds up maps over the \fullyfactored{} layout most when
elements are narrow and need more arithmetic.

\section{Related Work}
\label{sec:related}
\gibbon{}~\cite{ecoop17-gibbon} serializes ADT values into a single contiguous buffer and compiles traversals to operate directly on it, which improves spatial locality.
\oldcalc{}~\cite{LoCal, Vollmer2021Thesis} introduces a type-safe language to write traversals over serialized ADTs by abstracting pointers as locations, 
and adds indirection pointers, which allows sharing of serialized data, and random-access pointers to access fields of an ADT in $\mathcal{O}(1)$.
Later work parallelizes traversals over such representations~\cite{Gibbon2}.
Serialization formats such as Cap'n Proto~\cite{capnproto} also read data in place, but through accessors generated for a fixed schema rather than compiled traversals over general ADTs.
Our work lifts the single-buffer requirement of \gibbon{} and \oldcalc{}, so that one value can span several buffers.

Several typed languages make data layout explicit and type-safe.
Typed assembly language~\cite{morrisett1999tal} brings types to machine code and the memory it accesses, region calculi~\cite{regioncalcs} track the regions in which objects are allocated in the types, and Cyclone~\cite{cyclone-pldi} lets C programmers control data representation and memory management safely.

Prior \AoS{}-to-\SoA{} transformations target array-like data structures.
Rompf et al.~\cite{Rompf2013OptimizingDS} support them in an extensible compiler framework, and Radtke and Weinzierl~\cite{radtke2025annotationguidedaostosoaconversionsgpu} apply them to annotated C++ programs.
\colobus{} applies the transformation to recursive ADTs and generates the code that traverses them.

Other work optimizes the layout of recursive data structures.
Marmoset~\cite{Marmoset} chooses a serialization order for the fields of an ADT, but still places each value in a single buffer.
For pointer-based structures, Chilimbi et al.\ cluster tree nodes in memory~\cite{Chilimbi1999} and reorder and split struct fields~\cite{Chilimbi1999b}, Truong et al.~\cite{Truong1998} group the same field of many heap-allocated structs together, and automatic pool allocation~\cite{Lattner2005} gives each data structure a memory pool of its own.
However, these transformations retain pointers, which serialization largely removes.

Earlier work vectorizes tree traversals by running many independent traversals in lockstep, for N-body simulations~\cite{makino90} and for general tree-traversal workloads on SIMD hardware~\cite{ren13cgo, ren14taco} and GPUs~\cite{goldfarb13sc}.
GPU layout optimizers restructure array-based data in GPU kernels, guided by graphs of the relationships between fields~\cite{AutomaticDLOptForGPUs} or by the memory-access patterns observed at run time~\cite{Leo}.
In contrast, \colobus{} vectorizes map traversals over arbitrary recursive ADTs written in a functional language: factoring turns a map over one field into a loop over a homogeneous buffer.

LLAMA~\cite{LLAMA} is a C++ library that decouples data layout from computation, so a program can be retuned for new hardware by changing only its layout description.
\colobus{} instead takes a layout pragma for a datatype, generates a~\fullyfactored{} serialized representation and the traversals over it automatically.

\section{Conclusion}

We introduced the \fullyfactored{} layout, which stores the fields of a serialized value in separate buffers, so that traversals skip the fields they do not read and maps over scalar fields can be vectorized.
\ourcalc{} is a type-safe language for programs over this layout, with a type-safety proof mechanized in Lean~4, and \colobus{} implements the layout in \gibbon{}, together with mutable cursors and optimizations for dependence-free maps.

Several directions remain open. Letting each datatype declare its own tree of paths would bring the \unfactored{} and \hybridfactored{} layouts into \ourcalc{}: with only the empty path $\ngEmptyPath$, a datatype takes the \unfactored{} layout, and with a chosen subset of paths, only the selected fields are factored. \colobus{} loopifies only annotated maps; loopifying folds
and more general recursive functions requires recognizing their recursion
scheme~\cite{meijer1991bananas}, which can be derived automatically from recursive
definitions~\cite{hu1996hylomorphisms}. Finally, the per-buffer loops suit GPU offloading via OpenMP.

\section{AI Use}

We used generative AI tools, Anthropic's Claude (Sonnet and Opus, versions 4.5 onward)
and OpenAI's GPT (versions 5 through 6), moving to newer versions as they were released
during this work. For the formalism, AI helped prove the auxiliary lemmas behind
progress, preservation and type-safety, simplify the metatheory, find counterexamples
that exposed flaws in earlier versions of the metatheory, and write the Lean~4
mechanization, whose headline results are the theorems of Section~\ref{sec:preservation}.
For the compiler, AI helped implement and harden parts of \colobus{}, including
loopification, stored scalar counts, selective buffer sharing, and vectorization at
different integer widths. For the evaluation, AI helped design the real-world benchmarks,
starting from the Treelogy suite~\cite{hegde2017treelogy}, so that their datatypes and
memory-access patterns resemble those of production software. AI also helped write the
benchmarking scripts, which collect the running times and hardware-counter measurements,
and the scripts that render the tables and figures.

\bibliography{refs}

\iftoggle{EXTND}{
\newpage
\onecolumn
\appendix
\renewcommand{\floatpagefraction}{0.75}
\makeatletter
\setlength{\@fptop}{0pt}
\setlength{\@fpsep}{\floatsep}
\setlength{\@fpbot}{0pt plus 1fil}
\makeatother
\newcommand{\HereInput}[1]{\begingroup
  \let\colobusfloattable\table \let\endcolobusfloattable\endtable
  \renewenvironment{table}[1][]{\colobusfloattable[H]}{\endcolobusfloattable}%
  \input{#1}\endgroup}

The appendices follow the order of the paper.
Appendices~\ref{sec:appendix-formal}--\ref{sec:appendix-proof-structure} complete the
formal development of Section~\ref{sec:lang}: the full rules and a second worked example,
the well-formedness definitions, and the structure of the type-safety proof.
Appendix~\ref{sec:appendix-compiler} adds compiler material to
Section~\ref{sec:compiler-implementation}, and
Appendices~\ref{sec:appendix-eval} and~\ref{sec:appendix-clang} give the measurements
behind Section~\ref{sec:eval}, compiled with GCC and with Clang.

\section{Additional Formal Material}
\label{sec:appendix-formal}

This appendix repeats the grammar of Section~\ref{sec:lang}, gives the
typing and reduction rules that the section describes, and works through a second example, \lstinline|buildTree|,
whose self-recursive constructor also exercises $\gramkw{after}$.
Appendix~\ref{sec:well-formedness} defines the well-formedness judgments the
rules and the metatheory rely on, and
Appendix~\ref{sec:appendix-proof-structure} gives the structure of the
type-safety proof.

Section~\ref{sec:static-semantics} explains the rules that enforce the
order of writes; the remaining typing rules work as follows.
\textsc{T-Var} and \textsc{T-Concrete-Loc} type a variable or a location value
at a complete location. \textsc{T-Let} adds the bound variable to $\TENV$ and
its location to $\SENV$. \textsc{T-LetRegion} binds a fresh region at its
annotated type $\TYP_{\mathrm{ann}}$ and marks it unstarted, and
\textsc{T-LetLoc-Start} then binds its first location, setting the region's
state to $\ngConstFam{\ngLoc}$ and making $\ngLoc$ pending at every path.
\textsc{T-LetLoc-After} steps past a complete, consumable value and, like
\textsc{T-LetLoc-Advance-Tags}, moves $\AENV$ over the descendant set.
\textsc{T-Case} requires exactly one clause for each constructor of the
scrutinee's type, and \textsc{T-Pattern} places a clause's factored binders at
the scrutinee's position extended by their keys and its self-recursive binders
at the scrutinee's position. \textsc{T-FunctionDef} checks a body starting from
state $\ngConstFam{\ngLoc}$ for the result's atom $\ngLoc$, with $\ngLoc$
pending at every path of the result region.
\socalfullrules

\begin{figure*}[tp]
  \footnotesize
  \centering
  \input{formal_grammar_meta}

  \socalgrammarcolumns

  \normalsize
  \caption{Grammar of \ourcalc{}. This is a copy of Figure~\ref{fig:grammar}.
    \captionscrunch}
  \phantomsection
  \label{fig:grammar2}
  \label{fig:grammar2-env}
\end{figure*}

\begin{figure*}[tp]
  \centering
  \socalrulestack[\textwidth]{%
    \socaljudgment{Typing}{\socalJTyping}%
    \socalpairfit{\input{paper-formalism/paper-fragments/rule-tvar-tconcreteloc}}\socalrulesep
    \socalwhereright{\input{paper-formalism/paper-fragments/rule-tlet}}\socalrulesep
    \input{paper-formalism/paper-fragments/rule-tcase}\socalrulesep
    \socalwhereright{\input{paper-formalism/paper-fragments/rule-tletregion}}\socalgroupsep
    \socaljudgment{Function Declarations}{\socalJFun}%
    \socalwhereright{\input{paper-formalism/paper-fragments/rule-tfunctiondef}}
  }
  \caption{Static typings of \ourcalc{}: variables, $\gramkw{let}$, regions,
    $\gramkw{case}$ and function declarations.
    \captionscrunch}
  \phantomsection
  \label{fig:types12}
\end{figure*}

\begin{figure*}[tp]
  \centering
  \socalrulestack[\textwidth]{%
    \socaljudgment{Typing}{\socalJTyping}%
    \socalwhereright{\input{paper-formalism/paper-fragments/rule-tletloc-start}}\socalrulesep
    \input{paper-formalism/paper-fragments/rule-tletloc-advtags}\socalrulesep
    \input{paper-formalism/paper-fragments/rule-tletloc-after}\socalrulesep
    \input{paper-formalism/paper-fragments/rule-tletloc-advfield}
  }
  \caption{Static typings of \ourcalc{}: the location binders.
    \captionscrunch}
  \phantomsection
  \label{fig:types7-appendix}
\end{figure*}

\begin{figure*}[tp]
  \centering
  \socalrulestack[\textwidth]{%
    \socaljudgment{Typing}{\socalJTyping}%
    \input{paper-formalism/paper-fragments/rule-tdataconstructor}\socalgroupsep
    \socaljudgment{Patterns}{\socalJPattern}%
    \input{paper-formalism/paper-fragments/rule-tpattern}
  }
  \caption{Static typings of \ourcalc{}: the constructor rule and patterns.
    \captionscrunch}
  \phantomsection
  \label{fig:types2-appendix}
  \label{fig:types12b}
\end{figure*}

\begin{figure*}[tp]
  \centering
  \socalrulestack[\textwidth]{%
    \socaljudgment{Typing}{\socalJTyping}%
    \input{paper-formalism/paper-fragments/rule-tapp}\socalgroupsep
    \socaljudgment{Top-Level Programs}{\socalJProgram}%
    \input{paper-formalism/paper-fragments/rule-tprogram}
  }
  \caption{Static typings of \ourcalc{}:
    function application and whole programs.}
  \phantomsection
  \label{fig:types4-appendix}
\end{figure*}

\begin{figure*}[tp]
  \centering
  \socalrulestack[\textwidth]{%
    \socaljudgment{Reduction}{\socalJStep}%
    \input{paper-formalism/paper-fragments/gr-evalctx}\socalrulesep
    \socalwhereright{\input{paper-formalism/paper-fragments/rule-dletval-dletregion}}\socalrulesep
    \socalwhereright{\input{paper-formalism/paper-fragments/rule-dletloc-start}}\socalrulesep
    \input{paper-formalism/paper-fragments/rule-dctx}\socalrulesep
    \socalwhereright{\input{paper-formalism/paper-fragments/rule-ddataconstructor}}\socalrulesep
    \socalwhereright{\input{paper-formalism/paper-fragments/rule-dletloc-advtags}}\socalrulesep
    \socalwhereright{\input{paper-formalism/paper-fragments/rule-dletloc-after}}\socalrulesep
    \input{paper-formalism/paper-fragments/rule-dletloc-advfield}
  }
  \caption{Dynamic semantics of \ourcalc{}: $\gramkw{let}$, regions, location
    binders and constructors.
    \captionscrunch}
  \phantomsection
  \label{fig:dynamic12}
\end{figure*}

\begin{figure*}[tp]
  \centering
  \socalrulestack[\textwidth]{%
    \socaljudgment{Reduction}{\socalJStep}%
    \input{paper-formalism/paper-fragments/rule-dcase}\socalrulesep
    \socalwhereright{\input{paper-formalism/paper-fragments/rule-dapp}}\socalgroupsep
    \socaljudgment{Value Well-Formedness}{\vwf{\STOR}{\ngPos}{c}{\TYP}{c'}}%
    \input{paper-formalism/paper-fragments/rule-vwfdata}
  }
  \caption{Dynamic semantics of \ourcalc{}: case analysis, calls,
    and value well-formedness. \captionscrunch}
  \phantomsection
  \label{fig:dynamic2}
\end{figure*}

\subsection{A Second Worked Example: \lstinline|buildTree|}%
\label{sec:appendix-buildtree}

\begin{figure}[tp]
\centering
\begin{lstlisting}[style=localblock]
buildtree : (*@$\forall_{\ngLoc^{\ngReg}}$@*) . Int -> Tree @ (*@$\ngLoc^{\ngReg}$@*)
buildtree n =
  if n <= 0
  then -- a Leaf carries one factored field: write it, then step its cursor
       let i : Int @ (*@$\ngLoc^{\ngReg.\ngFieldKey{\mathsf{Leaf}}{1}}$@*) = 1 in
       letloc (*@$\ngLoc_1^{\ngReg}$@*) = advanceField (*@$\ngLoc^{\ngReg}$@*) (Leaf,1) beyond (Int @ (*@$\ngLoc^{\ngReg.\ngFieldKey{\mathsf{Leaf}}{1}}$@*)) in
       (Leaf (*@$\ngLoc^{\ngReg}$@*) i)
  else -- a Node carries no factored field: just reserve its tag cell
       letloc (*@$\ngLoc_a^{\ngReg}$@*) = advanceTags (*@$\ngLoc^{\ngReg}$@*) for Node in
       -- build the left subtree in place, at the cell past the tag:
       let left : Tree @ (*@$\ngLoc_a^{\ngReg}$@*) = buildtree [(*@$\ngLoc_a^{\ngReg}$@*)] (n - 1) in
       -- the right subtree starts past every cursor the left subtree moved:
       letloc (*@$\ngLoc_b^{\ngReg}$@*) = after (Tree @ (*@$\ngLoc_a^{\ngReg}$@*)) in
       let right : Tree @ (*@$\ngLoc_b^{\ngReg}$@*) = buildtree [(*@$\ngLoc_b^{\ngReg}$@*)] (n - 1) in
       -- write the Node tag, claiming both children:
       (Node (*@$\ngLoc^{\ngReg}$@*) left right)
\end{lstlisting}
\caption{\lstinline|buildTree| in \ourcalc.}%
\label{fig:buildtree-socal}
\end{figure}

\begin{table*}[!tp]
  \begingroup
  \centering
  \formalgrammarkeywords
  \setlength{\tabcolsep}{0pt}
  \fontsize{8}{9.6}\selectfont
  \newcommand{\tCodeKw}{%
    \renewcommand{\gramkw}[1]{\mathord{{\socalkeepcolor{FormalCodeKw}\mathrm{##1}}}}%
    \renewcommand{\gramwd}[1]{\gramkw{##1}}}
  \renewcommand{\arraystretch}{1.45}
  \newcommand{\tq}{(\mathsf{Leaf},1)}
  \newcommand{\tL}{\ngLoc^{\ngReg}}
  \newcommand{\tLq}{\ngLoc^{\ngReg.q}}
  \newcommand{\tLa}{\ngLoc_a^{\ngReg}}
  \newcommand{\tLaq}{\ngLoc_a^{\ngReg.q}}
  \newcommand{\tLb}{\ngLoc_b^{\ngReg}}
  \newcommand{\tLbq}{\ngLoc_b^{\ngReg.q}}
  \newcommand{\tLone}{\ngLoc_1^{\ngReg}}
  \newcommand{\tLtwo}{\ngLoc_2^{\ngReg}}
  \newcommand{\aL}{$\ngLoc$}
  \newcommand{\aLa}{$\ngLoc_a$}
  \newcommand{\aLb}{$\ngLoc_b$}
  \newcommand{\aLone}{$\ngLoc_1$}
  \newcommand{\aLtwo}{$\ngLoc_2$}
  \newcommand{\tTree}{\mathsf{Tree}}
  \newcommand{\tNode}{\mathsf{Node}}
  \newcommand{\tLeaf}{\mathsf{Leaf}}
  \newcommand{\tGap}{\hspace{0.9em}}
  \newcommand{\tSet}[1]{$\{#1\}$}
  \newcommand{\tNone}{$\emptyset$}
  \caption{Type checking a program that builds $\tNode\;(\tLeaf\;1)\;(\tLeaf\;2)$
    at the location $\tL$ that starts region $\ngReg$ (Appendix~\ref{sec:appendix-buildtree}).
    Here $\TYP_{\ngReg} = \tTree$ and $q \equiv \tq$, so $\ngPathsOf{\tTree} =
    \{\ngEmptyPath,\, q\}$; the two $\AENV$ columns give the atom allocation has
    reached at each position.}
  \label{fig:types-example}
  \label{fig:types-example-cont}
  \begin{adjustbox}{center,max width=\textwidth}
    \begin{tabular}{@{}r@{\tGap}>{\tCodeKw}l@{\tGap}c@{\tGap}c@{\tGap}l@{\tGap}l@{}}
      & Code & $\AENV(\ngReg.\ngEmptyPath)$ & $\AENV(\ngReg.q)$
        & $\NENV$ & $\UENV$ \\ \hline
      \emph{seed} &
        &
        \aL & \aL &
        \tSet{\tL,\, \tLq} &
        \tNone \\
      $1$ &
        $\letloc{\tLa}{\ngAdvTags{\tL}{\tNode}}{}$ &
        \aLa & \aLa &
        \tSet{\tL,\, \tLa,\, \tLaq} &
        \tNone \\
      $2$ &
        $\letpack{i_1 : \Int@\tLaq}{1}{}$ &
        \aLa & \aLa &
        \tSet{\tL,\, \tLa} &
        \tSet{\tLaq} \\
      $3$ &
        $\letloc{\tLone}{\ngAdvField{\tLa}{\tLeaf}{1}{\Int@\tLaq}}{}$ &
        \aLone & \aLa &
        \tSet{\tL,\, \tLa,\, \tLone} &
        \tSet{\tLaq} \\
      $4$ &
        $\letpack{x : \tTree@\tLa}{\datacon{\tLeaf}{\tLa}{i_1}}{}$ &
        \aLa & \aLa &
        \tSet{\tL} &
        \tSet{\tLa} \\
      $5$ &
        $\letloc{\tLb}{\ngAfterTy{\tTree@\tLa}}{}$ &
        \aLb & \aLb &
        \tSet{\tL,\, \tLb,\, \tLbq} &
        \tSet{\tLa} \\
      $6$ &
        $\letpack{i_2 : \Int@\tLbq}{2}{}$ &
        \aLb & \aLb &
        \tSet{\tL,\, \tLb} &
        \tSet{\tLa,\, \tLbq} \\
      $7$ &
        $\letloc{\tLtwo}{\ngAdvField{\tLb}{\tLeaf}{1}{\Int@\tLbq}}{}$ &
        \aLtwo & \aLb &
        \tSet{\tL,\, \tLb,\, \tLtwo} &
        \tSet{\tLa,\, \tLbq} \\
      $8$ &
        $\letpack{y : \tTree@\tLb}{\datacon{\tLeaf}{\tLb}{i_2}}{}$ &
        \aLb & \aLb &
        \tSet{\tL} &
        \tSet{\tLa,\, \tLb} \\
      $9$ &
        $\datacon{\tNode}{\tL}{(x,\, y)}$ &
        \aL & \aL &
        \tNone &
        \tSet{\tL} \\
    \end{tabular}
  \end{adjustbox}
  \endgroup
\end{table*}

Section~\ref{sec:lang} types \lstinline|buildList|, whose values have one self-recursive field.
\lstinline|buildTree| allocates a perfectly balanced \lstinline|Tree|, whose \lstinline|Node| has two, and so also exercises $\gramkw{after}$:
\begin{code}
buildtree : Int -> Tree
buildtree n = if n <= 0 then Leaf 1 else Node (buildtree (n - 1)) (buildtree (n - 1))
\end{code}
\figref{fig:buildtree-socal} shows it in \ourcalc{}, compiled for the \fullyfactored{} layout.
The tree occupies two positions in $\ngReg$: $\ngReg$ itself, whose buffer holds the tags, and $\ngReg.\ngFieldKey{\mathsf{Leaf}}{1}$, whose buffer holds the integers of \lstinline|Leaf| nodes.
In the \lstinline|Leaf| branch, the program writes the integer at the factored position, steps past it with $\gramkw{advanceField}$, and writes the \lstinline|Leaf| tag.
In the \lstinline|Node| branch, $\gramkw{advanceTags}$ reserves the cell for the \lstinline|Node| tag and binds the location just past it, $\ngLoc_a^{\ngReg}$, where the left subtree is built in place.
The right subtree starts at $\ngLoc_b^{\ngReg}$, which $\gramkw{after}$ places past the end of the left subtree in every buffer it occupies.
Finally, the program writes the \lstinline|Node| tag at $\ngLoc^{\ngReg}$, completing the node over its two children.

Table~\ref{fig:types-example} type checks one unfolding of this recursion, a straight-line program that builds $\mathsf{Node}\;\allowbreak(\mathsf{Leaf}\;1)\;\allowbreak(\mathsf{Leaf}\;2)$ at a location $\ngLoc^{\ngReg}$ that a $\gramkw{start}$ has begun.
The $\gramkw{letloc}$s record
\[
  \begin{array}{@{}r@{\;}l@{}}
    \CENV = \{ & \ngLoc^{\ngReg} \mapsto \ngStartRegTy{\ngReg}{\mathsf{Tree}},\;
                 \ngLoc_a^{\ngReg} \mapsto \ngAdvTags{\ngLoc^{\ngReg}}{\mathsf{Node}}, \\
               & \ngLoc_1^{\ngReg} \mapsto \ngAdvField{\ngLoc_a^{\ngReg}}{\mathsf{Leaf}}{1}{\Int@\ngLoc_a^{\ngReg.q}}, \\
               & \ngLoc_b^{\ngReg} \mapsto \ngAfterTy{\mathsf{Tree}@\ngLoc_a^{\ngReg}}, \\
               & \ngLoc_2^{\ngReg} \mapsto \ngAdvField{\ngLoc_b^{\ngReg}}{\mathsf{Leaf}}{1}{\Int@\ngLoc_b^{\ngReg.q}}\;\} .
  \end{array}
\]
Steps~1 and~5 bind an atom that occupies both positions of the new subtree (\textsc{T-LetLoc-Advance-Tags} and \textsc{T-LetLoc-After}), so each adds two locations to $\NENV$ and moves $\AENV$ at both $\ngEmptyPath$ and $q$; step~1 also drops $\ngLoc^{\ngReg.q}$, since a \lstinline|Node| has no field at $q$.
Step~3, by contrast, adds a single location and moves $\AENV$ at $\ngEmptyPath$ only (\textsc{T-LetLoc-Advance-Field}).
Writing a payload (steps~2 and~6) moves its location from $\NENV$ to $\UENV$.
Each constructor write (steps~4, 8 and~9, \textsc{T-DataConstructor}) removes its destination and chain from $\NENV$, claims its fields from $\UENV$, adds the destination to $\UENV$, and resets $\AENV$ to the destination's atom.
The run ends with $\NENV = \emptyset$, $\UENV = \{\ngLoc^{\ngReg}\}$ and $\AENV$ back at the constant family it started with, the state in which a program's main expression must end.

\section{Well-Formedness}
\label{sec:well-formedness}

Type safety (Section~\ref{sec:type-safety}) is stated for well-formed
configurations (Definition~\ref{def:cfgwf}). This appendix builds that notion up from its parts: auxiliary definitions, the allocation front and the write log; well-formedness of values, of location constraints and of the allocation state; the statements these definitions cite; well-formedness of the store and the store view; and finally well-formedness of configurations and of a program's initial configuration, with the step renaming and multistep reduction that the theorems use. Every definition is the formalism's own. The remaining definitions that the others cite, which the paper does not otherwise need, follow at the end.

\begingroup
\small
\setlength{\emergencystretch}{3em}

\subsection{Auxiliary definitions}
\label{sec:wf-auxiliaries}
\label{sec:store-model-resolution}

\begin{definition}[Constructor and field lookup]
\label{def:ctors-typeoffield}
$\ctors{\TYP}$ is the set of pairs $(\DC\,\overharpoon{\TYP'})$
declared for $\TYP$, and $\typeoffield(\DC,j) = \TYP'_j$ where
$(\DC\,\overharpoon{\TYP'}) \in \ctors{\TYP}$ for the unique $\TYP$
declaring $\DC$.
\end{definition}
\begin{definition}[Factors-to]
\label{def:packs}
Write $\tau \packs \tau'$, read $\tau$ \emph{factors to} $\tau'$, when
some constructor of $\tau$ declares a field of type $\tau'$ that is not
$\tau$ itself. Datatype well-formedness (Definition~\ref{def:wfdecl}) requires
$\packs$ to be acyclic.
\end{definition}
\begin{definition}[Regions and location atoms]
\label{def:region-location-sorts}
A region is named by a region variable, written $\ngReg$. A
\emph{location atom}, written $\ngLoc$, is a bare name, drawn from a
countably infinite alphabet, with no structure of its own. Every
location is \emph{built on} an atom, whatever position it stands at.
In program text an atom always carries a position; the metatheory's
notation also names atoms bare, in substitutions and freshness
conditions. $\ngLoc$ and $\ngReg$ are metavariables over static names:
names that occur in program text.
\[
  \begin{array}{@{}l@{\quad}l@{\quad}l@{}}
    \ngLoc & \in \textup{Location Atoms}
      & \textit{static} \\
    \ngReg & \in \textup{Region Variables}
      & \textit{static}
  \end{array}
\]
\end{definition}
\begin{definition}[Factored positions]
\label{def:factored}
Write $\ngFactored{\TYP}$ for $\TYP$'s factored positions: the keys
$\ngFieldKey{\DC}{j}$ with $(\DC\,\overharpoon{\TYP'}) \in \ctors{\TYP}$
and $\TYP'_j \neq \TYP$.
\end{definition}
\begin{definition}[Paths of a type]
\label{def:pathsof}
Write $\ngPathsOf{\TYP}$ for the paths that lead from $\TYP$ through its factored fields: the
empty path, together with $\ngFieldKey{\DC}{j}.\ngPath$ for each
factored position $\ngFieldKey{\DC}{j}$ of $\TYP$ and each
$\ngPath \in \ngPathsOf{\typeoffield(\DC,j)}$. The set is finite
because $\packs$ is acyclic and finite (Definition~\ref{def:packs};
there are finitely many declarations).
\end{definition}
\begin{definition}[Type at a path]
\label{def:typeat}
Write $\ngTypeAt{\TYP}{\ngPath}$ for the type at path $\ngPath$ from
$\TYP$:
\[
  \ngTypeAt{\TYP}{\ngEmptyPath} = \TYP
  \qquad
  \ngTypeAt{\TYP}{\ngFieldKey{\DC}{j}.\ngPath} =
    \ngTypeAt{\typeoffield(\DC,j)}{\ngPath}
  \quad\text{$\ngFieldKey{\DC}{j}$ a factored position of $\TYP$.}
\]
The side condition is part of the equation and confines the second
clause to the factored positions of $\TYP$, so
$\ngTypeAt{\TYP}{\ngPath}$ is defined at exactly the paths of $\TYP$, $\ngPathsOf{\TYP}$. The recursion terminates because the
path is finite and each step consumes one key.
\end{definition}
\begin{definition}[Datatype-declaration well-formedness]
\label{def:wfdecl}
$\wfdecl\;\overharpoon{\DD}$ holds iff:
\begin{enumerate}
\item \textbf{Name uniqueness.} Each datatype name is declared by at
  most one declaration of $\overharpoon{\DD}$, and each constructor
  name $\DC$ appears in exactly one declaration, and exactly once
  there. So $\DC$ alone determines its datatype, its field types, and
  its arity.
\item \textbf{Type completeness.} Every datatype name appearing as a
  field type is declared by $\overharpoon{\DD}$. With item~1 this
  makes $\typeoffield(\DC,j)$ a declared type at every declared
  constructor $\DC$ and every $1 \le j \le$ the arity of $\DC$.
\item \textbf{Field ordering.} Within every constructor
  $\DC\;\TYP'_1\ldots\TYP'_\litnum$ of a datatype $\TYP$, the fields
  factoring to another datatype precede the self-recursive ones.
  Writing $m \le \litnum$ for the boundary, factored fields occupy
  $1..m$ and self-recursive fields $m{+}1..\litnum$. A declaration that
  interleaves them, such as
  $\mathtt{Node}\;\mathtt{Tree}\;\mathtt{P}\;\mathtt{Tree}$, is
  rejected; $\mathtt{Node}\;\mathtt{P}\;\mathtt{Tree}\;\mathtt{Tree}$
  is accepted.
\item \textbf{Acyclicity.} $\packs$ is acyclic.
\item \textbf{Inhabitation.} Every declaration declares at least one
  constructor, so $\ctors{\TYP}$ is non-empty at every declared $\TYP$.
\end{enumerate}
\end{definition}
\begin{definition}[Function lookup]
\label{def:function-lookup}
$\mathrm{Function}(f)$ is the scheme, parameter vector, and body that
$\overharpoon{\FD}$ declares for $f$, unique because a program
declares each function once.
\end{definition}
\begin{definition}[Region typing environment]
\label{def:region-typing-environment}
A region typing environment is given by
\[
  \begin{array}{@{}l@{\hspace{0.55em}}l@{\;\;}l@{\;\;}l@{}}
    \textup{Region Typing Environments} & \Theta & \gramdef &
      \set{\ngReg_1 \mapsto \TYP_1, \ldots, \ngReg_\litnum \mapsto \TYP_\litnum}
  \end{array}
\]
with $\litnum \ge 0$: the listing is finite, and each region name
occurs in it at most once. $\Theta$ is therefore a finite partial
function from region names to types, and $\ngDom\Theta$ is the set of
region names it lists. Write $\TYP_{\ngReg}$ for $\Theta(\ngReg)$, at
the region typing environment the surrounding statement fixes and only
where $\ngReg \in \ngDom\Theta$.
\end{definition}
\begin{definition}[Extension]
\label{def:region-typing-environment-order}
Write $\Theta \sqsubseteq \Theta'$ when $\ngDom\Theta \subseteq
\ngDom{\Theta'}$ and $\Theta'(\ngReg) = \Theta(\ngReg)$ at every
$\ngReg \in \ngDom\Theta$: $\Theta'$ agrees with $\Theta$ everywhere
$\Theta$ is defined, and may be defined at further regions besides.
\end{definition}

Figure~\ref{fig:grammar} gives layout cursors and the location map as plain
productions. In full, the layout cursors for a type $\TYP$ and the location
map are:
\[
  \begin{array}{@{}l@{}}
    \textup{For each type}\;\TYP,\;\;
    c \in \ngLayoutCursorT{\TYP} \;\;\gramdef\;\;
    \ngCurs{i}\,
      \overharpoon{(\DC,\,j,\,c_{\DC,j})}
    \\[1.2mm]
    \qquad\text{with one entry}\;
      c_{\DC,j} \in \ngLayoutCursorT{\typeoffield(\DC,j)}
      \;\text{at each}\;
      \ngFieldKey{\DC}{j} \in \ngFactored{\TYP}
      \;\text{and no other key.}
  \end{array}
\]
\[
  \MENV \;\gramdef\; \set{\ngLoc_1^{\ngPos_1} \mapsto c_1, \ldots}
  \quad\text{with}\; c_\ind \in
    \ngLayoutCursorT{\ngTypeAt{\TYP_{\ngReg_\ind}}{\ngPath_\ind}}
  \;\text{at}\; \ngPos_\ind = \ngReg_\ind.\ngPath_\ind,
\]
\[
  \text{and}\; \ngLoc_\ind \neq \ngLoc_\kappa \;\text{for}\; \ind \neq \kappa:
  \text{the keys' \emph{atoms} are pairwise distinct.}
\]
\begin{definition}[Cell address]
\label{def:cell-address}
For a position $\ngPos$ and index $i$, write $\ngCell{\ngPos}{i}$ for
the cell address the pair names: the $i$-th cell of the array at
position $\ngPos$.
\end{definition}
\begin{definition}[Cursor operations]
\label{def:cursor-ops}
\[
  \begin{array}{@{}l@{\;\;}l@{}}
    \ngZeroCursor{\TYP} & : \ngLayoutCursorT{\TYP} \\
    \ngProj{c}{\ngPath} & : \ngLayoutCursorT{\ngTypeAt{\TYP}{\ngPath}}
      \quad\text{for}\; c : \ngLayoutCursorT{\TYP},\;
        \ngPath \in \ngPathsOf{\TYP} \\
    \ngTagNext{c} & : \ngLayoutCursorT{\TYP}
      \quad\text{for}\; c : \ngLayoutCursorT{\TYP} \\
    \ngReplace{c}{\ngFieldKey{\DC}{j}}{g} & : \ngLayoutCursorT{\TYP} \\
    & \quad\text{for}\; c : \ngLayoutCursorT{\TYP},\;
        g : \ngLayoutCursorT{\typeoffield(\DC,j)},\;
        \ngFieldKey{\DC}{j} \in \ngFactored{\TYP} \\
    \ngCell{\ngPos}{c} & \text{, the cell}\;
      \ngCell{\ngPos}{c(\ngEmptyPath)}\;\text{, for a cursor read at}\;
      \ngPos
  \end{array}
\]
$\ngZeroCursor{\TYP}$ is the cursor with $0$ at every path.
$\ngTagNext{c}$ raises $c$'s own root index by one and carries every
other entry across unchanged. $\ngReplace{c}{\ngFieldKey{\DC}{j}}{g}$
is $c$ with the entry at key $\ngFieldKey{\DC}{j}$ replaced by $g$.
$\ngProj{c}{\ngPath}$ is the entry $c$'s own vector reaches by
following $\ngPath$ key by key -- each key of $\ngPath$ is one the
vector reached so far carries, which the formation of
$\ngLayoutCursorT{\cdot}$ guarantees at every
$\ngPath \in \ngPathsOf{\TYP}$. Write $c(\ngPath)$ for the root index
$\ngProj{c}{\ngPath}$ itself carries -- a natural number, not a
cursor, and so a different reading of $c$ than projection: $\ngProj
c\ngPath$ descends to a sub-cursor, while $c(\ngPath)$ reads off that
sub-cursor's own head index. The last line overloads
$\ngCell{\cdot}{\cdot}$: for a cursor $c$ read at a position $\ngPos$,
$\ngCell{\ngPos}{c}$ names the cell at $c$'s own root index, reducing
to Definition~\ref{def:cell-address}'s address at $c(\ngEmptyPath)$,
$c$'s index at the empty path.
\end{definition}
\begin{definition}[Zero cursor]
\label{def:zerocursor}
For a type $\TYP$, define $\ngZeroCursor{\TYP} \in \ngLayoutCursorT{\TYP}$
by
\[
  \ngZeroCursor{\TYP} \;=\; \ngCurs{0}\;
    \overharpoon{(\DC,\,j,\,\ngZeroCursor{\typeoffield(\DC,j)})}
\]
where the vector sends every $\ngFieldKey{\DC}{j} \in \ngFactored{\TYP}$
to $\ngZeroCursor{\typeoffield(\DC,j)}$. The recursion is the one the
formation of $\ngLayoutCursorT{\cdot}$ follows, and terminates for the
same reason: $\packs$ is acyclic and finite. The result is the cursor with $0$ at
every path of $\TYP$, so $\ngProj{\ngZeroCursor{\TYP}}{\ngPath}$ has
root index $0$ at every $\ngPath \in \ngPathsOf{\TYP}$.

A zero cursor names no position: a cursor is indices and shape alone,
and the position it is read at is supplied separately, by the location
it is bound to.
\end{definition}
\begin{definition}[Predecessor]
\label{def:poschain}
At a location-constraint environment $\CENV$, which every reading of
these relations names,
$\ngLoc'^{\ngPos}$ is a \emph{successor} of $\ngLoc^{\ngPos}$ -- and
$\ngLoc^{\ngPos}$ is a \emph{predecessor} of $\ngLoc'^{\ngPos}$ -- when
$\CENV(\ngLoc'^{\ngPos})$ is $\ngAdvTags{\ngLoc^{\ngPos}}{\DC}$, or
$\ngAfterTy{\TYP@\ngLoc^{\ngPos}}$, or
$\ngAdvField{\ngLoc^{\ngPos}}{\DC}{k}{\ngLocatedTy}$, for some $\DC$,
$k$, $\TYP$, and $\ngLocatedTy$. A \emph{position chain at $\ngPos$} is
a sequence $\ngLoc_0^{\ngPos},\ldots,\ngLoc_k^{\ngPos}$ in which each
entry is a successor of the one before it. At $k = 0$ the sequence is
the single entry $\ngLoc_0^{\ngPos}$, which is a position chain at
$\ngPos$ since it has no adjacent pair.
\end{definition}
\begin{definition}[Field chain]
\label{def:fieldchain}
Write $\ngFieldChain{\CENV}{\ngLoc_0^{\ngReg.\ngPath}}{\DC}{j}{\ngLoc'^{\ngReg.\ngPath}}$
for: $\ngLoc'^{\ngReg.\ngPath}$ is where $\DC$'s field chain from
$\ngLoc_0^{\ngReg.\ngPath}$ stands before its $j$-th field-advance
link. Precisely, this holds when there are locations
$\ngLoc_0^{\ngReg.\ngPath},\ldots,\ngLoc_{j-1}^{\ngReg.\ngPath}$ with
$\ngLoc'^{\ngReg.\ngPath} = \ngLoc_{j-1}^{\ngReg.\ngPath}$ and
\[
  \CENV(\ngLoc_k^{\ngReg.\ngPath}) \;=\;
    \ngAdvField{\ngLoc_{k-1}^{\ngReg.\ngPath}}{\DC}{k}
      {\typeoffield(\DC,k)@\ngLoc_0^{\ngReg.\ngPath.\ngFieldKey{\DC}{k}}}
\]
for every $1 \le k \le j-1$. At $j=1$ this range is empty: no $\CENV$
equation is required, the location sequence is $\ngLoc_0^{\ngReg.\ngPath}$
alone, and $\ngFieldChain{\CENV}{\ngLoc_0^{\ngReg.\ngPath}}{\DC}{1}{\ngLoc_0^{\ngReg.\ngPath}}$
holds: the chain before its first link is the starting location
itself.
\end{definition}
\begin{definition}[Field-advance links]
\label{def:fieldlinks}
Write
\[
  \ngFieldLinks{\CENV} \;=\;
    \set{\ngLoc_1^{\ngPos} \in \ngDom\CENV \;\mid\;
      \CENV(\ngLoc_1^{\ngPos})\;\text{is an}\;\gramkw{advanceField}\;\text{form}}
\]
for the locations whose defining location expression has the form
$\gramkw{advanceField}$.
\end{definition}
\begin{definition}[Chain root]
\label{def:chainroot}
$\ngChainRoot{\CENV}{\ngLoc^{\ngReg.\ngPath}}$ is defined exactly when
there is a finite sequence
$\ngLoc_0^{\ngReg.\ngPath},\ldots,\ngLoc_k^{\ngReg.\ngPath} =
\ngLoc^{\ngReg.\ngPath}$ with
\[
  \ngLoc_0^{\ngReg.\ngPath} \notin \ngFieldLinks{\CENV}
  \quad\text{(Definition~\ref{def:fieldlinks})}
  \qquad\text{and}\qquad
  \CENV(\ngLoc_i^{\ngReg.\ngPath}) =
    \ngAdvField{\ngLoc_{i-1}^{\ngReg.\ngPath}}{\DC_i}{k_i}{\ngLocatedTy_i}
\]
for every $1 \le i \le k$, for some $\DC_i$, $k_i$, and
$\ngLocatedTy_i$ at each index $i$; its value is then
$\ngLoc_0^{\ngReg.\ngPath}$.
\end{definition}
\begin{definition}[Resolving a location in $\MENV$]
\label{def:mresolve}
$\MENV(\ngLoc^{\ngReg.\ngPath})$ is defined when some
$\ngLoc^{\ngReg.\ngPath_0} \in \ngDom\MENV$ has
$\ngPath = \ngPath_0.\ngPath_1$ for some
$\ngPath_1 \in \ngPathsOf{\ngTypeAt{\TYP_{\ngReg}}{\ngPath_0}}$, and it
is then the projection
\[
  \MENV(\ngLoc^{\ngReg.\ngPath})
    \;=\; \ngProj{\MENV(\ngLoc^{\ngReg.\ngPath_0})}{\ngPath_1}
    \;\in\; \ngLayoutCursorT{\ngTypeAt{\ngTypeAt{\TYP_{\ngReg}}{\ngPath_0}}
                                {\ngPath_1}}
    \;=\; \ngLayoutCursorT{\ngTypeAt{\TYP_{\ngReg}}{\ngPath}} ,
\]
the last step by the path composition
$\ngTypeAt{\ngTypeAt{\TYP}{\ngPath_0}}{\ngPath_1} =
\ngTypeAt{\TYP}{\ngPath_0.\ngPath_1}$. The cell it
names is $\ngCell{\ngReg.\ngPath}{\cdot}$ at that projection's own
index.
\end{definition}
\begin{definition}[Allocated positions]
\label{def:allocpos}
For a location map $\MENV$, write
\[
  \ngAllocatedPositions\MENV \;=\;
    \set{\ngReg.\ngPath \;\mid\;
      \MENV\;\text{is defined at some}\;\ngLoc_x^{\ngReg.\ngPath}} ,
\]
and call its members the \emph{allocated positions} of $\MENV$.
\end{definition}

Write $\STOR : \ngWLog{\ngAllocatedPositions\MENV}$ for the concrete
runtime-store formation condition, instantiating the abstract judgment
$\STOR : \ngWLog{W}$ (Section~\ref{sec:worlds-static-envs}) at
$W = \ngAllocatedPositions\MENV$. Its first condition then makes every
event of $\STOR$ stand at a position $\MENV$ allocates, and its second,
unchanged from the abstract statement, still bars two events from ever
sharing one cell address.

\subsection{The allocation front}
\label{sec:wf-front}

\begin{definition}[The allocation-frontier cursor of a location]
\label{def:front}
At a store $\STOR$, a location map $\MENV$ and a pending-location set
$\NENV$: for a location $\ngLoc_x$ at which $\MENV$ is defined, standing at
$\ngReg.\ngPath$, write $\ngFront{\ngLoc_x}$ for the end cursor of
$\vwf{\STOR}{\ngReg.\ngPath}{\MENV(\ngLoc_x)}{\ngTypeAt{\TYP_{\ngReg}}{\ngPath}}{\cdot}$
when $\ngLoc_x \notin \NENV$ and such a traversal exists, and for
$\MENV(\ngLoc_x)$ otherwise.
\end{definition}

\subsection{The write log and allocation states}
\label{sec:wf-writelog}
\label{sec:worlds-static-envs}

A write event pairs a cell address with the constructor tag written there.
The store records these events in the order they occur. Stated
abstractly, for an arbitrary set $W$ of positions permitted to hold a
write:
\[
  \begin{array}{@{}l@{\hspace{0.55em}}l@{\;\;}l@{\;\;}l@{}}
    \textup{Write Events} & w & \gramdef &
      (\ngCell{\ngPos}{i},\DC) \\
    \textup{Stores} & \STOR & \gramdef &
      \varepsilon \gramor \STOR \cdot w
  \end{array}
\]
$\ngSnoc{\STOR}{a}{\DC}$ is the second production, appending one event
-- address $a$, tag $\DC$ -- to the end of $\STOR$. Write
$\ngAddrs{\STOR}$ for the set of addresses $\STOR$ already holds,
defined on that grammar by
\[
  \ngAddrs{\varepsilon} = \emptyset ,
  \qquad
  \ngAddrs{\ngSnoc{\STOR}{\ngCell{\ngPos}{i}}{\DC}}
    = \ngAddrs{\STOR} \cup \set{\ngCell{\ngPos}{i}} .
\]
Write $\STOR : \ngWLog{W}$ when both of the following hold.
\begin{enumerate}
  \item\label{wlog:positions} \emph{Positions are permitted.} Every
    event of $\STOR$ names an address $\ngCell{\ngPos}{i}$ with
    $\ngPos \in W$.
  \item\label{wlog:once} \emph{No address is written twice.} No address
    occurs in two events of $\STOR$.
\end{enumerate}

Figure~\ref{fig:grammar} writes an allocation function $\ngFam$ as a finite
map. In full, an allocation state is
\[
  \ngAP \;\gramdef\; \phi \gramor \ngFam
  \quad\text{with}\; \ngFam\;\text{a partial map from}\; \ngPathsOf{\TYP_{\ngReg}}
  \;\text{to}\; \ngAtom,\;\text{at the key}\; \ngReg
\]
For each region, $\AENV$ records either that the region has not yet
been started ($\phi$) or, once it has, a family $\ngFam$ that records
an atom at the paths of the region's type. An allocation
state carrying $\phi$ is called \emph{unstarted} and one carrying a
family \emph{started}.

Three conventions fix this representation. First,
$\phi$ is a distinguished marker, distinct from every allocation
function, every atom and every location, and it carries no location
atom and no region name.
Second, $\ngFam$ has values in $\ngAtom$, so $\phi$ is never one of its
results. In the input allocation environment of a derivation
witnessing configuration well-formedness (Definition~\ref{def:cfgwf})
$\ngFam$ is total on $\ngPathsOf{\TYP_{\ngReg}}$: a region is started
or it is not, and never partly. This totality is a formation condition
that definition asks of its witness, not a condition of every
typing judgment: a location substitution
(Definition~\ref{def:locsubst}) moves a family onto the paths below a
non-root position of another region, and, where that region had no
family of its own, the family is defined only on those paths. An
allocation environment is \emph{of its sort} at $\Theta$ when every
family stands at a region of $\ngDom\Theta$ and is total on
$\ngPathsOf{\TYP_{\ngReg}}$; the input allocation environment of a
derivation witnessing configuration well-formedness is of its sort
(Definition~\ref{def:cfgwf}); that of an arbitrary judgment need not be. Third, $\ngReg \notin \ngDom\AENV$ is a
third state, distinct from both: no entry stands at $\ngReg$ at all, and
the readings below are undefined there rather than $\phi$.
\begin{definition}[Reading $\AENV$]
\label{def:aenv-core}
For $\ngReg \in \ngDom\AENV$ and $\ngPath \in
\ngPathsOf{\TYP_{\ngReg}}$, write $\ngEntry{\AENV}{\ngReg.\ngPath}$, or
just $\AENV(\ngReg.\ngPath)$, for
\[
  \ngEntry{\AENV}{\ngReg.\ngPath} \;=\;
  \begin{cases}
    \phi & \text{if}\; \AENV(\ngReg) = \phi, \\
    \ngFam(\ngPath)^{\ngReg.\ngPath}
      & \text{if}\; \AENV(\ngReg) = \ngFam .
  \end{cases}
\]
The two cases are exclusive and exhaustive, $\phi$ being no allocation
function. At a $\phi$ entry the reading marks that no atom has been
recorded at $\ngReg.\ngPath$ yet, precisely because $\ngReg$ is
unstarted. The reading is partial: it is undefined at a region outside
$\ngDom\AENV$, at a path outside $\ngPathsOf{\TYP_{\ngReg}}$, and at a
path where a family that is not total (Section~\ref{sec:worlds-static-envs})
is undefined. An
equation $\AENV(\ngReg.\ngPath) = X$ written as a premise of a rule or
as a conjunct of a clause or a hypothesis asserts both that the left
side is defined and that it is $X$; this convention is used throughout
and is not repeated at each site, as for
$\ngChainRoot{\cdot}{\cdot}$ (Definition~\ref{def:chainroot}).
\end{definition}
A bare region and a position read $\AENV$ at two different arguments,
and the answers lie in two different sorts. At a bare region,
$\AENV(\ngReg)$ is the value the map itself holds there: $\phi$ or an
allocation function, a value of sort $\ngAFam{\ngReg}$. At a position,
$\AENV(\ngReg.\ngPath)$ is the observation above: $\phi$ or a location.
$\phi$ is the only value the two sorts share, an allocation function
being no location, so which reading is meant is fixed by the argument
--- a region, or a region and a path together --- and never by the
answer. The
root observation is accordingly written $\AENV(\ngReg.\ngEmptyPath)$ and
never $\AENV(\ngReg)$: they are two different lookups, and where
$\ngReg$ is started they do not even agree in sort.

Two allocation environments are accordingly compared at a position by
their \emph{observations}, and $\phi$ is one of the two things an
observation can be. Every \emph{agreement} condition this document
states between two allocation environments at a position asks that both
observations be defined there and be the same one --- either both
$\phi$, or one and the same location --- and is therefore met at a
position both read as $\phi$. The reading also
determines the map value it came from. At a $\ngReg \in \ngDom\AENV$
and a $\ngPath \in \ngPathsOf{\TYP_{\ngReg}}$ --- the arguments at which
the equation above answers at all --- $\AENV(\ngReg.\ngPath) = \phi$
exactly when $\AENV(\ngReg) = \phi$, and $\AENV(\ngReg.\ngPath)$ is a
location exactly when $\AENV(\ngReg) = \ngFam$ for some allocation
function $\ngFam$, the two cases of the equation above being exhaustive
and $\phi$ distinct from every location.

Write
$\ngRegionsOf\AENV = \ngDom\AENV$ for the set of regions at which
$\AENV$ has an entry (for any finite map $X$ on regions, we also say
that $X$ \emph{keys} $\ngReg$ when $\ngReg \in \ngDom X$), and
\[
  \ngPositionsOf\AENV \;=\;
    \set{\ngReg.\ngPath \;\mid\; \ngReg \in \ngRegionsOf\AENV,\;
      \ngPath \in \ngPathsOf{\TYP_{\ngReg}}} .
\]
for the set of positions the observation answers at.

The rules write $\AENV$ only through five operations.
\begin{enumerate}
\item \emph{Region insertion.} $\AENV[\ngReg \mapsto \phi]$, at
  a $\ngReg \notin \ngRegionsOf\AENV$.
\item \emph{Region removal.} $\AENV \setminus \ngReg$, dropping one key.
\item \emph{Start.} $\AENV[\ngReg \mapsto \ngConstFam\ngLoc]$,
  where $\ngConstFam\ngLoc$ is the constant family
  $\ngConstFam{\ngLoc}(\ngPath) = \ngLoc$ at every
  $\ngPath \in \ngPathsOf{\TYP_{\ngReg}}$.
\item \emph{Point update.} At an allocation state that is an allocation
  function $\ngFam$, and a $\ngPath \in \ngPathsOf{\TYP_{\ngReg}}$,
  $\ngFam[\ngPath \mapsto \ngLoc]$ is ordinary function update: the
  family sending $\ngPath$ to $\ngLoc$ and agreeing with $\ngFam$ at
  every other path of $\ngPathsOf{\TYP_{\ngReg}}$ at which $\ngFam$ is
  defined. It is total there whenever $\ngFam$ is, and the result is
  again an allocation state. The operation is
  \emph{undefined} at $\phi$: an unstarted region carries no family to
  update.
\item \emph{Descendant-set update.} For $\ngReg \in \ngDom\Theta$ and
  $\ngPath \in \ngPathsOf{\TYP_{\ngReg}}$, write
  \[
    \ngDesc{\ngReg}{\ngPath} \;=\;
      \set{\ngPath.\ngPath_c \;\mid\; \ngPath_c \in
        \ngPathsOf{\ngTypeAt{\TYP_{\ngReg}}{\ngPath}}}
  \]
  for the \emph{descendant set} of $\ngPath$: the paths $\ngPath$
  extends by any further path, including $\ngPath$ itself (taking
  $\ngPath_c = \ngEmptyPath$). Then, at an allocation state that is an
  allocation function $\ngFam$,
  \[
    \ngDescOverride{\ngFam}{\ngReg}{\ngPath}{\ngLoc}
      \;=\; \ngFam'
    \quad\text{where}\quad
    \ngFam'(\ngPath') =
    \begin{cases}
      \ngLoc & \text{if}\; \ngPath' \in \ngDesc{\ngReg}{\ngPath}, \\
      \ngFam(\ngPath') & \text{otherwise,}
    \end{cases}
  \]
  total on $\ngPathsOf{\TYP_{\ngReg}}$ whenever $\ngFam$ is. As at the point update,
  the override is undefined at $\phi$.
\end{enumerate}
\begin{definition}[Predecessor set]
\label{def:sources}
Each location expression names at most one location it is defined
from: a $\gramkw{start}$ form names none; $\ngAdvTags{\ngLoc'^{\ngPos}}{\DC}$,
$\ngAfterTy{\TYP@\ngLoc'^{\ngPos}}$, and
$\ngAdvField{\ngLoc'^{\ngPos}}{\DC}{j}{\ngLocatedTy}$ each name
$\ngLoc'^{\ngPos}$. Write
\[
  \ngSources{\CENV} \;=\;
    \set{\ngLoc'^{\ngPos} \mid \text{some}\; \CENV(\ngLoc_1^{\ngPos_1})
      \;\text{names}\; \ngLoc'^{\ngPos}\;\text{this way}}
\]
for the \emph{predecessor set} of $\CENV$.
\end{definition}

\subsection{Value well-formedness and derived views}
\label{sec:vwf-derived-views}

Section~\ref{sec:worlds-static-envs} defined the store as an
append-only log of address-tag events. This section defines what it
means for a stretch of that log, starting at one cursor, to hold a
well-formed serialized value of a given type, and the cursor
immediately past it. The value well-formedness judgment (VWF) describes
serialized-value traversal. A derivation of this judgment is called a
\emph{traversal}.

Write $\vwf{\STOR}{\ngPos}{c}{\TYP}{c'}$ to mean that $\STOR$ holds a
value of type $\TYP$ at the cursor $c$, read at the position $\ngPos$,
with end cursor $c'$. Both $c$ and $c'$ are members of
$\ngLayoutCursorT\TYP$, the judgment's own type index. The position is
a second index, not a component of $c$: a cursor carries indices and
cursor shape alone, so the judgment must separately be told which
position's arrays its root index reads into. The relation is partial:
a store, position, and cursor need not admit any derivation at all,
and its one rule, \ngRuleName{VWF-Data}, states exactly when one exists.

\subsection{Location-constraint realization}
\label{sec:wf-creal}

\begin{definition}[Location-constraint realization]
\label{def:creal}
$\crealj{\STOR}{\CENV}{\MENV}$ holds when, for every
$\ngLoc^{\ngReg.\ngPath} \in \ngDom\CENV$, $\MENV$ is defined at
$\ngLoc^{\ngReg.\ngPath}$, is defined at every argument location the
entry names, and the entry $\CENV(\ngLoc^{\ngReg.\ngPath})$ and the
binding $\MENV(\ngLoc^{\ngReg.\ngPath})$ agree as follows.
Here an argument location is the predecessor named by an
$\gramkw{advanceTags}$, $\gramkw{after}$ or $\gramkw{advanceField}$
entry. In the last form, the located type is read for its type alone:
its recorded location $\ngLoc_0^{\ngPos_0}$ need not be resolved by
$\MENV$, and realization imposes no condition on $\ngPos_0$.
\[
  \begin{array}{@{}l@{}}
    \CENV(\ngLoc^{\ngReg.\ngPath}) = \ngStartRegTy{\ngReg}{\TYP}: \quad
      \text{every cursor of}\; \MENV(\ngLoc^{\ngReg.\ngPath}) \;\text{is at index}\; 0. \\[2mm]
    \CENV(\ngLoc^{\ngReg.\ngPath}) = \ngAdvTags{\ngLoc'^{\ngReg.\ngPath}}{\DC}: \quad
      \MENV(\ngLoc^{\ngReg.\ngPath}) = \ngTagNext{\MENV(\ngLoc'^{\ngReg.\ngPath})}. \\[2mm]
    \CENV(\ngLoc^{\ngReg.\ngPath}) = \ngAfterTy{\TYP'@\ngLoc'^{\ngReg.\ngPath}}: \quad
      \vwf{\STOR}{\ngReg.\ngPath}{\MENV(\ngLoc'^{\ngReg.\ngPath})}{\TYP'}{\MENV(\ngLoc^{\ngReg.\ngPath})}. \\[2mm]
    \CENV(\ngLoc^{\ngReg.\ngPath}) = \ngAdvField{\ngLoc'^{\ngReg.\ngPath}}{\DC}{j}
        {\TYP'@\ngLoc_0^{\ngPos_0}}: \\
    \qquad \text{with}\; c = \MENV(\ngLoc'^{\ngReg.\ngPath}),\;\exists g.\;
          \vwf{\STOR}{\ngReg.\ngPath.\ngFieldKey{\DC}{j}}{c(\DC,j)}{\TYP'}{g}
      \;\text{and}\; \MENV(\ngLoc^{\ngReg.\ngPath}) = \ngReplace{c}{\ngFieldKey{\DC}{j}}{g}.
  \end{array}
\]
\end{definition}

\subsection{Well-formedness of allocation}
\label{sec:wf-allocwf}

\begin{definition}[Allocation well-formedness]
\label{def:allocwf}
$\allocwfj{\STOR}{\MENV}{\AENV}{\NENV}$ holds when item~\ref{awf:unstarted}
holds at every $\ngReg \in \ngRegionsOf\AENV$ and item~\ref{awf:entry}
holds at every $\ngReg.\ngPath \in \ngPositionsOf\AENV$
(Definition~\ref{def:aenv-core}):
\begin{enumerate}[label=\arabic*., ref=\arabic*]
\item\label{awf:unstarted} \emph{Unstarted region.} If $\AENV(\ngReg) =
  \ngUnstarted$, $\MENV$ resolves no location at any position
  $\ngReg.\ngPath$ with $\ngPath \in \ngPathsOf{\TYP_{\ngReg}}$, the
  positions that $\ngReg$'s datatype $\TYP_{\ngReg}$ determines:
  \[
    \nexists\, \ngLoc_x,\, \ngPath \in \ngPathsOf{\TYP_{\ngReg}}.\;\;
      \MENV\;\text{defined at}\; \ngLoc_x^{\ngReg.\ngPath} .
  \]
\item\label{awf:entry} \emph{Entry.} If $\AENV(\ngReg.\ngPath) = E$ is
  a location, then $\MENV$ resolves $E$. By
  Definition~\ref{def:aenv-core} that observation is
  $\ngFam(\ngPath)^{\ngReg.\ngPath}$, so $E$ already carries the
  position $\ngReg.\ngPath$ and is written here without a further one:
  \[
    \AENV(\ngReg.\ngPath) = E \;\implies\; \MENV\;\text{defined at}\;
      E ,
  \]
  and, writing $E$ for this entry throughout:
  \begin{enumerate}[label=(\alph*), ref=\theenumi(\alph*)]
    \item\label{awf:entry-written} \emph{Completed-or-unwritten entry.}
      If $E \notin \NENV$, then either a serialized-value traversal
      from $\MENV(E)$ exists at the position's own type, or no such
      traversal exists and the cell $\MENV(E)$ names is unwritten:
      \[
        E \notin \NENV \implies
        \Big(\exists c.\; \vwf{\STOR}{\ngReg.\ngPath}{\MENV(E)}
          {\ngTypeAt{\TYP_{\ngReg}}{\ngPath}}{c}\Big)
        \;\lor\;
        \big(\MENV(E)(\ngEmptyPath) \notin \ngDom{\STOR(\ngReg.\ngPath)}\big) .
      \]
    \item\label{awf:above-mark} \emph{Pending entry.} If $E \in \NENV$,
      the cell $\MENV(E)$ names has index strictly greater than
      $\ngMaxIdx{\ngReg.\ngPath}{\STOR}$:
      \[
        E \in \NENV \implies
        \MENV(E)(\ngEmptyPath) > \ngMaxIdx{\ngReg.\ngPath}{\STOR} .
      \]
  \end{enumerate}
\end{enumerate}
\end{definition}

\subsection{Statements the well-formedness definitions cite}
\label{sec:wf-cited}

\begin{definition}[Derivations and their parts]
\label{def:derivation}
\ngBind{\ngTyCtx}{\TENV}{\SENV}{\CENV}%
\ngBind{\ngAlCtx}{\AENV}{\NENV}{\UENV}%
\ngBind{\ngAlCtx'}{\AENV'}{\NENV'}{\UENV'}%
A \emph{derivation} of a judgment of this or any other form defined
below is a finite tree of rule instances: each premise of a rule
instance that is itself a judgment is the conclusion of one child of
that instance, and an instance with no such premises is a \emph{leaf}.
Read off that tree:
\begin{enumerate}
  \item\label{der:root} the \emph{root} of $D$ is its topmost rule
    instance, and the judgment that instance concludes is the one $D$
    derives;
  \item\label{der:node} a \emph{node} of $D$ is one particular
    occurrence of a rule instance in that tree --- a position in the
    tree, not a judgment: two nodes may conclude syntactically equal
    judgments and remain different nodes;
  \item\label{der:sub} the \emph{subderivation rooted at} a node $J$ of
    $D$ is the derivation consisting of that occurrence together with
    every node below it, and it derives $J$'s own conclusion;
  \item\label{der:derives} \emph{$D$ derives $J$} means that $D$'s root
    conclusion is $J$, and \emph{$J$ is derivable} means that some
    derivation derives $J$;
  \item\label{der:io} for a node concluding an expression judgment
    $\ngJT{\Theta}{\ngTyCtx}{\ngAlCtx}{\EXPR : \ngLocatedTy}{\ngAlCtx'}$, where
    $\ngShow{\ngTyCtx}$, $\ngShow{\ngAlCtx}$ and $\ngShow{\ngAlCtx'}$, its
    \emph{input components} are exactly $\TENV,\SENV,\CENV,\AENV,\NENV,\UENV$, its
    \emph{output components} exactly $\AENV',\NENV',\UENV'$, and its
    \emph{subject} is $\EXPR$; $\Theta$ is a parameter of the judgment,
    neither an input nor an output component;
  \item\label{der:theta} a node's \emph{region typing environment} is
    the region typing environment of the judgment it concludes, the
    $\Theta$ written first in it: the root's is that of the judgment
    $D$ derives, and a premise's is the one its rule writes in that
    premise;
  \item\label{der:whole} each of items~\ref{der:node} to~\ref{der:theta},
    said of a derivation rather than of a node, is said of its root
    node.
\end{enumerate}
\end{definition}
A judgment may have more than one derivation: the rules leave data
existentially chosen at several premises, and two proof trees choosing
differently derive the same judgment. So \emph{a} derivation of a
judgment is written wherever only derivability is known, and a witness
is bound by name wherever one is carried; \emph{the} derivation is
written only where a preceding statement has fixed which one is meant.
Unqualified, \emph{derivation} below means a derivation of static
typing, built from the rules of this section. The serialized-value
traversal judgment of Section~\ref{sec:vwf-derived-views} also has
derivations, which the document calls \emph{traversals}.

The typing context's $\TENV$, $\SENV$, and $\CENV$ are read, and a
premise may extend them for its own sub-derivation, but the judgment's
conclusion never reports a new value for any of the three: they are
lexical, not threaded. $\AENV$, $\NENV$, and $\UENV$ are this judgment's
input allocation context; $\AENV'$, $\NENV'$, and $\UENV'$ are its output
allocation context. The output triple is a static postcondition of
$\EXPR$, stated for the regions visible at the judgment: its intended
reading is the allocation state of those regions should evaluation of
$\EXPR$ reach a value, and it is connected to runs only through
configuration well-formedness and Preservation
(Sections~\ref{sec:config-wf} and~\ref{sec:preservation}). It is not a
description of the whole runtime allocation state, and it asserts
nothing about whether evaluation terminates. A
region $\EXPR$ binds itself is not visible at the judgment and does not
occur in $\AENV'$ or $\UENV'$ ($\textsc{T-LetRegion}$), although its
runtime region and storage remain allocated once its binder has been
reduced (Section~\ref{sec:dynamic-semantics}).
This judgment mentions neither $\MENV$ nor $\STOR$:
membership in $\SENV$ or $\UENV$ is a static fact this judgment reads
or produces, never evidence that a store cell, cursor, or write
already exists.

Every judgment of this section takes a region typing environment
$\Theta$ (Definition~\ref{def:region-typing-environment}) as its first
parameter, and in a rule $\TYP_{\ngReg}$ abbreviates $\Theta(\ngReg)$
at the $\Theta$ of the rule's conclusion. As there, $\TYP_{\ngReg}$
denotes a type only where $\ngReg \in \ngDom\Theta$, so a rule that
writes $\TYP_{\ngReg}$, in a premise or in its where-block, applies
only where $\ngReg \in \ngDom\Theta$. Every rule but one gives each of
its premises the $\Theta$ of its conclusion.
$\textsc{T-LetRegion}$ is the exception: its body premise takes its
conclusion's $\Theta$ extended by the bound region at its annotation
$\ngAnn$, the bound region not being in the conclusion's $\Theta$. So
each node of a derivation carries a $\Theta$ of its own
(item~\ref{der:theta}): a premise's $\Theta$ extends its conclusion's
(Definition~\ref{def:region-typing-environment-order}), it equals it
except at a $\textsc{T-LetRegion}$ node, and a node's $\Theta$ is the
root's extended by the regions of the $\gramkw{letregion}$ binders
enclosing it. A derivation is \emph{read at} $\Theta$ when its root's
region typing environment is $\Theta$, that is, when the judgment it
derives is written with $\Theta$. No typing rule removes an entry of
$\Theta$ or changes one.

Three further judgment forms appear, each fully defined at the one
rule that introduces it and named here only for their input and
output shape, alongside the expression judgment above. Each takes a
region typing environment $\Theta$ as its first parameter, as the
expression judgment does:
\[
  \ngBind{\ngTyCtx}{\TENV}{\SENV}{\CENV}
  \ngBind{\ngAlCtx}{\AENV}{\NENV}{\UENV}
  \ngBind{\ngAlCtx'}{\AENV'}{\NENV'}{\UENV'}
  \begin{array}{@{}l@{\quad}l@{\quad}l@{}}
    \textup{Judgment} & \textup{Reads} & \textup{Produces} \\
    \ngJPT{\Theta}{\TYP''@\ngLoc_s}{\ngTyCtx}{\ngAlCtx}{\pat : \ngLocatedTy}{\ngAlCtx'}
      & \Theta,\ngTyCtx,\ngAlCtx\;\text{and the scrutinee's located type}\;\TYP''@\ngLoc_s
      & \ngAlCtx' \\
    \ngJFunT{\Theta}{\FD} & \Theta\;\text{only: checks one closed declaration} & \text{nothing} \\
    \ngJProgT{\Theta}{top : \ngLocatedTy}{\ngAlCtx'}
      & \Theta\;\text{only: builds its own initial contexts}
      & \ngAlCtx'
  \end{array}
\]
The function-declaration judgment $\tcfun$ and the whole-program
judgment $\vdash_{\mathit{prog}}$ each build their own initial typing
and allocation contexts, $\ngTyCtx_0$ and $\ngAlCtx_0$, from scratch.
Neither builds a region typing environment: each is read at the
$\Theta$ it is given and types its body or main expression at that
same $\Theta$. A program's own initial region typing environment is
constructed outside the judgment, from the program and its bootstrap
data (Definition~\ref{def:program-region-typing-environment}), and
admission supplies it as the $\Theta$ of $\vdash_{\mathit{prog}}$
(Definition~\ref{def:admitted}).
$\tcfun$ checks one declaration and produces no output allocation
context; $\vdash_{\mathit{prog}}$ produces one, since a program's
main expression is typed by the expression judgment above and inherits
its output.
\begin{lemma}[Functionality and extension stability of value traversal]
\label{lem:vwf-fun}
Let $\wfdecl\;\overharpoon{\DD}$ hold.
Let $\STOR$ be any store.
\begin{enumerate}
  \item \emph{Functionality.} If $\STOR$ satisfies
    condition~\ref{wlog:once} (Definition~\ref{def:storeview}),
    $\vwf{\STOR}{\ngPos}{c}{\TYP}{c_1}$ and
    $\vwf{\STOR}{\ngPos}{c}{\TYP}{c_2}$, then $c_1 = c_2$; moreover a
    derivable instance of the judgment has exactly one derivation.
  \item \emph{Stability under store extension.} If
    $\vwf{\STOR}{\ngPos}{c}{\TYP}{c'}$, $\STOR'$ extends $\STOR$
    only at addresses $\STOR$ leaves unwritten, and $\STOR'$ satisfies
    condition~\ref{wlog:once}, then
    $\vwf{\STOR'}{\ngPos}{c}{\TYP}{c'}$.
\end{enumerate}
\end{lemma}
\begin{definition}[The read set of a traversal]
\label{def:readset}
For a derivation $D$ of $\vwf{\STOR}{\ngPos}{c}{\TYP}{c'}$, define
$\ngFootprint{D}$, its \emph{read set}, by: the root use of VWF-Data
contributes the one cell its own tag-read premise names, and $D$'s
read set is that cell together with the ordinary union of the read
sets of its $m$ factored subderivations and its $p$ self-recursive
ones. A cursor component that a use of VWF-Data only carries across
unchanged -- never reading a tag from it -- contributes nothing to the
read set.
\end{definition}
\begin{lemma}[Constraint realization makes predecessor roots well founded]
\label{lem:creal-wellfounded}
Let $\STOR$ satisfy condition~\ref{wlog:once} (Definition~\ref{def:storeview}), $\STOR$ being the store read below.
Let $\crealj{\STOR}{\CENV}{\MENV}$ hold. For a position
$\ngPos = \ngReg.\ngPath$ and a location $L^{\ngPos}$ at which
$\MENV$ is defined, write
\[
  \mu_{\ngPos}(L) \;=\;
    \sum_{\alpha \in \ngPathsOf{\ngTypeAt{\TYP_{\ngReg}}{\ngPath}}}
      \MENV(L^{\ngPos})(\alpha) ,
\]
a natural number. Then:
\begin{enumerate}
  \item\label{cwf:rank} if $\ngLoc'^{\ngPos}$ is a successor of
    $\ngLoc^{\ngPos}$ (Definition~\ref{def:poschain}) by an entry
    $\CENV$ records, then
    $\mu_{\ngPos}(\ngLoc) < \mu_{\ngPos}(\ngLoc')$;
  \item\label{cwf:acyclic} the predecessor edges $\CENV$ records are
    accordingly acyclic at every position;
  \item\label{cwf:total} $\ngChainRoot{\CENV}{L^{\ngPos}}$ is
    defined at every $L^{\ngPos}$ at which $\MENV$ is defined; and
  \item\label{cwf:unique} its value there is unique.
\end{enumerate}
\end{lemma}

\subsection{Store well-formedness}
\label{sec:wf-storewf}

\begin{definition}[Store well-formedness]
\label{def:storewf}
Two separate prerequisites govern this definition, and neither implies
the other.

\emph{Formation of the store.} The lookup $\STOR(\ngPos)(i)$, and every
traversal built from it, is defined wherever $\STOR$ satisfies
condition~\ref{wlog:once}. The written-index sets
$\ngDom{\STOR(\ngPos)}$ ask nothing. The remaining notation of
Definition~\ref{def:storeview} is defined only on a write log at
$\ngAllocatedPositions\MENV$, which implies that condition. Accordingly,
clauses~\ref{swf:complete}, \ref{swf:constraints},
\ref{swf:claims-walk}, \ref{swf:claims-disjoint}
and~\ref{swf:member-front}, and the three definitions nested in
clause~\ref{swf:claims}, are read where $\STOR$ satisfies
condition~\ref{wlog:once}. Clauses~\ref{swf:nursery}, \ref{swf:chains},
\ref{swf:entry-max}, \ref{swf:straddler} and~\ref{swf:nursery-cross}
ask no formation of the store. Clauses~\ref{swf:allocation},
\ref{swf:claims-cover}, \ref{swf:node-order} and~\ref{swf:cursor-cut},
clause~\ref{swf:claims} as a whole, and the judgment as a whole, are
read at a write log at $\ngAllocatedPositions\MENV$.

\emph{Well-formed declarations.} Clauses~\ref{swf:claims-cover},
\ref{swf:claims-disjoint}, \ref{swf:node-order}, \ref{swf:cursor-cut}
and~\ref{swf:member-front}, and the three definitions nested in
clause~\ref{swf:claims}, read end cursors, member read sets,
member-boundary cursors and allocation-frontier cursors
(Definition~\ref{def:front}). These are unique only under
$\wfdecl\;\overharpoon{\DD}$ (Lemma~\ref{lem:vwf-fun}(1)), so those
clauses, and the judgment as a whole, are read at declarations
satisfying it.

$\storewfj{\SENV}{\CENV}{\AENV}{\NENV}{\UENV}{\MENV}{\STOR}$ holds when
all of the following do.
\begin{enumerate}[label=\arabic*., ref=\arabic*]

\item\label{swf:complete} \textbf{Complete-location consistency.} Every
location in $\SENV$ resolves in $\MENV$ and begins a serialized-value
traversal at the type its position determines.
\[
  \forall\, \ngLoc^{\ngReg.\ngPath} \in \SENV.\;\;
    \MENV\;\text{defined at}\;\ngLoc^{\ngReg.\ngPath}
    \;\land\;
    \exists c.\; \vwf{\STOR}{\ngReg.\ngPath}{\MENV(\ngLoc^{\ngReg.\ngPath})}
      {\ngTypeAt{\TYP_{\ngReg}}{\ngPath}}{c} .
\]

\item\label{swf:constraints} \textbf{Location constraints realized.}
$\CENV$'s record of how each location was introduced agrees with
$\MENV$ and $\STOR$ (Definition~\ref{def:creal}).
\[
  \crealj{\STOR}{\CENV}{\MENV} .
\]

\item\label{swf:allocation} \textbf{Allocation state consistent.}
$\AENV$'s entries agree with $\MENV$ and $\STOR$
(Definition~\ref{def:allocwf}).
\[
  \allocwfj{\STOR}{\MENV}{\AENV}{\NENV} .
\]

\item\label{swf:nursery} \textbf{Pending locations name unwritten cells.}
Every pending location resolves, at a cell $\STOR$ has not yet
written.
\[
  \forall\, \ngLoc^{\ngReg.\ngPath} \in \NENV.\;
    \exists c.\; \MENV(\ngLoc^{\ngReg.\ngPath}) = c
    \;\land\; c(\ngEmptyPath) \notin \ngDom{\STOR(\ngReg.\ngPath)} .
\]

\item\label{swf:claims} \textbf{Consumable values account for the store.}
Every consumable value resolves at a written root cell and has a
traversal; every written cell lies under some consumable value's
traversal; distinct consumable values read disjoint cells.
\begin{enumerate}[label=(\alph*), ref=\theenumi(\alph*)]

\item\label{swf:claims-walk} \emph{Every consumable value resolves and
traverses.}
\[
  \begin{aligned}
  \forall\, \ngLoc_u^{\ngReg.\ngPath} \in \UENV.\;\;
    & \MENV\;\text{defined at}\;\ngLoc_u^{\ngReg.\ngPath}
    \;\land\;
    \MENV(\ngLoc_u^{\ngReg.\ngPath})(\ngEmptyPath) \in \ngDom{\STOR(\ngReg.\ngPath)}
    \;\land{} \\
    & \exists c_u.\; \vwf{\STOR}{\ngReg.\ngPath}{\MENV(\ngLoc_u^{\ngReg.\ngPath})}
      {\ngTypeAt{\TYP_{\ngReg}}{\ngPath}}{c_u} .
  \end{aligned}
\]
This traversal's end cursor and read set are named next
(Definitions~\ref{def:endof} and~\ref{def:memberreadset}), before the
two remaining subclauses.

\begin{definition}[Value end cursor]
\label{def:endof}
For $\ngLoc_u^{\ngReg.\ngPath} \in \UENV$, clause~\ref{swf:claims-walk}
supplies a traversal
$\vwf{\STOR}{\ngReg.\ngPath}{\MENV(\ngLoc_u)}{\ngTypeAt{\TYP_{\ngReg}}{\ngPath}}{c_u}$,
whose end cursor $c_u$ is $\ngLoc_u$'s own; write
$\ngEndOf{\ngLoc_u} = c_u$. Functionality of serialized-value traversal
(Lemma~\ref{lem:vwf-fun}(1)) makes $c_u$ unique, so $\ngEndOf{\ngLoc_u}$
is well defined.
\end{definition}

\begin{definition}[Member read set]
\label{def:memberreadset}
For $\ngLoc_u^{\ngReg.\ngPath} \in \UENV$, let $D_u$ be the unique
derivation (Lemma~\ref{lem:vwf-fun}(1)) that clause~\ref{swf:claims-walk}
supplies for $\ngLoc_u$.
Write $\ngFootprint{\ngLoc_u} = \ngFootprint{D_u}$
(Definition~\ref{def:readset}) for $\ngLoc_u$'s \emph{member read set}.
\end{definition}

\begin{definition}[Member-boundary cursor]
\label{def:memberboundary}
For an atom $\ngLoc_x$ at a position $\ngPos$, with
$\ngLoc_x^{\ngPos} \in \NENV \cup \UENV$, write $\ngMemFront{\ngLoc_x}$ for
that member's \emph{member-boundary cursor}, read at the same $\STOR$,
$\MENV$, $\NENV$ and $\UENV$ the clauses above are:
\[
  \ngMemFront{\ngLoc_x} =
  \begin{cases}
    \MENV(\ngLoc_x), & \ngLoc_x \in \NENV, \\
    \ngEndOf{\ngLoc_x}, & \ngLoc_x \in \UENV .
  \end{cases}
\]
\end{definition}
The two cases of Definition~\ref{def:memberboundary} are disjoint at
every location $\MENV$ resolves: clause~\ref{swf:nursery} makes a
pending member's root cell unwritten, and clause~\ref{swf:claims-walk}
makes a consumable member's root cell written, so no location is both.

\item\label{swf:claims-cover} \emph{Every written cell lies under some
traversal.}
\[
  \begin{aligned}
    \forall\, \ngPos \in \ngDom\STOR,\; i \in \ngDom{\STOR(\ngPos)}.\;\;
    \exists\, & \ngLoc_u^{\ngReg.\ngPath} \in \UENV,\;
      \ngPath' \in \ngPathsOf{\ngTypeAt{\TYP_{\ngReg}}{\ngPath}}.\;\;
    \ngPos = \ngReg.\ngPath.\ngPath' \;\land{} \\
    & \MENV(\ngLoc_u^{\ngReg.\ngPath})(\ngPath') \le i <
      \ngEndOf{\ngLoc_u^{\ngReg.\ngPath}}(\ngPath') .
  \end{aligned}
\]

\item\label{swf:claims-disjoint} \emph{Distinct members read disjoint
cells.}
\[
  \forall\, \ngLoc_{u,1}^{\ngPos_1}, \ngLoc_{u,2}^{\ngPos_2} \in \UENV.\;\;
    \ngLoc_{u,1}^{\ngPos_1} \neq \ngLoc_{u,2}^{\ngPos_2} \implies
    \ngFootprint{\ngLoc_{u,1}} \cap \ngFootprint{\ngLoc_{u,2}} = \emptyset .
\]
\end{enumerate}

\item\label{swf:chains} \textbf{Pending-location chain coherence.}
Three conjuncts hold at once: pending locations at one position name
one cell exactly when an $\gramkw{advanceField}$ chain joins them; a
pending location is the recorded argument of at most one entry of
$\CENV$, counted over all of $\ngDom\CENV$ and not its own position
alone; and a pending allocation entry is the argument of none.
\[
  \begin{array}{@{}l@{}}
  \forall\, \ngLoc_x^{\ngReg.\ngPath}, \ngLoc_y^{\ngReg.\ngPath} \in \NENV,\;
    \text{both resolved by}\;\MENV: \\[0.8mm]
  \qquad \MENV(\ngLoc_x)(\ngEmptyPath) = \MENV(\ngLoc_y)(\ngEmptyPath) \\[0.8mm]
  \qquad\quad \iff\;
    \text{some position chain}\;
    \ngLoc_0^{\ngReg.\ngPath},\ldots,\ngLoc_k^{\ngReg.\ngPath}\;
    \text{at}\;\ngReg.\ngPath\;
    (\text{Definition~\ref{def:poschain}}) \\[0.8mm]
  \qquad\qquad\quad
    \text{has every link an}\;\gramkw{advanceField}\;
    \text{predecessor link and}\;\set{\ngLoc_0,\ngLoc_k} = \set{\ngLoc_x,\ngLoc_y};
    \\[2.4mm]
  \forall\, \ngLoc_x^{\ngReg.\ngPath} \in \NENV:\;\;
    \big|\set{\ngLoc'^{\ngPos'} \in \ngDom\CENV \;\mid\;
      \CENV(\ngLoc'^{\ngPos'})\;\text{names}\; \ngLoc_x^{\ngReg.\ngPath}}\big|
    \le 1 ; \\[2.4mm]
  \forall\, \ngReg.\ngPath\;\text{with}\; \AENV(\ngReg.\ngPath) = E:\;\;
    E \in \NENV \implies E \notin \ngSources{\CENV} .
  \end{array}
\]

\item\label{swf:entry-max} \textbf{Allocation entry is maximal.} At one
position, every pending member's root index is at most the allocation
entry's.
\[
  \begin{aligned}
    &\forall\, \ngReg.\ngPath\;\text{with}\;\AENV(\ngReg.\ngPath) = E,\;
      \text{a location}: \\
    &\quad \forall\, \ngLoc_x \in \NENV\;\text{standing at}\;\ngReg.\ngPath,\;
      \text{resolved by}\;\MENV:\quad
      \MENV(\ngLoc_x)(\ngEmptyPath) \;\le\; \MENV(E)(\ngEmptyPath) .
  \end{aligned}
\]

\item\label{swf:node-order} \textbf{Constructor tag follows its
contents in write order.} Let $D_u$ be the traversal
clause~\ref{swf:claims-walk} supplies for a consumable member $\ngLoc_u$. Its
\emph{node subderivations} are $D_u$ itself and, at any depth, every
subderivation of it (item~\ref{der:sub} of
Definition~\ref{def:derivation}),
each a use of $\textsc{VWF-Data}$, the serialized-value traversal
judgment's only rule; that rule's remaining premises are side
conditions rather than traversal judgments and root no subderivation.
Each use's own tag-read premise $\STOR(\ngPos')(i') = \DC'$ names its
\emph{root cell} $a_0 = \ngCell{\ngPos'}{i'}$, written last of the
cells that use reads.
\[
  \begin{aligned}
    &\forall\, \ngLoc_u \in \UENV,\;\text{every node subderivation}\;D'\;
      \text{of}\;D_u,\;\text{with root cell}\;a_0 : \\
    &\quad \forall\, a \in \ngFootprint{D'} \setminus \set{a_0} :\quad
      \ngStampOf{\STOR}{a_0} \;>\; \ngStampOf{\STOR}{a} .
  \end{aligned}
\]

\item\label{swf:cursor-cut} \textbf{Boundary-cursor write-order
separation.} Write $c = \ngMemFront{\ngLoc_x}$
(Definition~\ref{def:memberboundary}) for the member-boundary cursor of
a pending or consumable member $\ngLoc_x$ standing at $\ngReg.\ngPath$. Then
$c$ cuts the write order: what lies below $c$ at a path was written
before anything at or above $c$ at any extension of that path.
\[
  \begin{aligned}
    &\forall\, \ngLoc_x \in \NENV \cup \UENV\;\text{standing at}\;
      \ngReg.\ngPath,\;\text{resolved by}\;\MENV,\;\;
      c = \ngMemFront{\ngLoc_x} : \\
    &\quad \forall\, \ngPath_1, \ngPath_2 \in
      \ngPathsOf{\ngTypeAt{\TYP_{\ngReg}}{\ngPath}}\;\text{with}\;
      \ngPath_1\;\text{a prefix of}\;\ngPath_2, \\
    &\quad \forall\, i_1 \in \ngDom{\STOR(\ngReg.\ngPath.\ngPath_1)}\;
      \text{with}\;i_1 < c(\ngPath_1),\;\;
      \forall\, i_2 \in \ngDom{\STOR(\ngReg.\ngPath.\ngPath_2)}\;
      \text{with}\;i_2 \ge c(\ngPath_2) : \\
    &\quad\qquad
      \ngStampOf{\STOR}{\ngCell{\ngReg.\ngPath.\ngPath_1}{i_1}} \;<\;
      \ngStampOf{\STOR}{\ngCell{\ngReg.\ngPath.\ngPath_2}{i_2}} .
  \end{aligned}
\]

\item\label{swf:straddler} \textbf{At most the chain root spans a
position.} A location \emph{spans} its position when its own atom
stands in $\NENV$ again at a strict descendant of that position --- a
non-empty path of the position's own type. The subscript names the
pending set the membership is tested in.
\[
  \operatorname{Spans}_{\NENV}(\ngLoc_x, \ngReg.\ngPath) \iff
  \exists\, \ngPath' \neq \ngEmptyPath.\;
    \ngPath' \in \ngPathsOf{\ngTypeAt{\TYP_{\ngReg}}{\ngPath}}
    \land \ngLoc_x^{\ngReg.\ngPath.\ngPath'} \in \NENV .
\]
At a position whose entry is a location, any pending member standing
there that spans it must be the allocation entry's chain root.
\[
  \begin{aligned}
    &\forall\, \ngReg.\ngPath\;\text{with}\; \AENV(\ngReg.\ngPath) = E,\;
      \text{a location}: \\
    &\quad \forall\, \ngLoc_x^{\ngReg.\ngPath} \in \NENV :\quad
      \operatorname{Spans}_{\NENV}(\ngLoc_x, \ngReg.\ngPath) \;\implies\;
      \ngLoc_x^{\ngReg.\ngPath} = \ngChainRoot{\CENV}{E} .
  \end{aligned}
\]
The consequent asserts that $\ngChainRoot{\CENV}{E}$ is defined and
equals $\ngLoc_x^{\ngReg.\ngPath}$, as Definition~\ref{def:chainroot}'s
convention reads such an equation; both operands are displayed located,
the predicate above having read $\ngLoc_x$ at a second position as
well. The quantifier's own qualifier secures that definedness:
item~\ref{awf:entry} is guarded at an entry that is a location, which
is exactly the entry this clause quantifies at, so
clause~\ref{swf:allocation} resolves $E$ in $\MENV$ at its own
position; clause~\ref{swf:constraints} is
$\crealj{\STOR}{\CENV}{\MENV}$; and
Lemma~\ref{lem:creal-wellfounded} then answers at $E$. Neither of those
two clauses reads $\ngChainRoot{\CENV}{\cdot}$, so nothing here rests
on this clause itself.

\item\label{swf:member-front} \textbf{Member boundary lies within the
allocation frontier.} At a position whose entry is a location, a member
standing there is held within the entry path by path, its
member-boundary index at each path of the position's own type compared
with the entry's allocation-frontier index there.
\[
  \begin{aligned}
    &\forall\, \ngReg.\ngPath\;\text{with}\; \AENV(\ngReg.\ngPath) = E,\;
      \text{a location}: \\
    &\quad \forall\, \ngLoc_x \in \NENV \cup \UENV\;\text{standing at}\;
      \ngReg.\ngPath,\;\text{resolved by}\;\MENV, \\
    &\quad \forall\, \ngPath' \in
      \ngPathsOf{\ngTypeAt{\TYP_{\ngReg}}{\ngPath}} :\quad
      \ngMemFront{\ngLoc_x}(\ngPath') \;\le\; \ngFront{E}(\ngPath') .
  \end{aligned}
\]
Both readings stay inside their domains.
$\ngMemFront{\ngLoc_x}$ is read at a member $\MENV$ resolves, where
Definition~\ref{def:memberboundary} answers in either case; and
$\ngFront{E}$ is read at an entry this clause asks to be a location,
which item~\ref{awf:entry} then resolves in $\MENV$, where
Definition~\ref{def:front} is total. Both results lie in
$\ngLayoutCursorT{\ngTypeAt{\TYP_{\ngReg}}{\ngPath}}$ and so carry an
entry at every path of the position's own type.

\item\label{swf:nursery-cross} \textbf{Pending cursors do not reverse
order below a strict lead.} If one pending cursor has a smaller index
at a path, its index is no larger than the other cursor's at any
extension of that path.
\[
  \begin{aligned}
    &\forall\, \ngLoc_x^{\ngReg.\ngPath}, \ngLoc_y^{\ngReg.\ngPath} \in \NENV,\;
      \text{both resolved by}\;\MENV, \\
    &\quad \forall\, \ngPath_1, \ngPath_2 \in
      \ngPathsOf{\ngTypeAt{\TYP_{\ngReg}}{\ngPath}}\;\text{with}\;
      \ngPath_1\;\text{a prefix of}\;\ngPath_2 : \\[1mm]
    &\quad\qquad \MENV(\ngLoc_x)(\ngPath_1) < \MENV(\ngLoc_y)(\ngPath_1)
      \;\implies\;
      \MENV(\ngLoc_x)(\ngPath_2) \le \MENV(\ngLoc_y)(\ngPath_2) .
  \end{aligned}
\]
Equality at $\ngPath_1$ establishes no conclusion at $\ngPath_2$.

\end{enumerate}
\end{definition}

\subsection{The store view}
\label{sec:wf-storeview}

\begin{definition}[The store's derived view]
\label{def:storeview}
Write $\STOR_n$ for the $n$-th event of a store $\STOR$,
$0 \le n < \ngLen{\STOR}$. Where $\STOR$ satisfies
condition~\ref{wlog:once} of the write-log judgment --- no address
occurs in two of its events --- write
\[
  \STOR(\ngPos)(i) = \DC
    \;\iff\; \exists n.\; \STOR_n = (\ngCell{\ngPos}{i},\DC) ;
\]
that condition leaves at most one event at each address, so this is a
partial function of the address, undefined at every address no event
names. For any store, write
\[
  \ngDom{\STOR(\ngPos)} \;=\;
    \set{i \;\mid\; \ngCell{\ngPos}{i} \in \ngAddrs\STOR}
\]
for the indices written at $\ngPos$; it reads the events alone and asks
no condition at all. The following notation is defined from $\MENV$
and $\STOR$ and from nothing else, for
$\STOR : \ngWLog{\ngAllocatedPositions\MENV}$:
\[
  \begin{array}{@{}l@{\;\;}l@{}}
    \ngDom\STOR &=\; \ngAllocatedPositions\MENV \\[0.6mm]
    \ngStampOf{\STOR}{\ngCell{\ngPos}{i}} = n
      &\iff\; \exists \DC.\; \STOR_n = (\ngCell{\ngPos}{i},\DC) \\[0.6mm]
    \ngCells\STOR &=\; \ngLen{\STOR} \\[0.6mm]
    \ngMaxIdx{\ngPos}{\STOR} &=\;
      \max\big(\set{-1} \cup \ngDom{\STOR(\ngPos)}\big)
  \end{array}
\]
\end{definition}

\subsection{Configurations and program entry}
\label{sec:wf-configurations}
\label{sec:config-wf}
\label{sec:metatheory}

Fix datatype declarations $\overharpoon{\DD}$ and function declarations
$\overharpoon{\FD}$ for the remainder of this section, together with
the bootstrap data $\ngReg,\ngLoc^{\ngReg},\TYP_b$
(Definition~\ref{def:program-region-typing-environment}).
$\TYP_b$ is fixed data of these statements, not a lookup. By contrast,
$\TYP_{\ngReg}$ is $\Theta(\ngReg)$ at whatever $\Theta$ a statement
below is read at. The two agree exactly where that $\Theta$ carries
$\ngReg$ at $\TYP_b$. Where a statement needs this agreement, adequacy
asks for it (Definition~\ref{def:region-environment-adequacy}, root
agreement~\ref{adeq:root-bootstrap}). Where a statement needs to name the
bootstrap region apart from a bound $\ngReg$ of its own it writes
$\ngReg_b$, which is this same $\ngReg$. Assume $\wfdecl\;\overharpoon{\DD}$, a
condition on $\overharpoon{\DD}$ alone
(Definition~\ref{def:wfdecl}), and take it fixed for the remainder of
the section. Each statement carries $\ngJFunT{\Theta}{\FD_\ind}$, for every
$\ind \le \litnum = |\overharpoon{\FD}|$, at that statement's own
$\Theta$; $\textsc{T-FunctionDef}$'s where-block reads
$\ngPathsOf{\ngReg}$ there. It is a fact of the declarations and that
environment together. No rule of $\ngStepsto$ changes the declarations,
but $\textsc{D-LetRegion}$ extends $\Theta$ to some $\Theta'$, so the proof of
Theorem~\ref{thm:preservation} re-establishes $\ngJFunT{\Theta'}{\FD_\ind}$ at the
extended environment by Lemma~\ref{lem:world-weaken}. No runtime expression is fixed by this convention: every
$\EXPR$ below (or a decorated variant, $\EXPR_1$, $\EXPR''$, and so
on) is that one definition's or theorem's own variable, ranging over
whatever expression a reduction has produced, since $\ngStepsto$
replaces it at every step.
\begin{lemma}[Region-environment weakening]
\label{lem:world-weaken}
Let $\Theta \sqsubseteq \Theta'$
(Definition~\ref{def:region-typing-environment-order}), and let $J$
range over: any expression, value, or pattern typing judgment; any
declaration, function, or program judgment; any formation condition on
$\AENV$ or $\MENV$; any cursor projection or $\MENV$-resolution; any
serialized-value traversal; allocation well-formedness; every clause
of store well-formedness; discharge; global binder uniqueness and
scope discipline; and environment-name containment. Write $J$ at
$\Theta$ for $J$ read at the region typing environment $\Theta$: for a
typing, function or program judgment, $J$ with $\Theta$ written as its
region typing environment; for each of the others, $J$ with the
$\Theta$ its definition reads. Assume:
\begin{enumerate}
  \item $J$ holds at $\Theta$, by some derivation where $J$ has one;
  \item unless $J$ is an expression, value or pattern typing judgment,
    $\Theta$ is adequate for $J$;
  \item where $J$ reads the store, $\STOR$ satisfies the formation its
    notation needs: condition~\ref{wlog:once}
    (Definition~\ref{def:storeview}) for a serialized-value traversal,
    $\STOR : \ngWLog{\ngAllocatedPositions\MENV}$ for allocation
    well-formedness, and, for a clause of store well-formedness, the
    formation that Definition~\ref{def:storewf}'s opening assigns it;
  \item where $J$ is one of the five clauses that
    Definition~\ref{def:storewf} reads under
    $\wfdecl\;\overharpoon{\DD}$, that condition holds as well;
  \item no region of $\ngDom{\Theta'} \setminus \ngDom\Theta$ is bound by
    a $\gramkw{letregion}$ of $J$'s subject, where $J$ is a typing
    judgment; of the declaration's body, where $J$ is a function
    judgment $\tcfun$; or of its main expression or any declaration's
    body, where $J$ is a program judgment.
\end{enumerate}
Then $J$ holds at $\Theta'$, by the same derivation read at
$\Theta'$: the same rule at every node, and every node's judgment the
original's except for its region typing environment, which is
$\Theta'$ in place of $\Theta$, extended by the same
$\gramkw{letregion}$ binders enclosing the node
(Definition~\ref{def:derivation}, item~\ref{der:theta}).
\end{lemma}

Reduction relates \emph{configurations}
\[
  \ngCfg{\Theta}{\STOR}{\MENV}{\EXPR} .
\]
$\Theta$ is the region typing environment the configuration carries
(Definition~\ref{def:region-typing-environment}). A run starts at the
program's own region typing environment
(Definition~\ref{def:program-region-typing-environment}), and each
region a step allocates extends it. $\STOR$ is the append-only store log
(Section~\ref{sec:worlds-static-envs}); $\MENV$ is the location map
(Section~\ref{sec:store-model-resolution}); $\EXPR$ is the expression
being evaluated. The angle brackets group the four arguments of the
reduction relations; they are not a construct of the term language.
$\Theta$ fixes the type of each region the configuration may name, and
configuration well-formedness (Section~\ref{sec:config-wf}) types
$\EXPR$ at $\Theta$. A configuration holds $\Theta$ but none of the
static environments $\TENV,\SENV,\CENV,\AENV,\NENV,\UENV$: those are
components of typing derivations, and configuration well-formedness
asks for some derivation typing $\EXPR$ at $\Theta$ and reads them
there. We call
$\STOR;\MENV;\EXPR$ the configuration's \emph{runtime state}; a
predicate that reads a region typing environment beside it, such as
adequacy (Definition~\ref{def:region-environment-adequacy}), is stated
for a runtime state at a given $\Theta$. The datatype and
function declarations $\overharpoon{\DD}$ and $\overharpoon{\FD}$ are
fixed, ambient program data, the same throughout a run, and are not
components of a configuration.

\emph{Contextual reduction} is
\[
  \ngStepCfg{\Theta}{\STOR}{\MENV}{\EXPR}{\Theta'}{\STOR'}{\MENV'}{\EXPR'} ,
\]
one step of a run, from the source configuration
$\ngCfg{\Theta}{\STOR}{\MENV}{\EXPR}$ to the target configuration
$\ngCfg{\Theta'}{\STOR'}{\MENV'}{\EXPR'}$. Nine rules define
\emph{basic contraction} (equivalently, \emph{basic reduction}),
\[
  \ngStepCfgB{\Theta}{\STOR}{\MENV}{r}{\Theta'}{\STOR'}{\MENV'}{r'} ,
\]
each stated at a whole \emph{redex} $r$ -- a term that the left-hand side of one rule
matches -- rather than an arbitrary expression; its right-hand $r'$ is the \emph{contractum}. A tenth
rule, $\textsc{D-Ctx}$, is the only one concluding contextual reduction:
it lifts one basic contraction under an evaluation context
(Section~\ref{sec:terms-values-programs}), so every step of a run is
one $\textsc{D-Ctx}$ instance over one basic contraction. Basic
contraction has one further parameter, which its configurations do not
show: the whole expression $\ngCtxFill{\ngCtx}{r}$ of the
$\textsc{D-Ctx}$ instance that lifts it. Its fresh-name premises are
read against that expression, not against the redex alone
(Definition~\ref{def:fresh-dynamic}). A statement about a basic
contraction on its own either names the expression its premises are
read against or holds of the contraction under whichever expression it
is derived; neither form asserts that the contraction is derivable
under another expression.

\begin{definition}[Freshness]
\label{def:fresh-core}
Let $a$ be a name: a location atom, a region, or a term variable. For
finite syntactic objects $F_1,\ldots,F_k$ -- terms, types, maps, sets,
or vectors of function declarations $\overharpoon{\FD}$ -- write
$\ngFresh{a}{F_1,\ldots,F_k}$, read ``$a$ is fresh for
$F_1,\ldots,F_k$'', to mean $a$ occurs, free or bound, in
none of them. Occurrence is structural; bound occurrences count. A
location atom $\ngLoc$ occurs wherever a location $\ngLoc^{\ngPos}$
built on it does: as a map key or inside a map's value, inside a
runtime value $\ngLocVal{\ngLoc^{\ngPos}}$, in any annotation or
location expression, at the binding occurrence of a $\gramkw{letloc}$
or of a $\gramkw{case}$ clause's annotation, and in a located type
$\TYP@\ngLoc^{\ngPos}$. A region $\ngReg$ occurs as the region
component of any position -- hence in a key of $\AENV$ and in an
address of $\STOR$ -- and at the binder of a $\gramkw{letregion}$. The
store is among the places a region occurs because its addresses are
built from positions; a layout cursor is not, carrying indices and
shape and no name of its own. Where prose names a location
$\ngLoc^{\ngPos}$ rather than a bare atom, it means freshness of that
location's own atom $\ngLoc$ -- a position carries no name of its own
and is never freshened.
\end{definition}
\begin{definition}[Fresh for a transition]
\label{def:fresh-dynamic}
Let $\textsc{D-Ctx}$ lift a basic contraction of the redex $r$ under
the evaluation context $\ngCtx$, from the source configuration
$\ngCfg{\Theta}{\STOR}{\MENV}{\ngCtxFill{\ngCtx}{r}}$, at the function
declarations $\overharpoon{\FD}$. A name
(a location atom, a region, or a term variable) is \emph{fresh for the
transition} when
\[
  \ngFresh{a}{\STOR,\;\MENV,\;\ngCtxFill{\ngCtx}{r},\;
    \overharpoon{\FD}}
  \quad\text{and}\quad a \notin \ngDom\Theta
\]
(Definition~\ref{def:fresh-core}): the list $\textsc{D-LetRegion}$
writes out. In every dynamic rule, a name picked \emph{fresh} without
further qualification -- $\ngLoc_f$ in the four
$\gramkw{letloc}$ rules, $\overharpoon{\ngLoc_f}$ in
$\textsc{D-Case}$ -- is fresh for the transition, and so is every name
$\mathrm{Freshen}$ picks in $\textsc{D-App}$. When one rule picks
several names at once, they are pairwise distinct, chosen in a fixed
order, and each later pick also avoids the names already chosen. A name
fresh for the transition is also called \emph{fresh for the
configuration} $\ngCfg{\Theta}{\STOR}{\MENV}{\ngCtxFill{\ngCtx}{r}}$,
the declarations being fixed beside it.
\end{definition}
\begin{definition}[Freshen]
\label{def:freshen}
Let $\fvar\,\overharpoon{\var} = \EXPR$ be a declaration of
$\overharpoon{\FD}$. The \emph{binding occurrences} of $\EXPR$ are:
every $\gramkw{letloc}$'s bound location atom; every
$\gramkw{letregion}$'s bound region; every $\gramkw{case}$ clause's
bound location atoms; and every bound term variable, whether a
$\gramkw{let}$'s or a clause's. A scheme's own quantified locations
and region are not among them: no quantifier is a binder syntactically
contained in $\EXPR$, and $\textsc{D-App}$'s own substitution $\sigma$
instantiates those separately (below).

Taking the binding occurrences of $\EXPR$ in a fixed order, choose for
each one a fresh name of the same sort -- fresh, per the convention
above, for the current transition, and pairwise distinct from every
name already chosen for an earlier binding occurrence of this same
application. $\mathrm{Freshen}(\EXPR)$ is $\EXPR$ with, at each
binding occurrence, the occurrence itself and every occurrence of the
bound name standing in that one binder's own lexical scope -- a
$\gramkw{letloc}$'s or $\gramkw{letregion}$'s continuation, a
$\gramkw{case}$ clause's own branch, or a term-variable binder's own
scope -- replaced by its chosen name, and no other occurrence touched.
The sequence of chosen names, in the fixed order of the binding
occurrences, is the \emph{supply} of that application.
\end{definition}
\begin{definition}[Atom renaming]
\label{def:atom-renaming}
An atom renaming $\subst{\cdot}{\ngLoc}{\ngLoc_f}$ replaces atom
$\ngLoc$ by $\ngLoc_f$ at every position $\ngLoc$ occurs at, and
leaves every region, path, and position otherwise unchanged.
\end{definition}
\begin{definition}[Region renaming]
\label{def:region-renaming}
A region renaming $\ngReg \mapsto \ngReg_f$ replaces region $\ngReg$
by another region $\ngReg_f$: at every position whose region is $\ngReg$,
$\ngReg.\ngPath$ becomes $\ngReg_f.\ngPath$, for every $\ngPath$; and
the region typing environment gives $\ngReg_f$ whatever it gives
$\ngReg$: $\TYP_{\ngReg_f} = \TYP_{\ngReg}$ where
$\ngReg \in \ngDom\Theta$, and nothing where $\ngReg \notin
\ngDom\Theta$. $\Theta$'s entry at $\ngReg$ itself is kept.
\end{definition}
\begin{definition}[Clause-complete derivation]
\label{def:clause-complete}
A derivation of the expression judgment, or of
\ngRuleName{T-Pattern}'s clause judgment, is \emph{clause-complete} when
every \ngRuleName{T-Case} instance in it, at its root or at any node beneath,
has a non-empty clause vector, $\litnum \ge 1$.
\end{definition}
\begin{lemma}[Atom renaming]
\label{lem:rename-atom}
Let $\ngLoc_f$ be a location atom occurring in none of
$\EXPR,\TENV,\SENV,\CENV,\AENV,\NENV,\UENV$, and suppose no binder of
$\EXPR$ binds $\ngLoc$. If a clause-complete derivation
(Definition~\ref{def:clause-complete}) derives
$\ngJT{\Theta}{\ngTup{\TENV}{\SENV}{\CENV}}{\ngTup{\AENV}{\NENV}{\UENV}}
  {\EXPR : \ngLocatedTy}{\ngTup{\AENV'}{\NENV'}{\UENV'}}$ then
\[
  \ngJLongT{\Theta}
    {\ngTup{\subst{\TENV}{\ngLoc}{\ngLoc_f}}{\subst{\SENV}{\ngLoc}{\ngLoc_f}}
      {\subst{\CENV}{\ngLoc}{\ngLoc_f}}}
    {\ngTup{\subst{\AENV}{\ngLoc}{\ngLoc_f}}{\subst{\NENV}{\ngLoc}{\ngLoc_f}}
      {\subst{\UENV}{\ngLoc}{\ngLoc_f}}}
    {\subst{\EXPR}{\ngLoc}{\ngLoc_f} : \subst{\ngLocatedTy}{\ngLoc}{\ngLoc_f}}
    {\ngTup{\subst{\AENV'}{\ngLoc}{\ngLoc_f}}{\subst{\NENV'}{\ngLoc}{\ngLoc_f}}
      {\subst{\UENV'}{\ngLoc}{\ngLoc_f}}} ,
\]
the atom renaming of Definition~\ref{def:atom-renaming} applied
pointwise to each term and each environment.
\end{lemma}
\begin{definition}[Global binder uniqueness and scope discipline]
\label{def:singlybound}
A term $\EXPR$ satisfies \emph{global binder uniqueness and scope
discipline against} declarations $\overharpoon{\FD}$ and a bootstrap
location $\ngLoc^{\ngReg}$ when the following all hold.

\emph{Uniqueness.} No two binders drawn together from: every
$\gramkw{letloc}$ in $\EXPR$; every location a $\gramkw{case}$ clause
binds in its annotation, in $\EXPR$; every location a declaration's
scheme quantifies; every $\gramkw{letloc}$ or $\gramkw{case}$-clause
binder standing in a declaration's body; and $\ngLoc^{\ngReg}$ itself,
bind one location atom. The same is asked of region names, whose
binders are every $\gramkw{letregion}$ in $\EXPR$, every region a
declaration's scheme quantifies, every $\gramkw{letregion}$ standing in
a declaration's body, and the bootstrap region $\ngReg$.

\emph{Main-expression scope.} For every location atom or region a
binder of $\EXPR$ binds -- a $\gramkw{letloc}$, a $\gramkw{letregion}$,
or a $\gramkw{case}$ clause's annotation -- that binder's own binding
occurrence is permitted, and every other occurrence of the name in
$\EXPR$ stands inside that binder's own body: a $\gramkw{letloc}$'s or a
$\gramkw{letregion}$'s continuation, or the clause's own branch.

\emph{Declaration scope.} A declaration binds names of two kinds,
scoped differently. A location or region its scheme quantifies has the
whole declaration as its scope, scheme and body together: its own
binding occurrence is permitted, every other occurrence of the name
stands within that declaration -- within its scheme's argument and
result located types and within its body alike -- and it occurs in
$\EXPR$ nowhere and in another declaration nowhere. A $\gramkw{letloc}$,
$\gramkw{letregion}$, or $\gramkw{case}$-clause binder standing in a
declaration's body is scoped exactly as in the main expression, read
within that one declaration: its own binding occurrence is permitted,
every other occurrence of the name within that declaration stands
inside that binder's own body, and, as for a scheme-quantified name, it
occurs in $\EXPR$ nowhere and in another declaration nowhere. It is not
given scope over the declaration's body as a whole, only over its own.

\emph{Bootstrap scope.} $\ngLoc^{\ngReg}$ and the bootstrap region
$\ngReg$ have the whole main expression $\EXPR$ as their scope; the
definition asks nothing of their occurring, or not occurring, inside a
declaration.

The condition constrains only the names these binders take. It says
nothing about a name no binder of $\EXPR$ or of a declaration takes,
which is why it survives the steps that consume a binder. A call's own
instantiated copy of a declaration carries fresh names in its binders'
place, not the declaration's own, so the
condition is asked of the program as written, once, rather than of
every instantiated copy. A program satisfies global binder uniqueness
and scope discipline when its main expression does, against its own
declarations and its own bootstrap.
\end{definition}
\begin{definition}[Environment-name containment]
\label{def:grounded}
The static components $\TENV,\SENV,\CENV,\AENV,\NENV,\UENV$ satisfy
\emph{environment-name containment} for a runtime state
$\STOR;\MENV;\EXPR$ when every location atom and region occurring in
them also occurs in $\MENV$, in $\EXPR$, or in the declarations:
\[
  \operatorname{names}_{\ell,R}(\TENV,\SENV,\CENV,\AENV,\NENV,\UENV)
  \;\subseteq\;
  \operatorname{names}_{\ell,R}(\MENV,\EXPR,\overharpoon{\DD},\overharpoon{\FD}) .
\]
\end{definition}
\begin{definition}[Adequate region typing environment]
\label{def:region-environment-adequacy}
\emph{Adequacy of a term.} Write $\Theta \vdash_{\mathit{adq}} \EXPR$
when the expression $\EXPR$ is adequate at $\Theta$, read with its
binders. It is defined by recursion on $\EXPR$, from the following
readings of the forms a term writes directly:
\begin{itemize}
  \item a location $\ngLoc^{\ngReg.\ngPath}$ written outside a located
    type --- a $\gramkw{letloc}$ binder's location, a constructor's
    destination, a call's location vector, $\ngLocVal{\ngLoc}$, the
    argument of a location expression --- is adequate at $\Theta$ when
    $\Theta(\ngReg)$ is defined and is a declared datatype;
  \item a value is adequate when it is a variable, or a location value
    whose location is adequate;
  \item a located type $\TYP@\ngLoc^{\ngReg.\ngPath}$ is adequate at
    $\Theta$ when its location is, and $\ngPath \in
    \ngPathsOf{\TYP_{\ngReg}}$ and $\ngTypeAt{\TYP_{\ngReg}}{\ngPath} =
    \TYP$;
  \item a location expression $\ngStartRegTy{\ngReg}{\TYP}$ is adequate
    at $\Theta$ when $\Theta(\ngReg) = \TYP$ and $\TYP$ is declared; an
    $\gramkw{advanceTags}$ form when its argument location is; an
    $\gramkw{after}$ form when its located type is; and an
    $\gramkw{advanceField}$ form when its argument location and its
    located type are.
\end{itemize}
Then:
\begin{enumerate}
  \item\label{adeq:letregion} $\Theta \vdash_{\mathit{adq}}
    \letregty{\ngReg}{\TYP}{\EXPR}$ when $\ngReg \notin \ngDom\Theta$,
    $\TYP$ is a declared datatype, and
    $\Theta[\ngReg \mapsto \TYP] \vdash_{\mathit{adq}} \EXPR$;
  \item\label{adeq:other} every other form is adequate at $\Theta$
    when every location, value, located type and location expression
    written directly in it is adequate at $\Theta$, and each of its
    immediate subexpressions is adequate at $\Theta$: a $\gramkw{let}$'s
    annotation and its two subexpressions, a $\gramkw{letloc}$'s
    location, location expression and body, a constructor
    application's destination and fields, a call's location vector and
    arguments, and a $\gramkw{case}$'s scrutinee and, at each clause,
    its binders' located types and its body.
\end{enumerate}
$\gramkw{letregion}$ is the one form that binds a region, and the only
one whose body is read at a different $\Theta$.

\emph{Adequacy for a judgment.} $\Theta$ is \emph{adequate} for $J$
when all five conditions below hold. Write
$\operatorname{regions}(J)$ for the regions occurring in $J$'s
\emph{free components}: its selected static components, the result of
a judgment of $J$, its runtime store and location map, and the schemes
of the declarations beside it --- not its region typing environment,
which is what adequacy is asked of; not its subject, and not a
declaration's body, which the second condition reads with their
binders.

\emph{Coverage.} $\Theta$ is defined at every region in
$\operatorname{regions}(J)$, and at the bootstrap region $\ngReg_b$
wherever a statement fixes bootstrap data beside $J$.

\emph{Subjects.} $\Theta \vdash_{\mathit{adq}} \EXPR$ for $J$'s own
subject $\EXPR$, and $\Theta \vdash_{\mathit{adq}} \EXPR_{\FD}$ for the
body $\EXPR_{\FD}$ of every declaration beside $J$.

\emph{Root agreement.} $\Theta(\ngReg) = \TYP$ at each of the
following, these being the free forms that name a region and a type
with no path between them:
\begin{enumerate}
  \item\label{adeq:root-start} a $\ngStartRegTy{\ngReg}{\TYP}$ entry of
    $\CENV$ among $J$'s selected static components;
  \item\label{adeq:root-scheme} a function scheme beside $J$, at each
    argument pair $(\ngReg_\ind,\TYP_\ind)$ and at its result pair
    $(\ngReg,\TYP)$; and
  \item\label{adeq:root-bootstrap} the pair $(\ngReg_b,\TYP_b)$,
    wherever a statement fixes bootstrap data beside $J$.
\end{enumerate}

\emph{Path agreement.} At every located type
$\TYP@\ngLoc^{\ngReg.\ngPath}$ standing in $J$'s variable environment
$\TENV$, in an entry of its $\CENV$, or as the result of a judgment of
$J$:
\[
  \ngPath \in \ngPathsOf{\TYP_{\ngReg}}
  \qquad\text{and}\qquad
  \ngTypeAt{\TYP_{\ngReg}}{\ngPath} = \TYP .
\]

\emph{Declaredness.} $\Theta(\ngReg)$ is a datatype name some
declaration of $\overharpoon{\DD}$ declares, at every region in
$\operatorname{regions}(J)$, and at the bootstrap region $\ngReg_b$
wherever a statement fixes bootstrap data beside $J$.
\end{definition}
\begin{definition}[Static/runtime invariant]
\label{def:wfinv}
Write
\[
  \Theta;\; \TENV;\SENV;\CENV;\AENV;\NENV;\UENV
    \vdash_{\mathit{inv}} \MENV;\STOR;\EXPR
\]
when all four of the following hold, the last two being read only
where the first two do.
\begin{enumerate}
  \item\label{wfinv:wlog} $\STOR : \ngWLog{\ngAllocatedPositions\MENV}$
    (Definition~\ref{def:storeview}).
  \item\label{wfinv:wfdecl} $\wfdecl\;\overharpoon{\DD}$ (Definition~\ref{def:wfdecl}),
    at the declarations fixed beside the judgment.
  \item\label{wfinv:storewf}
    $\storewfj{\SENV}{\CENV}{\AENV}{\NENV}{\UENV}{\MENV}{\STOR}$
    (Definition~\ref{def:storewf}).
  \item\label{wfinv:adequate} $\Theta$ is adequate
    (Definition~\ref{def:region-environment-adequacy}) for
    $\TENV,\SENV,\CENV,\AENV,\NENV,\UENV$, for the runtime state
    $\STOR;\MENV;\EXPR$, and for the declarations and the bootstrap
    region, location and type fixed beside them.
\end{enumerate}
Where $\TENV = \emptyset$, write
\[
  \Theta;\; \SENV;\CENV;\AENV;\NENV;\UENV
    \vdash_{\mathit{inv}} \MENV;\STOR;\EXPR
\]
for the same judgment.
\end{definition}
\begin{definition}[Discharged]
\label{def:settled}
Let $\NENV_f$ and $\UENV_f$ be two supplied sets of locations.
A derivation of the expression judgment with input pending-location
set $\NENV$ is \emph{discharged against} $(\NENV_f,\UENV_f)$ when
\[
  \NENV \cap \NENV_f = \emptyset
  \qquad\text{and}\qquad
  \ngPath'' = \ngEmptyPath \;\text{for every}\;
  \ngLoc''^{{\ngReg}''.\ngPath''} \in \UENV_f .
\]
Thus none of its input pending locations belongs to $\NENV_f$,
and every location in $\UENV_f$ stands at its region's root.
A derivation is \emph{discharged outright} when it is discharged
against its own output pair $(\NENV',\UENV')$.
\end{definition}
\begin{definition}[Well-typed configuration]
\label{def:cfgwf}
For a datatype $\TYP$ and a configuration
$\ngCfg{\Theta}{\STOR}{\MENV}{\EXPR}$, write
\[
  \vdash \ngCfg{\Theta}{\STOR}{\MENV}{\EXPR} : \TYP
\]
and say the configuration is \emph{well typed at $\TYP$}, relative to
the declarations
$\overharpoon{\DD},\overharpoon{\FD}$ and the bootstrap region,
location and type this section fixes, when
\begin{enumerate}
  \item\label{cfgwf:wfdecl} $\wfdecl\;\overharpoon{\DD}$ holds; and
  \item\label{cfgwf:tcfun} $\ngJFunT{\Theta}{\FD_\ind}$ holds for
    every $\ind \le \litnum$;
\end{enumerate}
and there are $\SENV,\CENV,\AENV,\NENV,\UENV,\AENV',\NENV',\UENV'$, a
location atom $\ngLoc_{\mathrm{res}}$, a region $\ngReg_r$, a path
$\ngPath_r$ and a derivation $D$ of
\[
  \ngJT{\Theta}{\ngTup{\emptyset}{\SENV}{\CENV}}{\ngTup{\AENV}{\NENV}{\UENV}}
  {\EXPR : \TYP@\ngLoc_{\mathrm{res}}^{\ngReg_r.\ngPath_r}}{\ngTup{\AENV'}{\NENV'}{\UENV'}}\]
such that all five of the following hold. (The witnesses
$\ngLoc_{\mathrm{res}}$ and $\ngReg_r$ are unrelated to the bootstrap
atom and region $\ngLoc^{\ngReg}$; they may coincide with them.)
\begin{enumerate}[resume]
  \item\label{cfgwf:settled} $D$ is discharged outright
    (Definition~\ref{def:settled}).
  \item\label{cfgwf:singlybound} $\EXPR$ and $\overharpoon{\FD}$
    satisfy global binder uniqueness and scope discipline relative to
    the bootstrap location $\ngLoc^{\ngReg}$
    (Definition~\ref{def:singlybound}).
  \item\label{cfgwf:grounded} $\emptyset,\SENV,\CENV,\AENV,\NENV,\UENV$
    satisfies environment-name containment
    (Definition~\ref{def:grounded}): every location atom and region
    they name occurs in $\MENV$, in $\EXPR$, or in the declarations.
  \item\label{cfgwf:invariant}
    $\Theta;\; \SENV;\CENV;\AENV;\NENV;\UENV
      \vdash_{\mathit{inv}} \MENV;\STOR;\EXPR$
    (Definition~\ref{def:wfinv}), the static/runtime invariant at
    $\TENV = \emptyset$: store well-formedness at those five components
    with $\MENV$ and $\STOR$, and adequacy of $\Theta$ for them, for the
    runtime state, and for the declarations and the bootstrap data
    beside them.
  \item\label{cfgwf:boot} $\MENV$ binds the bootstrap location
    $\ngLoc^{\ngReg}$: $\MENV(\ngLoc^{\ngReg})$ is defined. A name
    picked fresh for the configuration is then never the bootstrap atom.
\end{enumerate}
Every object named in the definition is of its sort, as
Section~\ref{sec:worlds-static-envs} fixes it at $\Theta$. In
particular $\Theta$, $\SENV$, $\CENV$, $\AENV$, $\NENV$, $\UENV$ and
$\MENV$ are finite. Each cursor of $\MENV$ has the type at its key's
position, and the atoms of $\MENV$'s keys are pairwise distinct.
Each family of $\AENV$ is defined exactly on the paths of its region's
type. These formation conditions are part of the judgment, and a
reduction step re-establishes them with the rest
(Theorem~\ref{thm:preservation}).
Write $\ngCfgWF{\TYP}{\Theta}{\STOR}{\MENV;\EXPR}$ for the same
judgment.
\end{definition}
\begin{definition}[Step-ready]
\label{def:step-ready}
A derivation of
\[
  \ngJT{\Theta}{\ngTup{\emptyset}{\SENV}{\CENV}}{\ngTup{\AENV}{\NENV}{\UENV}}
  {\EXPR : \TYP@\ngLoc_{\mathrm{res}}^{\ngReg_r.\ngPath_r}}{\ngTup{\AENV'}{\NENV'}{\UENV'}}\]
-- $\ngLoc_{\mathrm{res}}$ a location atom, $\ngReg_r$ a region,
$\ngPath_r$ a path, none
fixed by the ambient context of Section~\ref{sec:metatheory} -- is
\emph{step-ready} at $\Theta$ when $\wfdecl\;\overharpoon{\DD}$
holds, $\ngJFunT{\Theta}{\FD_\ind}$ holds for every $\ind \le \litnum$,
$\STOR : \ngWLog{\ngAllocatedPositions\MENV}$
(Definition~\ref{def:storeview}), and, read only where those three
hold, clauses
\ref{swf:complete}, \ref{swf:constraints} and \ref{swf:nursery} of
Definition~\ref{def:storewf} hold of
$\SENV,\CENV,\AENV,\NENV,\UENV,\MENV,\STOR$, and $\Theta$ is adequate
(Definition~\ref{def:region-environment-adequacy}) for the derivation
together with the runtime state $\STOR;\MENV;\EXPR$, exactly as
item~\ref{wfinv:adequate} of Definition~\ref{def:wfinv} reads it
inside item~\ref{cfgwf:invariant} of Definition~\ref{def:cfgwf}.
\end{definition}
\begin{definition}[Program region typing environment]
\label{def:program-region-typing-environment}
Let $P = \overharpoon{\DD};\overharpoon{\FD};\EXPR$ be a program and let
\[
  b \;=\; (\ngReg_b,\;\ngLoc_b,\;\TYP_b)
\]
be a region variable, a location atom and a datatype declared in
$\overharpoon{\DD}$, supplied
beside $P$: \emph{bootstrap data}. They are not part of $P$, whose
grammar carries declarations, function declarations and an expression
and no bootstrap triple; $\ngReg_b$ and $\ngLoc_b$ are the bootstrap
region and location a $\textsc{T-Program}$ derivation is to take, and
$\TYP_b$ is the datatype that region is to be carried at. Write
$\mathcal{B}(P;b)$ for the finite relation between region names and
types holding exactly:
\begin{enumerate}
  \item $(\ngReg,\TYP)$ for every scheme's result region $\ngReg$ at
    its result type $\TYP$;
  \item $(\ngReg_\ind,\TYP_\ind)$ for every scheme's argument region
    at that argument's type; and
  \item $(\ngReg_b,\TYP_b)$, from $b$.
\end{enumerate}
The first two families are read off $P$'s text and the third is
$b$'s own pair, so no family reads a region typing environment. A
$\gramkw{letregion}$ annotation is not a family: a region a
$\gramkw{letregion}$ binds is not in the domain of the $\Theta$ its
binder is typed at ($\textsc{T-LetRegion}$), and the body reads the
annotation from the binder itself.
$P$ has \emph{consistent region annotations at $b$} when
$\mathcal{B}(P;b)$ is single-valued:
\[
  (\ngReg,\TYP_1) \in \mathcal{B}(P;b) \;\land\;
  (\ngReg,\TYP_2) \in \mathcal{B}(P;b)
  \;\implies\; \TYP_1 = \TYP_2 .
\]
For such a $P$ and $b$, the \emph{program region typing environment}
$\Theta_{P;b}$
is $\mathcal{B}(P;b)$ read as the finite partial function it then is, and
so is a region typing environment of
Definition~\ref{def:region-typing-environment}'s grammar, its listing
carrying one entry for each region $\mathcal{B}(P;b)$ relates. Where $\mathcal{B}(P;b)$
is not single-valued, $\Theta_{P;b}$ is undefined, and a program that
wants one region name at two types writes two region names.
\end{definition}
\begin{definition}[Declared program annotations]
\label{def:declared-annotations}
$P = \overharpoon{\DD};\overharpoon{\FD};\EXPR$ has \emph{declared
annotations at $b$} when every datatype name standing in one of the
following is declared by some declaration of $\overharpoon{\DD}$:
\begin{enumerate}
  \item\label{da:letregion} a $\letregty{\ngReg}{\TYP}{\cdot}$
    annotation occurring in $\EXPR$ or in a declaration's body;
  \item\label{da:start} a $\ngStartRegTy{\ngReg}{\TYP}$ location
    expression occurring in $\EXPR$ or in a declaration's body;
  \item\label{da:scheme} a scheme's argument types and its result
    type; and
  \item\label{da:bootstrap} $\TYP_b$, the datatype of $b$.
\end{enumerate}
\end{definition}
\begin{lemma}[Initial configuration well-formedness]
\label{lem:initial-cfgwf}
Let $P = \overharpoon{\DD};\overharpoon{\FD};\EXPR_0$ and let
$b = (\ngReg,\ngLoc,\TYP_b)$ be this section's bootstrap data; let $P$
have consistent region annotations at $b$
(Definition~\ref{def:program-region-typing-environment}) and declared
annotations at $b$ (Definition~\ref{def:declared-annotations}); let $\EXPR_0$
and $\overharpoon{\FD}$ satisfy global binder uniqueness and scope
discipline against the bootstrap location $\ngLoc^{\ngReg}$
(Definition~\ref{def:singlybound}); and let a
$\ngRuleName{T-Program}$ derivation of
\[
  \ngJProgT{\Theta_{P;b}}{P : \TYP@\ngLoc_{\mathrm{res}}^{\ngReg_r.\ngPath_r}}
    {\ngTup{\AENV'}{\NENV'}{\UENV'}}
\]
be given, for some $\AENV'$, $\NENV'$ and $\UENV'$, taking $\ngReg$ and
$\ngLoc^{\ngReg}$ as its bootstrap region and location; its
main-expression premise stands at the same located type. Define
\[
  \MENV_0 \;=\; \set{\ngLoc^{\ngReg} \mapsto \ngZeroCursor{\TYP_b}} ,
  \qquad
  \STOR_0 \;=\; \varepsilon .
\]
($\ngZeroCursor{\TYP_b}$ is defined: the $\ngRuleName{T-Program}$
derivation has the premise $\wfdecl\;\overharpoon{\DD}$, whose
acyclicity condition makes Definition~\ref{def:zerocursor}'s recursion
terminate.) Then
\[
  \vdash \ngCfg{\Theta_{P;b}}{\STOR_0}{\MENV_0}{\EXPR_0} : \TYP ,
\]
and in particular, by item~\ref{wfinv:wlog} of
Definition~\ref{def:wfinv} inside item~\ref{cfgwf:invariant} of
Definition~\ref{def:cfgwf},
$\STOR_0$ is a write log at $\ngAllocatedPositions{\MENV_0}$.
\end{lemma}
\begin{definition}[Rule-induced renaming]
\label{def:step-renaming}
For a basic contraction, under whichever enclosing expression it is
derived, the renaming $\rho$ it induces is:
\[
  \begin{array}{@{}l@{\qquad}l@{}}
    \textsc{D-Let-Val},\ \textsc{D-Case},\
      \textsc{D-DataConstructor},\ \textsc{D-App}
      & \text{the identity} \\[1mm]
    \textsc{D-LetRegion}
      & \text{the region renaming}\;
        \ngReg \mapsto \ngReg_f\;
        (\text{Definition~\ref{def:region-renaming}}) \\[1mm]
    \text{the four}\;\gramkw{letloc}\;\text{rules}
      & \text{the atom renaming}\;
        \ngLoc \mapsto \ngLoc_f\;
        (\text{Lemma~\ref{lem:rename-atom}})
  \end{array}
\]
carrying in each case the rule's own bound name to the fresh one its
freshness premise chooses. At $\textsc{D-Ctx}$, $\rho$ is the renaming the
inner contraction induces. Application is written postfix, as in
$\NENV_{\mathrm{tgt}}\rho$.
\end{definition}
\begin{definition}[Multistep reduction]
\label{def:multistep}
Write
$\ngStepCfgStar{\Theta}{\STOR}{\MENV}{\EXPR}{\Theta'}{\STOR'}{\MENV'}{\EXPR'}$
for the reflexive-transitive closure of contextual reduction
(Section~\ref{sec:dynamic-semantics}); like that relation, it is a
relation on configurations. It is the smallest relation closed under
two constructors.
\[
  \inferrule*[lab=MS-Refl]{ }{
    \ngStepCfgStar{\Theta}{\STOR}{\MENV}{\EXPR}{\Theta}{\STOR}{\MENV}{\EXPR}
  }
\]
\[
  \inferrule*[lab=MS-Step]{
    \ngStepCfg{\Theta}{\STOR}{\MENV}{\EXPR}{\Theta_1}{\STOR_1}{\MENV_1}{\EXPR_1}
    \\
    \ngStepCfgStar{\Theta_1}{\STOR_1}{\MENV_1}{\EXPR_1}
      {\Theta_2}{\STOR_2}{\MENV_2}{\EXPR_2}
  }{
    \ngStepCfgStar{\Theta}{\STOR}{\MENV}{\EXPR}
      {\Theta_2}{\STOR_2}{\MENV_2}{\EXPR_2}
  }
\]
\end{definition}

\subsection{Additional definitions}
\label{sec:wf-auto-deps}
\begin{definition}[Admitted program]
\label{def:admitted}
A program $P$ is \emph{admitted} at bootstrap data $b$ when it has
consistent region annotations at $b$
(Definition~\ref{def:program-region-typing-environment}), declared
annotations at $b$ (Definition~\ref{def:declared-annotations}), global
binder uniqueness and scope discipline
(Definition~\ref{def:singlybound}), and a $\ngRuleName{T-Program}$
derivation of
$\ngJProgT{\Theta_{P;b}}{P : \ngLocatedTy}{\ngTup{\AENV'}{\NENV'}{\UENV'}}$
for some located type $\ngLocatedTy$ and some $\AENV'$, $\NENV'$ and
$\UENV'$: the program judgment at the region typing environment
$\Theta_{P;b}$.
\end{definition}
\begin{definition}[Bound-name avoidance]
\label{def:avoids}
Fix a finite object built from one of: a map entry $k \mapsto v$, in
which case both $k$ and $v$ are checked; a member of a set-valued
environment such as $\SENV$, $\NENV$, or $\UENV$, in which case the
member itself is; or the source or the target of a location or region
substitution, in which case the location or region itself is. A
location atom or a region \emph{occurs} in the object checked when
Definition~\ref{def:fresh-core}'s occurrence relation says so. A path
carries no location atom and no region of its own: the region a
position names belongs to the position, not to a path that extends it.

The object \emph{avoids} $\EXPR$ when no location atom or region
occurring in it is bound by a $\gramkw{letloc}$, a $\gramkw{letregion}$,
or a $\gramkw{case}$ clause's annotation of $\EXPR$.
\end{definition}
\begin{definition}[Evaluation focus]
\label{def:evalfocus}
For an expression $\EXPR$, its \emph{evaluation focus} $\ngFocus{\EXPR}$
is the $\EXPR_0$ with $\EXPR = \ngCtxFill{\ngCtx}{\EXPR_0}$, for the
unique $\ngCtx$ as large as the grammar above allows.
\end{definition}
\begin{definition}[Inert allocation-family restriction]
\label{def:inert-restriction}
Let $D_0$ derive
$\ngJT{\Theta}{\ngTup{\TENV}{\SENV}{\CENV}}{\ngTup{\AENV}{\NENV}{\UENV}}
  {\EXPR : \ngLocatedTy}{\ngTup{\AENV'}{\NENV'}{\UENV'}}$, let $Q$ be a finite set of
regions, and let $\MENV$ be a location map. $D$ is the
\emph{$Q$-restriction of $D_0$ against $\MENV$} when every $\ngReg \in Q$
satisfies
\begin{enumerate}
  \item\label{inert:unstarted} $\AENV(\ngReg) = \phi$ --- the value the
    map holds at that key, which is a different reading from the
    positional observation $\phi$ that
    Definition~\ref{def:aenv-core} then returns at each path of the
    family;
  \item\label{inert:absent} $\ngReg$ occurs in none of $\EXPR$,
    $\TENV$, $\SENV$, $\NENV$, $\UENV$ or $\ngLocatedTy$, and at no
    key and no recorded argument of an entry of $\CENV$ --- the one
    location of $\CENV$ not asked is an
    $\ngAdvField{\ngLoc'^{\ngPos}}{\DC}{j}{\ngLocatedTy'}$ entry's
    located-type location, the location of $\ngLocatedTy'$, which the
    typing rules match in $\CENV$ and never read through $\AENV$
    (Lemma~\ref{lem:restriction-premise}); and
  \item\label{inert:unresolved} $\MENV$ resolves no
    $L^{\ngReg.\ngPath}$, at any atom $L$ and any path $\ngPath$;
\end{enumerate}
and $D$ is what Lemma~\ref{lem:strengthen-a}, applied once per member
of $Q$, leaves: the same subject, the same rule at every node, the same
region typing environment at every node, the same
$\TENV,\SENV,\CENV,\NENV,\UENV$, the same located result type and the
same output pending-location and consumable-value sets, with every
allocation environment of the derivation tree --- the input $\AENV$ and
the output $\AENV'$ alike --- carrying no key in $Q$.
\end{definition}
\begin{definition}[Location substitution]
\label{def:locsubst}
A location substitution $\sigma$ is given by a vector of \emph{root
locations} $\overharpoon{\ngLoc^{\ngReg}}$ --- each an atom
$\ngLoc_\ind$ paired with a region $\ngReg_\ind$ at the empty path, the
atoms pairwise distinct --- together with a vector of \emph{targets}
$\overharpoon{\ngLoc'^{{\ngReg}'.\ngPath'}}$ of the same length: root
location $\ngLoc_\ind^{\ngReg_\ind}$ is replaced by target
$\ngLoc_\ind'^{{\ngReg}_\ind'.\ngPath_\ind'}$. Call $\ngReg_\ind$ one
of $\sigma$'s \emph{domain regions} and $\ngLoc_\ind$ one of its
\emph{distinguished atoms} at that region. Several roots may share one
domain region; the definition below is stated for that case and
specializes to one root per region.

$\sigma$ is \emph{well formed} when two conditions hold of its two
vectors.
\begin{enumerate}
  \item\label{ls:prefix} \emph{One target prefix per domain region.}
    $\ngReg_\ind = \ngReg_\kappa \implies
     {\ngReg}_\ind'.\ngPath_\ind' = {\ngReg}_\kappa'.\ngPath_\kappa'$.
    Write $\ngPos_\sigma(\ngReg)$ for that one target position, at each
    domain region $\ngReg$.
  \item\label{ls:atom} \emph{One target atom per distinguished atom.}
    $\ngLoc_\ind = \ngLoc_\kappa \implies
     \ngLoc_\ind' = \ngLoc_\kappa'$. The roots' own atoms are pairwise
    distinct, so this is a condition on a repeated occurrence of one
    root in the two vectors.
\end{enumerate}
At a well-formed $\sigma$, define for each domain region $\ngReg$ the
\emph{atom map} $a_{\ngReg}$ on location atoms:
\[
  a_{\ngReg}(\ngLoc_\ind) = \ngLoc_\ind'
  \quad\text{at each root}\; \ngLoc_\ind^{\ngReg}\;\text{of}\;\sigma
  \;\text{standing at}\;\ngReg ,
  \qquad
  a_{\ngReg}(\ngLoc'') = \ngLoc''
  \quad\text{at every other atom} .
\]
Item~\ref{ls:atom} makes $a_{\ngReg}$ a function, and the roots'
pairwise-distinct atoms make its two clauses agree nowhere. The action
on a location is then
\[
  \ngLoc^{\ngReg.\ngPath}\sigma \;=\;
    a_{\ngReg}(\ngLoc)^{\ngPos_\sigma(\ngReg).\ngPath}
  \quad\text{at a domain region}\;\ngReg ,
  \qquad
  \ngLoc^{\ngPos}\sigma \;=\; \ngLoc^{\ngPos}
  \quad\text{at every other region} ,
\]
the residual path $\ngPath$ carried unchanged beneath the target
prefix. The induced actions are these and no others.
\begin{itemize}
  \item \emph{On atoms.} At a domain region $\ngReg$, the atom of a
    location standing there is rewritten by $a_{\ngReg}$; an atom
    standing at no domain region is untouched. An atom is rewritten
    relative to the region it stands at, never absolutely: one atom
    may be distinguished at one domain region and fixed at another.
  \item \emph{On positions.} $\ngPos\sigma$ is
    $\ngPos_\sigma(\ngReg).\ngPath$ at $\ngPos = \ngReg.\ngPath$ for a
    domain region $\ngReg$, and $\ngPos$ itself otherwise. The
    positional action reads item~\ref{ls:prefix} alone and is
    independent of every atom map.
  \item \emph{On locations.} As displayed above: position and atom
    together.
  \item \emph{On terms.} $\EXPR\sigma$ rewrites every location
    occurring in $\EXPR$ --- in a $\gramkw{letloc}$'s bound location and
    its location expression, in a $\gramkw{case}$ clause's annotations,
    in a $\gramkw{datacon}$'s destination, in an application's location
    vector, and in every located-type annotation --- and leaves every
    term variable, constructor name and function name alone.
  \item \emph{On located types.} $(\TYP@\ngLoc^{\ngPos})\sigma =
    \TYP@(\ngLoc^{\ngPos}\sigma)$: the type component is untouched.
  \item \emph{On static environments.} $\TENV\sigma$ rewrites each
    binding's located type; $\SENV\sigma$, $\NENV\sigma$ and
    $\UENV\sigma$ are set images under the action on locations;
    $\CENV\sigma$ rewrites each key and each location its value names;
    and $\AENV\sigma$ is read region by region. The action on allocation
    environments is defined at a \emph{prefix-separated} $\sigma$: two
    distinct domain regions whose target positions stand in one region
    carry target paths neither of which is a prefix of the other, so
    that each position lies at or below at most one domain region's
    target position. A region $\ngReg$ that
    $\sigma$ moves to the position ${\ngReg}'.\ngPath'$ loses its key.
    A target region ${\ngReg}'$ that $\sigma$ does not itself move
    holds a family, started whatever $\ngReg$ and ${\ngReg}'$ held. For
    each domain region $\ngReg$ whose target position is
    ${\ngReg}'.\ngPath'$, at a path $\ngPath'.\ngPath''$ it observes the
    rewritten location wherever $\ngReg$'s family observes a location at
    $\ngReg.\ngPath''$, and nothing otherwise; at a path below no domain
    region's target position it holds what $\AENV$'s own family at
    ${\ngReg}'$ holds, if $\AENV$ has one there, and nothing otherwise. Every other region
    keeps its entry. A
    set image is defined whether or not $\sigma$ is injective; where
    $\sigma$ identifies two members, the image holds one.
\end{itemize}
Well-formedness makes $\sigma$ a function. It does \emph{not} make
$\sigma$ injective: two
distinct roots at two different domain regions may carry one target
location, and two domain regions may carry target prefixes one of
which extends the other. Such a $\sigma$ acts on locations, terms,
located types, variable environments and location sets as above, but it is not
prefix-separated and has no action on allocation environments: at a
path below both prefixes, two domain regions' families would each supply
the one observation there. Where a result needs injectivity --- on
locations, on positions, or on both --- or the allocation action, it
obtains the condition separately.

\emph{Relocations.} A \emph{relocation} is a map $f$ on locations
together with a map $g$ on positions, a set of \emph{moved atoms}, a set
of \emph{moved regions} and a set of \emph{target atoms}, such that
$f$ moves a location's position by $g$ alone ($\ngLoc^{\ngPos}f$ stands
at $g(\ngPos)$); $g$ carries a residual path unchanged
($g(\ngPos.\ngPath) = g(\ngPos).\ngPath$) and fixes every position of a
region that is not moved; $f$ fixes every location whose atom and whose
region are both unmoved, keeps every atom that is not moved, and sends
every moved atom to a target atom. A relocation acts on terms, located
types, variable environments and location sets as a location substitution
does, location by location; it is not asked to be injective, and it is
given no action on location-constraint or allocation environments. The
action of a well-formed location substitution is a relocation, with its
distinguished atoms moved, its domain regions moved, and its roots' and
targets' atoms as target atoms.
\end{definition}
\begin{definition}[Reduct package]
\label{def:reduct-package}
Fix declarations $\overharpoon{\DD},\overharpoon{\FD}$ and a bootstrap
location $\ngLoc^{\ngReg}$. Let $D_{\mathrm{step}}$ derive
\[
  \ngJT{\Theta}{\ngTup{\TENV}{\SENV}{\CENV}}{\ngTup{\AENV}{\NENV}{\UENV}}
  {\EXPR : \ngLocatedTy}{\ngTup{\AENV_o}{\NENV_o}{\UENV_o}}\]
at a region typing environment $\Theta$, discharged against
$\NENV_{\mathrm{tgt}},\UENV_{\mathrm{tgt}}$; let $\ngCtx_{\mathrm{out}}$
be an evaluation context, so that $\EXPR$ stands in the whole
expression $\ngCtxFill{\ngCtx_{\mathrm{out}}}{\EXPR}$; and let
$\ngStepCfg{\Theta}{\STOR}{\MENV}{\EXPR}{\Theta''}{\STOR''}{\MENV''}{\EXPR''}$
lifting the basic contraction $B$, with induced renaming $\rho$
(Definition~\ref{def:step-renaming}). Write $Q_B$ for the set of
regions the contraction allocates: $Q_B = \set{\ngReg_f}$ when $B$
is \ngRuleName{D-LetRegion} with fresh region $\ngReg_f$, and
$Q_B = \emptyset$ at every other contraction. The \emph{reduct package}
holds when there are
$\SENV'',\CENV'',\AENV'',\NENV'',\UENV''$, $\AENV'''$ and a set
$\mathcal{V}$ of locations, the \emph{additional output consumable
locations}, with:
\begin{enumerate}
  \item\label{rp:theta} $\Theta \sqsubseteq \Theta''$
    (Definition~\ref{def:region-typing-environment-order});
  \item\label{rp:typing} a derivation of
    $\ngJT{\Theta''}{\ngTup{\TENV}{\SENV''}{\CENV''}}{\ngTup{\AENV''}{\NENV''}{\UENV''}}
  {\EXPR'' :
     \ngLocatedTy\,\rho}{\ngTup{\AENV'''}{\NENV_o\rho}{\UENV_o\rho \cup \mathcal{V}}}$,
    at the same located type up to $\rho$;
  \item\label{rp:outputs} that derivation's output pending-location and
    consumable-value sets are $\NENV_o\rho$ and
    $\UENV_o\rho \cup \mathcal{V}$, as item~\ref{rp:typing} displays
    them, where every member of $\mathcal{V}$ stands at the root of a
    region of $Q_B$, and no member of $\UENV_o\rho$ stands at a region
    of $Q_B$; so $\mathcal{V} = \emptyset$ at every contraction but
    \ngRuleName{D-LetRegion};
  \item\label{rp:settled} it is discharged against
    $\NENV_{\mathrm{tgt}}\rho$ and
    $\UENV_{\mathrm{tgt}}\rho \cup \mathcal{V}$
    (Definition~\ref{def:settled});
  \item\label{rp:grounded} $\TENV,\SENV'',\CENV'',\AENV'',\NENV'',\UENV''$
    satisfy environment-name containment for
    $\STOR'';\MENV'';\ngCtxFill{\ngCtx_{\mathrm{out}}}{\EXPR''}$, the
    whole reduct expression (Definition~\ref{def:grounded});
  \item\label{rp:singly} $\EXPR''$ satisfies global binder uniqueness and
    scope discipline against the same
    $\overharpoon{\FD}$ and $\ngLoc^{\ngReg}$
    (Definition~\ref{def:singlybound});
  \item\label{rp:invariant}
    $\Theta'';\; \TENV;\SENV'';\CENV'';\AENV'';\NENV'';\UENV''
      \vdash_{\mathit{inv}} \MENV'';\STOR'';\EXPR''$
    (Definition~\ref{def:wfinv}), at the same $\TENV$
    item~\ref{rp:typing} exhibits: store well-formedness of those five
    components with $\MENV''$ and $\STOR''$, and adequacy of $\Theta''$
    for them, for the runtime state and for the declarations and
    bootstrap data beside them;
  \item\label{rp:senv} $\SENV'' \supseteq \SENV\rho$;
  \item\label{rp:cenv} $\CENV''$ agrees with $\CENV\rho$ at every key
    the latter has, and carries at most one key besides, the location
    $B$ binds if $B$ binds one, at the entry $B$'s rule records there
    --- the count being of keys and taken against $\CENV\rho$, and a
    contextual $\ngRuleName{T-Let}$ layer adding no second key and
    changing no entry, since it returns the inner package's own
    $\CENV''$, so the clause holds of it as of the inner package;
  \item\label{rp:aenv} $\AENV'''$ agrees with $\AENV_o\rho$ at every
    position at which $\NENV_o\rho$ holds a location and at every
    position $\ngReg.\ngEmptyPath$ that is a key of $\AENV$, agreement
    meaning that both observations are defined there and are the same
    observation (Definition~\ref{def:aenv-core}): either both are
    $\phi$, or both are one and the same location; and
    every entry $\AENV'''$ holds names only atoms occurring in
    $\TENV,\SENV,\CENV,\AENV,\NENV,\UENV$, in $\EXPR$, or among the
    names the contraction freshens; and, the \emph{cone clause},
    $\AENV'''$ observes as $\AENV_o\rho$ (the two give the same
    observation wherever either is defined; unlike agreement, both may
    be undefined) at every position except these:
    \begin{itemize}
      \item at a position of a region $\ngReg_f$ of $Q_B$, where the
        clause asserts nothing of $\AENV'''$ and asks only that
        $\AENV_o\rho(\ngReg_f)$ be undefined, the redex's output having
        been masked there;
      \item where $B$ is \ngRuleName{D-App}, with output location
        $\ngLoc_{out}^{\ngPos_{out}}$ at
        $\ngPos_{out} = \ngReg'_{out}.\ngPath_0$, at a position $q$ of
        $\mathcal{C}^\circ = \mathcal{C} \setminus \set{\ngPos_{out}}$,
        where $\mathcal{C} = \set{\ngReg'_{out}.\ngPath_c \mid
        \ngPath_c \in \ngDesc{\ngReg'_{out}}{\ngPath_0}}$, $\AENV'''$
        may observe $\ngLoc_{out}^{\,q}$ instead;
    \end{itemize}
    and where, at \ngRuleName{D-App}, $\AENV'''$ and $\AENV_o\rho$
    disagree at some position of $\mathcal{C}^\circ$, the output
    $\AENV_o\rho$ and $\NENV_o\rho$ satisfy the three \emph{cone
    conditions} relative to $\AENV_\bullet$, the input allocation
    environment of the node $D_B$ item~\ref{rp:pending} names. For an
    allocation environment $\AENV_1$ and a pending-location set
    $\NENV_1$, these are:
    \begin{enumerate}
      \item\label{rp:cone-pending} $\NENV_1$ holds no location at any
        position of $\mathcal{C}^\circ$;
      \item\label{rp:cone-out} $\ngLoc_{out}^{\ngPos_{out}} \notin
        \NENV_1$; and
      \item\label{rp:cone-agree} $\AENV_1$ observes as $\AENV_\bullet$
        at every position of $\mathcal{C}$.
    \end{enumerate}
  \item\label{rp:pending} $\ngPendUp{B}{D_B}{\NENV_B''}$
    (Definition~\ref{def:pending-update}), where $\ngCtx$ is the
    evaluation context of the $\ngRuleName{D-Ctx}$ instance the
    reduction lifts, $D_B$ is the node of $D_{\mathrm{step}}$ the
    $\ngRuleName{T-Let}$ first-premise descent of $\ngCtx$'s own layer
    count reaches (Definition~\ref{def:let-descent}), and $\NENV_B''$
    is the input pending-location set of the node
    item~\ref{rp:typing}'s derivation's descent of that same length
    reaches; and
  \item\label{rp:forward} the reduction writes forward
    (Definition~\ref{def:writes-forward}); and
  \item\label{rp:local} every new entry of $\SENV''$, $\CENV''$,
    $\AENV''$, $\AENV'''$, $\NENV''$ and $\UENV''$ --- compared against
    $\SENV\rho$, $\CENV\rho$, $\AENV\rho$, $\AENV_o\rho$, $\NENV\rho$
    and $\UENV\rho$ respectively --- and every member of $\mathcal{V}$
    is locally provided
    (Definition~\ref{def:local-provenance}) at
    $D_{\mathrm{step}}$'s own subject $\EXPR$.
\end{enumerate}
\end{definition}
\begin{definition}[Rooted typing derivation]
\label{def:rooted}
$D_{\mathrm{step}}$ is \emph{rooted in $D_{\mathrm{root}}$ against
$\MENV$}, written
\[
  D_{\mathrm{step}} \sqsubseteq^{\MENV}_{\mathrm{rt}} D_{\mathrm{root}} ,
\]
when there are a derivation $D_0$ and a finite set of regions $Q$ with
\begin{enumerate}
  \item\label{rt:descent} $D_0 \preceq_{\gramkw{let}} D_{\mathrm{root}}$
    (Definition~\ref{def:let-descent}), so that $D_0$ is the
    subderivation of $D_{\mathrm{root}}$ rooted at a node reached from
    $D_{\mathrm{root}}$'s own root by following zero or more
    $\ngRuleName{T-Let}$ first expression premises;
  \item\label{rt:restrict} $D_{\mathrm{step}}$ the $Q$-restriction of
    $D_0$ against $\MENV$ (Definition~\ref{def:inert-restriction});
  \item\label{rt:discharged} $D_{\mathrm{root}}$ discharged outright
    (Definition~\ref{def:settled}); and
  \item\label{rt:binders} $D_{\mathrm{root}}$'s subject satisfying
    global binder uniqueness and scope discipline against the
    declarations and the bootstrap location
    (Definition~\ref{def:singlybound}).
\end{enumerate}
\end{definition}
\begin{definition}[T-Let first-premise descent]
\label{def:let-descent}
For derivations of the expression judgment, write
$D' \preceq_{\gramkw{let}} D$ when
$D'$ is the subderivation of $D$ rooted
at a node reached from $D$'s own root by following zero
or more links, each link passing from a $\ngRuleName{T-Let}$ node to
the child concluding that node's first expression premise
(Definition~\ref{def:derivation}\ref{der:sub}).
\end{definition}
\begin{definition}[Local provenance]
\label{def:local-provenance}
Fix a subject expression $\EXPR$ and the renaming $\rho$ a basic
contraction induces. A location atom or a region is \emph{locally
provided} when it occurs in $\EXPR$ or is one of the names that
contraction freshens. Each
comparison below fixes what counts as a \emph{new entry} of its own
component, and asks that every location atom and every region occurring
in one be locally provided. Atoms and regions are asked for separately
throughout: a location contributes both its atom and its position's
region.
\begin{enumerate}
  \item\label{loc:sets} \emph{The three sets.} A new entry of
    $\SENV''$ against $\SENV\rho$, of $\NENV''$ against $\NENV\rho$, and
    of $\UENV''$ against $\UENV\rho$, is a member of the first not in
    the second. Its names are its own atom and its position's region.
  \item\label{loc:cenv} \emph{The location-constraint environment.} A
    new entry of $\CENV''$ against $\CENV\rho$ is a key of
    $\ngDom{\CENV''}$ that is not a key of $\ngDom{\CENV\rho}$, taken
    together with the value $\CENV''$ holds there;
    item~\ref{rp:cenv} already forbids an old key at a changed value,
    so no such pair arises. Its names are the key's own atom and
    region, and every atom and region its value writes: the argument
    location of an $\gramkw{advanceTags}$, $\gramkw{after}$ or
    $\gramkw{advanceField}$ form, the region and type of a
    $\gramkw{start}$ form, and the located type the last two carry,
    whose own location counts among them.
  \item\label{loc:aenv} \emph{The allocation environment.} A new entry
    of $\AENV''$ against $\AENV\rho$ is of either of two kinds: a region
    key of $\ngDom{\AENV''}$ that is not a key of $\ngDom{\AENV\rho}$,
    whose names are that region together with the atoms of the family
    value held there --- $\ngUnstarted$ holding none --- or a position
    at which both are defined and observe different locations
    (Definition~\ref{def:aenv-core}), whose names are that position's
    region and the atom $\AENV''$ observes.
  \item\label{loc:aenv3} \emph{The exhibited allocation environment.} A
    new entry of $\AENV'''$ against $\AENV_o\rho$ is of either of two
    kinds, read as in item~\ref{loc:aenv}: a position at which the two
    observe different locations, or a region key of $\AENV'''$ that is
    not a key of $\AENV_o\rho$, whose names are that region together
    with the atoms of the family value held there.
  \item\label{loc:revealed} \emph{The additional output consumable locations.}
    Every member of the set $\mathcal{V}$ of
    Definition~\ref{def:reduct-package} is a new entry of the reduct's
    output consumable-value set against $\UENV_o\rho$. Its names are
    its own atom and its position's region.
\end{enumerate}
Item~\ref{loc:revealed} is vacuous at every contraction but
$\ngRuleName{D-LetRegion}$, where $\mathcal{V}$ may be non-empty
(item~\ref{rp:outputs} of Definition~\ref{def:reduct-package}).

\end{definition}
\begin{definition}[Pending-location update]
\label{def:pending-update}
Let $B$ be a basic contraction, derived under whichever enclosing
expression, and let $D_B$ derive the expression
judgment whose subject is $B$'s own --- the redex --- with input
pending-location set $\NENV_B$ and output set $\NENV_{B,o}$, and let
clauses~\ref{swf:constraints} and~\ref{swf:nursery} of
Definition~\ref{def:storewf} hold of $D_B$'s own input components with
$\MENV$ and $\STOR$, $\STOR$ satisfying condition~\ref{wlog:once} (Definition~\ref{def:storeview}). For a set
$\NENV_B''$, write $\ngPendUp{B}{D_B}{\NENV_B''}$ when the row $B$'s
rule selects holds. Every name a row uses is read either at $B$ or at
$D_B$'s own last rule instance, and at no other node.
\[
  \begin{array}{@{}l@{\quad}l@{}}
    \ngRuleName{D-Let-Val} & \NENV_B'' = \NENV_B \\
    \ngRuleName{D-LetRegion} & \NENV_B'' = \NENV_B \\
    \ngRuleName{D-Case} & \NENV_B'' = \NENV_B \\
    \ngRuleName{D-LetLoc-Start} &
      \NENV_B'' = \NENV_B \cup
        \set{\ngLoc_f^{\ngReg.\ngPath} \mid
          \ngPath \in \ngPathsOf{\TYP_{\ngReg}}} \\
    \ngRuleName{D-LetLoc-Advance-Field} &
      \NENV_B'' = \NENV_B \cup \set{\ngLoc_f^{\ngReg.\ngPath}} \\
    \ngRuleName{D-LetLoc-Advance-Tags} &
      \NENV_B'' = \Big(\NENV_B \cup
        \set{\ngLoc_f^{\ngReg.\ngPath.\ngPath_c} \mid
          \ngPath_c \in \ngPathsOf{\ngTypeAt{\TYP_{\ngReg}}{\ngPath}}}\Big)
        - \NENV_r \\
    \ngRuleName{D-LetLoc-After} &
      \NENV_B'' = \NENV_B \cup
        \set{\ngLoc_f^{\ngReg.\ngPath.\ngPath_c} \mid
          \ngPath_c \in \ngPathsOf{\ngTypeAt{\TYP_{\ngReg}}{\ngPath}}} \\
    \ngRuleName{D-App} & \NENV_B'' = \NENV_B \cup F \\
    \ngRuleName{D-DataConstructor} & \NENV_B'' = \NENV_{B,o}
  \end{array}
\]
$\ngLoc_f$ is the atom $B$ freshens and $\ngReg.\ngPath$ the position
$B$'s own $\gramkw{letloc}$ binds at, both read at $B$. $\NENV_r$ is
the set $D_B$'s own $\ngRuleName{T-LetLoc-Advance-Tags}$ where-block
removes, and it is not a datum of $B$: it is built on
$\ngLoc_0 = \ngChainRoot{\CENV}{\ngLoc'^{\ngReg.\ngPath}}$ at $D_B$'s
own location-constraint environment, which the dynamic rule does not
carry. That chain root is defined, and the two clauses asked above are
what make it so: clause~\ref{swf:constraints} is
$\crealj{\STOR}{\CENV}{\MENV}$, and
Lemma~\ref{lem:creal-wellfounded} answers at every location $\MENV$
resolves, the rule's own $\ngLoc'^{\ngReg.\ngPath} \in \NENV$ premise
and clause~\ref{swf:nursery} putting the argument among them.

Those two clauses are all this definition asks, and it asks them for
that one row. The other eight read nothing beyond $B$ itself, $D_B$'s
own judgment and the declarations: five of them return
$\NENV_B$ unchanged or extended by a family built on the atom $B$
freshens at the position $B$'s own $\gramkw{letloc}$ binds, both read
at $B$; $\ngRuleName{D-App}$'s reads $B$'s location vector and
$\fvar$'s scheme; and $\ngRuleName{D-DataConstructor}$'s reads
$\NENV_{B,o}$, a component of $D_B$'s own judgment. Both clauses
descend: they are read at the six input components
Lemma~\ref{lem:descent} carries down unchanged.
$F$ is
the output family $D_B$'s own $\ngRuleName{T-App}$ where-block names,
\[
  F \;=\; \set{\ngLoc_{out}^{\ngPos_{out}.\ngPath} \;\mid\;
    \ngPath \in \ngPathsOf{\TYP}} ,
\]
at the call's own output location --- the caller's, not a callee
formal --- and $\rho$ is the identity at
$\ngRuleName{D-App}$, so no name in the row is renamed. Unlike
$\NENV_r$, $F$ is recoverable from $B$ alone together with the
declarations: $B$'s location vector has $\ngLoc_{out}^{\ngPos_{out}}$
as its last entry, and $\fvar$'s scheme, which the declarations fix,
gives $\TYP$ and hence $\ngPathsOf\TYP$.
\end{definition}
\begin{definition}[Writing forward]
\label{def:writes-forward}
Let $\ngStepCfg{\Theta}{\STOR}{\MENV}{\EXPR}{\Theta''}{\STOR''}{\MENV''}{\EXPR''}$
lift $B$, and let $\CENV$ be the location-constraint environment of the
pre-reduction state. The reduction \emph{writes forward} when, for every
event of $\STOR''$ that is not an event of $\STOR$ (Lemma~\ref{lem:step-appends}
making these exactly the events appended) with address $\ngCell{\ngReg.\ngPath}{i}$ and with
$\ngLoc_w^{\ngReg.\ngPath}$ the destination the contraction writes at,
and for every construction ancestor $\ngLoc_y^{\ngReg.\ngPath_0}$ of
$\ngLoc_w^{\ngReg.\ngPath}$ in $\CENV$
(Definition~\ref{def:construction-ancestry}) \emph{that $\MENV$
resolves} (Definition~\ref{def:mresolve}), writing
$\ngPath = \ngPath_0.\ngPath_1$,
\[
  \ngProj{\MENV(\ngLoc_y^{\ngReg.\ngPath_0})}{\ngPath_1}(\ngEmptyPath)
  \;\le\; i ,
\]
both indices read in $\STOR(\ngReg.\ngPath)$.
\end{definition}
\begin{definition}[Construction ancestry]
\label{def:construction-ancestry}
Let $\ngLoc_x^{\ngReg.\ngPath}$ be a location. A \emph{predecessor
chain from $\ngLoc_x$} is a sequence
$\ngLoc_{\mathrm{anc},0}^{\ngReg.\ngPath},\ldots,
\ngLoc_{\mathrm{anc},n}^{\ngReg.\ngPath}$ ($n \ge 0$) with
$\ngLoc_{\mathrm{anc},0} = \ngLoc_x$ and, for $0 \le k < n$,
$\ngLoc_{\mathrm{anc},k}^{\ngReg.\ngPath}$ a successor of
$\ngLoc_{\mathrm{anc},k+1}^{\ngReg.\ngPath}$
(Definition~\ref{def:poschain}); $n = 0$ gives the one-element chain
$\ngLoc_{\mathrm{anc},0} = \ngLoc_x$ alone. An atom $\ngLoc_y$ at a
position $\ngPos_y$ is a \emph{construction ancestor} of
$\ngLoc_x^{\ngReg.\ngPath}$ when, for some predecessor chain
$\ngLoc_{\mathrm{anc},0},\ldots,\ngLoc_{\mathrm{anc},n}$ from
$\ngLoc_x$, either
$\ngLoc_y^{\ngPos_y} = \ngLoc_{\mathrm{anc},n}^{\ngReg.\ngPath}$, or the
location $\ngLoc_y^{\ngPos_y}$ is built on $\ngLoc_{\mathrm{anc},n}$'s
own atom, standing at a proper prefix position of $\ngReg.\ngPath$:
$\ngLoc_y^{\ngPos_y} = \ngLoc_{\mathrm{anc},n}^{\ngReg.\ngPath_0}$ for
some $\ngPath_0$ with $\ngPath = \ngPath_0.\ngPath_1$ and
$\ngPath_1 \neq \ngEmptyPath$.

Every construction ancestor arises this way: zero or more predecessor
links taken backward from $\ngLoc_x$, ending at some
$\ngLoc_{\mathrm{anc},n}$, followed by at most one move to the location
built on $\ngLoc_{\mathrm{anc},n}$'s own atom, standing at a strictly
shorter, prefix position -- never the other order, and never more than
one such move. Taking $n = 0$ and no prefix move makes $\ngLoc_x$ its
own construction ancestor. This is a static relation on locations,
induced by $\CENV$'s own predecessor links and by position prefixes
alone, read at one fixed location-constraint environment; it records no history of when
an entry was introduced.
\end{definition}

\endgroup

\section{Structure of the Type-Safety Proof}\label{sec:appendix-proof-structure}

Section~\ref{sec:type-safety} states progress, preservation and type safety for
\ourcalc{}. This appendix gives the lemmas their proofs rest on, each with a
sketch of its proof; the supplementary material contains every proof in full.
Type safety (Theorem~\ref{thm:type-safety}) follows from the well-formedness
of a program's initial configuration (Lemma~\ref{lem:initial-cfgwf}) and from
a general form of type safety, which in turn composes preservation over a run
of steps and then applies progress (Section~\ref{sec:appendix-proof-safety}).
Progress reduces to a core lemma about typing derivations that a configuration
can step from (Section~\ref{sec:appendix-proof-progress}). Preservation reduces
to a core lemma that dispatches each basic reduction to the rule that typed its
redex (Section~\ref{sec:appendix-proof-preservation}); both cores read a typing
derivation backwards through an inversion lemma, and preservation substitutes
values and locations into derivations
(Section~\ref{sec:appendix-proof-substitution}).

\begingroup
\small
\setlength{\emergencystretch}{3em}

\subsection{Type safety from progress and preservation}\label{sec:appendix-proof-safety}

\begin{lemma}[Type safety, general]
\label{lem:type-safety-general}
Let $\vdash \ngCfg{\Theta_0}{\STOR_0}{\MENV_0}{\EXPR_0} : \TYP$
hold (Definition~\ref{def:cfgwf}) and let
\[
  \ngStepCfgStar{\Theta_0}{\STOR_0}{\MENV_0}{\EXPR_0}
    {\Theta_k}{\STOR_k}{\MENV_k}{\EXPR_k} .
\]
Then $\Theta_0 \sqsubseteq \Theta_k$, and
$\vdash \ngCfg{\Theta_k}{\STOR_k}{\MENV_k}{\EXPR_k} : \TYP$, and
either $\EXPR_k$ is a $\VAL$, or there are
$\STOR_{k+1},\MENV_{k+1},\EXPR_{k+1}$ and a $\Theta_{k+1}$ with
$\Theta_k \sqsubseteq \Theta_{k+1}$ and
\[
  \ngStepCfg{\Theta_k}{\STOR_k}{\MENV_k}{\EXPR_k}
    {\Theta_{k+1}}{\STOR_{k+1}}{\MENV_{k+1}}{\EXPR_{k+1}} .
\]
\end{lemma}
\begin{proof}[Proof sketch]
The lemma combines two results and needs no induction of its own.
Lemma~\ref{lem:preservation-compose}, applied to the lemma's well-formedness hypothesis and the multistep reduction, gives $\Theta_0 \sqsubseteq \Theta_k$ and
$\vdash \ngCfg{\Theta_k}{\STOR_k}{\MENV_k}{\EXPR_k} : \TYP$.
Theorem~\ref{thm:progress}, applied to that reached configuration, gives the
last conclusion: $\EXPR_k$ is a $\VAL$, or the configuration steps to one whose
region typing environment extends $\Theta_k$.
\end{proof}

\begin{lemma}[Multistep preservation]
\label{lem:preservation-compose}
Let $\vdash \ngCfg{\Theta_0}{\STOR_0}{\MENV_0}{\EXPR_0} : \TYP$
hold (Definition~\ref{def:cfgwf}) and let
\[
  \ngStepCfgStar{\Theta_0}{\STOR_0}{\MENV_0}{\EXPR_0}
    {\Theta_n}{\STOR_n}{\MENV_n}{\EXPR_n} .
\]
Then
\[
  \Theta_0 \sqsubseteq \Theta_n
  \qquad\text{and}\qquad
  \vdash \ngCfg{\Theta_n}{\STOR_n}{\MENV_n}{\EXPR_n} : \TYP .
\]
\end{lemma}
\begin{proof}[Proof sketch]
By induction on the derivation of the multistep relation
(Definition~\ref{def:multistep}). At \textsc{MS-Refl} the configuration is
carried across unchanged, so well-formedness is the hypothesis itself and
$\Theta_0 \sqsubseteq \Theta_0$ holds by reflexivity. At \textsc{MS-Step},
Theorem~\ref{thm:preservation} applied to the first reduction gives
$\Theta_0 \sqsubseteq \Theta_1$ and well-formedness of the intermediate
configuration $\ngCfg{\Theta_1}{\STOR_1}{\MENV_1}{\EXPR_1}$. That
well-formedness is exactly the hypothesis the induction hypothesis needs for
the remaining, shorter multistep derivation, which yields
$\Theta_1 \sqsubseteq \Theta_n$ and well-formedness of the final
configuration. Transitivity of $\sqsubseteq$
(Definition~\ref{def:region-typing-environment-order}) composes the two
orderings.
\end{proof}

\subsection{Progress}\label{sec:appendix-proof-progress}

\begin{lemma}[Progress, step-ready form]
\label{lem:progress-core}
Let a derivation of
$\ngJT{\Theta}{\ngTup{\emptyset}{\SENV}{\CENV}}{\ngTup{\AENV}{\NENV}{\UENV}}
  {\EXPR : \TYP@\ngLoc_{\mathrm{res}}^{\ngReg_r.\ngPath_r}}{\ngTup{\AENV'}{\NENV'}{\UENV'}}$
be step-ready at
$\Theta$. Then either $\EXPR$ is a $\VAL$, or there are
$\STOR'',\MENV'',\EXPR''$ and $\Theta''$ with
$\Theta \sqsubseteq \Theta''$ and
$\ngStepCfg{\Theta}{\STOR}{\MENV}{\EXPR}{\Theta''}{\STOR''}{\MENV''}{\EXPR''}$.
\end{lemma}
\begin{proof}[Proof sketch]
We prove a stronger claim: a step-ready non-value has the form
$\ngCtxFill{\ngCtx}{r}$, and the redex $r$ contracts at some
$\Theta'' \sqsupseteq \Theta$ whatever enclosing term $W$ its freshness
premises are checked against. These are the only premises of the basic rules
(Section~\ref{sec:dynamic-semantics}) that read the enclosing term, and fresh
names always exist because $\STOR$, $\MENV$, $W$, the declarations and
$\ngDom\Theta$ are finite. The lemma is then the case $W = \EXPR$, lifted by
\textsc{D-Ctx}. The proof is by induction on the typing derivation, with
Lemma~\ref{lem:inversion} fixing the last rule from $\EXPR$'s outermost form.
\textsc{T-Var} cannot occur because the variable environment is empty, and
\textsc{T-Concrete-Loc} concludes at a value. In the \textsc{T-Let} case,
either \textsc{D-Let-Val} fires, or Lemma~\ref{lem:descent} shows that the bound expression's sub-derivation has the conclusion's six inputs; with the adequacy of $\Theta$, which step-readiness carries separately, it is step-ready at the same $\Theta$; the induction hypothesis then
supplies the redex, and the quantification over $W$ lets it step inside the
larger term. \textsc{T-LetRegion}, \textsc{T-LetLoc-Start} and \textsc{T-App}
look up nothing in $\STOR$ and resolve no location in $\MENV$: their dynamic
premises are fresh names, the start rule's type equation, a zero cursor that
exists under $\wfdecl$, and, for a call, a declaration, matching arities and
single-valued substitutions guaranteed by \textsc{T-App}'s premises and
$\tcfun$. In the remaining cases the store well-formedness clauses that
step-readiness carries supply what the dynamic rules read. The resolution
conjunct of clause~\ref{swf:nursery} defines $\MENV$ at the predecessor of a tag
or field advance and at a constructor's destination, and that clause's second
conjunct shows the destination address is unwritten.
Clause~\ref{swf:complete} supplies the traversal that \textsc{D-LetLoc-After}
reads, and the one that \textsc{D-LetLoc-Advance-Field} reads at the chain
root's field, which clause~\ref{swf:constraints} identifies with the
predecessor's cursor. For \textsc{T-Case}, clause~\ref{swf:complete} provides a
binding and a traversal at the scrutinee location; Lemma~\ref{lem:vwf-inv}
reads that traversal back to a written tag $\DC$ and derivable field
traversals, and \textsc{T-Case}'s coverage and distinctness premises select
exactly one clause for $\DC$.
\end{proof}

\begin{lemma}[Descent]
\label{lem:descent}
Let
\[
  \ngJT{\Theta}{\ngTup{\TENV}{\SENV}{\CENV}}{\ngTup{\AENV}{\NENV}{\UENV}}
  {\ngCtxFill{\ngCtx}{\EXPR_0} : \ngLocatedTy}{\ngTup{\AENV'}{\NENV'}{\UENV'}} .\]
Then the derivation contains a judgment for $\EXPR_0$ itself, at the
same $\Theta$ and with the same six inputs $\ngTup{\TENV}{\SENV}{\CENV};\,\ngTup{\AENV}{\NENV}{\UENV}$ as the conclusion
above. If $\ngCtx = [\,]$, its outputs are the conclusion's,
$\ngTup{\AENV'}{\NENV'}{\UENV'}$. Otherwise its outputs are those of the first
premise of $\ngCtx$'s \emph{innermost} $\gramkw{let}$, the one whose
bound expression is $\EXPR_0$; this is $\ngCtx$'s last layer. That
$\gramkw{let}$ need not be the innermost $\gramkw{let}$ of the subject.
The lemma asks nothing of $\ngCtx$'s size, so $\EXPR_0$ may itself be a
$\gramkw{let}$ when $\ngCtx$ is not taken as large as the grammar
allows. (Definition~\ref{def:evalfocus} takes $\ngCtx$ as large as
possible, and its decomposition is one of those this lemma applies
to.)

Two slots of the derivation carry that triple as $\EXPR_0$'s own
output and no third does: that $\gramkw{let}$'s first premise's own
output slot, and its second premise's input slot, which is what the
$\gramkw{let}$'s own continuation is read at. No slot further out is
one of them --- an enclosing $\gramkw{let}$ writes its \emph{second}
premise's outputs at its own output slot, and its first premise's
output slot carries the outputs of typing that premise's whole subject,
which is larger than $\EXPR_0$. This is a claim about which slots, not
about which values: two judgments of one derivation may carry equal
triples, and where a continuation leaves the allocation state alone
they do.
\end{lemma}
\begin{proof}[Proof sketch]
By induction on the structure of $\ngCtx$, whose grammar has two productions,
$[\,]$ and $\letpack{\var:\ngLocatedTy'}{\ngCtx'}{\EXPR_2}$. For
$\ngCtx = [\,]$ the conclusion itself is the judgment for $\EXPR_0$. In the
$\gramkw{let}$ case, Lemma~\ref{lem:inversion} makes the last rule
\textsc{T-Let}, whose first premise types $\ngCtxFill{\ngCtx'}{\EXPR_0}$ at the
conclusion's own six inputs unchanged. The induction hypothesis at the smaller
context $\ngCtx'$ then yields the judgment for $\EXPR_0$ at those inputs. Its
outputs are the first premise's when $\ngCtx' = [\,]$; otherwise they are those
of $\ngCtx'$'s innermost $\gramkw{let}$, and wrapping one more $\gramkw{let}$
outside leaves that layer innermost. The two-slot claim follows from
\textsc{T-Let}'s own shape: it reads its second premise at its first premise's
outputs, an enclosing $\gramkw{let}$ writes its second premise's outputs at its
own output slot, and its first premise's output slot types a subject strictly
larger than $\EXPR_0$.
\end{proof}

\begin{lemma}[Reading a traversal back]
\label{lem:vwf-inv}
Let $\wfdecl\;\overharpoon{\DD}$ hold.
Let $\STOR$ satisfy condition~\ref{wlog:once} (Definition~\ref{def:storeview}), $\STOR$ being the store read below.
If $\vwf{\STOR}{\ngPos}{\ngCurs{i}\,\overharpoon{CF}}{\TYP}{c'}$ is
derivable, then $\STOR(\ngPos)(i)$ is written; writing $\DC$ for the
tag it holds, there is a unique $\overharpoon{\TYP'}$ with
$(\DC\,\overharpoon{\TYP'}) \in \ctors{\TYP}$; the vector
$\overharpoon{CF}$ has an entry at $\ngFieldKey{\DC}{j}$ for each
$j \le m$, each read by the premise at $d_{j-1}$, every earlier entry
replacing its own key alone; and the traversals
\[
  \vwf{\STOR}{\ngPos.\ngFieldKey{\DC}{j}}{d_{j-1}(\DC,j)}{\TYP'_j}{g_j}
  \qquad\text{and}\qquad
  \vwf{\STOR}{\ngPos}{f_{k-1}}{\TYP}{f_k}
\]
that $\ngRuleName{VWF-Data}$ (Section~\ref{sec:vwf-derived-views}) carries
as premises are themselves derivable, at the $d_j$ and $f_k$ that rule
computes.
\end{lemma}
\begin{proof}[Proof sketch]
No induction is needed: \textsc{VWF-Data}
(Section~\ref{sec:vwf-derived-views}) is the judgment's only rule, and its
conclusion fixes the position, cursor and field vector it is read at. So any
derivation of the judgment is an instance of that rule at $\ngPos$, $i$ and
$\overharpoon{CF}$, and the traversals in the statement are that instance's
premises. The rule's first premise reads the tag at $\STOR(\ngPos)(i)$, so that
cell is written. Uniqueness of $\DC$ and $\overharpoon{\TYP'}$ follows from
$\wfdecl$ (Definition~\ref{def:wfdecl}). The vector has an entry at each factored key $\ngFieldKey{\DC}{j}$, $j \leq m$, because the corresponding premise reads $d_{j-1}(\DC,j)$ at that key.
\end{proof}

\subsection{Preservation}\label{sec:appendix-proof-preservation}

\begin{lemma}[Strengthened preservation]
\label{lem:preservation-core}
Fix declarations $\overharpoon{\DD},\overharpoon{\FD}$ and a bootstrap
location $\ngLoc^{\ngReg}$, with $\wfdecl\;\overharpoon{\DD}$ and
$\ngJFunT{\Theta}{\FD_\ind}$ for every $\ind \le \litnum$, at the region
typing environment $\Theta$ named next, each at a
clause-complete body derivation (Definition~\ref{def:clause-complete}). Let
$D_{\mathrm{root}}$ be a clause-complete derivation of the expression
judgment at a region typing environment $\Theta$, and let $D_{\mathrm{step}}$ derive
\[
  \ngJT{\Theta}{\ngTup{\TENV}{\SENV}{\CENV}}{\ngTup{\AENV}{\NENV}{\UENV}}
  {\EXPR : \ngLocatedTy}{\ngTup{\AENV_o}{\NENV_o}{\UENV_o}}\]
at that same $\Theta$, with
$D_{\mathrm{step}} \sqsubseteq^{\MENV}_{\mathrm{rt}} D_{\mathrm{root}}$
(Definition~\ref{def:rooted}): that relation asks $D_{\mathrm{root}}$
discharged outright (Definition~\ref{def:settled}) and its subject
under global binder uniqueness and scope discipline against those
declarations and that bootstrap location
(Definition~\ref{def:singlybound}), and supplies an actual descendant
$D_0 \preceq_{\gramkw{let}} D_{\mathrm{root}}$ of which
$D_{\mathrm{step}}$ is the $Q$-restriction against $\MENV$
(Definition~\ref{def:inert-restriction}), so that
Lemma~\ref{lem:descent} reads $D_0$'s six input components as
$D_{\mathrm{root}}$'s own and $D_{\mathrm{step}}$'s differ from them at
no key outside $Q$; let
\[
  \Theta;\; \TENV;\SENV;\CENV;\AENV;\NENV;\UENV
    \vdash_{\mathit{inv}} \MENV;\STOR;\EXPR
\]
hold (Definition~\ref{def:wfinv}); let
$\TENV,\SENV,\CENV,\AENV,\NENV,\UENV$ satisfy environment-name
containment (Definition~\ref{def:grounded}) for
$\STOR;\MENV;\ngCtxFill{\ngCtx_{\mathrm{out}}}{\EXPR}$, a whole
expression in which $\EXPR$ stands under the evaluation context
$\ngCtx_{\mathrm{out}}$ (the empty context when $\EXPR$ is that whole
expression);
let the location of every located type in the range of $\TENV$ be a
member of $\SENV$; let $\MENV$ bind the bootstrap location
$\ngLoc^{\ngReg}$; let $D_{\mathrm{root}}$'s input allocation
environment and $\MENV$ be of their sorts at $\Theta$
(Section~\ref{sec:worlds-static-envs});
and let $D_{\mathrm{step}}$ be discharged against
$\NENV_{\mathrm{tgt}},\UENV_{\mathrm{tgt}}$, with
\begin{enumerate}
  \item\label{core:bound} every location of $\NENV_{\mathrm{tgt}}$
    built on an atom a $\gramkw{letloc}$ or a $\gramkw{case}$-clause
    annotation of $\EXPR$ binds standing in $\NENV_o$; and
  \item\label{core:callout} every location of $\NENV_{\mathrm{tgt}}$
    built on the output atom of an application standing in $\EXPR$
    standing in $\NENV_o$.
\end{enumerate}
Finally let
\[
  \ngStepCfg{\Theta}{\STOR}{\MENV}{\EXPR}{\Theta''}{\STOR''}{\MENV''}{\EXPR''}
\]
hold, at the evaluation context $\ngCtx$ its one $\ngRuleName{D-Ctx}$
instance selects and over the basic contraction that instance lifts,
with $\rho$ the renaming that contraction induces
(Definition~\ref{def:step-renaming}), with $\NENV_{\mathrm{tgt}}$
holding no location built on an atom that contraction freshens, and
with no name that contraction freshens occurring in
$\ngCtxFill{\ngCtx_{\mathrm{out}}}{\EXPR}$. Then
the reduct package
(Definition~\ref{def:reduct-package}) holds at $D_{\mathrm{step}}$ and
that reduction, at $\ngCtx_{\mathrm{out}}$, with that same $\rho$.
\end{lemma}
\begin{proof}[Proof sketch]
The proof is by structural induction on the evaluation context $\ngCtx$ that the
single \textsc{D-Ctx} instance selects, measured by its number of
$\gramkw{let}$ layers. The reduction derivation itself has no proper
subderivation to recurse into, and every layer lifts the same basic
contraction, so $\Theta''$, $\STOR''$, $\MENV''$ and the renaming $\rho$
(Definition~\ref{def:step-renaming}) are fixed once for all cases. At the
empty context, Lemma~\ref{lem:inversion} fixes the typing rule from the redex's
outermost form, and each of the nine basic contractions has its own preservation lemma, whose reduct package (Definition~\ref{def:reduct-package}) the core takes unchanged. The package holds a typing derivation of the reduct at $\Theta''$, with the outputs renamed by $\rho$; store well-formedness (Definition~\ref{def:storewf}) and allocation-environment well-formedness (Definition~\ref{def:allocwf}) for the updated $\AENV$, $\NENV$, $\UENV$ and store; the adequacy of $\Theta''$, which completes the static/runtime invariant (Definition~\ref{def:wfinv}); and the agreement facts that the enclosing layers need. The core proof supplies each case lemma's case-specific hypotheses. The target pending set holds no location built on the atom a $\gramkw{letloc}$ binds, because the lemma's own condition places such locations in the output pending set, which the rule's premise keeps them out of. It holds none built on a call's output atom, by the matching condition, \textsc{T-App}'s where-clause and discharge (Definition~\ref{def:settled}). At a $\gramkw{let}$ layer,
Lemma~\ref{lem:inversion} makes the last rule \textsc{T-Let} and the hypotheses
descend to its first premise. The proof first drops from that premise the
unstarted regions named only in the continuation, and re-establishes
rootedness for the smaller derivation. The induction hypothesis then gives the
result for the inner reduction, and the preservation lemma for \textsc{D-Ctx}
reassembles \textsc{T-Let}: it restores the dropped unstarted regions and
transports the continuation's derivation to the inner reduct's new output
allocation environment, at the same renaming.
\end{proof}

\begin{lemma}[Inversion]
\label{lem:inversion}
Let
\[
  \ngJT{\Theta}{\ngTup{\TENV}{\SENV}{\CENV}}{\ngTup{\AENV}{\NENV}{\UENV}}
  {\EXPR : \ngLocatedTy}{\ngTup{\AENV'}{\NENV'}{\UENV'}} .\]
The last rule of the derivation is fixed by $\EXPR$'s own outermost
form, one of the seven productions of the expression grammar
(Section~\ref{sec:terms-values-programs}):
\[
\begin{array}{@{}l@{\qquad}l@{}}
  \var & \ngRuleName{T-Var} \\
  \ngLocVal{\ngLoc'^{\ngPos'}} & \ngRuleName{T-Concrete-Loc} \\
  \letpack{\var:\ngLocatedTy'}{\EXPR_1}{\EXPR_2} & \ngRuleName{T-Let} \\
  \letregty{\ngReg'}{\TYP'}{\EXPR_1} & \ngRuleName{T-LetRegion} \\
  \letloc{\ngLoc'^{\ngPos'}}{\ngStartRegTy{\ngReg'}{\TYP'}}{\EXPR_1}
    & \ngRuleName{T-LetLoc-Start} \\
  \letloc{\ngLoc'^{\ngPos'}}{\ngAfterTy{\ngLocatedTy'}}{\EXPR_1}
    & \ngRuleName{T-LetLoc-After} \\
  \letloc{\ngLoc'^{\ngPos'}}{\ngAdvTags{\ngLoc''^{\ngPos''}}{\DC}}{\EXPR_1}
    & \ngRuleName{T-LetLoc-Advance-Tags} \\
  \letloc{\ngLoc'^{\ngPos'}}
    {\ngAdvField{\ngLoc''^{\ngPos''}}{\DC}{j}{\ngLocatedTy'}}{\EXPR_1}
    & \ngRuleName{T-LetLoc-Advance-Field} \\
  \datacon{\DC}{\ngLoc'^{\ngPos'}}{\overharpoon{\VAL}}
    & \ngRuleName{T-DataConstructor} \\
  \case{\VAL}{\overharpoon{\pat}} & \ngRuleName{T-Case} \\
  \fapp{\overharpoon{\ngLoc'^{\ngPos'}}}{\overharpoon{\VAL}}
    & \ngRuleName{T-App}
\end{array}
\]
and matches that rule's own conclusion: the rule's metavariables
instantiate so its conclusion is exactly this judgment and its
premises hold under that instantiation. This is matching, not
substitution -- an output slot is written as the input metavariable
itself at some rules and as a computed expression at others, so the
match yields a binding at the former and an equation at the latter.

Read the output slots rule by rule. \ngRuleName{T-Var} and
\ngRuleName{T-Concrete-Loc} repeat their own inputs, so the match gives
$\AENV' = \AENV$, $\NENV' = \NENV$, $\UENV' = \UENV$ directly.
\ngRuleName{T-DataConstructor} writes an expression at all three output
slots, and \ngRuleName{T-App} at its pending-location and
consumable-value slots (its $\AENV$ slot repeating the input); at each
such slot the match gives the equation the rule states, and this is
the only sense in which either rule ``computes'' an output.
\ngRuleName{T-LetRegion} writes its body premise's own outputs with its
own region masked: if that premise concludes at outputs
$\ngTup{\AENV_1}{\NENV_1}{\UENV_1}$, the match gives
$\AENV' = \AENV_1 \setminus \ngReg'$, $\NENV' = \NENV_1$ and
$\UENV' = \UENV_1 \setminus \ngReg'$, the members of $\UENV_1$ at a
region other than $\ngReg'$. The remaining six rules copy a premise's
own outputs into the conclusion verbatim, so the match binds them and
computes nothing. At \ngRuleName{T-Case} the copied triple is the one
every clause premise shares. A \ngRuleName{T-Case} with no clause has
no premise, and then its output triple is unconstrained.

There is one thing the output slots never expose. A $\gramkw{letloc}$ rule's
own pending-location set, discharge included, is an input it hands to
its continuation premise; the conclusion's output pending-location set
is whatever that premise returns. A tag advance's intermediate
pending-location set is therefore invisible in the judgment's own
$\NENV'$, and an argument that needs it must read the premise, not the
conclusion.

Every premise but one is read at environments computable from the
conclusion's own inputs and the rule's where-clause. The exception is
\ngRuleName{T-Let}, whose second premise is read at the first premise's own
outputs; those are existentially quantified and the conclusion does not
name them. \ngRuleName{T-Let}'s \emph{first} premise is not an exception: it
receives the conclusion's own six inputs unchanged. Every premise takes the
conclusion's own region typing environment $\Theta$, except
\ngRuleName{T-LetRegion}'s body premise, which takes
$\Theta[\ngReg' \mapsto \TYP']$.
\end{lemma}
\begin{proof}[Proof sketch]
Fix the derivation; its last rule instance concludes the given judgment. That
rule is determined by case analysis on $\EXPR$'s outermost form, not by
induction. The expression productions are pairwise distinct and there is no
subsumption or weakening rule, so only rules whose conclusion has $\EXPR$'s
form can come last. The two productions that carry several rules, values
(\textsc{T-Var}, \textsc{T-Concrete-Loc}) and $\gramkw{letloc}$ (four rules),
are separated by the value grammar and by the four location-expression forms,
so the rule is unique, and it exists because it is the derivation's own last
rule. The output-slot claims are read off each rule's conclusion: a slot
written as a metavariable binds it, and a slot written as an expression yields
that equation. Every premise is read at environments fixed by the conclusion
and the where-clause, except \textsc{T-Let}'s second premise, which is read at
the first premise's outputs. Only \textsc{T-LetRegion}'s body premise changes
$\Theta$, to $\Theta[\ngReg' \mapsto \TYP']$
(Definition~\ref{def:derivation}).
\end{proof}

Three smaller results recur in the preservation proofs and in the definitions
they rely on: a reduction only appends to the store, restricting a derivation
commutes with its premises, and an unstarted region that a judgment never
names can be removed from its allocation environment. Their proofs are in the
supplementary material.

\begin{lemma}[A reduction appends at most one event]
\label{lem:step-appends}
Let $\ngStepCfg{\Theta}{\STOR}{\MENV}{\EXPR}{\Theta''}{\STOR''}{\MENV''}{\EXPR''}$.
Then $\STOR$ is an initial segment of $\STOR''$, and $\STOR''$ has at
most one event $\STOR$ lacks.
\end{lemma}
\begin{lemma}[Restriction at a premise]
\label{lem:restriction-premise}
Let $D_0$ be clause-complete (Definition~\ref{def:clause-complete}),
and let $D$ be the $Q$-restriction of $D_0$ against $\MENV$. Then $D$ and
$D_0$ carry the same subject and the same rule at every node --- so
Lemma~\ref{lem:inversion} determines one rule at each, and an instance
may be inverted at $D$ exactly where it may be inverted at $D_0$ ---
and, at every node and every expression premise of the rule standing
there, the subderivation of $D$ at that premise is the
$Q$-restriction, against that same $\MENV$, of the subderivation of
$D_0$ at the corresponding premise.
\end{lemma}
\begin{lemma}[Removing an unused unstarted region]
\label{lem:strengthen-a}
Let $\ngJT{\Theta}{\ngTup{\TENV}{\SENV}{\CENV}}{\ngTup{\AENV}{\NENV}{\UENV}}
  {\EXPR : \ngLocatedTy}{\ngTup{\AENV'}{\NENV'}{\UENV'}}$, let $\ngReg$ occur in none
of $\EXPR,\TENV,\SENV,\NENV,\UENV$ and $\ngLocatedTy$, nor at any key or
recorded argument of an entry of $\CENV$ (an
$\gramkw{advanceField}$ entry's located-type location is not
asked), and let
$\AENV(\ngReg) = \ngUnstarted$. Then
\[
  \ngJT{\Theta}{\ngTup{\TENV}{\SENV}{\CENV}}{\ngTup{\AENV \setminus \ngReg}{\NENV}{\UENV}}
  {\EXPR : \ngLocatedTy}{\ngTup{\AENV' \setminus \ngReg}{\NENV'}{\UENV'}}\]
is derivable, where $\AENV \setminus \ngReg$ removes the key $\ngReg$
-- the whole family entry, not one position of it -- from $\AENV$
(Definition~\ref{def:aenv-core}), and $\AENV' \setminus \ngReg$ removes
that same key from $\AENV'$.
\end{lemma}

\subsection{Substitution}\label{sec:appendix-proof-substitution}

\begin{lemma}[Value substitution]
\label{lem:subst-val}
Let $\ngJT{\Theta}{\ngTup{\TENV[\var \mapsto \TYP_1@\ngLoc_1^{\ngPos_1}]}{\SENV}{\CENV}}{\ngTup{\AENV}{\NENV}{\UENV}}
  {\EXPR : \ngLocatedTy}{\ngTup{\AENV'}{\NENV'}{\UENV'}}$
and let $\VAL$ be a value with
$\ngJT{\Theta}{\ngTup{\TENV}{\SENV}{\CENV}}{\ngTup{\AENV}{\NENV}{\UENV}}
  {\VAL : \TYP_1@\ngLoc_1^{\ngPos_1}}{\ngTup{\AENV}{\NENV}{\UENV}}$, and let the
location of every located type in the range of $\TENV$ be a member of
$\SENV$. Then
$\ngJT{\Theta}{\ngTup{\TENV}{\SENV}{\CENV}}{\ngTup{\AENV}{\NENV}{\UENV}}
  {\subst{\EXPR}{\var}{\VAL} : \ngLocatedTy}{\ngTup{\AENV'}{\NENV'}{\UENV'}}$.
\end{lemma}
\begin{proof}[Proof sketch]
By induction on the derivation for $\EXPR$. \textsc{T-Var} is the only rule
that reads a term variable. At the substituted occurrence of $\var$, its
premise $\TENV(\var) = \TYP_1@\ngLoc_1^{\ngPos_1}$ is exactly the located type
at which $\VAL$ is typed, so $\VAL$'s derivation takes the place of that
\textsc{T-Var} instance; because a value rule, like \textsc{T-Var}, leaves
$\AENV$, $\NENV$ and $\UENV$ unchanged, the outputs are the ones the
surrounding derivation expects. $\VAL$'s derivation stays valid at the
occurrence: $\SENV$ only grows between $\var$'s binder and its use, and
$\VAL$'s derivation reads $\SENV$ only through a positive membership premise;
beneath a $\gramkw{letregion}$ the leaf is rebuilt at the extended region
typing environment (Definition~\ref{def:derivation}). A binder of $\EXPR$ that
would capture a free variable of $\VAL$ is first renamed to a fresh variable.
The freshness premises of \textsc{T-LetRegion}, the $\gramkw{letloc}$ rules and \textsc{T-Pattern} still hold. Dropping $\var$'s entry removes names, and the entry of $\TENV$ at $\var$ that the drop may uncover has its location in $\SENV$ by the lemma's hypothesis, so every name a rule binds is already fresh for it.
\end{proof}

\begin{lemma}[Location substitution]
\label{lem:subst-loc}
Let $\sigma$ be a location substitution (Definition~\ref{def:locsubst})
with the one root location $\ngLoc^{\ngReg}$ and the one target
$\ngLoc'^{\,{\ngReg}'.\ngPath'}$, and let a clause-complete derivation
(Definition~\ref{def:clause-complete}) derive
$\ngJT{\Theta}{\ngTup{\TENV}{\SENV}{\CENV}}{\ngTup{\AENV}{\NENV}{\UENV}}
  {\EXPR : \ngLocatedTy}{\ngTup{\AENV'}{\NENV'}{\UENV'}}$, satisfying:
\begin{itemize}
  \item \emph{A fresh target region.} ${\ngReg}'$ occurs in none of
    $\EXPR,\TENV,\SENV,\CENV,\AENV,\NENV,\UENV$.
  \item \emph{The target atom.} $\ngLoc' = \ngLoc$, or $\ngLoc'$
    occurs in none of $\EXPR,\TENV,\SENV,\CENV,\AENV,\NENV,\UENV$.
  \item \emph{Avoidance.} The target $\ngLoc'^{\,{\ngReg}'.\ngPath'}$
    avoids $\EXPR$ (Definition~\ref{def:avoids}).
  \item \emph{No region-starting forms at $\ngReg$.} $\EXPR$ contains
    no $\gramkw{letregion}$ binding $\ngReg$ and no $\gramkw{start}$
    binding a location at $\ngReg$.
  \item \emph{The root case for $\ngRuleName{T-LetLoc-After}$.} Every
    $\ngRuleName{T-LetLoc-After}$ node of the derivation standing at
    $\ngReg$'s own root position has a location of its own input
    pending-location set there.
  \item \emph{Type agreement.}
    $\ngTypeAt{\TYP_{{\ngReg}'}}{\ngPath'} = \TYP_{\ngReg}$.
  \item \emph{A family at $\ngReg$.} $\AENV$ holds a family, not
    $\phi$, at $\ngReg$.
\end{itemize}
Then
\[
  \ngJT{\Theta}{\ngTup{\TENV\sigma}{\SENV\sigma}{\CENV\sigma}}{\ngTup{\AENV\sigma}{\NENV\sigma}{\UENV\sigma}}
  {\EXPR\sigma :
  \ngLocatedTy\,\sigma}{\ngTup{\AENV'\sigma}{\NENV'\sigma}{\UENV'\sigma}} ,\]
an environment rewritten pointwise, at keys and values alike, and an
allocation environment read region by region: $\ngReg$'s entry moves to
${\ngReg}'$, whose family observes $a_{\ngReg}(\ngFam(\ngPath''))$ at
${\ngReg}'.\ngPath'.\ngPath''$ wherever $\ngReg$'s family $\ngFam$
observes $\ngFam(\ngPath'')$ at $\ngReg.\ngPath''$, and nothing
elsewhere; every other region keeps its entry.
\end{lemma}
\begin{proof}[Proof sketch]
By induction on the derivation, checking each rule by reading its premises
through $\sigma$, which sends each location $\ngLoc''^{\ngReg.\ngPath''}$ to
$a_{\ngReg}(\ngLoc'')^{{\ngReg}'.\ngPath'.\ngPath''}$ and fixes every location
at another region. The key fact is that no judgment of the derivation names
${\ngReg}'$, nor $\ngLoc'$ unless $\ngLoc' = \ngLoc$: a premise's inputs come
from its conclusion's inputs and subject, and an output names no region or location atom that its judgment's subject and inputs do not carry. This last property needs
clause-completeness (Definition~\ref{def:clause-complete}), since a
\textsc{T-Case} with no clauses could otherwise conclude at outputs naming two
locations that $\sigma$ identifies. Hence $\sigma$ is injective on the
positions and locations read at every judgment, which preserves membership in
$\NENV$ and $\UENV$, the negative premises, $\ngFieldLinks{\CENV}$
(Definition~\ref{def:fieldlinks}), $\ngSources{\CENV}$
(Definition~\ref{def:sources}), chains, and every set difference a rule forms.
Positional premises carry over by type agreement, since
$\ngTypeAt{\cdot}{\cdot}$ composes along concatenated paths
(Definition~\ref{def:typeat}), and the freshness premises survive: a bound region is neither $\ngReg$, at which $\AENV$ holds a family, nor ${\ngReg}'$, which $\EXPR$ does not name, and avoidance keeps every bound atom distinct from $\ngLoc'$.
\end{proof}

\endgroup

\FloatBarrier
\section{Additional Compiler Material}
\label{sec:appendix-compiler}

This appendix adds four pieces to Section~\ref{sec:compiler-implementation}:
how a datatype's declaration determines the region tree of its factored
layout, how \colobus{} adapts \gibbon{}'s pointers to that layout, how \colobus{} lowers \ourcalc{} to \lamcur{}, and a program written
with \lamcur{}'s mutable cursors.

\subsection{Turning datatype descriptions into location structures}
\begin{algorithm}[tp]
\caption{Computing a datatype's region tree}
\label{alg:build-soa-loc}
\small
\begin{algorithmic}[1]
\State \textbf{Input:} datatype $\TYP$, environment $\Delta$
\State \textbf{Output:} the paths $\ngPathsOf{\TYP}$ of $\TYP$'s region tree
\State \textbf{Notation:} $\DC$: Data Constructor,\; $j$: Field Index,\; $\sigma$: Field Type.
\Function{Paths}{$\TYP,\Delta$}
\State $\Pi \gets \set{\ngEmptyPath}$
\ForAll{$(\DC,fields) \in \Call{ConstructorsOf}{\TYP,\Delta}$} \Comment{$[(\DC,[(j,\sigma)])]$}
\ForAll{$(j,\sigma) \in fields$}
\If{$\sigma \neq \TYP$} \Comment{$\ngFieldKey{\DC}{j} \in \ngFactored{\TYP}$; skip self-recursive fields}
\ForAll{$\ngPath \in \Call{Paths}{\sigma,\Delta}$}
\State $\Pi \gets \Pi \cup \set{\ngFieldKey{\DC}{j}.\ngPath}$
\EndFor
\EndIf
\EndFor
\EndFor
\State \Return $\Pi$ \Comment{finite: $\packs$ is acyclic}
\EndFunction
\end{algorithmic}
\end{algorithm}

Algorithm~\ref{alg:build-soa-loc} shows formally how a datatype determines its factored layout.
Given a datatype $\tau$ and environment $\Delta$, \lstinline|ConstructorsOf| returns constructors and their fields as $(K,(j,\sigma))$ pairs.
\textsc{Paths} computes $\tau$'s region tree: the empty path and, for each field whose type differs from $\tau$, that field's key followed by every path of the field's own type. Self-recursive fields add no path, since they stay at their parent's position.
\textsc{T-Program} and \textsc{T-FunctionDef} start a region with the constant family $\ngConstFam{\ngLoc}$ of Section~\ref{sec:static-semantics}, which sends every path of this tree to one fresh atom $\ngLoc$; the positions, not fresh names, distinguish the locations.

The pseudocode covers the \fullyfactored{} case, in which a field of another datatype is factored in turn. In \colobus{}, such a field keeps the layout of its own datatype, so declaring that datatype \unfactored{} yields the \hybridfactored{} layout of Section~\ref{sec:colobus-input}.

\subsection{Pointers in the Factored Layout}%
\label{sec:pointers}

In fully serialized data, reaching a field can take time linear in the size of the data serialized before it, even when the computation does not need that data.
To reach the right subtree of a binary tree in a serialized layout, for example, a traversal must first step over the left subtree.
Gibbon~\cite{LoCal} solves this problem with random-access pointers, which give constant-time access to the fields of a value.
To share values instead of copying them, it uses indirection pointers.
\gibbon{} also uses redirection pointers to link the chunks of a growing region.
This subsection describes how we adapt all three to the \fullyfactored{} layout.

\paragraph{Redirection pointers.}
\gibbon{} allocates a region as a sequence of chunks.
The first chunk is small; when a write might run past the end of the current chunk, \gibbon{} allocates a new chunk and links it from the current one with a \textit{redirection pointer}.
Each new chunk is twice the size of the previous one, up to a fixed maximum.
In the \unfactored{} layout a traversal reads a single buffer. Thus, on reaching a redirection pointer it simply jumps to the new chunk.
A factored value spans several buffers.
Thus, when any one of a value's buffers needs a new chunk, \colobus{} redirects the tag buffer and all of the value's scalar-field buffers together.
A traversal therefore checks for a redirection only in the tag buffer, and on finding one follows the redirection in every other buffer at once.
Redirecting together may leave a chunk partly unused, but since chunks grow geometrically we expect this fragmentation to be small.

\paragraph{Indirection pointers.}
\textit{Indirection} pointers let a new value share existing data instead of copying it.
In the \unfactored{} layout, sharing is encoded by an additional constructor, \lstinline|IndirectionE| \cil{Cursor}, that points to the shared value.
In the \fullyfactored{} layout, a shared value has one cursor per buffer, so the indirection carries all of them: \lstinline|IndirectionE| \cil{Cursor[N]}, where \lstinline|N| is the number of buffers the value occupies.
The indirection is written in the tag buffer alone.
A traversal that finds the indirection tag in the tag buffer simply continues from the \lstinline|N| stored cursors.
With garbage collection enabled, the region holding the tag buffer takes a reference to each of the \lstinline|N| shared buffers, so the collector keeps every one of them alive while the indirection is reachable.

\paragraph{Random-access pointers.}
A traversal that needs only some fields of a value may still have to step through the serialized fields before them~\cite{LoCal}.
In the \unfactored{} layout, \gibbon{} avoids this cost by storing, in each node, pointers to the fields that come after the first packed field.
\colobus{} applies the same idea to the \fullyfactored{} layout.
Its random-access nodes are stored in the tag buffer, and for a factored field they hold one pointer per buffer of that field, so a traversal can jump directly to a field whose data lies in other buffers.

\subsection{From \ourcalc{} to \lamcur}%
\label{sec:socal-to-nocal}

\begin{figure}
  \centering
  \input{formal_nocal_grammar.tex}
  \caption{Grammar of \lamcur{} (an extension of \ourcalc{}). Operations added by \colobus{} are shaded.}%
  \label{fig:nocal-grammar}
\end{figure}

After \colobus{} constructs location-annotated \ourcalc{} code, it converts it to \lamcur{} (Figure~\ref{fig:nocal-grammar}), a lower-level IR in which locations become \textit{cursors} into serialized buffers.
The \ourcalc{} formalism is deliberately abstract; the lowering to \lamcur{} is where the implementation chooses a concrete, efficient representation.

The lowering of a location is type directed.
A \unfactored{} location lowers to a single cursor, as in \gibbon{}.
A \fullyfactored{} location lowers to a cursor array, \lstinline{CursorArray N}, where \lstinline|N| is the statically known number of buffers.
The array follows the structure of the datatype.
Index 0 holds the cursor of the tag buffer.
Each field then contributes its cursors in order: one for a field in a single buffer, and the whole array of a factored field, inlined at consecutive indices.
Each buffer therefore has a fixed index, known at compile time.

The location IR of the implementation has three operations on factored locations, together with matching operations on regions.
\lstinline{GenSoALoc} builds a factored location from the location of the tag buffer and one location per factored field; it lowers to \lstinline{makeCursorArray}, which builds a cursor array with the tag cursor first.
\lstinline{GetDataConLocSoA} selects the location of the tag buffer and lowers to \lstinline{indexCursorArray} at index 0.
\lstinline{GetFieldLocSoA} selects the location of one field, identified by its constructor and position.
For a field in a single buffer it lowers to \lstinline{indexCursorArray} at that field's index; for a factored field it builds a new cursor array from the field's consecutive indices.
Other operations on single cursors, such as reading and writing tags and scalars, map directly to existing \gibbon{} primitives.

For mutable cursors, \lamcur{} adds the type \lstinline{MutCursor}, a reference to a cursor, with the operations \lstinline{addrOfCursor}, \lstinline{derefMutCursor}, \lstinline{writeCursorMutable}, and \lstinline{bumpCursorMutable}.
Generated code updates cursor positions in place through these references instead of binding fresh cursor variables.
A cursor array needs no separate mutable form, since it is already passed by reference and updated in place.
Section~\ref{sec:nocal-example} shows \lstinline{sumTree} in \lamcur{}.

The optimizations of Section~\ref{sec:factored-opts} add three more groups of primitives: counted and chunk-walking loops (\lstinline{for} and \lstinline{whileCursor}), the selective indirections that share individual buffers, and vector operations, which the code generator emits as SIMD code.

\subsection{Example of working with mutable cursors in \lamcur}%
\label{sec:nocal-example}

\begin{figure}
\begin{code}
sumTree :: CursorArray 2 -> Int
sumTree arr =
  let dcur     = addrOfCursor (indexCursorArray arr 0) in
  let lcur     = addrOfCursor (indexCursorArray arr 1) in
  let (tag, _) = readTag (derefMutCursor dcur) in
  let _        = bumpCursorMutable dcur 1 in
  switch tag of
    Leaf -> let (n, _) = readScalar Int (derefMutCursor lcur) in
            let _      = bumpCursorMutable lcur (sizeof Int) in
            n
    Node -> let a = sumTree arr in
            let b = sumTree arr in
            a + b
\end{code}
\caption{\lstinline|sumTree| in \lamcur with mutable cursors}%
\label{fig:sumtree-nocal}%
\label{fig:nocal-example}%
\end{figure}

The code in Figure~\ref{fig:nocal-example} shows \lstinline{sumTree} in \lamcur{} and gives an example of various cursor-related operations in \lamcur.
The \lstinline|Tree| datatype with the factored layout uses two buffers: one for data-constructor tags and one for \lstinline|Leaf| payloads.
Therefore, the lowered function consumes a \lstinline|CursorArray 2|.
The function binds mutable references to both cursors at entry and bumps them in place as it traverses the tree.
Recursive calls receive the same cursor array, so cursor state is threaded through the traversal by in-place updates rather than by allocating fresh cursor bindings at each step.

\FloatBarrier
\section{Additional Evaluation Results}\label{sec:appendix-eval}

\input{tables/pldi/_preamble}

This appendix gives the measurements behind Section~\ref{sec:eval} in full,
for the C compiled by GCC; Appendix~\ref{sec:appendix-clang} repeats them with
Clang. Section~\ref{sec:appendix-eval-legend} explains how to read the tables.
The whole-suite results (Section~\ref{sec:appendix-eval-suite}) compare the
layouts program by program, and the comparisons with other languages
(Sections~\ref{sec:appendix-tree-sweep} and~\ref{sec:appendix-ghc-mlton}) place
\colobus{} beside mainstream compilers. Section~\ref{sec:appendix-papi} explains how the hardware counters were measured, and the per-program results (Section~\ref{sec:appendix-eval-programs}) describe each benchmark and give its timings pass by pass, with the counters of the two programs whose cache behavior Section~\ref{sec:eval} analyzes. The supplementary material holds every table of this
run, including the per-pass effect of each optimization and the hardware
counters of every program.

\subsection{Reading the tables}\label{sec:appendix-eval-legend}

Every column of a timing table is one configuration of the compiler, named by
a symbol (Table~\ref{tab:pldi-legend}). The base letter is the layout, $A$ for
the \unfactored{} layout (array of structs) and $S$ for the \fullyfactored{}
layout (struct of arrays). The subscripts name how \colobus{} generates code:
$r$ recursive traversal, $i$ immutable cursors, $m$ mutable cursors, $\ell$
loopified, $b$ selective buffer sharing and $v$ SIMD vectorization.
The superscript names a C-compiler setting that differs from the default: $-t$
disables tail-call optimization and $-av$ disables the C auto-vectorizer, while
$+av$ marks a configuration that has an $-av$ twin. \gibbon{},
$A_{ri}^{\scriptscriptstyle +av}$, is shaded in every table that shows it. The
Pointer column compiles the AoS source to a pointer-based layout, one heap
object per node; it is not a serialized layout and is shown for reference.
Every configuration was compiled for the same SIMD target, 256-bit AVX2 (\texttt{-mavx2}), which the SIMD pass and the C compiler's auto-vectorizer both use. Every configuration frees the region chunks of each timed iteration (\texttt{--reclaim-iterate-regions}), so memory use does not grow with the iteration count. Some inputs were resized from their defaults: Add1TreeInt16 (depth = 34; default 33), Add1TreeInt32 (depth = 34; default 33), Add1TreeInt64 (depth = 34; default 33), Add1TreeInt8 (depth = 34; default 33), MonoTree (depth = 25; default 23), reduceNestedList (n (outer list elements) = 10,000; default 1,000,000).

Times are the median time of one iteration of a pass, in milliseconds. In each
row the fastest configuration is \textcolor{gibbonfast}{green} and the slowest
\textcolor{gibbonslow}{red}. $A^{\min}/S^{\min}$ divides a row's fastest
\unfactored{} time by its fastest \fullyfactored{} time; a value above $1$
means the \fullyfactored{} layout is faster. A cell without a time says why:
`*' did not compile, `-' compiled but did not run to completion, `**' ran but
gave a wrong result, `$\ddagger$' exhausted the C stack, and `?' was not
measured. A `$\ddagger$' is expected for configurations that recurse once per
element: mutable cursors with tail calls turn the same traversal into a loop,
and the inputs are sized for the loop. Wide tables are split, the \unfactored{}
and pointer-based configurations first and the \fullyfactored{} ones in a
companion table.

Fold tables add two columns: \textbf{Uses}, the fields a pass binds out of the
datatype's total, and \textbf{Dead\%}, the fraction it never binds. A pass may bind a
field without reading it: \PassName{verifyIR} (Section~\ref{sec:eval-compiler}) binds
all nine fields of its datatype but reads only the opcode and the tail. Their
columns are the recursive configurations, since a fold has nothing to
loopify. Map tables add $\Sigma_b$, the buffers a pass leaves unmodified out of
the \fullyfactored{} value's total, read from the pass's annotations: it
bounds what selective buffer sharing can reuse.
Table~\ref{tab:benchmark-characteristics} lists each benchmark's datatype: its
fields, its buffers in the \fullyfactored{} layout, and how much of it the
passes leave untouched.

  \begingroup
    \setlength{\textfloatsep}{4pt plus 1pt minus 1pt}%
    \setlength{\floatsep}{3pt plus 1pt minus 1pt}%
    \setlength{\intextsep}{3pt plus 1pt minus 1pt}%
    \setlength{\abovecaptionskip}{3pt}%
    \setlength{\belowcaptionskip}{2pt}%
    \begin{table}[htbp]
\centering
\caption{Configuration symbols.}
\label{tab:pldi-legend}
\small\gibbonnumfont
\setlength{\tabcolsep}{3pt}
\gibbonfit{%
\begin{tabular}{@{}c p{0.9\linewidth}@{}}
\toprule
\textbf{Symbol} & \textbf{Configuration} \\
\midrule
$A_{ri}^{\scriptscriptstyle -t}$ & AoS, recursive traversal, immutable cursors, C tail-call optimization disabled \\
$A_{ri}^{\scriptscriptstyle -av}$ & AoS, recursive traversal, immutable cursors, C auto-vectorization disabled \\
\gibbonvanilla $A_{ri}^{\scriptscriptstyle +av}$ & \gibbonvanilla AoS, recursive traversal, immutable cursors, C auto-vectorization enabled \\
$A_{rm}^{\scriptscriptstyle -t}$ & AoS, recursive traversal, mutable cursors, C tail-call optimization disabled \\
$A_{rm}^{\scriptscriptstyle -av}$ & AoS, recursive traversal, mutable cursors, C auto-vectorization disabled \\
$A_{rm}^{\scriptscriptstyle +av}$ & AoS, recursive traversal, mutable cursors, C auto-vectorization enabled \\
$A_{\ell}^{\scriptscriptstyle -av}$ & AoS, loopified, C auto-vectorization disabled \\
\bottomrule
\end{tabular}}
\end{table}%
  \endgroup
  \begingroup
    \setlength{\textfloatsep}{4pt plus 1pt minus 1pt}%
    \setlength{\floatsep}{3pt plus 1pt minus 1pt}%
    \setlength{\intextsep}{3pt plus 1pt minus 1pt}%
    \setlength{\abovecaptionskip}{3pt}%
    \setlength{\belowcaptionskip}{2pt}%
    \begin{table}[htbp]
\centering
\caption{Benchmark characteristics.}
\label{tab:benchmark-characteristics}
\small
\gibbonfit{%
\begin{tabular}{llrrrrrr}
\toprule
Benchmark & ADT & Fields & Buffers & Folds & Maps & Uses & Dead \\
\midrule
\texttt{ColorOctree} & \texttt{ColorOctree} & 25 & 18 & 4 & 2 & 2--13 & 92\% \\
\texttt{Compiler} & \texttt{IR} & 9 & 8 & 9 & 2 & 2--9 & 78\% \\
\texttt{DBQuery} & \texttt{Query} & 15 & 13 & 6 & 2 & 3--15 & 80\% \\
\texttt{DecisionTree} & \texttt{DTree} & 7 & 6 & 9 & 0 & 2--3 & 71\% \\
\texttt{DecisionTreeClassify} & \texttt{DTree} & 7 & 6 & 2 & 0 & 4--5 & 43\% \\
\texttt{DomTree} & \texttt{DOM} & 15 & 14 & 4 & 2 & 3--15 & 80\% \\
\texttt{KDTree} & \texttt{KDTree} & 17 & 16 & 6 & 0 & 11--14 & 35\% \\
\texttt{LinearListReduction} & \texttt{List} & 11 & 11 & 3 & 3 & 2--11 & 82\% \\
\texttt{List} & \texttt{List} & 2 & 2 & 3 & 1 & 1--2 & 50\% \\
\texttt{MonoTree} & \texttt{Tree} & 3 & 2 & 2 & 1 & 3 & 0\% \\
\texttt{ObjectGraph} & \texttt{Heap} & 5 & 4 & 7 & 2 & 3--5 & 40\% \\
\texttt{OctTree} & \texttt{Octree} & 16 & 9 & 6 & 2 & 8--16 & 50\% \\
\texttt{PiecewiseFunctions} & \texttt{PW} & 8 & 7 & 6 & 2 & 3--8 & 62\% \\
\texttt{reduceNestedList} & \texttt{List} & 3 & 3 & 1 & 0 & 2 & 33\% \\
\texttt{TernaryTree} & \texttt{Tree} & 5 & 3 & 1 & 1 & 5 & 0\% \\
\texttt{Trie} & \texttt{Trie} & 10 & 9 & 5 & 2 & 3--10 & 70\% \\
\bottomrule
\end{tabular}}
\end{table}%
  \endgroup

\subsection{Whole-suite results}\label{sec:appendix-eval-suite}

Tables~\ref{tab:summary} and~\ref{tab:summary-vs-vanilla} summarize each
program by the sum of its passes' median times, end to end and separately for
fold and map passes. Table~\ref{tab:summary} compares the fastest
\unfactored{} configuration, loopified
($A_{\ell}^{\scriptscriptstyle +av}$), with the fastest \fullyfactored{} one
($S_{\ell bv}^{\scriptscriptstyle +av}$): what the layout itself is worth when
both sides have every optimization that applies. Table~\ref{tab:summary-vs-vanilla}
puts \gibbon{} on the \unfactored{} side, so its speedups are what
\colobus{} gains over the compiler it extends. A speedup above $1$ means the
\fullyfactored{} configuration is faster, and \textbf{bold} marks one above
$1.1$. A pass sum can hide a pass that goes the other way; the per-program
tables show each pass. The OctTree row sums the separate OctTree programs, and
ColorOctree's passes also appear on a row of their own. \textbf{ADT fields} is
the number of fields of the benchmark's datatype and \textbf{SoA bufs} the
number of buffers it occupies in the \fullyfactored{} layout.
Tables~\ref{tab:cursor_comparison_times} and~\ref{tab:cursor_comparison_speedups}
compare mutable and immutable cursors in both layouts by whole-program time,
the build included.

  \begingroup
    \setlength{\textfloatsep}{4pt plus 1pt minus 1pt}%
    \setlength{\floatsep}{3pt plus 1pt minus 1pt}%
    \setlength{\intextsep}{3pt plus 1pt minus 1pt}%
    \setlength{\abovecaptionskip}{3pt}%
    \setlength{\belowcaptionskip}{2pt}%
    \begin{table}[htbp]
\centering
\caption{Pass-sum time per program (ms), loopified AoS against the fully optimized SoA configuration.}
\label{tab:summary}
\small\gibbonnumfont
\setlength{\tabcolsep}{3pt}
\gibbonfit{%
\begin{tabular}{l c c r r r r r r r r r}
\toprule
\textbf{Program} & \textbf{ADT} & \textbf{SoA} & \multicolumn{3}{c}{\textbf{End-to-end}} & \multicolumn{3}{c}{\textbf{Fold passes}} & \multicolumn{3}{c}{\textbf{Map passes}} \\
\cmidrule(lr){4-6}\cmidrule(lr){7-9}\cmidrule(lr){10-12}
 & fields & bufs & $A_{\ell}^{\scriptscriptstyle +av}$ (ms) & $S_{\ell bv}^{\scriptscriptstyle +av}$ (ms) & $A_{\ell}^{\scriptscriptstyle +av}$/$S_{\ell bv}^{\scriptscriptstyle +av}$ & $A_{\ell}^{\scriptscriptstyle +av}$ (ms) & $S_{\ell bv}^{\scriptscriptstyle +av}$ (ms) & $A_{\ell}^{\scriptscriptstyle +av}$/$S_{\ell bv}^{\scriptscriptstyle +av}$ & $A_{\ell}^{\scriptscriptstyle +av}$ (ms) & $S_{\ell bv}^{\scriptscriptstyle +av}$ (ms) & $A_{\ell}^{\scriptscriptstyle +av}$/$S_{\ell bv}^{\scriptscriptstyle +av}$ \\
\midrule
OctTree & 16 & 9 & 1038.00 & 642.80 & \textbf{1.61$\times$} & 443.70 & 392.40 & \textbf{1.13$\times$} & 594.00 & 250.40 & \textbf{2.37$\times$} \\
Compiler & 9 & 8 & 308.70 & 60.90 & \textbf{5.07$\times$} & 159.30 & 41.20 & \textbf{3.86$\times$} & 149.50 & 19.70 & \textbf{7.59$\times$} \\
DBQuery & 15 & 13 & 295.00 & 143.30 & \textbf{2.06$\times$} & 157.80 & 124.20 & \textbf{1.27$\times$} & 137.20 & 19.10 & \textbf{7.20$\times$} \\
DecisionTree & 7 & 6 & 132.50 & 64.20 & \textbf{2.06$\times$} & 132.50 & 64.20 & \textbf{2.06$\times$} & -- & -- & -- \\
DecisionTreeClassify & 7 & 6 & 171.50 & 185.50 & 0.92$\times$ & 171.50 & 185.50 & 0.92$\times$ & -- & -- & -- \\
DomTree & 15 & 14 & 211.80 & 122.20 & \textbf{1.73$\times$} & 156.50 & 64.60 & \textbf{2.42$\times$} & 55.30 & 57.60 & 0.96$\times$ \\
\bottomrule
\end{tabular}}
\end{table}%
  \endgroup
  \begingroup
    \setlength{\textfloatsep}{4pt plus 1pt minus 1pt}%
    \setlength{\floatsep}{3pt plus 1pt minus 1pt}%
    \setlength{\intextsep}{3pt plus 1pt minus 1pt}%
    \setlength{\abovecaptionskip}{3pt}%
    \setlength{\belowcaptionskip}{2pt}%
    \begin{table}[htbp]
\centering
\caption{Pass-sum time per program (ms), \gibbon{} against the fully optimized SoA configuration.}
\label{tab:summary-vs-vanilla}
\small\gibbonnumfont
\setlength{\tabcolsep}{3pt}
\gibbonfit{%
\begin{tabular}{l c c r r r r r r r r r}
\toprule
\textbf{Program} & \textbf{ADT} & \textbf{SoA} & \multicolumn{3}{c}{\textbf{End-to-end}} & \multicolumn{3}{c}{\textbf{Fold passes}} & \multicolumn{3}{c}{\textbf{Map passes}} \\
\cmidrule(lr){4-6}\cmidrule(lr){7-9}\cmidrule(lr){10-12}
 & fields & bufs & \gibbonvanilla $A_{ri}^{\scriptscriptstyle +av}$ (ms) & $S_{\ell bv}^{\scriptscriptstyle +av}$ (ms) & $A_{ri}^{\scriptscriptstyle +av}$/$S_{\ell bv}^{\scriptscriptstyle +av}$ & \gibbonvanilla $A_{ri}^{\scriptscriptstyle +av}$ (ms) & $S_{\ell bv}^{\scriptscriptstyle +av}$ (ms) & $A_{ri}^{\scriptscriptstyle +av}$/$S_{\ell bv}^{\scriptscriptstyle +av}$ & \gibbonvanilla $A_{ri}^{\scriptscriptstyle +av}$ (ms) & $S_{\ell bv}^{\scriptscriptstyle +av}$ (ms) & $A_{ri}^{\scriptscriptstyle +av}$/$S_{\ell bv}^{\scriptscriptstyle +av}$ \\
\midrule
OctTree & 16 & 9 & \gibbonvanilla 1237.00 & 642.80 & \textbf{1.92$\times$} & \gibbonvanilla 543.30 & 392.40 & \textbf{1.38$\times$} & \gibbonvanilla 693.70 & 250.40 & \textbf{2.77$\times$} \\
Compiler & 9 & 8 & \gibbonvanilla 1179.00 & 60.90 & \textbf{19.35$\times$} & \gibbonvanilla 739.50 & 41.20 & \textbf{17.93$\times$} & \gibbonvanilla 439.80 & 19.70 & \textbf{22.33$\times$} \\
DBQuery & 15 & 13 & \gibbonvanilla 377.60 & 143.30 & \textbf{2.63$\times$} & \gibbonvanilla 216.60 & 124.20 & \textbf{1.74$\times$} & \gibbonvanilla 161.00 & 19.10 & \textbf{8.45$\times$} \\
DecisionTree & 7 & 6 & \gibbonvanilla 299.90 & 64.20 & \textbf{4.67$\times$} & \gibbonvanilla 299.90 & 64.20 & \textbf{4.67$\times$} & \gibbonvanilla -- & -- & -- \\
DecisionTreeClassify & 7 & 6 & \gibbonvanilla 179.10 & 185.50 & 0.97$\times$ & \gibbonvanilla 179.10 & 185.50 & 0.97$\times$ & \gibbonvanilla -- & -- & -- \\
DomTree & 15 & 14 & \gibbonvanilla 346.60 & 122.20 & \textbf{2.84$\times$} & \gibbonvanilla 286.60 & 64.60 & \textbf{4.43$\times$} & \gibbonvanilla 60.00 & 57.60 & 1.04$\times$ \\
\bottomrule
\end{tabular}}
\end{table}%
  \endgroup
  \begingroup
    \setlength{\textfloatsep}{4pt plus 1pt minus 1pt}%
    \setlength{\floatsep}{3pt plus 1pt minus 1pt}%
    \setlength{\intextsep}{3pt plus 1pt minus 1pt}%
    \setlength{\abovecaptionskip}{3pt}%
    \setlength{\belowcaptionskip}{2pt}%
    \begin{table}[htbp]
\centering
\caption{Whole-program time (ms) with mutable and immutable cursors.}
\label{tab:cursor_comparison_times}
\small
\gibbonfit{%
\begin{tabular}{p{5.5cm} r r r r}
\toprule
\textbf{Program} & $A_{rm}^{\scriptscriptstyle +av}$ & $A_{ri}^{\scriptscriptstyle +av}$ & $S_{rm}^{\scriptscriptstyle +av}$ & $S_{ri}$ \\
\midrule
Compiler & 7291.89 & 25886.29 & \textbf{5474.52} & -- \\
DBQuery & \textbf{6978.33} & 8536.69 & 7042.90 & 24351.06 \\
DecisionTree & 2862.36 & 6385.00 & \textbf{1432.13} & 26317.26 \\
DecisionTreeClassify & \textbf{3770.88} & 3954.00 & 4067.79 & 6070.69 \\
DomTree & 4965.40 & 7800.60 & \textbf{3430.47} & 24162.54 \\
KDTree & 4061.04 & 5563.08 & \textbf{3587.87} & 12092.04 \\
LinearListReduction & \textbf{17487.09} & 39260.26 & 17668.01 & -- \\
reduceNestedList & 129.30 & 524.57 & \textbf{123.68} & 248.95 \\
List & \textbf{9143.34} & -- & 9958.25 & -- \\
MonoTree & 5819.32 & 11605.71 & \textbf{5793.01} & 18427.48 \\
ObjectGraph & 6183.45 & 10404.33 & \textbf{6051.36} & 19075.95 \\
OctTree & \textbf{15205.72} & 16469.85 & 17586.36 & -- \\
PiecewiseFunctions\_norm2Estimate & 728.93 & 1478.99 & \textbf{636.92} & 7305.29 \\
PiecewiseFunctions\_truncateTolViolations & 654.58 & 1176.14 & \textbf{415.26} & 5886.95 \\
PiecewiseFunctions\_compressMass & 675.69 & 1125.94 & \textbf{417.82} & 6272.88 \\
PiecewiseFunctions\_autorefineMaxLevel & 740.36 & 1132.38 & \textbf{571.64} & 6567.35 \\
PiecewiseFunctions\_pmapCutHistogram & 691.86 & 1399.33 & \textbf{453.09} & 6000.36 \\
PiecewiseFunctions\_lbDeuxLoadProxy & 712.27 & 1193.52 & \textbf{546.28} & 6319.94 \\
PiecewiseFunctions\_addConstPW & \textbf{3069.89} & 3280.35 & 3382.24 & 11537.47 \\
PiecewiseFunctions\_diffPW & \textbf{3193.65} & 3380.02 & 3520.03 & 12587.43 \\
TernaryTree & \textbf{2443.64} & 2994.54 & 2444.46 & 14012.56 \\
Trie & 5222.13 & 6886.51 & \textbf{5110.62} & 21842.05 \\
ColorOctree & \textbf{13643.25} & 14854.04 & 20087.62 & -- \\
\bottomrule
\end{tabular}}
\end{table}%
  \endgroup
  \begingroup
    \setlength{\textfloatsep}{4pt plus 1pt minus 1pt}%
    \setlength{\floatsep}{3pt plus 1pt minus 1pt}%
    \setlength{\intextsep}{3pt plus 1pt minus 1pt}%
    \setlength{\abovecaptionskip}{3pt}%
    \setlength{\belowcaptionskip}{2pt}%
    \begin{table}[htbp]
\centering
\caption{Speedups between the cursor configurations of Table~\ref{tab:cursor_comparison_times}.}
\label{tab:cursor_comparison_speedups}
\small
\gibbonfit{%
\begin{tabular}{p{5.5cm} r r r r}
\toprule
\textbf{Program} & $A_{rm}^{\scriptscriptstyle +av}$/$S_{rm}^{\scriptscriptstyle +av}$ & $A_{ri}^{\scriptscriptstyle +av}$/$A_{rm}^{\scriptscriptstyle +av}$ & $A_{ri}^{\scriptscriptstyle +av}$/$S_{rm}^{\scriptscriptstyle +av}$ & $A_{ri}^{\scriptscriptstyle +av}$/$S_{ri}$ \\
\midrule
Compiler & \textbf{1.33$\times$} & \textbf{3.55$\times$} & \textbf{4.73$\times$} & -- \\
DBQuery & 0.99$\times$ & \textbf{1.22$\times$} & \textbf{1.21$\times$} & 0.35$\times$ \\
DecisionTree & \textbf{2.00$\times$} & \textbf{2.23$\times$} & \textbf{4.46$\times$} & 0.24$\times$ \\
DecisionTreeClassify & 0.93$\times$ & 1.05$\times$ & 0.97$\times$ & 0.65$\times$ \\
DomTree & \textbf{1.45$\times$} & \textbf{1.57$\times$} & \textbf{2.27$\times$} & 0.32$\times$ \\
KDTree & \textbf{1.13$\times$} & \textbf{1.37$\times$} & \textbf{1.55$\times$} & 0.46$\times$ \\
LinearListReduction & 0.99$\times$ & \textbf{2.25$\times$} & \textbf{2.22$\times$} & -- \\
\bottomrule
\end{tabular}}
\end{table}%
  \endgroup

\subsection{Comparison with other languages}\label{sec:appendix-tree-sweep}

Figure~\ref{fig:tree-sweep} redoes Figure~4 of \citet{ecoop17-gibbon}, which
compared hand-written packed C with pointer-based C and mainstream languages on
three traversals of a complete binary tree: building it (\texttt{buildTree}), adding a constant to every leaf
(\texttt{add1Tree}) and summing the leaves (\texttt{sumTree}). We run the same
three traversals at tree depths 10 to 26, for \gibbon{}, for the
\unfactored{} and \fullyfactored{} configurations of \colobus{}, for a
pointer-based build, and for the same program in GHC, MLton, OCaml, Rust,
Racket, Java and Chez Scheme. The comparisons below are against the fastest
\colobus{} configuration at each depth.

On \texttt{add1Tree}, \colobus{} is $3.2$--$4.9\times$ faster than \gibbon{} and faster than every other language at every depth, by $1.44$--$122\times$. On \texttt{sumTree}, it is $3.5$--$5.0\times$ faster than \gibbon{} and faster than every other language at most depths: Java at depth 18 is faster. At every other depth, \colobus{} is the fastest. On \texttt{buildTree}, it is $1.05$--$1.12\times$ faster than \gibbon{} and faster than every other language at most depths: Racket at depths 10--16, Chez Scheme at depths 10, 11 and 14--16, OCaml at depth 13 and GHC at depths 14 and 15 are faster. From depth 17 on, \colobus{} is the fastest.

\begin{figure*}[tp]
  \centering
  \includegraphics{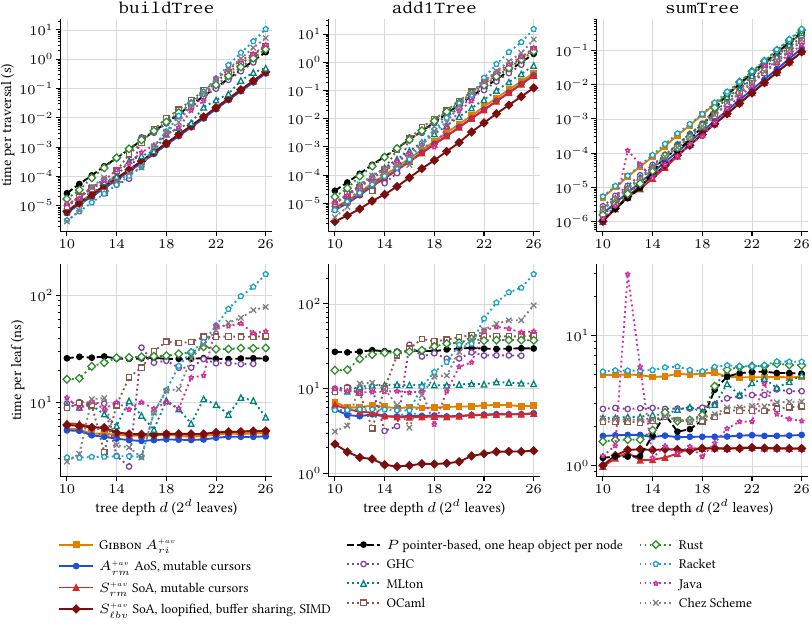}
  \caption{\texttt{buildTree}, \texttt{add1Tree} and \texttt{sumTree} on a complete binary tree of depth $d$ ($2^d$ leaves, $d = 10, \ldots, 26$): Figure~4 of \citet{ecoop17-gibbon} redone with the configurations of \colobus{} (symbols as in Table~\ref{tab:pldi-legend}). Top row: median time per traversal; bottom row: the same time per leaf. Each point is the median of 21 iterations in each of 3 interleaved rounds on one core of an Intel Core i7-12700K. Every point's result was checked against an independent model of the program. The generated C is compiled with GCC 16.2.0 for AVX2. The other languages are GHC 9.14.1 (\texttt{-O2}), MLton 20241230, OCaml 5.5.1 (\texttt{ocamlopt -O3}, flambda), Rust 1.99.0 (\texttt{-C opt-level=3 -C target-cpu=x86-64-v3}), Racket 9.3 CS, OpenJDK 27 and Chez Scheme 10.4.1 (optimize-level 3); each runs at its default runtime settings, and a run was stopped once its resident memory exceeded 69\,GB. Missing points exceeded the memory limit: GHC, \texttt{buildTree} and \texttt{add1Tree} at depth 26. The pointer-based build never frees memory, so its iterations were capped to fit: for \texttt{buildTree} and \texttt{add1Tree} it ran 14, 7, 3 and 1 iterations at depths 23, 24, 25 and 26.}
  \label{fig:tree-sweep}
\end{figure*}

\subsection{Comparison with GHC and MLton}\label{sec:appendix-ghc-mlton}

Table~\ref{tab:comparison_ghc_mlton} compares \colobus{} with GHC and MLton on
the programs of the suite. GHC and MLton run hand-written ports of each program,
and every program runs at its own default input size, which differs from the
sizes of the other tables; \PassName{reduceNestedList}, which has no port, is
left out. The \colobus{} columns are the \unfactored{} (AoS) and
\fullyfactored{} (SoA) layouts with mutable cursors. Each time is the sum of a
program's per-pass median times for one iteration, so building the input is
not included.

\begin{table}[htbp]
\centering
\caption{Runtime comparison of \colobus{} (AoS/SoA), GHC, and MLton. Times are total median per iteration (ms). \textbf{Bold} marks the fastest time for each program. The Geomean row covers the 29 program(s) every column measured.}
\label{tab:comparison_ghc_mlton}
\small
\gibbonfit{%
\begin{tabular}{l r r r r}
\toprule
\textbf{Program} & \textbf{\colobus{}-AoS} & \textbf{\colobus{}-SoA} & \textbf{GHC} & \textbf{MLton} \\
\midrule
Compiler & 315.60 & \textbf{221.10} & 1286.00 & 433.70 \\
DBQuery & \textbf{308.60} & 312.10 & 614.80 & 460.40 \\
DecisionTree & 124.10 & \textbf{63.70} & 410.30 & 336.00 \\
DecisionTreeClassify & 171.20 & 189.00 & 324.40 & \textbf{87.00} \\
DomTree & 210.40 & \textbf{141.60} & 428.00 & 146.50 \\
KDTree & 181.20 & \textbf{162.60} & 492.90 & 246.30 \\
LinearListReduction & 702.00 & 690.10 & 5646.00 & \textbf{553.00} \\
List & 397.30 & \textbf{380.60} & 13020.00 & 1550.00 \\
MonoTree & 66.00 & \textbf{65.30} & 590.80 & 169.70 \\
ObjectGraph & 269.70 & \textbf{264.10} & 671.60 & 366.50 \\
Oct:sumMass & 32.20 & \textbf{26.60} & 67.30 & 37.90 \\
Oct:sumEnergy & 36.00 & \textbf{30.10} & 171.30 & 51.30 \\
Oct:countActive & 33.30 & \textbf{27.30} & 79.90 & 40.00 \\
Oct:countParticles & 33.00 & \textbf{26.40} & 34.60 & 37.10 \\
Oct:barnesHutPotential & \textbf{41.90} & 44.70 & 169.30 & 52.70 \\
Oct:fmmPotential & 96.80 & \textbf{96.10} & 550.70 & 281.90 \\
Oct:scaleEnergy & \textbf{144.50} & 181.50 & 1183.00 & 170.30 \\
Oct:clearFlags & 154.00 & 180.00 & 288.80 & \textbf{83.10} \\
PiecewiseFunctions\_norm2Estimate & 24.90 & \textbf{22.70} & 61.80 & 33.40 \\
PiecewiseFunctions\_truncateTolViolations & 22.00 & \textbf{11.40} & 63.30 & 43.60 \\
PiecewiseFunctions\_compressMass & 22.70 & \textbf{11.60} & 56.10 & 39.70 \\
PiecewiseFunctions\_autorefineMaxLevel & 25.50 & \textbf{19.20} & 63.40 & 31.30 \\
PiecewiseFunctions\_pmapCutHistogram & 24.10 & \textbf{13.10} & 77.80 & 52.90 \\
PiecewiseFunctions\_lbDeuxLoadProxy & 24.90 & \textbf{17.70} & 60.60 & 30.00 \\
PiecewiseFunctions\_addConstPW & 119.00 & 130.90 & 182.70 & \textbf{87.90} \\
PiecewiseFunctions\_diffPW & 121.70 & 132.20 & 502.60 & \textbf{94.10} \\
TernaryTree & 109.40 & \textbf{102.50} & 311.40 & 187.50 \\
Trie & 216.60 & \textbf{209.00} & 509.60 & 291.80 \\
ColorOctree & 536.40 & 740.40 & 2930.00 & \textbf{474.00} \\
\midrule
\textbf{Geomean} & 93.80 & \textbf{81.40} & 310.40 & 123.10 \\
\bottomrule
\end{tabular}}
\end{table}

\subsection{Hardware counters}\label{sec:appendix-papi}

Section~\ref{sec:eval} explains two results with hardware counters: the
\PassName{Compiler} passes and \PassName{LinearListReduction}'s
\PassName{reduction} pass. Their counter tables appear with those programs'
results in Section~\ref{sec:appendix-eval-programs}; the supplementary material
holds the counters of every program.
The counts come from PAPI, read around
the timed region of each iteration, and each cell is the median over the
iterations of that pass. They were measured in separate runs, with binaries
that read the counters, so a count and a time never come from the same
executable. The machine's performance core counts four programmable events
besides cycles and instructions, so every counter executable ran twice, once
for each half of the counters.

Each table gives, per pass and configuration, cycles, retired instructions,
retired loads that missed the L1 data cache, the L2 and the last-level cache,
L1 instruction-cache misses, loads that missed every TLB level, and the cache
lines filled into the L2, which counts demand loads, stores and hardware
prefetches alike. In each row the fewest is \textcolor{gibbonfast}{green} and
the most \textcolor{gibbonslow}{red}. A counter whose every pass stays below
1000 counts per iteration is measurement noise and is left out.

\FloatBarrier
\subsection{The C auto-vectorizer over the factored layout}\label{sec:appendix-autovec}

The C auto-vectorizer does not always help. With mutable cursors, it slows every
\fullyfactored{} pass of \PassName{KDTree} by $8$--$49\%$ ($S_{rm}^{\scriptscriptstyle +av}$ against
$S_{rm}^{\scriptscriptstyle -av}$), but leaves the \unfactored{} passes within 2\% of their time without it. The two layouts
give GCC different code to vectorize. Over the \unfactored{} layout, GCC's basic-block (SLP)
vectorizer finds little: most pass functions contain no vector instructions, and none uses
256-bit registers. Over the \fullyfactored{} layout, a pass updates an array of cursors, one
per buffer, and the SLP vectorizer packs groups of scalar operations, such as these cursor
updates, into 128- and 256-bit vector code in every pass. Disabling only the SLP vectorizer
brings every \fullyfactored{} pass to within 1\% of its time with the auto-vectorizer off.
Hardware counters read around each pass reveal two distinct costs.

In \PassName{nearestDist}, \PassName{countInRange}, \PassName{sumMassInRange} and
\PassName{photonMappingBenchmark}, the vectorized statements lie mostly off the path a
query takes; \PassName{countInRange} vectorizes only the code that follows an
indirection, which the \PassName{KDTree} input does not contain. The cost is in the function's frame:
because the function uses 256-bit registers, GCC realigns the stack on entry, saves
two more registers and executes \texttt{vzeroupper} before every call and return.
Every recursive call pays this cost, so these passes execute 7--12\% more instructions
and take 8--24\% more cycles; limiting GCC to 128-bit vectors removes this cost.

In \PassName{twoPointCorrelation} and \PassName{pointCloudNeighborhood}, the vector
code is on the hot path, where it makes up 45\% and 17\% of the executed instructions.
It moves values from scalar registers into vector lanes and back, and these moves most
plausibly stall on the single execution port that performs them (port~5). The passes
execute only 6\% and 4\% more instructions, but complete 18\% to 28\% fewer of them per cycle. With 128-bit vectors \PassName{pointCloudNeighborhood}
recovers, but \PassName{twoPointCorrelation}, which still routes each value through a
vector lane, stays $1.32\times$ slower.

Two further experiments rebuild the bit-width programs of Section~\ref{sec:eval-width}
with one compiler flag changed. In the
loopified configuration, GCC's loop vectorizer alone delivers the C auto-vectorizer's full
gain ($S_{\ell}^{\scriptscriptstyle -av}$ against $S_{\ell}^{\scriptscriptstyle +av}$).
Disabling only its SLP vectorizer leaves \PassName{add1Tree} at 8 bits and the
arithmetic kernel at 8 and 16 bits unchanged. Disabling only its loop vectorizer makes
\PassName{add1Tree} $1.53\times$ slower, and the kernel $4.46\times$ slower at 8 bits
and $2.42\times$ slower at 16 bits. Compiled for SSE2, whose registers are 128 bits
wide, the SIMD pass makes the kernel $2.17\times$ slower at 8 bits and
$1.82\times$ slower at 16 bits than with AVX2, but leaves \PassName{add1Tree} at 8 bits
about as fast ($1.04\times$).

The lesson complements Section~\ref{sec:eval-width}: over the \fullyfactored{} layout,
vectorizing loops over whole buffers speeds up maps, and GCC's loop vectorizer provides
that gain, whereas its SLP vectorizer, applied to recursive passes, only adds overhead.

\FloatBarrier
\subsection{Per-program results}\label{sec:appendix-eval-programs}

Each subsection below describes one benchmark, lists its passes as its tables
name them, and gives its fold and map tables. The order follows
Section~\ref{sec:eval}: the two case studies first, then the other real-world
programs, then the synthetic programs. The supplementary material adds, for
every program, the effect of each optimization pass by pass, as a percentage
change in time.

\subsubsection{KDTree and Compiler}\label{sec:appendix-eval-case-studies}

Section~\ref{sec:eval} describes both programs and discusses their tables. The
hardware counters of \PassName{Compiler} follow its timing tables; they show
the cache misses that \PassName{verifyIR} avoids over the \fullyfactored{} layout.

  \begingroup
    \setlength{\textfloatsep}{4pt plus 1pt minus 1pt}%
    \setlength{\floatsep}{3pt plus 1pt minus 1pt}%
    \setlength{\intextsep}{3pt plus 1pt minus 1pt}%
    \setlength{\abovecaptionskip}{3pt}%
    \setlength{\belowcaptionskip}{2pt}%
    \begin{table}[htbp]
\centering
\caption{Fold passes of \texttt{KDTree}: median time per pass (ms), AoS and pointer-based configurations.}
\label{tab:pldi-fold-KDTree}
\label{tab:KDTree}
\scriptsize\gibbonnumfont
\setlength{\tabcolsep}{3pt}
\gibbonfit{%
\begin{tabular}{l c c r r r r r r r}
\toprule
\textbf{Pass} & & &\multicolumn{6}{c}{\textbf{AoS}}&\multicolumn{1}{c}{\textbf{Pointer}} \\
\cmidrule(lr){4-9}\cmidrule(lr){10-10}
 & \textbf{Uses} & \textbf{Dead\%} & $A_{ri}^{\scriptscriptstyle -t}$ & $A_{ri}^{\scriptscriptstyle -av}$ & \gibbonvanilla $A_{ri}^{\scriptscriptstyle +av}$ & $A_{rm}^{\scriptscriptstyle -t}$ & $A_{rm}^{\scriptscriptstyle -av}$ & $A_{rm}^{\scriptscriptstyle +av}$ & $P$ \\
\midrule
Find nearest Neighbour & 13/17 & 24\% & 40.85 & 28.89 & \gibbonvanilla 41.29 & 28.92 & 28.83 & 28.88 & 64.72 \\
countInRange tight\_box & 13/17 & 24\% & 41.24 & 28.46 & \gibbonvanilla 41.70 & 28.63 & 28.57 & 28.54 & 62.50 \\
sumMassInRange & 14/17 & 18\% & 41.03 & 28.47 & \gibbonvanilla 41.49 & 28.40 & 28.16 & 28.24 & 62.20 \\
twoPointCorrelation bin\_8\_16 & 11/17 & 35\% & 42.19 & 31.09 & \gibbonvanilla 42.62 & 31.30 & 30.47 & 31.09 & 66.89 \\
pointCloudNeighborhood & 11/17 & 35\% & 41.30 & 27.67 & \gibbonvanilla 41.28 & 27.41 & 27.20 & 27.37 & 59.18 \\
photonMappingBenchmark & 12/17 & 29\% & 48.01 & 44.39 & \gibbonvanilla 48.43 & 41.88 & 41.92 & 41.97 & 79.71 \\
\bottomrule
\end{tabular}}
\end{table}

\begin{table}[htbp]
\centering
\caption{Fold passes of \texttt{KDTree}: median time per pass (ms), SoA configurations.}
\label{tab:pldi-fold-KDTree-soa}
\scriptsize\gibbonnumfont
\setlength{\tabcolsep}{3pt}
\gibbonfit{%
\begin{tabular}{l c c r r r r r r}
\toprule
\textbf{Pass} & & &\multicolumn{5}{c}{\textbf{SoA}}& \\
\cmidrule(lr){4-8}
 & \textbf{Uses} & \textbf{Dead\%} & $S_{ri}^{\scriptscriptstyle -t}$ & $S_{ri}$ & $S_{rm}^{\scriptscriptstyle -t}$ & $S_{rm}^{\scriptscriptstyle -av}$ & $S_{rm}^{\scriptscriptstyle +av}$ & $A^{\min}$/$S^{\min}$ \\
\midrule
Find nearest Neighbour & 13/17 & 24\% & \textcolor{gibbonslow}{103.40} & 103.30 & 23.91 & \textcolor{gibbonfast}{21.31} & 23.68 & \textbf{1.35$\times$} \\
countInRange tight\_box & 13/17 & 24\% & 84.24 & \textcolor{gibbonslow}{84.59} & 23.23 & \textcolor{gibbonfast}{19.14} & 22.84 & \textbf{1.49$\times$} \\
sumMassInRange & 14/17 & 18\% & 77.92 & \textcolor{gibbonslow}{78.03} & 24.84 & \textcolor{gibbonfast}{20.69} & 22.27 & \textbf{1.36$\times$} \\
twoPointCorrelation bin\_8\_16 & 11/17 & 35\% & 104.80 & \textcolor{gibbonslow}{104.90} & 28.74 & \textcolor{gibbonfast}{19.61} & 29.13 & \textbf{1.55$\times$} \\
pointCloudNeighborhood & 11/17 & 35\% & 103.40 & \textcolor{gibbonslow}{103.70} & 21.40 & \textcolor{gibbonfast}{16.72} & 21.58 & \textbf{1.63$\times$} \\
photonMappingBenchmark & 12/17 & 29\% & 86.91 & \textcolor{gibbonslow}{86.92} & 42.48 & \textcolor{gibbonfast}{38.94} & 42.61 & 1.08$\times$ \\
\bottomrule
\end{tabular}}
\end{table}%
  \endgroup
\FloatBarrier

  \begingroup
    \setlength{\textfloatsep}{4pt plus 1pt minus 1pt}%
    \setlength{\floatsep}{3pt plus 1pt minus 1pt}%
    \setlength{\intextsep}{3pt plus 1pt minus 1pt}%
    \setlength{\abovecaptionskip}{3pt}%
    \setlength{\belowcaptionskip}{2pt}%
    \begin{table}[htbp]
\centering
\caption{Fold passes of \texttt{Compiler}: median time per pass (ms), AoS and pointer-based configurations.}
\label{tab:pldi-fold-Compiler}
\label{tab:Compiler}
\scriptsize\gibbonnumfont
\setlength{\tabcolsep}{3pt}
\gibbonfit{%
\begin{tabular}{l c c r r r r r r r}
\toprule
\textbf{Pass} & & &\multicolumn{6}{c}{\textbf{AoS}}&\multicolumn{1}{c}{\textbf{Pointer}} \\
\cmidrule(lr){4-9}\cmidrule(lr){10-10}
 & \textbf{Uses} & \textbf{Dead\%} & $A_{ri}^{\scriptscriptstyle -t}$ & $A_{ri}^{\scriptscriptstyle -av}$ & \gibbonvanilla $A_{ri}^{\scriptscriptstyle +av}$ & $A_{rm}^{\scriptscriptstyle -t}$ & $A_{rm}^{\scriptscriptstyle -av}$ & $A_{rm}^{\scriptscriptstyle +av}$ & $P$ \\
\midrule
verifyIR & 9/9 & 0\% & 84.37 & \textcolor{gibbonslow}{93.97} & \gibbonvanilla 81.14 & 61.90 & 13.67 & 13.22 & 28.40 \\
instCountPass & 2/9 & 78\% & 80.46 & \textcolor{gibbonslow}{94.52} & \gibbonvanilla 80.61 & 61.72 & 13.37 & 12.91 & 29.90 \\
blockCountPass & 2/9 & 78\% & 80.23 & \textcolor{gibbonslow}{94.81} & \gibbonvanilla 80.32 & 61.75 & 13.34 & 12.90 & 28.23 \\
memoryOpStatsPass & 3/9 & 67\% & 80.52 & \textcolor{gibbonslow}{94.59} & \gibbonvanilla 80.49 & 64.27 & 13.44 & 13.01 & 29.75 \\
castInstCountPass & 3/9 & 67\% & 80.53 & \textcolor{gibbonslow}{93.97} & \gibbonvanilla 80.47 & 63.08 & 13.26 & 12.80 & 28.58 \\
branchStatsPass & 2/9 & 78\% & 80.53 & \textcolor{gibbonslow}{94.63} & \gibbonvanilla 80.49 & 63.08 & 13.22 & 12.79 & 28.20 \\
latencyModelPass & 3/9 & 67\% & 80.51 & \textcolor{gibbonslow}{92.81} & \gibbonvanilla 80.50 & 63.12 & 13.32 & 12.88 & 28.43 \\
has cycle & 4/9 & 56\% & 94.61 & \textcolor{gibbonslow}{94.78} & \gibbonvanilla 94.76 & 61.74 & 55.63 & 55.56 & 41.28 \\
throughputModelPass & 3/9 & 67\% & 80.59 & \textcolor{gibbonslow}{93.97} & \gibbonvanilla 80.69 & 63.12 & 13.40 & 12.96 & 28.47 \\
\bottomrule
\end{tabular}}
\end{table}

\begin{table}[htbp]
\centering
\caption{Fold passes of \texttt{Compiler}: median time per pass (ms), SoA configurations.}
\label{tab:pldi-fold-Compiler-soa}
\scriptsize\gibbonnumfont
\setlength{\tabcolsep}{3pt}
\gibbonfit{%
\begin{tabular}{l c c r r r r r r}
\toprule
\textbf{Pass} & & &\multicolumn{5}{c}{\textbf{SoA}}& \\
\cmidrule(lr){4-8}
 & \textbf{Uses} & \textbf{Dead\%} & $S_{ri}^{\scriptscriptstyle -t}$ & $S_{ri}$ & $S_{rm}^{\scriptscriptstyle -t}$ & $S_{rm}^{\scriptscriptstyle -av}$ & $S_{rm}^{\scriptscriptstyle +av}$ & $A^{\min}$/$S^{\min}$ \\
\midrule
verifyIR & 9/9 & 0\% & $\ddagger$ & $\ddagger$ & 44.42 & \textcolor{gibbonfast}{3.42} & 3.44 & \textbf{3.87$\times$} \\
instCountPass & 2/9 & 78\% & $\ddagger$ & $\ddagger$ & 42.92 & \textcolor{gibbonfast}{1.92} & 1.92 & \textbf{6.72$\times$} \\
blockCountPass & 2/9 & 78\% & $\ddagger$ & $\ddagger$ & 42.71 & \textcolor{gibbonfast}{1.92} & 1.92 & \textbf{6.73$\times$} \\
memoryOpStatsPass & 3/9 & 67\% & $\ddagger$ & $\ddagger$ & 46.37 & \textcolor{gibbonfast}{3.82} & 3.82 & \textbf{3.41$\times$} \\
castInstCountPass & 3/9 & 67\% & $\ddagger$ & $\ddagger$ & 45.12 & \textcolor{gibbonfast}{3.39} & 3.40 & \textbf{3.77$\times$} \\
branchStatsPass & 2/9 & 78\% & $\ddagger$ & $\ddagger$ & 45.14 & \textcolor{gibbonfast}{3.43} & 3.43 & \textbf{3.73$\times$} \\
latencyModelPass & 3/9 & 67\% & $\ddagger$ & $\ddagger$ & 44.17 & 3.03 & \textcolor{gibbonfast}{3.00} & \textbf{4.30$\times$} \\
has cycle & 4/9 & 56\% & $\ddagger$ & $\ddagger$ & 45.31 & 17.27 & \textcolor{gibbonfast}{17.25} & \textbf{3.22$\times$} \\
throughputModelPass & 3/9 & 67\% & $\ddagger$ & $\ddagger$ & 44.14 & 3.04 & \textcolor{gibbonfast}{3.02} & \textbf{4.29$\times$} \\
\bottomrule
\end{tabular}}
\end{table}%
  \endgroup
  \begingroup
    \setlength{\textfloatsep}{4pt plus 1pt minus 1pt}%
    \setlength{\floatsep}{3pt plus 1pt minus 1pt}%
    \setlength{\intextsep}{3pt plus 1pt minus 1pt}%
    \setlength{\abovecaptionskip}{3pt}%
    \setlength{\belowcaptionskip}{2pt}%
    \begin{table}[htbp]
\centering
\caption{Map passes of \texttt{Compiler}: median time per pass (ms), AoS and pointer-based configurations.}
\label{tab:pldi-map-Compiler}
\tiny\gibbonnumfont
\setlength{\tabcolsep}{3pt}
\gibbonfit{%
\begin{tabular}{l c c r r r r r r r r r}
\toprule
\textbf{Pass} & & &\multicolumn{8}{c}{\textbf{AoS}}&\multicolumn{1}{c}{\textbf{Pointer}} \\
\cmidrule(lr){4-11}\cmidrule(lr){12-12}
 & \textbf{$\Sigma_b$} & \textbf{$\Sigma_b\%$} & $A_{ri}^{\scriptscriptstyle -t}$ & $A_{ri}^{\scriptscriptstyle -av}$ & \gibbonvanilla $A_{ri}^{\scriptscriptstyle +av}$ & $A_{rm}^{\scriptscriptstyle -t}$ & $A_{rm}^{\scriptscriptstyle -av}$ & $A_{rm}^{\scriptscriptstyle +av}$ & $A_{\ell}^{\scriptscriptstyle -av}$ & $A_{\ell}^{\scriptscriptstyle +av}$ & $P$ \\
\midrule
targetReturnPass & 7/8 & 88\% & 201.90 & 204.00 & \gibbonvanilla 217.30 & 213.70 & 80.27 & 78.68 & 76.31 & 75.56 & \textcolor{gibbonslow}{236.20} \\
stripSideEffectsPass & 7/8 & 88\% & 208.20 & 205.00 & \gibbonvanilla 222.50 & 175.90 & 76.47 & 76.18 & 74.12 & 73.90 & \textcolor{gibbonslow}{223.90} \\
\bottomrule
\end{tabular}}
\end{table}

\begin{table}[htbp]
\centering
\caption{Map passes of \texttt{Compiler}: median time per pass (ms), SoA configurations.}
\label{tab:pldi-map-Compiler-soa}
\tiny\gibbonnumfont
\setlength{\tabcolsep}{3pt}
\gibbonfit{%
\begin{tabular}{l c c r r r r r r r r r r r r}
\toprule
\textbf{Pass} & & &\multicolumn{11}{c}{\textbf{SoA}}& \\
\cmidrule(lr){4-14}
 & \textbf{$\Sigma_b$} & \textbf{$\Sigma_b\%$} & $S_{ri}^{\scriptscriptstyle -t}$ & $S_{ri}$ & $S_{rm}^{\scriptscriptstyle -t}$ & $S_{rm}^{\scriptscriptstyle -av}$ & $S_{rm}^{\scriptscriptstyle +av}$ & $S_{\ell}^{\scriptscriptstyle -av}$ & $S_{\ell}^{\scriptscriptstyle +av}$ & $S_{\ell b}^{\scriptscriptstyle -av}$ & $S_{\ell b}^{\scriptscriptstyle +av}$ & $S_{\ell bv}^{\scriptscriptstyle -av}$ & $S_{\ell bv}^{\scriptscriptstyle +av}$ & $A^{\min}$/$S^{\min}$ \\
\midrule
targetReturnPass & 7/8 & 88\% & $\ddagger$ & $\ddagger$ & 193.50 & 92.89 & 91.44 & 100.40 & 99.37 & 10.45 & 9.96 & 10.48 & \textcolor{gibbonfast}{9.89} & \textbf{7.64$\times$} \\
stripSideEffectsPass & 7/8 & 88\% & $\ddagger$ & $\ddagger$ & 193.50 & 89.25 & 91.59 & 102.20 & 102.70 & 9.88 & 9.83 & 9.86 & \textcolor{gibbonfast}{9.80} & \textbf{7.54$\times$} \\
\bottomrule
\end{tabular}}
\end{table}%
  \endgroup
\FloatBarrier

  \begingroup
    \setlength{\textfloatsep}{4pt plus 1pt minus 1pt}%
    \setlength{\floatsep}{3pt plus 1pt minus 1pt}%
    \setlength{\intextsep}{3pt plus 1pt minus 1pt}%
    \setlength{\abovecaptionskip}{3pt}%
    \setlength{\belowcaptionskip}{2pt}%
    \begin{table}[htbp]
\centering
\caption{Hardware counters of \texttt{Compiler}: median count per pass and iteration, AoS and pointer-based configurations.}
\label{tab:Compiler_pldi_counters_by_config}
\tiny\gibbonnumfont
\setlength{\tabcolsep}{3pt}
\gibbonfit{%
\begin{tabular}{l r r r r r r r r r}
\toprule
\textbf{Pass} &\multicolumn{8}{c}{\textbf{AoS}}&\multicolumn{1}{c}{\textbf{Pointer}} \\
\cmidrule(lr){2-9}\cmidrule(lr){10-10}
 & $A_{ri}^{\scriptscriptstyle -t}$ & $A_{ri}^{\scriptscriptstyle -av}$ & \gibbonvanilla $A_{ri}^{\scriptscriptstyle +av}$ & $A_{rm}^{\scriptscriptstyle -t}$ & $A_{rm}^{\scriptscriptstyle -av}$ & $A_{rm}^{\scriptscriptstyle +av}$ & $A_{\ell}^{\scriptscriptstyle -av}$ & $A_{\ell}^{\scriptscriptstyle +av}$ & $P$ \\
\midrule
\multicolumn{10}{l}{\textit{Cycles}} \\
verifyIR & 4.2e+08 & \textcolor{gibbonslow}{4.8e+08} & \gibbonvanilla 4.1e+08 & 3.1e+08 & 7.0e+07 & 6.8e+07 & 6.8e+07 & 6.8e+07 & 1.5e+08 \\
instCountPass & 4.2e+08 & \textcolor{gibbonslow}{4.8e+08} & \gibbonvanilla 4.1e+08 & 3.1e+08 & 6.9e+07 & 6.7e+07 & 6.6e+07 & 6.7e+07 & 1.5e+08 \\
blockCountPass & 4.4e+08 & \textcolor{gibbonslow}{4.9e+08} & \gibbonvanilla 4.1e+08 & 3.2e+08 & 6.8e+07 & 6.7e+07 & 6.6e+07 & 6.7e+07 & 1.5e+08 \\
memoryOpStatsPass & 4.2e+08 & \textcolor{gibbonslow}{4.8e+08} & \gibbonvanilla 4.1e+08 & 3.3e+08 & 6.9e+07 & 6.8e+07 & 6.7e+07 & 6.8e+07 & 1.5e+08 \\
castInstCountPass & 4.2e+08 & \textcolor{gibbonslow}{4.8e+08} & \gibbonvanilla 4.3e+08 & 3.2e+08 & 6.8e+07 & 6.6e+07 & 6.6e+07 & 6.7e+07 & 1.5e+08 \\
\bottomrule
\end{tabular}}
\end{table}

\begin{table}[htbp]
\centering
\caption{Hardware counters of \texttt{Compiler}: median count per pass and iteration, AoS and pointer-based configurations (continued).}
\label{tab:Compiler_pldi_counters_by_config-part2}
\tiny\gibbonnumfont
\setlength{\tabcolsep}{3pt}
\gibbonfit{%
\begin{tabular}{l r r r r r r r r r}
\toprule
\textbf{Pass} &\multicolumn{8}{c}{\textbf{AoS}}&\multicolumn{1}{c}{\textbf{Pointer}} \\
\cmidrule(lr){2-9}\cmidrule(lr){10-10}
 & $A_{ri}^{\scriptscriptstyle -t}$ & $A_{ri}^{\scriptscriptstyle -av}$ & \gibbonvanilla $A_{ri}^{\scriptscriptstyle +av}$ & $A_{rm}^{\scriptscriptstyle -t}$ & $A_{rm}^{\scriptscriptstyle -av}$ & $A_{rm}^{\scriptscriptstyle +av}$ & $A_{\ell}^{\scriptscriptstyle -av}$ & $A_{\ell}^{\scriptscriptstyle +av}$ & $P$ \\
\midrule
\multicolumn{10}{l}{\textit{Instructions}} \\
verifyIR & 1.9e+08 & \textcolor{gibbonslow}{2.0e+08} & \gibbonvanilla 1.9e+08 & 1.3e+08 & 8.5e+07 & 8.5e+07 & 8.5e+07 & 8.5e+07 & \textcolor{gibbonfast}{7.4e+07} \\
instCountPass & 1.6e+08 & \textcolor{gibbonslow}{1.7e+08} & \gibbonvanilla 1.6e+08 & 8.9e+07 & 6.3e+07 & 6.3e+07 & 6.3e+07 & 6.3e+07 & \textcolor{gibbonfast}{5.5e+07} \\
blockCountPass & 1.6e+08 & \textcolor{gibbonslow}{1.7e+08} & \gibbonvanilla 1.6e+08 & 8.4e+07 & 5.8e+07 & 5.8e+07 & 5.8e+07 & 5.8e+07 & 5.0e+07 \\
memoryOpStatsPass & 1.9e+08 & \textcolor{gibbonslow}{2.0e+08} & \gibbonvanilla 1.9e+08 & 1.4e+08 & 9.2e+07 & 9.2e+07 & 9.2e+07 & 9.2e+07 & \textcolor{gibbonfast}{8.4e+07} \\
castInstCountPass & 1.8e+08 & \textcolor{gibbonslow}{1.9e+08} & \gibbonvanilla 1.8e+08 & 1.3e+08 & 8.6e+07 & 8.6e+07 & 8.6e+07 & 8.6e+07 & \textcolor{gibbonfast}{7.8e+07} \\
\bottomrule
\end{tabular}}
\end{table}

\begin{table}[htbp]
\centering
\caption{Hardware counters of \texttt{Compiler}: median count per pass and iteration, AoS and pointer-based configurations (continued).}
\label{tab:Compiler_pldi_counters_by_config-part3}
\tiny\gibbonnumfont
\setlength{\tabcolsep}{3pt}
\gibbonfit{%
\begin{tabular}{l r r r r r r r r r}
\toprule
\textbf{Pass} &\multicolumn{8}{c}{\textbf{AoS}}&\multicolumn{1}{c}{\textbf{Pointer}} \\
\cmidrule(lr){2-9}\cmidrule(lr){10-10}
 & $A_{ri}^{\scriptscriptstyle -t}$ & $A_{ri}^{\scriptscriptstyle -av}$ & \gibbonvanilla $A_{ri}^{\scriptscriptstyle +av}$ & $A_{rm}^{\scriptscriptstyle -t}$ & $A_{rm}^{\scriptscriptstyle -av}$ & $A_{rm}^{\scriptscriptstyle +av}$ & $A_{\ell}^{\scriptscriptstyle -av}$ & $A_{\ell}^{\scriptscriptstyle +av}$ & $P$ \\
\midrule
\multicolumn{10}{l}{\textit{L1D load misses (retired)}} \\
verifyIR & \textcolor{gibbonslow}{4.0e+06} & 3.4e+06 & \gibbonvanilla 3.3e+06 & 2.2e+06 & 3.3e+06 & 3.3e+06 & 3.3e+06 & 3.3e+06 & 3.4e+06 \\
instCountPass & \textcolor{gibbonslow}{4.3e+06} & 3.5e+06 & \gibbonvanilla 3.6e+06 & 3.4e+06 & 4.0e+06 & 4.0e+06 & 4.0e+06 & 4.0e+06 & 3.7e+06 \\
blockCountPass & \textcolor{gibbonslow}{4.6e+06} & 3.6e+06 & \gibbonvanilla 3.6e+06 & 3.4e+06 & 4.0e+06 & 4.0e+06 & 4.0e+06 & 4.0e+06 & 3.7e+06 \\
memoryOpStatsPass & \textcolor{gibbonslow}{3.9e+06} & 3.3e+06 & \gibbonvanilla 3.3e+06 & 3.0e+06 & 1.4e+06 & 1.5e+06 & 1.5e+06 & 1.5e+06 & 3.2e+06 \\
castInstCountPass & \textcolor{gibbonslow}{4.0e+06} & 3.4e+06 & \gibbonvanilla 3.3e+06 & 3.0e+06 & 2.1e+06 & 2.1e+06 & 2.1e+06 & 2.1e+06 & 2.9e+06 \\
\bottomrule
\end{tabular}}
\end{table}

\begin{table}[htbp]
\centering
\caption{Hardware counters of \texttt{Compiler}: median count per pass and iteration, AoS and pointer-based configurations (continued).}
\label{tab:Compiler_pldi_counters_by_config-part4}
\tiny\gibbonnumfont
\setlength{\tabcolsep}{3pt}
\gibbonfit{%
\begin{tabular}{l r r r r r r r r r}
\toprule
\textbf{Pass} &\multicolumn{8}{c}{\textbf{AoS}}&\multicolumn{1}{c}{\textbf{Pointer}} \\
\cmidrule(lr){2-9}\cmidrule(lr){10-10}
 & $A_{ri}^{\scriptscriptstyle -t}$ & $A_{ri}^{\scriptscriptstyle -av}$ & \gibbonvanilla $A_{ri}^{\scriptscriptstyle +av}$ & $A_{rm}^{\scriptscriptstyle -t}$ & $A_{rm}^{\scriptscriptstyle -av}$ & $A_{rm}^{\scriptscriptstyle +av}$ & $A_{\ell}^{\scriptscriptstyle -av}$ & $A_{\ell}^{\scriptscriptstyle +av}$ & $P$ \\
\midrule
\multicolumn{10}{l}{\textit{L1I misses}} \\
verifyIR & 3.6e+02 & 4.0e+02 & \gibbonvanilla 4.6e+02 & 3.4e+02 & 3.3e+02 & 2.0e+02 & 3.0e+02 & 2.4e+02 & 8.8e+01 \\
instCountPass & 7.6e+02 & 7.0e+02 & \gibbonvanilla 7.0e+02 & 4.0e+02 & 4.4e+02 & 3.9e+02 & 4.6e+02 & 3.2e+02 & 5.3e+01 \\
blockCountPass & 6.5e+02 & 6.6e+02 & \gibbonvanilla 5.5e+02 & 3.9e+02 & 3.7e+02 & 3.4e+02 & 4.7e+02 & 2.7e+02 & 8.5e+01 \\
memoryOpStatsPass & 4.7e+02 & 4.5e+02 & \gibbonvanilla 4.6e+02 & 4.9e+02 & 3.9e+02 & 3.0e+02 & 3.5e+02 & 2.2e+02 & 7.1e+01 \\
castInstCountPass & 6.8e+02 & 8.1e+02 & \gibbonvanilla 6.8e+02 & 6.0e+02 & 4.0e+02 & 3.5e+02 & 3.7e+02 & 2.6e+02 & 5.0e+01 \\
branchStatsPass & 4.5e+02 & 6.4e+02 & \gibbonvanilla 4.5e+02 & 6.2e+02 & 4.5e+02 & 3.8e+02 & 3.8e+02 & 3.8e+02 & 6.2e+01 \\
latencyModelPass & 4.4e+02 & 5.2e+02 & \gibbonvanilla 6.7e+02 & 7.4e+02 & 3.4e+02 & 3.1e+02 & 3.2e+02 & 2.8e+02 & 5.2e+01 \\
has cycle & 7.4e+02 & 7.7e+02 & \gibbonvanilla 6.5e+02 & 7.3e+02 & 5.0e+02 & 4.2e+02 & 4.7e+02 & 4.0e+02 & 6.3e+01 \\
throughputModelPass & 6.0e+02 & 5.6e+02 & \gibbonvanilla 4.9e+02 & 5.4e+02 & 3.8e+02 & 3.4e+02 & 3.3e+02 & 3.9e+02 & 7.8e+01 \\
targetReturnPass & 6.0e+04 & 5.3e+04 & \gibbonvanilla 6.1e+04 & 8.1e+04 & 5.8e+04 & 5.9e+04 & 4.2e+04 & 4.6e+04 & \textcolor{gibbonslow}{1.6e+05} \\
stripSideEffectsPass & 1.2e+05 & 1.3e+05 & \gibbonvanilla 1.2e+05 & 1.1e+05 & 1.2e+05 & 1.2e+05 & 1.0e+05 & 1.2e+05 & 1.5e+05 \\
\bottomrule
\end{tabular}}
\end{table}

\begin{table}[htbp]
\centering
\caption{Hardware counters of \texttt{Compiler}: median count per pass and iteration, AoS and pointer-based configurations (continued).}
\label{tab:Compiler_pldi_counters_by_config-part5}
\tiny\gibbonnumfont
\setlength{\tabcolsep}{3pt}
\gibbonfit{%
\begin{tabular}{l r r r r r r r r r}
\toprule
\textbf{Pass} &\multicolumn{8}{c}{\textbf{AoS}}&\multicolumn{1}{c}{\textbf{Pointer}} \\
\cmidrule(lr){2-9}\cmidrule(lr){10-10}
 & $A_{ri}^{\scriptscriptstyle -t}$ & $A_{ri}^{\scriptscriptstyle -av}$ & \gibbonvanilla $A_{ri}^{\scriptscriptstyle +av}$ & $A_{rm}^{\scriptscriptstyle -t}$ & $A_{rm}^{\scriptscriptstyle -av}$ & $A_{rm}^{\scriptscriptstyle +av}$ & $A_{\ell}^{\scriptscriptstyle -av}$ & $A_{\ell}^{\scriptscriptstyle +av}$ & $P$ \\
\midrule
\multicolumn{10}{l}{\textit{Data TLB misses}} \\
verifyIR & 3.6e+02 & 1.9e+04 & \gibbonvanilla \textcolor{gibbonslow}{5.6e+04} & 4.6e+04 & 1.2e+02 & 1.5e+02 & 1.3e+02 & 1.5e+02 & \textcolor{gibbonfast}{9.9e+01} \\
instCountPass & 4.6e+02 & 1.9e+04 & \gibbonvanilla \textcolor{gibbonslow}{6.9e+04} & 5.0e+04 & 1.5e+02 & 2.2e+02 & 1.6e+02 & 1.8e+02 & \textcolor{gibbonfast}{8.1e+01} \\
blockCountPass & 3.9e+02 & 1.8e+04 & \gibbonvanilla \textcolor{gibbonslow}{6.7e+04} & 5.0e+04 & 1.5e+02 & 2.1e+02 & 2.0e+02 & 2.0e+02 & \textcolor{gibbonfast}{5.7e+01} \\
memoryOpStatsPass & 4.4e+02 & 1.9e+04 & \gibbonvanilla \textcolor{gibbonslow}{5.6e+04} & 5.0e+04 & 1.1e+02 & 1.8e+02 & 1.5e+02 & 1.6e+02 & \textcolor{gibbonfast}{6.3e+01} \\
castInstCountPass & 3.8e+02 & 2.1e+04 & \gibbonvanilla \textcolor{gibbonslow}{5.8e+04} & 5.0e+04 & 1.4e+02 & 1.7e+02 & 1.5e+02 & 1.5e+02 & \textcolor{gibbonfast}{6.8e+01} \\
\bottomrule
\end{tabular}}
\end{table}

\begin{table}[htbp]
\centering
\caption{Hardware counters of \texttt{Compiler}: median count per pass and iteration, AoS and pointer-based configurations (continued).}
\label{tab:Compiler_pldi_counters_by_config-part6}
\tiny\gibbonnumfont
\setlength{\tabcolsep}{3pt}
\gibbonfit{%
\begin{tabular}{l r r r r r r r r r}
\toprule
\textbf{Pass} &\multicolumn{8}{c}{\textbf{AoS}}&\multicolumn{1}{c}{\textbf{Pointer}} \\
\cmidrule(lr){2-9}\cmidrule(lr){10-10}
 & $A_{ri}^{\scriptscriptstyle -t}$ & $A_{ri}^{\scriptscriptstyle -av}$ & \gibbonvanilla $A_{ri}^{\scriptscriptstyle +av}$ & $A_{rm}^{\scriptscriptstyle -t}$ & $A_{rm}^{\scriptscriptstyle -av}$ & $A_{rm}^{\scriptscriptstyle +av}$ & $A_{\ell}^{\scriptscriptstyle -av}$ & $A_{\ell}^{\scriptscriptstyle +av}$ & $P$ \\
\midrule
\multicolumn{10}{l}{\textit{L2 load misses (retired)}} \\
verifyIR & 1.4e+06 & 1.1e+06 & \gibbonvanilla 1.3e+06 & 1.1e+06 & 2.5e+06 & 2.5e+06 & 2.5e+06 & \textcolor{gibbonslow}{2.5e+06} & 6.8e+05 \\
instCountPass & 1.7e+06 & 1.2e+06 & \gibbonvanilla 1.5e+06 & 2.0e+06 & 3.1e+06 & \textcolor{gibbonslow}{3.1e+06} & 3.1e+06 & 3.1e+06 & 6.5e+05 \\
blockCountPass & 1.7e+06 & 1.3e+06 & \gibbonvanilla 1.5e+06 & 2.0e+06 & 3.1e+06 & 3.1e+06 & 3.1e+06 & \textcolor{gibbonslow}{3.1e+06} & 6.4e+05 \\
memoryOpStatsPass & 1.4e+06 & 1.1e+06 & \gibbonvanilla 1.3e+06 & \textcolor{gibbonslow}{1.6e+06} & 9.8e+05 & 1.0e+06 & 1.0e+06 & 1.0e+06 & 4.6e+05 \\
castInstCountPass & 1.4e+06 & 1.1e+06 & \gibbonvanilla 1.3e+06 & \textcolor{gibbonslow}{1.6e+06} & 1.5e+06 & 1.6e+06 & 1.6e+06 & 1.6e+06 & 4.9e+05 \\
\bottomrule
\end{tabular}}
\end{table}

\begin{table}[htbp]
\centering
\caption{Hardware counters of \texttt{Compiler}: median count per pass and iteration, AoS and pointer-based configurations (continued).}
\label{tab:Compiler_pldi_counters_by_config-part7}
\tiny\gibbonnumfont
\setlength{\tabcolsep}{3pt}
\gibbonfit{%
\begin{tabular}{l r r r r r r r r r}
\toprule
\textbf{Pass} &\multicolumn{8}{c}{\textbf{AoS}}&\multicolumn{1}{c}{\textbf{Pointer}} \\
\cmidrule(lr){2-9}\cmidrule(lr){10-10}
 & $A_{ri}^{\scriptscriptstyle -t}$ & $A_{ri}^{\scriptscriptstyle -av}$ & \gibbonvanilla $A_{ri}^{\scriptscriptstyle +av}$ & $A_{rm}^{\scriptscriptstyle -t}$ & $A_{rm}^{\scriptscriptstyle -av}$ & $A_{rm}^{\scriptscriptstyle +av}$ & $A_{\ell}^{\scriptscriptstyle -av}$ & $A_{\ell}^{\scriptscriptstyle +av}$ & $P$ \\
\midrule
\multicolumn{10}{l}{\textit{LLC load misses (retired)}} \\
verifyIR & 1.4e+06 & 1.1e+06 & \gibbonvanilla 1.3e+06 & 1.0e+06 & 2.4e+06 & 2.5e+06 & 2.5e+06 & \textcolor{gibbonslow}{2.5e+06} & 6.8e+05 \\
instCountPass & 1.6e+06 & 1.2e+06 & \gibbonvanilla 1.4e+06 & 1.9e+06 & 3.0e+06 & \textcolor{gibbonslow}{3.1e+06} & 3.0e+06 & 3.0e+06 & 6.3e+05 \\
blockCountPass & 1.7e+06 & 1.2e+06 & \gibbonvanilla 1.4e+06 & 1.9e+06 & 3.0e+06 & 3.0e+06 & 3.0e+06 & \textcolor{gibbonslow}{3.1e+06} & 6.2e+05 \\
memoryOpStatsPass & 1.4e+06 & 1.1e+06 & \gibbonvanilla 1.2e+06 & \textcolor{gibbonslow}{1.6e+06} & 9.7e+05 & 9.9e+05 & 9.9e+05 & 1.0e+06 & 4.4e+05 \\
castInstCountPass & 1.4e+06 & 1.1e+06 & \gibbonvanilla 1.3e+06 & \textcolor{gibbonslow}{1.6e+06} & 1.5e+06 & 1.5e+06 & 1.6e+06 & 1.6e+06 & 4.7e+05 \\
\bottomrule
\end{tabular}}
\end{table}

\begin{table}[htbp]
\centering
\caption{Hardware counters of \texttt{Compiler}: median count per pass and iteration, AoS and pointer-based configurations (continued).}
\label{tab:Compiler_pldi_counters_by_config-part8}
\tiny\gibbonnumfont
\setlength{\tabcolsep}{3pt}
\gibbonfit{%
\begin{tabular}{l r r r r r r r r r}
\toprule
\textbf{Pass} &\multicolumn{8}{c}{\textbf{AoS}}&\multicolumn{1}{c}{\textbf{Pointer}} \\
\cmidrule(lr){2-9}\cmidrule(lr){10-10}
 & $A_{ri}^{\scriptscriptstyle -t}$ & $A_{ri}^{\scriptscriptstyle -av}$ & \gibbonvanilla $A_{ri}^{\scriptscriptstyle +av}$ & $A_{rm}^{\scriptscriptstyle -t}$ & $A_{rm}^{\scriptscriptstyle -av}$ & $A_{rm}^{\scriptscriptstyle +av}$ & $A_{\ell}^{\scriptscriptstyle -av}$ & $A_{\ell}^{\scriptscriptstyle +av}$ & $P$ \\
\midrule
\multicolumn{10}{l}{\textit{L2 lines in}} \\
verifyIR & 1.8e+07 & \textcolor{gibbonslow}{2.1e+07} & \gibbonvanilla 1.8e+07 & 1.2e+07 & 5.1e+06 & 5.1e+06 & 5.1e+06 & 5.1e+06 & 7.5e+06 \\
instCountPass & 1.8e+07 & \textcolor{gibbonslow}{2.1e+07} & \gibbonvanilla 1.8e+07 & 1.2e+07 & 5.1e+06 & 5.1e+06 & 5.1e+06 & 5.1e+06 & 7.5e+06 \\
blockCountPass & 1.8e+07 & \textcolor{gibbonslow}{2.1e+07} & \gibbonvanilla 1.8e+07 & 1.2e+07 & 5.1e+06 & 5.1e+06 & 5.1e+06 & 5.1e+06 & 7.5e+06 \\
memoryOpStatsPass & 1.8e+07 & \textcolor{gibbonslow}{2.1e+07} & \gibbonvanilla 1.8e+07 & 1.2e+07 & 5.1e+06 & 5.1e+06 & 5.1e+06 & 5.1e+06 & 7.5e+06 \\
castInstCountPass & 1.8e+07 & \textcolor{gibbonslow}{2.1e+07} & \gibbonvanilla 1.8e+07 & 1.2e+07 & 5.1e+06 & 5.1e+06 & 5.1e+06 & 5.1e+06 & 7.5e+06 \\
\bottomrule
\end{tabular}}
\end{table}

\begin{table}[htbp]
\centering
\caption{Hardware counters of \texttt{Compiler}: median count per pass and iteration, SoA configurations.}
\label{tab:Compiler_pldi_counters_by_config-soa}
\tiny\gibbonnumfont
\setlength{\tabcolsep}{3pt}
\gibbonfit{%
\begin{tabular}{l r r r r r r r r r r r}
\toprule
\textbf{Pass} &\multicolumn{11}{c}{\textbf{SoA}} \\
\cmidrule(lr){2-12}
 & $S_{ri}^{\scriptscriptstyle -t}$ & $S_{ri}$ & $S_{rm}^{\scriptscriptstyle -t}$ & $S_{rm}^{\scriptscriptstyle -av}$ & $S_{rm}^{\scriptscriptstyle +av}$ & $S_{\ell}^{\scriptscriptstyle -av}$ & $S_{\ell}^{\scriptscriptstyle +av}$ & $S_{\ell b}^{\scriptscriptstyle -av}$ & $S_{\ell b}^{\scriptscriptstyle +av}$ & $S_{\ell bv}^{\scriptscriptstyle -av}$ & $S_{\ell bv}^{\scriptscriptstyle +av}$ \\
\midrule
\multicolumn{12}{l}{\textit{Cycles}} \\
verifyIR & -- & -- & 2.3e+08 & \textcolor{gibbonfast}{1.7e+07} & 1.8e+07 & 1.7e+07 & 1.8e+07 & 1.7e+07 & 1.9e+07 & 1.7e+07 & 1.7e+07 \\
instCountPass & -- & -- & 2.2e+08 & 9.9e+06 & 1.2e+07 & 9.8e+06 & 9.9e+06 & \textcolor{gibbonfast}{9.8e+06} & 9.8e+06 & 9.8e+06 & 9.9e+06 \\
blockCountPass & -- & -- & 2.2e+08 & 9.8e+06 & \textcolor{gibbonfast}{9.7e+06} & 9.8e+06 & 9.7e+06 & 9.8e+06 & 9.8e+06 & 9.8e+06 & 9.7e+06 \\
memoryOpStatsPass & -- & -- & 2.4e+08 & 2.0e+07 & 1.9e+07 & 1.9e+07 & 1.9e+07 & 1.9e+07 & 1.9e+07 & \textcolor{gibbonfast}{1.9e+07} & 1.9e+07 \\
castInstCountPass & -- & -- & 2.3e+08 & 1.7e+07 & 1.7e+07 & 1.7e+07 & 1.7e+07 & 1.7e+07 & \textcolor{gibbonfast}{1.7e+07} & 1.7e+07 & 1.7e+07 \\
\bottomrule
\end{tabular}}
\end{table}

\begin{table}[htbp]
\centering
\caption{Hardware counters of \texttt{Compiler}: median count per pass and iteration, SoA configurations (continued).}
\label{tab:Compiler_pldi_counters_by_config-soa-part2}
\tiny\gibbonnumfont
\setlength{\tabcolsep}{3pt}
\gibbonfit{%
\begin{tabular}{l r r r r r r r r r r r}
\toprule
\textbf{Pass} &\multicolumn{11}{c}{\textbf{SoA}} \\
\cmidrule(lr){2-12}
 & $S_{ri}^{\scriptscriptstyle -t}$ & $S_{ri}$ & $S_{rm}^{\scriptscriptstyle -t}$ & $S_{rm}^{\scriptscriptstyle -av}$ & $S_{rm}^{\scriptscriptstyle +av}$ & $S_{\ell}^{\scriptscriptstyle -av}$ & $S_{\ell}^{\scriptscriptstyle +av}$ & $S_{\ell b}^{\scriptscriptstyle -av}$ & $S_{\ell b}^{\scriptscriptstyle +av}$ & $S_{\ell bv}^{\scriptscriptstyle -av}$ & $S_{\ell bv}^{\scriptscriptstyle +av}$ \\
\midrule
\multicolumn{12}{l}{\textit{Instructions}} \\
verifyIR & -- & -- & 1.4e+08 & 8.6e+07 & 8.6e+07 & 8.6e+07 & 8.6e+07 & 8.6e+07 & 8.6e+07 & 8.6e+07 & 8.6e+07 \\
instCountPass & -- & -- & 8.9e+07 & 5.7e+07 & 5.7e+07 & 5.7e+07 & 5.7e+07 & 5.7e+07 & 5.7e+07 & 5.7e+07 & 5.7e+07 \\
blockCountPass & -- & -- & 8.4e+07 & 4.5e+07 & \textcolor{gibbonfast}{4.5e+07} & 4.5e+07 & 4.5e+07 & 4.5e+07 & 4.5e+07 & 4.5e+07 & 4.5e+07 \\
memoryOpStatsPass & -- & -- & 1.4e+08 & 9.1e+07 & 9.1e+07 & 9.1e+07 & 9.1e+07 & 9.1e+07 & 9.1e+07 & 9.1e+07 & 9.1e+07 \\
castInstCountPass & -- & -- & 1.4e+08 & 8.6e+07 & 8.6e+07 & 8.6e+07 & 8.6e+07 & 8.6e+07 & 8.6e+07 & 8.6e+07 & 8.6e+07 \\
branchStatsPass & -- & -- & 1.4e+08 & 8.6e+07 & 8.6e+07 & 8.6e+07 & 8.6e+07 & 8.6e+07 & 8.6e+07 & 8.6e+07 & 8.6e+07 \\
latencyModelPass & -- & -- & 1.1e+08 & 6.9e+07 & 6.9e+07 & 6.9e+07 & 6.9e+07 & 6.9e+07 & 6.9e+07 & 6.9e+07 & 6.9e+07 \\
has cycle & -- & -- & 1.7e+08 & 1.5e+08 & 1.5e+08 & 1.5e+08 & 1.5e+08 & 1.5e+08 & 1.5e+08 & 1.5e+08 & 1.5e+08 \\
throughputModelPass & -- & -- & 1.1e+08 & 6.9e+07 & 6.9e+07 & 6.9e+07 & 6.9e+07 & 6.9e+07 & 6.9e+07 & 6.9e+07 & 6.9e+07 \\
targetReturnPass & -- & -- & 7.4e+08 & 7.0e+08 & 7.0e+08 & 5.1e+08 & 5.1e+08 & 3.5e+07 & 1.0e+07 & 3.5e+07 & \textcolor{gibbonfast}{1.0e+07} \\
stripSideEffectsPass & -- & -- & 6.7e+08 & 6.5e+08 & 6.5e+08 & 5.0e+08 & 5.0e+08 & \textcolor{gibbonfast}{2.8e+05} & 2.8e+05 & 3.1e+05 & 3.1e+05 \\
\bottomrule
\end{tabular}}
\end{table}

\begin{table}[htbp]
\centering
\caption{Hardware counters of \texttt{Compiler}: median count per pass and iteration, SoA configurations (continued).}
\label{tab:Compiler_pldi_counters_by_config-soa-part3}
\tiny\gibbonnumfont
\setlength{\tabcolsep}{3pt}
\gibbonfit{%
\begin{tabular}{l r r r r r r r r r r r}
\toprule
\textbf{Pass} &\multicolumn{11}{c}{\textbf{SoA}} \\
\cmidrule(lr){2-12}
 & $S_{ri}^{\scriptscriptstyle -t}$ & $S_{ri}$ & $S_{rm}^{\scriptscriptstyle -t}$ & $S_{rm}^{\scriptscriptstyle -av}$ & $S_{rm}^{\scriptscriptstyle +av}$ & $S_{\ell}^{\scriptscriptstyle -av}$ & $S_{\ell}^{\scriptscriptstyle +av}$ & $S_{\ell b}^{\scriptscriptstyle -av}$ & $S_{\ell b}^{\scriptscriptstyle +av}$ & $S_{\ell bv}^{\scriptscriptstyle -av}$ & $S_{\ell bv}^{\scriptscriptstyle +av}$ \\
\midrule
\multicolumn{12}{l}{\textit{L1D load misses (retired)}} \\
verifyIR & -- & -- & 3.0e+05 & \textcolor{gibbonfast}{1.5e+04} & 1.9e+04 & 1.8e+04 & 1.9e+04 & 1.9e+04 & 2.8e+04 & 2.0e+04 & 1.6e+04 \\
instCountPass & -- & -- & 2.1e+05 & 1.4e+04 & 1.7e+04 & 1.3e+04 & 1.4e+04 & 1.2e+04 & 1.7e+04 & \textcolor{gibbonfast}{1.1e+04} & 1.8e+04 \\
blockCountPass & -- & -- & 2.1e+05 & 5.5e+03 & 1.5e+04 & 6.0e+03 & 1.6e+04 & 4.3e+03 & 7.2e+03 & \textcolor{gibbonfast}{3.4e+03} & 6.8e+03 \\
memoryOpStatsPass & -- & -- & 2.9e+05 & 3.6e+04 & \textcolor{gibbonfast}{2.3e+04} & 3.2e+04 & 3.5e+04 & 3.5e+04 & 3.8e+04 & 3.5e+04 & 3.6e+04 \\
castInstCountPass & -- & -- & 3.0e+05 & 2.8e+04 & 2.4e+04 & \textcolor{gibbonfast}{1.7e+04} & 2.2e+04 & 2.1e+04 & 1.8e+04 & 2.4e+04 & 2.1e+04 \\
\bottomrule
\end{tabular}}
\end{table}

\begin{table}[htbp]
\centering
\caption{Hardware counters of \texttt{Compiler}: median count per pass and iteration, SoA configurations (continued).}
\label{tab:Compiler_pldi_counters_by_config-soa-part4}
\tiny\gibbonnumfont
\setlength{\tabcolsep}{3pt}
\gibbonfit{%
\begin{tabular}{l r r r r r r r r r r r}
\toprule
\textbf{Pass} &\multicolumn{11}{c}{\textbf{SoA}} \\
\cmidrule(lr){2-12}
 & $S_{ri}^{\scriptscriptstyle -t}$ & $S_{ri}$ & $S_{rm}^{\scriptscriptstyle -t}$ & $S_{rm}^{\scriptscriptstyle -av}$ & $S_{rm}^{\scriptscriptstyle +av}$ & $S_{\ell}^{\scriptscriptstyle -av}$ & $S_{\ell}^{\scriptscriptstyle +av}$ & $S_{\ell b}^{\scriptscriptstyle -av}$ & $S_{\ell b}^{\scriptscriptstyle +av}$ & $S_{\ell bv}^{\scriptscriptstyle -av}$ & $S_{\ell bv}^{\scriptscriptstyle +av}$ \\
\midrule
\multicolumn{12}{l}{\textit{L1I misses}} \\
verifyIR & -- & -- & 1.1e+02 & 6.7e+01 & 7.9e+01 & 4.5e+01 & 5.5e+01 & 6.0e+01 & 6.9e+01 & 6.5e+01 & 6.9e+01 \\
instCountPass & -- & -- & 6.6e+01 & 2.0e+01 & 7.0e+01 & 3.3e+01 & 3.9e+01 & 2.3e+01 & 5.4e+01 & 4.5e+01 & 5.5e+01 \\
blockCountPass & -- & -- & 1.1e+02 & 3.1e+01 & 5.8e+01 & 5.2e+01 & 4.4e+01 & 1.6e+01 & 3.3e+01 & 4.8e+01 & 3.1e+01 \\
memoryOpStatsPass & -- & -- & 5.6e+01 & 7.5e+01 & 7.3e+01 & 5.4e+01 & 6.3e+01 & 2.5e+01 & 4.2e+01 & 4.5e+01 & 4.6e+01 \\
castInstCountPass & -- & -- & 1.1e+02 & 2.9e+01 & 5.7e+01 & 4.0e+01 & 5.5e+01 & 3.4e+01 & 6.1e+01 & 4.3e+01 & 3.9e+01 \\
branchStatsPass & -- & -- & 8.7e+01 & 5.6e+01 & 5.7e+01 & 6.0e+01 & 5.6e+01 & 3.3e+01 & 6.5e+01 & 6.8e+01 & 6.0e+01 \\
latencyModelPass & -- & -- & 1.6e+02 & 2.8e+01 & 6.7e+01 & 5.1e+01 & 4.1e+01 & 2.3e+01 & 5.2e+01 & 5.5e+01 & 5.6e+01 \\
has cycle & -- & -- & 1.5e+02 & 1.4e+02 & 1.4e+02 & 1.5e+02 & 1.5e+02 & 1.4e+02 & 1.2e+02 & 1.3e+02 & 1.4e+02 \\
throughputModelPass & -- & -- & 4.5e+01 & 6.0e+01 & 5.1e+01 & 6.3e+01 & 6.5e+01 & 3.4e+01 & 3.9e+01 & 3.9e+01 & 4.2e+01 \\
targetReturnPass & -- & -- & 1.0e+05 & 1.0e+05 & 1.1e+05 & 7.9e+04 & 7.8e+04 & 1.2e+04 & 1.3e+04 & 1.3e+04 & \textcolor{gibbonfast}{1.2e+04} \\
stripSideEffectsPass & -- & -- & 1.8e+05 & 2.1e+05 & \textcolor{gibbonslow}{2.1e+05} & 1.3e+05 & 1.3e+05 & 2.8e+04 & 2.8e+04 & \textcolor{gibbonfast}{2.6e+04} & 3.6e+04 \\
\bottomrule
\end{tabular}}
\end{table}

\begin{table}[htbp]
\centering
\caption{Hardware counters of \texttt{Compiler}: median count per pass and iteration, SoA configurations (continued).}
\label{tab:Compiler_pldi_counters_by_config-soa-part5}
\tiny\gibbonnumfont
\setlength{\tabcolsep}{3pt}
\gibbonfit{%
\begin{tabular}{l r r r r r r r r r r r}
\toprule
\textbf{Pass} &\multicolumn{11}{c}{\textbf{SoA}} \\
\cmidrule(lr){2-12}
 & $S_{ri}^{\scriptscriptstyle -t}$ & $S_{ri}$ & $S_{rm}^{\scriptscriptstyle -t}$ & $S_{rm}^{\scriptscriptstyle -av}$ & $S_{rm}^{\scriptscriptstyle +av}$ & $S_{\ell}^{\scriptscriptstyle -av}$ & $S_{\ell}^{\scriptscriptstyle +av}$ & $S_{\ell b}^{\scriptscriptstyle -av}$ & $S_{\ell b}^{\scriptscriptstyle +av}$ & $S_{\ell bv}^{\scriptscriptstyle -av}$ & $S_{\ell bv}^{\scriptscriptstyle +av}$ \\
\midrule
\multicolumn{12}{l}{\textit{Data TLB misses}} \\
verifyIR & -- & -- & 2.5e+04 & 1.5e+03 & 1.5e+03 & 1.5e+03 & 1.5e+03 & 1.5e+03 & 1.5e+03 & 1.5e+03 & 1.5e+03 \\
instCountPass & -- & -- & 2.4e+04 & 6.5e+02 & 6.4e+02 & 6.6e+02 & 6.3e+02 & 6.2e+02 & 6.8e+02 & 6.3e+02 & 6.7e+02 \\
blockCountPass & -- & -- & 2.4e+04 & 5.9e+02 & 6.1e+02 & 5.7e+02 & 6.2e+02 & 6.0e+02 & 6.1e+02 & 5.8e+02 & 5.8e+02 \\
memoryOpStatsPass & -- & -- & 2.6e+04 & 1.5e+03 & 1.5e+03 & 1.5e+03 & 1.5e+03 & 1.5e+03 & 1.5e+03 & 1.5e+03 & 1.5e+03 \\
castInstCountPass & -- & -- & 2.5e+04 & 1.5e+03 & 1.5e+03 & 1.5e+03 & 1.5e+03 & 1.5e+03 & 1.5e+03 & 1.5e+03 & 1.5e+03 \\
branchStatsPass & -- & -- & 2.6e+04 & 1.5e+03 & 1.5e+03 & 1.5e+03 & 1.5e+03 & 1.5e+03 & 1.5e+03 & 1.5e+03 & 1.5e+03 \\
latencyModelPass & -- & -- & 2.6e+04 & 1.5e+03 & 1.5e+03 & 1.5e+03 & 1.5e+03 & 1.5e+03 & 1.5e+03 & 1.5e+03 & 1.5e+03 \\
has cycle & -- & -- & 2.6e+04 & 1.8e+04 & 1.8e+04 & 1.8e+04 & 1.8e+04 & 1.8e+04 & 1.8e+04 & 1.8e+04 & 1.8e+04 \\
throughputModelPass & -- & -- & 2.5e+04 & 1.5e+03 & 1.5e+03 & 1.5e+03 & 1.5e+03 & 1.5e+03 & 1.5e+03 & 1.5e+03 & 1.5e+03 \\
targetReturnPass & -- & -- & \textcolor{gibbonslow}{1.4e+05} & 7.4e+03 & 7.8e+03 & 7.7e+03 & 7.6e+03 & 9.8e+02 & 2.0e+03 & 9.8e+02 & 2.2e+03 \\
stripSideEffectsPass & -- & -- & \textcolor{gibbonslow}{1.4e+05} & 6.4e+03 & 6.6e+03 & 3.3e+03 & 3.0e+03 & 1.1e+04 & 1.1e+04 & 1.1e+04 & 1.1e+04 \\
\bottomrule
\end{tabular}}
\end{table}

\begin{table}[htbp]
\centering
\caption{Hardware counters of \texttt{Compiler}: median count per pass and iteration, SoA configurations (continued).}
\label{tab:Compiler_pldi_counters_by_config-soa-part6}
\tiny\gibbonnumfont
\setlength{\tabcolsep}{3pt}
\gibbonfit{%
\begin{tabular}{l r r r r r r r r r r r}
\toprule
\textbf{Pass} &\multicolumn{11}{c}{\textbf{SoA}} \\
\cmidrule(lr){2-12}
 & $S_{ri}^{\scriptscriptstyle -t}$ & $S_{ri}$ & $S_{rm}^{\scriptscriptstyle -t}$ & $S_{rm}^{\scriptscriptstyle -av}$ & $S_{rm}^{\scriptscriptstyle +av}$ & $S_{\ell}^{\scriptscriptstyle -av}$ & $S_{\ell}^{\scriptscriptstyle +av}$ & $S_{\ell b}^{\scriptscriptstyle -av}$ & $S_{\ell b}^{\scriptscriptstyle +av}$ & $S_{\ell bv}^{\scriptscriptstyle -av}$ & $S_{\ell bv}^{\scriptscriptstyle +av}$ \\
\midrule
\multicolumn{12}{l}{\textit{L2 load misses (retired)}} \\
verifyIR & -- & -- & 4.5e+04 & \textcolor{gibbonfast}{3.8e+03} & 4.7e+03 & 4.4e+03 & 4.8e+03 & 4.7e+03 & 4.8e+03 & 4.8e+03 & 4.1e+03 \\
instCountPass & -- & -- & 3.9e+04 & 1.3e+03 & 1.3e+03 & 1.3e+03 & \textcolor{gibbonfast}{1.3e+03} & 1.3e+03 & 1.3e+03 & 1.3e+03 & 1.3e+03 \\
blockCountPass & -- & -- & 3.9e+04 & 1.2e+03 & 1.3e+03 & 1.2e+03 & 1.3e+03 & 1.0e+03 & 1.3e+03 & \textcolor{gibbonfast}{1.0e+03} & 1.2e+03 \\
memoryOpStatsPass & -- & -- & 4.5e+04 & 3.5e+03 & \textcolor{gibbonfast}{3.1e+03} & 3.3e+03 & 3.2e+03 & 3.4e+03 & 3.5e+03 & 3.4e+03 & 3.5e+03 \\
castInstCountPass & -- & -- & 4.5e+04 & \textcolor{gibbonfast}{2.8e+03} & 3.8e+03 & 3.0e+03 & 3.1e+03 & 3.2e+03 & 2.9e+03 & 3.2e+03 & 3.2e+03 \\
\bottomrule
\end{tabular}}
\end{table}

\begin{table}[htbp]
\centering
\caption{Hardware counters of \texttt{Compiler}: median count per pass and iteration, SoA configurations (continued).}
\label{tab:Compiler_pldi_counters_by_config-soa-part7}
\tiny\gibbonnumfont
\setlength{\tabcolsep}{3pt}
\gibbonfit{%
\begin{tabular}{l r r r r r r r r r r r}
\toprule
\textbf{Pass} &\multicolumn{11}{c}{\textbf{SoA}} \\
\cmidrule(lr){2-12}
 & $S_{ri}^{\scriptscriptstyle -t}$ & $S_{ri}$ & $S_{rm}^{\scriptscriptstyle -t}$ & $S_{rm}^{\scriptscriptstyle -av}$ & $S_{rm}^{\scriptscriptstyle +av}$ & $S_{\ell}^{\scriptscriptstyle -av}$ & $S_{\ell}^{\scriptscriptstyle +av}$ & $S_{\ell b}^{\scriptscriptstyle -av}$ & $S_{\ell b}^{\scriptscriptstyle +av}$ & $S_{\ell bv}^{\scriptscriptstyle -av}$ & $S_{\ell bv}^{\scriptscriptstyle +av}$ \\
\midrule
\multicolumn{12}{l}{\textit{LLC load misses (retired)}} \\
verifyIR & -- & -- & 4.1e+04 & 3.7e+03 & 4.5e+03 & 4.0e+03 & 4.5e+03 & 4.4e+03 & 4.4e+03 & 4.4e+03 & \textcolor{gibbonfast}{3.7e+03} \\
instCountPass & -- & -- & 3.5e+04 & \textcolor{gibbonfast}{0.0e+00} & 7.0e+00 & 2.0e+00 & 4.0e+00 & \textcolor{gibbonfast}{0.0e+00} & 1.0e+00 & 2.0e+00 & 4.0e+00 \\
blockCountPass & -- & -- & 3.5e+04 & \textcolor{gibbonfast}{0.0e+00} & \textcolor{gibbonfast}{0.0e+00} & \textcolor{gibbonfast}{0.0e+00} & 2.0e+00 & \textcolor{gibbonfast}{0.0e+00} & 2.0e+00 & 1.0e+00 & \textcolor{gibbonfast}{0.0e+00} \\
memoryOpStatsPass & -- & -- & 4.1e+04 & 2.8e+03 & \textcolor{gibbonfast}{2.4e+03} & 2.5e+03 & 2.4e+03 & 2.7e+03 & 2.8e+03 & 2.6e+03 & 2.7e+03 \\
castInstCountPass & -- & -- & 4.2e+04 & \textcolor{gibbonfast}{2.0e+03} & 3.0e+03 & 2.1e+03 & 2.3e+03 & 2.4e+03 & 2.1e+03 & 2.4e+03 & 2.4e+03 \\
\bottomrule
\end{tabular}}
\end{table}

\begin{table}[htbp]
\centering
\caption{Hardware counters of \texttt{Compiler}: median count per pass and iteration, SoA configurations (continued).}
\label{tab:Compiler_pldi_counters_by_config-soa-part8}
\tiny\gibbonnumfont
\setlength{\tabcolsep}{3pt}
\gibbonfit{%
\begin{tabular}{l r r r r r r r r r r r}
\toprule
\textbf{Pass} &\multicolumn{11}{c}{\textbf{SoA}} \\
\cmidrule(lr){2-12}
 & $S_{ri}^{\scriptscriptstyle -t}$ & $S_{ri}$ & $S_{rm}^{\scriptscriptstyle -t}$ & $S_{rm}^{\scriptscriptstyle -av}$ & $S_{rm}^{\scriptscriptstyle +av}$ & $S_{\ell}^{\scriptscriptstyle -av}$ & $S_{\ell}^{\scriptscriptstyle +av}$ & $S_{\ell b}^{\scriptscriptstyle -av}$ & $S_{\ell b}^{\scriptscriptstyle +av}$ & $S_{\ell bv}^{\scriptscriptstyle -av}$ & $S_{\ell bv}^{\scriptscriptstyle +av}$ \\
\midrule
\multicolumn{12}{l}{\textit{L2 lines in}} \\
verifyIR & -- & -- & 4.0e+06 & \textcolor{gibbonfast}{8.4e+05} & 8.5e+05 & 8.4e+05 & 8.4e+05 & 8.4e+05 & 8.4e+05 & 8.5e+05 & 8.4e+05 \\
instCountPass & -- & -- & 3.3e+06 & 1.2e+05 & 1.2e+05 & \textcolor{gibbonfast}{1.2e+05} & 1.2e+05 & 1.2e+05 & 1.2e+05 & 1.2e+05 & 1.3e+05 \\
blockCountPass & -- & -- & 3.3e+06 & 1.2e+05 & 1.2e+05 & \textcolor{gibbonfast}{1.2e+05} & 1.2e+05 & 1.2e+05 & 1.2e+05 & 1.2e+05 & 1.2e+05 \\
memoryOpStatsPass & -- & -- & 4.1e+06 & 8.6e+05 & 8.6e+05 & \textcolor{gibbonfast}{8.6e+05} & 8.6e+05 & 8.6e+05 & 8.6e+05 & 8.7e+05 & 8.6e+05 \\
castInstCountPass & -- & -- & 4.0e+06 & \textcolor{gibbonfast}{8.4e+05} & 8.4e+05 & 8.4e+05 & 8.4e+05 & 8.4e+05 & 8.4e+05 & 8.4e+05 & 8.4e+05 \\
\bottomrule
\end{tabular}}
\end{table}%
  \endgroup
\FloatBarrier

\subsubsection{DBQuery}
\PassName{DBQuery} models a database query-plan tree, the representation that
cost-based optimizers and execution planners use~\cite{selinger1979access}. As in
the iterator execution model~\cite{graefe1994volcano}, internal nodes are
operators such as joins and filters, and leaves are sequential or index scans.
Query-plan trees are also the input of database benchmarking and
performance-prediction workloads~\cite{zhao2024comparative}. The folds aggregate
cost and cardinality bottom-up (\PassName{sumCost}, \PassName{sumRows}), count
joins (\PassName{countJoins}), estimate memory pressure (\PassName{sumMemory},
\PassName{hashJoinPressure}) and measure filter selectivity
(\PassName{filterSelectivitySkew}); the maps update plan metadata
(\PassName{scaleCosts}, \PassName{clearQueryFlags}). Results are in
\TableJumpRef{tab:DBQuery}.
  \begingroup
    \setlength{\textfloatsep}{4pt plus 1pt minus 1pt}%
    \setlength{\floatsep}{3pt plus 1pt minus 1pt}%
    \setlength{\intextsep}{3pt plus 1pt minus 1pt}%
    \setlength{\abovecaptionskip}{3pt}%
    \setlength{\belowcaptionskip}{2pt}%
    \input{tables/pldi/pldi-fold-DBQuery}%
  \endgroup
  \begingroup
    \setlength{\textfloatsep}{4pt plus 1pt minus 1pt}%
    \setlength{\floatsep}{3pt plus 1pt minus 1pt}%
    \setlength{\intextsep}{3pt plus 1pt minus 1pt}%
    \setlength{\abovecaptionskip}{3pt}%
    \setlength{\belowcaptionskip}{2pt}%
    \input{tables/pldi/pldi-map-DBQuery}%
  \endgroup
\FloatBarrier

\subsubsection{DecisionTree}
Decision trees are a standard classification model~\cite{breiman1984cart}.
\PassName{DecisionTree} models a binary decision tree used for inference and
model inspection: internal nodes hold a feature, a threshold and two subtrees,
and leaves hold a label and sample statistics. Its passes are all folds:
structural traversals (\PassName{countNodes}, \PassName{countLeaves},
\PassName{treeDepth}) and statistics and diagnostics (\PassName{sumImpurity},
\PassName{sumSamples}, \PassName{countFeatureUses}, \PassName{countSmallLeaves},
\PassName{sumPathLengths}, \PassName{inferenceCost}).
\PassName{DecisionTreeClassify} runs classification queries over the same tree
(\PassName{classify Depth}, \PassName{classify Batch}); each query follows a single
root-to-leaf path. Results are in \TableJumpRef{tab:DecisionTree}.
  \begingroup
    \setlength{\textfloatsep}{4pt plus 1pt minus 1pt}%
    \setlength{\floatsep}{3pt plus 1pt minus 1pt}%
    \setlength{\intextsep}{3pt plus 1pt minus 1pt}%
    \setlength{\abovecaptionskip}{3pt}%
    \setlength{\belowcaptionskip}{2pt}%
    \input{tables/pldi/pldi-fold-DecisionTree}%
  \endgroup
\FloatBarrier

  \begingroup
    \setlength{\textfloatsep}{4pt plus 1pt minus 1pt}%
    \setlength{\floatsep}{3pt plus 1pt minus 1pt}%
    \setlength{\intextsep}{3pt plus 1pt minus 1pt}%
    \setlength{\abovecaptionskip}{3pt}%
    \setlength{\belowcaptionskip}{2pt}%
    \input{tables/pldi/pldi-fold-DecisionTreeClassify}%
  \endgroup
\FloatBarrier

\subsubsection{DomTree}
\PassName{DomTree} models a document and render tree like those of browser layout
and UI rendering engines~\cite{Orchard, meyerovich2010fast}. \lstinline|Element|
nodes hold style and layout metadata and their children, and \lstinline|Text|
nodes hold text metrics. The folds summarize the layout (\PassName{SumArea},
\PassName{find max Bottom}, \PassName{count styled}, \PassName{sumTextWidth}), and
the maps transform it (\PassName{computeWidths}, \PassName{scaleLayout}). Results
are in \TableJumpRef{tab:DomTree}.
  \begingroup
    \setlength{\textfloatsep}{4pt plus 1pt minus 1pt}%
    \setlength{\floatsep}{3pt plus 1pt minus 1pt}%
    \setlength{\intextsep}{3pt plus 1pt minus 1pt}%
    \setlength{\abovecaptionskip}{3pt}%
    \setlength{\belowcaptionskip}{2pt}%
    \input{tables/pldi/pldi-fold-DomTree}%
  \endgroup
  \begingroup
    \setlength{\textfloatsep}{4pt plus 1pt minus 1pt}%
    \setlength{\floatsep}{3pt plus 1pt minus 1pt}%
    \setlength{\intextsep}{3pt plus 1pt minus 1pt}%
    \setlength{\abovecaptionskip}{3pt}%
    \setlength{\belowcaptionskip}{2pt}%
    \input{tables/pldi/pldi-map-DomTree}%
  \endgroup
\FloatBarrier

\subsubsection{ObjectGraph}
\PassName{ObjectGraph} models a heap object graph, as traversed by garbage
collectors and memory profilers. Each node holds an object's identity, size, children and other metadata. The folds account for memory
(\PassName{totalHeapSize}, \PassName{countMarked}, \PassName{countLargeItems},
\PassName{liveBytes}, \PassName{deadBytes}, \PassName{countSurvivors},
\PassName{sumObjIds}), and the maps mimic collector phases
(\PassName{sweepUnmarked}, \PassName{touchHotObjects}). Results are in
\TableJumpRef{tab:ObjectGraph}.
  \begingroup
    \setlength{\textfloatsep}{4pt plus 1pt minus 1pt}%
    \setlength{\floatsep}{3pt plus 1pt minus 1pt}%
    \setlength{\intextsep}{3pt plus 1pt minus 1pt}%
    \setlength{\abovecaptionskip}{3pt}%
    \setlength{\belowcaptionskip}{2pt}%
    \input{tables/pldi/pldi-fold-ObjectGraph}%
  \endgroup
  \begingroup
    \setlength{\textfloatsep}{4pt plus 1pt minus 1pt}%
    \setlength{\floatsep}{3pt plus 1pt minus 1pt}%
    \setlength{\intextsep}{3pt plus 1pt minus 1pt}%
    \setlength{\abovecaptionskip}{3pt}%
    \setlength{\belowcaptionskip}{2pt}%
    \input{tables/pldi/pldi-map-ObjectGraph}%
  \endgroup
\FloatBarrier

\subsubsection{OctTree}
Octrees partition three-dimensional space hierarchically; they are used in N-body simulation~\cite{barnes1986hierarchical} and geometric modeling~\cite{Meagher1982}. \PassName{OctTree} models the octree of a
hierarchical N-body or multipole computation, with constructors
\PassName{Cell}, \PassName{Particle} and \PassName{EmptyOct}. Its folds compute
physical aggregates and potentials (\PassName{sumMass}, \PassName{sumEnergy},
\PassName{countActive}, \PassName{countParticles},
\PassName{barnesHutPotential}, \PassName{fmmPotential}), and its maps update
particles (\PassName{scaleEnergy}, \PassName{clearFlags}).
\PassName{ColorOctree} uses an adapted octree for color
quantization~\cite{Gervautz1990}. Its folds estimate the quantized palette and
its error (\PassName{paletteEntriesQuantized}, \PassName{quantizationErrorProxy},
\PassName{reduceColorCount}, \PassName{closestColor}), and its maps convert
colors and mark palette leaves (\PassName{toYCoCg}, \PassName{markPaletteLeaves}).
Results are in \TableJumpRef{tab:OctTree}.
  \begingroup
    \setlength{\textfloatsep}{4pt plus 1pt minus 1pt}%
    \setlength{\floatsep}{3pt plus 1pt minus 1pt}%
    \setlength{\intextsep}{3pt plus 1pt minus 1pt}%
    \setlength{\abovecaptionskip}{3pt}%
    \setlength{\belowcaptionskip}{2pt}%
    \input{tables/pldi/pldi-fold-OctTree}%
  \endgroup
  \begingroup
    \setlength{\textfloatsep}{4pt plus 1pt minus 1pt}%
    \setlength{\floatsep}{3pt plus 1pt minus 1pt}%
    \setlength{\intextsep}{3pt plus 1pt minus 1pt}%
    \setlength{\abovecaptionskip}{3pt}%
    \setlength{\belowcaptionskip}{2pt}%
    \input{tables/pldi/pldi-map-OctTree}%
  \endgroup
\FloatBarrier

  \begingroup
    \setlength{\textfloatsep}{4pt plus 1pt minus 1pt}%
    \setlength{\floatsep}{3pt plus 1pt minus 1pt}%
    \setlength{\intextsep}{3pt plus 1pt minus 1pt}%
    \setlength{\abovecaptionskip}{3pt}%
    \setlength{\belowcaptionskip}{2pt}%
    \input{tables/pldi/pldi-fold-ColorOctree}%
  \endgroup
  \begingroup
    \setlength{\textfloatsep}{4pt plus 1pt minus 1pt}%
    \setlength{\floatsep}{3pt plus 1pt minus 1pt}%
    \setlength{\intextsep}{3pt plus 1pt minus 1pt}%
    \setlength{\abovecaptionskip}{3pt}%
    \setlength{\belowcaptionskip}{2pt}%
    \input{tables/pldi/pldi-map-ColorOctree}%
  \endgroup
\FloatBarrier

\subsubsection{PiecewiseFunctions}
\PassName{PiecewiseFunctions}~\cite{Orchard, SRajbhandari} represents functions as
adaptive k-d trees, as in scientific simulation; it is modeled on the
MADNESS\footnote{\url{https://github.com/m-a-d-n-e-s-s/madness}} framework, whose
code base we scanned for folds and maps. Leaves hold local coefficients, a scale
and an error estimate, and internal nodes hold the split dimension, the split
value and refinement metadata. The folds are inspired by MADNESS (\PassName{norm2Estimate}, \PassName{truncateTolViolations}, \PassName{compressMass}, \PassName{autorefineMaxLevel}, \PassName{pmapCutHistogram}, \PassName{lbDeuxLoadProxy}), and the maps are \PassName{addConstPW} and \PassName{diffPW}. Results are in
\TableJumpRef{tab:PiecewiseFunctions}.
  \begingroup
    \setlength{\textfloatsep}{4pt plus 1pt minus 1pt}%
    \setlength{\floatsep}{3pt plus 1pt minus 1pt}%
    \setlength{\intextsep}{3pt plus 1pt minus 1pt}%
    \setlength{\abovecaptionskip}{3pt}%
    \setlength{\belowcaptionskip}{2pt}%
    \input{tables/pldi/pldi-fold-PiecewiseFunctions}%
  \endgroup
  \begingroup
    \setlength{\textfloatsep}{4pt plus 1pt minus 1pt}%
    \setlength{\floatsep}{3pt plus 1pt minus 1pt}%
    \setlength{\intextsep}{3pt plus 1pt minus 1pt}%
    \setlength{\abovecaptionskip}{3pt}%
    \setlength{\belowcaptionskip}{2pt}%
    \input{tables/pldi/pldi-map-PiecewiseFunctions}%
  \endgroup
\FloatBarrier

\subsubsection{Trie}
A trie~\cite{Trie-ref} stores keys by their prefixes. \PassName{Trie} represents
one as a binary tree, with constructors \PassName{TNode}, \PassName{TLeaf} and
\PassName{TEmpty}, as in prefix search and autocompletion. The folds query the
index (\PassName{sumPrefixFreq}, \PassName{countTerminals},
\PassName{sumSubtreeHints}, \PassName{autocompleteTopKProxy},
\PassName{countLazyNodes}), and the maps update its statistics
(\PassName{decayTrieStats}, \PassName{resetTraversalState}). Results are in
\TableJumpRef{tab:Trie}.
  \begingroup
    \setlength{\textfloatsep}{4pt plus 1pt minus 1pt}%
    \setlength{\floatsep}{3pt plus 1pt minus 1pt}%
    \setlength{\intextsep}{3pt plus 1pt minus 1pt}%
    \setlength{\abovecaptionskip}{3pt}%
    \setlength{\belowcaptionskip}{2pt}%
    \input{tables/pldi/pldi-fold-Trie}%
  \endgroup
  \begingroup
    \setlength{\textfloatsep}{4pt plus 1pt minus 1pt}%
    \setlength{\floatsep}{3pt plus 1pt minus 1pt}%
    \setlength{\intextsep}{3pt plus 1pt minus 1pt}%
    \setlength{\abovecaptionskip}{3pt}%
    \setlength{\belowcaptionskip}{2pt}%
    \input{tables/pldi/pldi-map-Trie}%
  \endgroup
\FloatBarrier

\subsubsection{LinearListReduction}\label{sec:appendix-eval-llr}
\PassName{LinearListReduction} is a list whose cells carry eleven scalar fields,
more than any one pass reads. Its folds (\PassName{reduction},
\PassName{maxField}, \PassName{countHeavy}) each read two of the eleven, so the
\unfactored{} layout steps over the other nine in every cell while the
\fullyfactored{} layout never loads their buffers. Its maps
(\PassName{bumpFirst}, \PassName{shiftTriple}, \PassName{touchMost}) write a new
list, so they copy the fields they do not change unless selective buffer sharing
reuses those buffers (Section~\ref{sec:factored-opts}). Results are in
\TableJumpRef{tab:LinearListReduction}, followed by the hardware counters that
Section~\ref{sec:eval} cites for \PassName{reduction}.
  \begingroup
    \setlength{\textfloatsep}{4pt plus 1pt minus 1pt}%
    \setlength{\floatsep}{3pt plus 1pt minus 1pt}%
    \setlength{\intextsep}{3pt plus 1pt minus 1pt}%
    \setlength{\abovecaptionskip}{3pt}%
    \setlength{\belowcaptionskip}{2pt}%
    \input{tables/pldi/pldi-fold-LinearListReduction}%
  \endgroup
  \begingroup
    \setlength{\textfloatsep}{4pt plus 1pt minus 1pt}%
    \setlength{\floatsep}{3pt plus 1pt minus 1pt}%
    \setlength{\intextsep}{3pt plus 1pt minus 1pt}%
    \setlength{\abovecaptionskip}{3pt}%
    \setlength{\belowcaptionskip}{2pt}%
    \input{tables/pldi/pldi-map-LinearListReduction}%
  \endgroup
\FloatBarrier

  \begingroup
    \setlength{\textfloatsep}{4pt plus 1pt minus 1pt}%
    \setlength{\floatsep}{3pt plus 1pt minus 1pt}%
    \setlength{\intextsep}{3pt plus 1pt minus 1pt}%
    \setlength{\abovecaptionskip}{3pt}%
    \setlength{\belowcaptionskip}{2pt}%
    \input{tables/pldi/LinearListReduction_pldi_counters_by_config}%
  \endgroup
\FloatBarrier

\subsubsection{ReduceNestedList}
\PassName{ReduceNestedList} is the nested list of the introduction: each cell
holds an integer and an inner list. Its one fold, \PassName{reduction}, reads the
integers and never the inner lists, which makes it a small test of skipping dead
data in a recursive datatype. Results are in \TableJumpRef{tab:reduceNestedList}.
  \begingroup
    \setlength{\textfloatsep}{4pt plus 1pt minus 1pt}%
    \setlength{\floatsep}{3pt plus 1pt minus 1pt}%
    \setlength{\intextsep}{3pt plus 1pt minus 1pt}%
    \setlength{\abovecaptionskip}{3pt}%
    \setlength{\belowcaptionskip}{2pt}%
    \input{tables/pldi/pldi-fold-reduceNestedList}%
  \endgroup
\FloatBarrier

\subsubsection{List}
\PassName{List} is the simplest baseline: a cons list with one map
(\PassName{add1 List}) and three folds (\PassName{length List},
\PassName{sumList}, \PassName{sumList tail recursive}). Results are in
\TableJumpRef{tab:List}.
  \begingroup
    \setlength{\textfloatsep}{4pt plus 1pt minus 1pt}%
    \setlength{\floatsep}{3pt plus 1pt minus 1pt}%
    \setlength{\intextsep}{3pt plus 1pt minus 1pt}%
    \setlength{\abovecaptionskip}{3pt}%
    \setlength{\belowcaptionskip}{2pt}%
    \input{tables/pldi/pldi-fold-List}%
  \endgroup
  \begingroup
    \setlength{\textfloatsep}{4pt plus 1pt minus 1pt}%
    \setlength{\floatsep}{3pt plus 1pt minus 1pt}%
    \setlength{\intextsep}{3pt plus 1pt minus 1pt}%
    \setlength{\abovecaptionskip}{3pt}%
    \setlength{\belowcaptionskip}{2pt}%
    \input{tables/pldi/pldi-map-List}%
  \endgroup
\FloatBarrier

\subsubsection{MonoTree}
\PassName{MonoTree} is a binary tree with integers only at its leaves, the
\lstinline|Tree| of Section~\ref{sec:background:c}. It has one map
(\PassName{add1Tree}) and two folds (\PassName{sumTree},
\PassName{sumTree TailRec}). Results are in \TableJumpRef{tab:MonoTree}.
  \begingroup
    \setlength{\textfloatsep}{4pt plus 1pt minus 1pt}%
    \setlength{\floatsep}{3pt plus 1pt minus 1pt}%
    \setlength{\intextsep}{3pt plus 1pt minus 1pt}%
    \setlength{\abovecaptionskip}{3pt}%
    \setlength{\belowcaptionskip}{2pt}%
    \input{tables/pldi/pldi-fold-MonoTree}%
  \endgroup
  \begingroup
    \setlength{\textfloatsep}{4pt plus 1pt minus 1pt}%
    \setlength{\floatsep}{3pt plus 1pt minus 1pt}%
    \setlength{\intextsep}{3pt plus 1pt minus 1pt}%
    \setlength{\abovecaptionskip}{3pt}%
    \setlength{\belowcaptionskip}{2pt}%
    \input{tables/pldi/pldi-map-MonoTree}%
  \endgroup
\FloatBarrier

\subsubsection{TernaryTree}
\PassName{TernaryTree} gives each internal node three children, to vary the
branching factor of the tree baselines. It has one map (\PassName{add 1 tree}) and
one fold (\PassName{sum tree}). Results are in \TableJumpRef{tab:TernaryTree}.
  \begingroup
    \setlength{\textfloatsep}{4pt plus 1pt minus 1pt}%
    \setlength{\floatsep}{3pt plus 1pt minus 1pt}%
    \setlength{\intextsep}{3pt plus 1pt minus 1pt}%
    \setlength{\abovecaptionskip}{3pt}%
    \setlength{\belowcaptionskip}{2pt}%
    \input{tables/pldi/pldi-fold-TernaryTree}%
  \endgroup
  \begingroup
    \setlength{\textfloatsep}{4pt plus 1pt minus 1pt}%
    \setlength{\floatsep}{3pt plus 1pt minus 1pt}%
    \setlength{\intextsep}{3pt plus 1pt minus 1pt}%
    \setlength{\abovecaptionskip}{3pt}%
    \setlength{\belowcaptionskip}{2pt}%
    \input{tables/pldi/pldi-map-TernaryTree}%
  \endgroup
\FloatBarrier

\subsubsection{Bit-width microbenchmarks}\label{sec:appendix-eval-remaining}

\PassName{Add1TreeInt}$N$ and \PassName{ArithmeticIntensityInt}$N$ run one map
pass over a tree of $N$-bit integers, for $N = 8$, 16, 32 and 64.
Table~\ref{tab:width} in
Section~\ref{sec:eval} summarizes them; their map tables follow.

  \begingroup
    \setlength{\textfloatsep}{4pt plus 1pt minus 1pt}%
    \setlength{\floatsep}{3pt plus 1pt minus 1pt}%
    \setlength{\intextsep}{3pt plus 1pt minus 1pt}%
    \setlength{\abovecaptionskip}{3pt}%
    \setlength{\belowcaptionskip}{2pt}%
    \input{tables/pldi/pldi-map-Add1TreeInt8}%
  \endgroup
\FloatBarrier

  \begingroup
    \setlength{\textfloatsep}{4pt plus 1pt minus 1pt}%
    \setlength{\floatsep}{3pt plus 1pt minus 1pt}%
    \setlength{\intextsep}{3pt plus 1pt minus 1pt}%
    \setlength{\abovecaptionskip}{3pt}%
    \setlength{\belowcaptionskip}{2pt}%
    \input{tables/pldi/pldi-map-Add1TreeInt16}%
  \endgroup
\FloatBarrier

  \begingroup
    \setlength{\textfloatsep}{4pt plus 1pt minus 1pt}%
    \setlength{\floatsep}{3pt plus 1pt minus 1pt}%
    \setlength{\intextsep}{3pt plus 1pt minus 1pt}%
    \setlength{\abovecaptionskip}{3pt}%
    \setlength{\belowcaptionskip}{2pt}%
    \input{tables/pldi/pldi-map-Add1TreeInt32}%
  \endgroup
\FloatBarrier

  \begingroup
    \setlength{\textfloatsep}{4pt plus 1pt minus 1pt}%
    \setlength{\floatsep}{3pt plus 1pt minus 1pt}%
    \setlength{\intextsep}{3pt plus 1pt minus 1pt}%
    \setlength{\abovecaptionskip}{3pt}%
    \setlength{\belowcaptionskip}{2pt}%
    \input{tables/pldi/pldi-map-Add1TreeInt64}%
  \endgroup
\FloatBarrier

  \begingroup
    \setlength{\textfloatsep}{4pt plus 1pt minus 1pt}%
    \setlength{\floatsep}{3pt plus 1pt minus 1pt}%
    \setlength{\intextsep}{3pt plus 1pt minus 1pt}%
    \setlength{\abovecaptionskip}{3pt}%
    \setlength{\belowcaptionskip}{2pt}%
    \input{tables/pldi/pldi-map-ArithmeticIntensityInt8}%
  \endgroup
\FloatBarrier

  \begingroup
    \setlength{\textfloatsep}{4pt plus 1pt minus 1pt}%
    \setlength{\floatsep}{3pt plus 1pt minus 1pt}%
    \setlength{\intextsep}{3pt plus 1pt minus 1pt}%
    \setlength{\abovecaptionskip}{3pt}%
    \setlength{\belowcaptionskip}{2pt}%
    \input{tables/pldi/pldi-map-ArithmeticIntensityInt16}%
  \endgroup
\FloatBarrier

  \begingroup
    \setlength{\textfloatsep}{4pt plus 1pt minus 1pt}%
    \setlength{\floatsep}{3pt plus 1pt minus 1pt}%
    \setlength{\intextsep}{3pt plus 1pt minus 1pt}%
    \setlength{\abovecaptionskip}{3pt}%
    \setlength{\belowcaptionskip}{2pt}%
    \input{tables/pldi/pldi-map-ArithmeticIntensityInt32}%
  \endgroup
\FloatBarrier

  \begingroup
    \setlength{\textfloatsep}{4pt plus 1pt minus 1pt}%
    \setlength{\floatsep}{3pt plus 1pt minus 1pt}%
    \setlength{\intextsep}{3pt plus 1pt minus 1pt}%
    \setlength{\abovecaptionskip}{3pt}%
    \setlength{\belowcaptionskip}{2pt}%
    \input{tables/pldi/pldi-map-ArithmeticIntensityInt64}%
  \endgroup
\FloatBarrier

\FloatBarrier
\section{Results with Clang}\label{sec:appendix-clang}

Section~\ref{sec:eval} and Appendix~\ref{sec:appendix-eval} compile the C that
\colobus{} generates with GCC~16.2.0. To see how much the results depend on the
C compiler, we repeated every measurement with Clang~23.1.3, on the same
machine, for AVX2, with the same configurations, inputs and measurement
procedure. This appendix gives the Clang results; the configuration symbols
are those of Table~\ref{tab:pldi-legend}, and every table names the compiler
in its caption.

\subsection{Clang against GCC}\label{sec:appendix-clang-vs-gcc}

Table~\ref{tab:compiler_comparison} divides each configuration's end-to-end
time under Clang by its time under GCC, so a value above 1 means the Clang
build is slower. The end-to-end time is the build plus every timed pass. The
two compilers ran as separate campaigns with otherwise identical settings. The
geometric-mean rows cover the programs where both compilers have a verified
time, and the fold and map rows sum only those passes; `--' marks a
configuration that did not verify under one of them.

  \begingroup
    \setlength{\textfloatsep}{4pt plus 1pt minus 1pt}%
    \setlength{\floatsep}{3pt plus 1pt minus 1pt}%
    \setlength{\intextsep}{3pt plus 1pt minus 1pt}%
    \setlength{\abovecaptionskip}{3pt}%
    \setlength{\belowcaptionskip}{2pt}%
    \begin{table}[htbp]
\centering
\caption{Clang time over GCC time, end to end, per program and configuration, AoS and pointer-based configurations.}
\label{tab:compiler_comparison}
\tiny\gibbonnumfont
\setlength{\tabcolsep}{3pt}
\gibbonfit{%
\begin{tabular}{l r r r r r r r r r}
\toprule
\textbf{Program} &\multicolumn{8}{c}{\textbf{AoS}}&\multicolumn{1}{c}{\textbf{Pointer}} \\
\cmidrule(lr){2-9}\cmidrule(lr){10-10}
 & $A_{ri}^{\scriptscriptstyle -t}$ & $A_{ri}^{\scriptscriptstyle -av}$ & \gibbonvanilla $A_{ri}^{\scriptscriptstyle +av}$ & $A_{rm}^{\scriptscriptstyle -t}$ & $A_{rm}^{\scriptscriptstyle -av}$ & $A_{rm}^{\scriptscriptstyle +av}$ & $A_{\ell}^{\scriptscriptstyle -av}$ & $A_{\ell}^{\scriptscriptstyle +av}$ & $P$ \\
\midrule
Compiler & 0.93 & 0.87 & \gibbonvanilla 0.91 & 0.85 & 0.99 & 1.01 & 1.01 & 1.01 & 1.07 \\
DBQuery & 1.04 & 1.16 & \gibbonvanilla 1.05 & 1.10 & 1.02 & 1.02 & 1.03 & 1.03 & 0.99 \\
DecisionTree & 1.02 & 1.67 & \gibbonvanilla 1.01 & 1.08 & 1.00 & 1.00 & 1.00 & 1.00 & 1.06 \\
DecisionTreeClassify & 1.66 & 1.01 & \gibbonvanilla 1.00 & 1.63 & 1.02 & 1.02 & 1.02 & 1.02 & 1.05 \\
DomTree & 1.03 & 1.28 & \gibbonvanilla 1.04 & 1.05 & 1.03 & 1.04 & 1.03 & 1.03 & 1.05 \\
KDTree & 1.01 & 1.08 & \gibbonvanilla 1.02 & 1.04 & 1.06 & 1.06 & 1.06 & 1.05 & 1.07 \\
LinearListReduction & 0.85 & 0.90 & \gibbonvanilla 0.84 & 0.81 & 0.99 & 1.00 & 1.00 & 1.00 & 1.05 \\
reduceNestedList & 0.92 & 0.93 & \gibbonvanilla 0.93 & 1.85 & 0.99 & 1.01 & 1.00 & 1.00 & 1.42 \\
List & -- & -- & \gibbonvanilla -- & -- & 1.05 & 1.05 & 1.04 & 1.04 & 1.39 \\
MonoTree & 1.07 & 1.47 & \gibbonvanilla 1.06 & 1.19 & 1.18 & 1.19 & 1.23 & 1.22 & 1.03 \\
ObjectGraph & 1.04 & 1.50 & \gibbonvanilla 1.04 & 1.07 & 1.01 & 1.01 & 1.04 & 1.05 & 1.00 \\
TernaryTree & 1.12 & 1.27 & \gibbonvanilla 1.12 & 1.07 & 0.93 & 0.95 & 1.05 & 1.01 & 1.07 \\
Trie & 1.00 & 1.14 & \gibbonvanilla 0.99 & 0.97 & 0.96 & 0.97 & 0.97 & 0.97 & 1.00 \\
ColorOctree & 1.07 & 1.10 & \gibbonvanilla 1.07 & 1.06 & 1.07 & 1.07 & 1.10 & 1.10 & 1.05 \\
Add1TreeInt8 & 1.19 & 1.27 & \gibbonvanilla 1.21 & 1.13 & 0.91 & 0.86 & 0.95 & 1.12 & 1.03 \\
Add1TreeInt16 & 1.14 & 1.23 & \gibbonvanilla 1.13 & 1.12 & 0.91 & 0.92 & 1.01 & 1.14 & 1.03 \\
Add1TreeInt32 & 1.16 & 1.15 & \gibbonvanilla 1.17 & 1.07 & 0.97 & 0.97 & 0.99 & 1.10 & 1.02 \\
Add1TreeInt64 & 1.11 & 1.18 & \gibbonvanilla 1.12 & 1.06 & 0.89 & 0.99 & 1.09 & 1.07 & 1.03 \\
ArithmeticIntensityInt8 & 1.29 & 1.06 & \gibbonvanilla 1.29 & 1.37 & 1.04 & 1.32 & 0.98 & 1.21 & 1.19 \\
ArithmeticIntensityInt16 & 1.95 & 0.98 & \gibbonvanilla 1.95 & 2.04 & 0.99 & 2.03 & 0.93 & 2.24 & 1.28 \\
ArithmeticIntensityInt32 & 1.30 & 0.98 & \gibbonvanilla 1.30 & 1.34 & 0.96 & 1.29 & 0.92 & 1.38 & 1.19 \\
ArithmeticIntensityInt64 & 0.96 & 0.99 & \gibbonvanilla 0.96 & 1.01 & 1.00 & 1.01 & 0.99 & 0.98 & 1.02 \\
OctTree & 1.16 & 1.10 & \gibbonvanilla 1.05 & 1.15 & 1.02 & 1.03 & 1.05 & 1.05 & 1.07 \\
PiecewiseFunctions & 1.00 & 1.21 & \gibbonvanilla 1.00 & 0.97 & 0.97 & 0.98 & 0.99 & 0.99 & 1.02 \\
\midrule
\textit{Geomean, end to end} & 1.11 & 1.14 & \gibbonvanilla 1.08 & 1.15 & 1.00 & 1.06 & 1.02 & 1.10 & 1.09 \\
\textit{Geomean, fold passes} & 1.07 & 1.38 & \gibbonvanilla 1.02 & 1.18 & 1.04 & 1.05 & 1.06 & 1.05 & 1.00 \\
\textit{Geomean, map passes} & 1.15 & 1.10 & \gibbonvanilla 1.14 & 1.11 & 1.00 & 1.10 & 1.01 & 1.13 & 1.09 \\
\bottomrule
\end{tabular}}
\end{table}

\begin{table}[htbp]
\centering
\caption{Clang time over GCC time, end to end, per program and configuration, SoA configurations.}
\label{tab:compiler_comparison-soa}
\tiny\gibbonnumfont
\setlength{\tabcolsep}{3pt}
\gibbonfit{%
\begin{tabular}{l r r r r r r r r r r r}
\toprule
\textbf{Program} &\multicolumn{11}{c}{\textbf{SoA}} \\
\cmidrule(lr){2-12}
 & $S_{ri}^{\scriptscriptstyle -t}$ & $S_{ri}$ & $S_{rm}^{\scriptscriptstyle -t}$ & $S_{rm}^{\scriptscriptstyle -av}$ & $S_{rm}^{\scriptscriptstyle +av}$ & $S_{\ell}^{\scriptscriptstyle -av}$ & $S_{\ell}^{\scriptscriptstyle +av}$ & $S_{\ell b}^{\scriptscriptstyle -av}$ & $S_{\ell b}^{\scriptscriptstyle +av}$ & $S_{\ell bv}^{\scriptscriptstyle -av}$ & $S_{\ell bv}^{\scriptscriptstyle +av}$ \\
\midrule
Compiler & -- & -- & 0.93 & 1.13 & 1.17 & 0.93 & 0.89 & 1.14 & 1.14 & 1.13 & 1.14 \\
DBQuery & 0.68 & 0.68 & 1.20 & 1.08 & 1.16 & 0.97 & 0.97 & 1.06 & 1.09 & 1.06 & 1.09 \\
DecisionTree & 0.48 & 0.48 & 1.29 & 1.36 & 1.75 & 1.36 & 1.75 & 1.35 & 1.86 & 1.37 & 1.73 \\
DecisionTreeClassify & 1.30 & 0.86 & 1.47 & 1.04 & 0.99 & 1.03 & 0.99 & 1.03 & 0.99 & 1.03 & 0.99 \\
DomTree & 0.64 & 0.65 & 1.15 & 1.14 & 1.17 & 1.11 & 1.09 & 1.11 & 1.10 & 1.11 & 1.10 \\
KDTree & 0.86 & 0.87 & 1.01 & 1.12 & 1.07 & 1.11 & 1.07 & 1.13 & 1.07 & 1.12 & 1.08 \\
LinearListReduction & -- & -- & 0.89 & 1.09 & 1.13 & 0.83 & 0.86 & 0.91 & 0.92 & 0.96 & 0.99 \\
reduceNestedList & 2.00 & 2.00 & 1.83 & 1.01 & 1.04 & 1.02 & 1.04 & 1.02 & 1.04 & 1.02 & 1.04 \\
List & -- & -- & -- & 1.04 & 1.02 & 1.06 & 1.08 & 1.06 & 1.07 & 1.08 & 1.06 \\
MonoTree & 0.92 & 0.92 & 1.24 & 1.16 & 1.38 & 1.16 & 1.32 & 1.17 & 1.34 & 1.18 & 1.34 \\
ObjectGraph & 0.88 & 0.87 & 1.01 & 1.24 & 1.19 & 1.27 & 1.28 & 1.32 & 1.32 & 1.31 & 1.31 \\
TernaryTree & 0.50 & 0.50 & 1.13 & 0.99 & 1.14 & 1.04 & 1.16 & 1.06 & 1.17 & 1.05 & 1.16 \\
Trie & 0.68 & 0.68 & 1.24 & 1.02 & 1.35 & 0.89 & 0.99 & 1.06 & 1.25 & 1.06 & 1.24 \\
ColorOctree & -- & -- & 1.18 & 1.11 & 1.14 & 1.01 & 0.98 & 1.07 & 1.02 & 1.09 & 1.04 \\
Add1TreeInt8 & 0.91 & 0.91 & 1.05 & 0.94 & 0.96 & 0.85 & 1.05 & 0.84 & 1.04 & 0.86 & 1.06 \\
Add1TreeInt16 & 0.92 & 0.93 & 1.38 & 0.95 & 1.33 & 0.88 & 1.20 & 0.82 & 1.21 & 0.88 & 1.20 \\
Add1TreeInt32 & 0.91 & 0.91 & 1.30 & 0.95 & 1.29 & 0.90 & 1.17 & 0.90 & 1.18 & 0.95 & 1.20 \\
Add1TreeInt64 & 0.94 & 0.94 & 1.24 & 0.95 & 1.22 & 0.91 & 1.13 & 0.92 & 1.14 & 0.93 & 1.13 \\
ArithmeticIntensityInt8 & 0.89 & 0.87 & 0.66 & 1.07 & 0.67 & 0.98 & 1.68 & 0.99 & 1.73 & 0.80 & 0.92 \\
ArithmeticIntensityInt16 & 0.93 & 0.93 & 1.14 & 0.96 & 1.13 & 0.97 & 1.63 & 0.96 & 1.69 & 0.95 & 1.19 \\
ArithmeticIntensityInt32 & 0.94 & 0.94 & 0.93 & 0.94 & 0.94 & 0.97 & 1.40 & 0.97 & 1.41 & 1.01 & 1.16 \\
ArithmeticIntensityInt64 & 1.22 & 1.22 & 1.35 & 0.97 & 1.33 & 0.97 & 1.24 & 0.97 & 1.25 & 0.97 & 1.25 \\
OctTree & -- & -- & 1.30 & 1.04 & 1.21 & 1.02 & 1.05 & 1.04 & 1.06 & 1.03 & 1.07 \\
PiecewiseFunctions & 0.52 & 0.52 & 1.29 & 1.02 & 1.43 & 1.05 & 1.17 & 1.09 & 1.25 & 1.09 & 1.26 \\
\midrule
\textit{Geomean, end to end} & 0.85 & 0.83 & 1.16 & 1.05 & 1.16 & 1.01 & 1.15 & 1.03 & 1.20 & 1.04 & 1.15 \\
\textit{Geomean, fold passes} & 0.65 & 0.63 & 1.10 & 1.33 & 1.30 & 1.33 & 1.29 & 1.34 & 1.29 & 1.34 & 1.28 \\
\textit{Geomean, map passes} & 0.93 & 0.93 & 1.09 & 1.04 & 1.14 & 0.93 & 1.05 & 0.97 & 1.11 & 0.98 & 0.99 \\
\bottomrule
\end{tabular}}
\end{table}
  \endgroup

Figure~\ref{fig:tree-sweep-clang} repeats the tree sweep of
Figure~\ref{fig:tree-sweep} for the \colobus{} configurations, built with each
compiler.

On \texttt{add1Tree}, a configuration built with Clang takes $0.86$--$1.03\times$ its time with GCC (furthest: $S_{\ell bv}^{\scriptscriptstyle +av}$ at depth 12, $0.86\times$); on \texttt{sumTree}, $0.72$--$1.29\times$ (furthest: $S_{\ell bv}^{\scriptscriptstyle +av}$ at depth 10, $1.29\times$); on \texttt{buildTree}, $0.97$--$1.10\times$ (furthest: $S_{rm}^{\scriptscriptstyle +av}$ at depth 10, $1.10\times$). A ratio above 1 means Clang is slower.

\begin{figure*}[tp]
  \centering
  \includegraphics{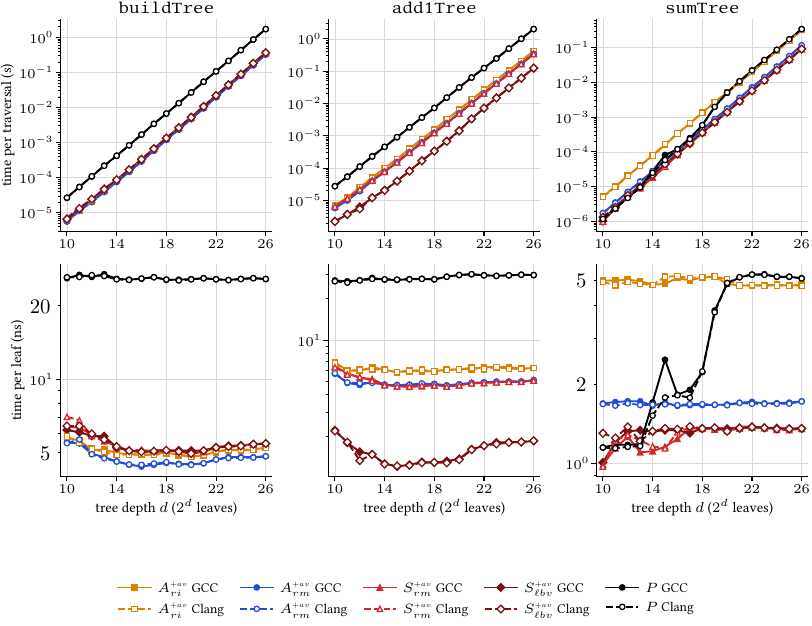}
  \caption{The tree sweep of Figure~\ref{fig:tree-sweep} for the configurations of \colobus{} and \gibbon{}, compiled with each C compiler: solid lines and filled markers GCC 16.2.0, dashed lines and open markers Clang 23.1.3, both for AVX2 (symbols as in Table~\ref{tab:pldi-legend}; $d = 10, \ldots, 26$). Top row: median time per traversal; bottom row: the same time per leaf. Each point is the median of 21 iterations in each of 3 interleaved rounds, checked against an independent model of the program.}
  \label{fig:tree-sweep-clang}
\end{figure*}

\subsection{End-to-end times}\label{sec:appendix-clang-stages}

\begin{figure}[tp]
  \centering
  \includegraphics{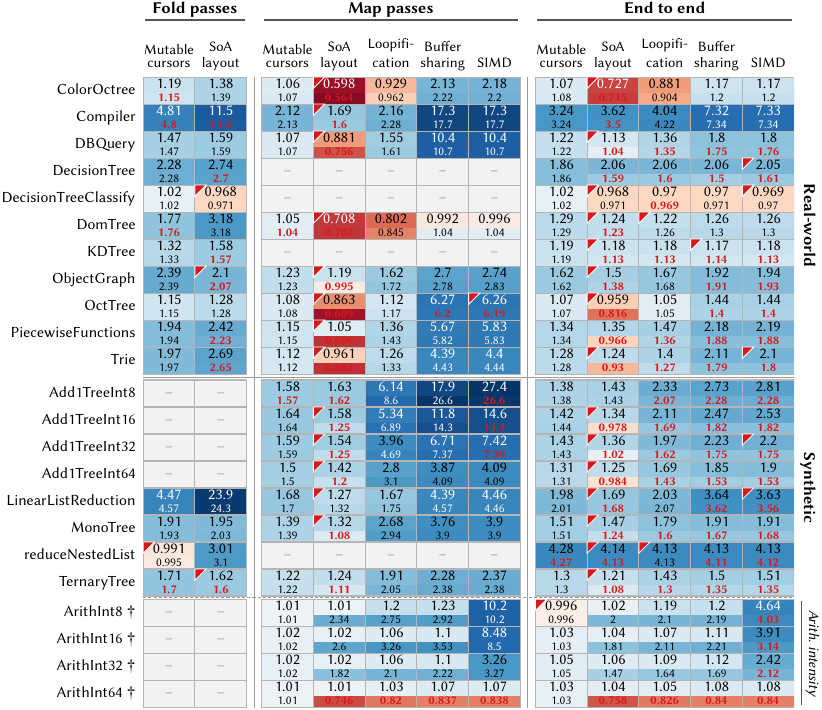}
  \caption{\emph{Clang~23.1.3.} Speedup over \gibbon{}
    ($A_{ri}^{\scriptscriptstyle +av}$) as each stage of the pipeline is added,
    as in Figure~\ref{fig:pldi-stages}, with the generated C compiled by Clang;
    \gibbon{} is compiled by Clang as well.}
  \label{fig:pldi-stages-clang}
\end{figure}

Figure~\ref{fig:pldi-stages-clang} is the Clang counterpart of
Figure~\ref{fig:pldi-stages}, and its cells are marked the same way.

\subsection{Bit widths}\label{sec:appendix-clang-widths}

Table~\ref{tab:width-clang} is the Clang counterpart of Table~\ref{tab:width}.

\input{tables/pldi/_preamble}
\begin{table}[t]
\centering
\caption{\emph{Clang 23.1.3.} The map passes \textit{add1Tree} and \textit{arithKernel} at four integer widths, with the columns of the main-text table.}
\label{tab:width-clang}
\footnotesize\gibbonnumfont
\setlength{\tabcolsep}{2.5pt}
\gibbonfit{%
\begin{tabular}{l rrrrrrrrrr cc}
\toprule
Width & \gibbon{} & Mutable & \shortstack{AoS\\loop} & SoA & \multicolumn{2}{c}{Loopified} & \multicolumn{2}{c}{\shortstack{Buffer\\sharing}} & \multicolumn{2}{c}{SIMD} & \multicolumn{2}{c}{Speedup} \\
\cmidrule(lr){2-2}\cmidrule(lr){3-3}\cmidrule(lr){4-4}\cmidrule(lr){5-5}\cmidrule(lr){6-7}\cmidrule(lr){8-9}\cmidrule(lr){10-11}\cmidrule(lr){12-13}
 & $A_{ri}^{\scriptscriptstyle +av}$ & $A_{rm}^{\scriptscriptstyle +av}$ & $A_{\ell}^{\scriptscriptstyle +av}$ & $S_{rm}^{\scriptscriptstyle +av}$ & $S_{\ell}^{\scriptscriptstyle -av}$ & $S_{\ell}^{\scriptscriptstyle +av}$ & $S_{\ell b}^{\scriptscriptstyle -av}$ & $S_{\ell b}^{\scriptscriptstyle +av}$ & $S_{\ell bv}^{\scriptscriptstyle -av}$ & $S_{\ell bv}^{\scriptscriptstyle +av}$ & \shortstack{SIMD\\gain} & \shortstack{vs.\\\gibbon{}} \\
\midrule
\multicolumn{13}{@{}l}{\textit{add1Tree}} \\
Int8 & 87.25 & 55.44 & 30.44 & 53.70 & 14.20 & 10.15 & 4.88 & 3.28 & 3.19 & 3.28 & 1.53$\times$ & 27.36$\times$ \\
Int16 & 90.68 & 55.41 & 34.79 & 72.76 & 16.98 & 13.17 & 7.68 & 6.36 & 6.19 & 6.30 & 1.24$\times$ & 14.64$\times$ \\
Int32 & 96.87 & 60.88 & 39.82 & 77.74 & 24.47 & 20.66 & 14.44 & 13.14 & 13.05 & 13.10 & 1.11$\times$ & 7.42$\times$ \\
Int64 & 108.80 & 72.35 & 51.82 & 90.42 & 38.90 & 35.09 & 28.11 & 26.59 & 26.59 & 26.58 & 1.06$\times$ & 4.09$\times$ \\
\midrule
\multicolumn{13}{@{}l}{\textit{arithKernel}} \\
Int8 & 534.30 & 528.40 & 464.10 & 228.20 & 442.80 & 193.80 & 435.10 & 182.60 & 52.39 & 52.40 & 8.31$\times$ & 10.20$\times$ \\
Int16 & 470.10 & 461.40 & 444.90 & 180.80 & 443.70 & 144.40 & 428.70 & 133.20 & 55.49 & 55.37 & 7.73$\times$ & 8.49$\times$ \\
Int32 & 484.30 & 474.00 & 455.60 & 266.60 & 455.00 & 230.30 & 438.70 & 218.10 & 148.40 & 148.10 & 2.96$\times$ & 3.27$\times$ \\
Int64 & 489.70 & 485.60 & 470.40 & 656.50 & 475.70 & 597.00 & 458.70 & 584.90 & 458.40 & 584.80 & 1.00$\times$ & 1.07$\times$ \\
\bottomrule
\end{tabular}}
\end{table}

\subsection{Whole-suite results}\label{sec:appendix-clang-suite}

Tables~\ref{tab:summary-clang} and~\ref{tab:summary-vs-vanilla-clang} are the
Clang counterparts of Tables~\ref{tab:summary} and~\ref{tab:summary-vs-vanilla},
and Tables~\ref{tab:cursor_comparison_times-clang}
and~\ref{tab:cursor_comparison_speedups-clang} of the cursor comparison; they
are read as in Section~\ref{sec:appendix-eval-suite}.
Every configuration was compiled for the same SIMD target, 256-bit AVX2 (\texttt{-mavx2}), which the SIMD pass and the C compiler's auto-vectorizer both use. Every configuration frees the region chunks of each timed iteration (\texttt{--reclaim-iterate-regions}), so memory use does not grow with the iteration count. Some inputs were resized from their defaults: Add1TreeInt16 (depth = 34; default 33), Add1TreeInt32 (depth = 34; default 33), Add1TreeInt64 (depth = 34; default 33), Add1TreeInt8 (depth = 34; default 33), MonoTree (depth = 25; default 23), reduceNestedList (n (outer list elements) = 10,000; default 1,000,000).

  \begingroup
    \setlength{\textfloatsep}{4pt plus 1pt minus 1pt}%
    \setlength{\floatsep}{3pt plus 1pt minus 1pt}%
    \setlength{\intextsep}{3pt plus 1pt minus 1pt}%
    \setlength{\abovecaptionskip}{3pt}%
    \setlength{\belowcaptionskip}{2pt}%
    \begin{table}[htbp]
\centering
\caption{\emph{Clang 23.1.3.} Pass-sum time per program (ms), loopified AoS against the fully optimized SoA configuration.}
\label{tab:summary-clang}
\small\gibbonnumfont
\setlength{\tabcolsep}{3pt}
\gibbonfit{%
\begin{tabular}{l c c r r r r r r r r r}
\toprule
\textbf{Program} & \textbf{ADT} & \textbf{SoA} & \multicolumn{3}{c}{\textbf{End-to-end}} & \multicolumn{3}{c}{\textbf{Fold passes}} & \multicolumn{3}{c}{\textbf{Map passes}} \\
\cmidrule(lr){4-6}\cmidrule(lr){7-9}\cmidrule(lr){10-12}
 & fields & bufs & $A_{\ell}^{\scriptscriptstyle +av}$ (ms) & $S_{\ell bv}^{\scriptscriptstyle +av}$ (ms) & $A_{\ell}^{\scriptscriptstyle +av}$/$S_{\ell bv}^{\scriptscriptstyle +av}$ & $A_{\ell}^{\scriptscriptstyle +av}$ (ms) & $S_{\ell bv}^{\scriptscriptstyle +av}$ (ms) & $A_{\ell}^{\scriptscriptstyle +av}$/$S_{\ell bv}^{\scriptscriptstyle +av}$ & $A_{\ell}^{\scriptscriptstyle +av}$ (ms) & $S_{\ell bv}^{\scriptscriptstyle +av}$ (ms) & $A_{\ell}^{\scriptscriptstyle +av}$/$S_{\ell bv}^{\scriptscriptstyle +av}$ \\
\midrule
OctTree & 16 & 9 & 1101.00 & 655.00 & \textbf{1.68$\times$} & 474.70 & 413.80 & \textbf{1.15$\times$} & 626.10 & 241.20 & \textbf{2.60$\times$} \\
Compiler & 9 & 8 & 315.80 & 86.70 & \textbf{3.64$\times$} & 159.60 & 67.70 & \textbf{2.36$\times$} & 156.20 & 19.00 & \textbf{8.21$\times$} \\
DBQuery & 15 & 13 & 318.10 & 171.20 & \textbf{1.86$\times$} & 168.10 & 155.60 & 1.08$\times$ & 150.00 & 15.50 & \textbf{9.66$\times$} \\
DecisionTree & 7 & 6 & 131.40 & 110.90 & \textbf{1.18$\times$} & 131.40 & 110.90 & \textbf{1.18$\times$} & -- & -- & -- \\
DecisionTreeClassify & 7 & 6 & 174.30 & 183.90 & 0.95$\times$ & 174.30 & 183.90 & 0.95$\times$ & -- & -- & -- \\
DomTree & 15 & 14 & 226.30 & 147.80 & \textbf{1.53$\times$} & 167.10 & 87.60 & \textbf{1.91$\times$} & 59.20 & 60.20 & 0.98$\times$ \\
KDTree & 17 & 16 & 200.10 & 167.40 & \textbf{1.20$\times$} & 200.10 & 167.40 & \textbf{1.20$\times$} & -- & -- & -- \\
LinearListReduction & 11 & 11 & 677.40 & 243.90 & \textbf{2.78$\times$} & 91.80 & 17.10 & \textbf{5.38$\times$} & 585.60 & 226.90 & \textbf{2.58$\times$} \\
reduceNestedList & 3 & 3 & 0.25 & 0.08 & \textbf{3.13$\times$} & 0.25 & 0.08 & \textbf{3.13$\times$} & -- & -- & -- \\
List & 2 & 2 & 389.40 & 326.10 & \textbf{1.19$\times$} & 146.90 & 138.50 & 1.06$\times$ & 242.50 & 187.50 & \textbf{1.29$\times$} \\
MonoTree & 3 & 2 & 291.50 & 222.40 & \textbf{1.31$\times$} & 167.40 & 161.60 & 1.04$\times$ & 124.10 & 60.90 & \textbf{2.04$\times$} \\
ObjectGraph & 5 & 4 & 254.70 & 209.80 & \textbf{1.21$\times$} & 132.10 & 145.80 & 0.91$\times$ & 122.60 & 64.00 & \textbf{1.91$\times$} \\
TernaryTree & 5 & 3 & 88.20 & 70.80 & \textbf{1.25$\times$} & 30.40 & 30.60 & 0.99$\times$ & 57.80 & 40.20 & \textbf{1.44$\times$} \\
Trie & 10 & 9 & 208.10 & 89.90 & \textbf{2.31$\times$} & 70.30 & 52.20 & \textbf{1.35$\times$} & 137.80 & 37.70 & \textbf{3.66$\times$} \\
PiecewiseFunctions & 8 & 7 & 376.10 & 174.80 & \textbf{2.15$\times$} & 156.10 & 127.00 & \textbf{1.23$\times$} & 220.10 & 47.80 & \textbf{4.60$\times$} \\
ColorOctree & 25 & 18 & 551.20 & 356.50 & \textbf{1.55$\times$} & 192.60 & 166.90 & \textbf{1.15$\times$} & 358.60 & 189.60 & \textbf{1.89$\times$} \\
\bottomrule
\end{tabular}}
\end{table}
  \endgroup
  \begingroup
    \setlength{\textfloatsep}{4pt plus 1pt minus 1pt}%
    \setlength{\floatsep}{3pt plus 1pt minus 1pt}%
    \setlength{\intextsep}{3pt plus 1pt minus 1pt}%
    \setlength{\abovecaptionskip}{3pt}%
    \setlength{\belowcaptionskip}{2pt}%
    \begin{table}[htbp]
\centering
\caption{\emph{Clang 23.1.3.} Pass-sum time per program (ms), \gibbon{} against the fully optimized SoA configuration.}
\label{tab:summary-vs-vanilla-clang}
\small\gibbonnumfont
\setlength{\tabcolsep}{3pt}
\gibbonfit{%
\begin{tabular}{l c c r r r r r r r r r}
\toprule
\textbf{Program} & \textbf{ADT} & \textbf{SoA} & \multicolumn{3}{c}{\textbf{End-to-end}} & \multicolumn{3}{c}{\textbf{Fold passes}} & \multicolumn{3}{c}{\textbf{Map passes}} \\
\cmidrule(lr){4-6}\cmidrule(lr){7-9}\cmidrule(lr){10-12}
 & fields & bufs & \gibbonvanilla $A_{ri}^{\scriptscriptstyle +av}$ (ms) & $S_{\ell bv}^{\scriptscriptstyle +av}$ (ms) & $A_{ri}^{\scriptscriptstyle +av}$/$S_{\ell bv}^{\scriptscriptstyle +av}$ & \gibbonvanilla $A_{ri}^{\scriptscriptstyle +av}$ (ms) & $S_{\ell bv}^{\scriptscriptstyle +av}$ (ms) & $A_{ri}^{\scriptscriptstyle +av}$/$S_{\ell bv}^{\scriptscriptstyle +av}$ & \gibbonvanilla $A_{ri}^{\scriptscriptstyle +av}$ (ms) & $S_{\ell bv}^{\scriptscriptstyle +av}$ (ms) & $A_{ri}^{\scriptscriptstyle +av}$/$S_{\ell bv}^{\scriptscriptstyle +av}$ \\
\midrule
OctTree & 16 & 9 & \gibbonvanilla 1290.00 & 655.00 & \textbf{1.97$\times$} & \gibbonvanilla 552.90 & 413.80 & \textbf{1.34$\times$} & \gibbonvanilla 737.00 & 241.20 & \textbf{3.06$\times$} \\
Compiler & 9 & 8 & \gibbonvanilla 1104.00 & 86.70 & \textbf{12.73$\times$} & \gibbonvanilla 767.80 & 67.70 & \textbf{11.34$\times$} & \gibbonvanilla 336.60 & 19.00 & \textbf{17.70$\times$} \\
DBQuery & 15 & 13 & \gibbonvanilla 413.00 & 171.20 & \textbf{2.41$\times$} & \gibbonvanilla 246.10 & 155.60 & \textbf{1.58$\times$} & \gibbonvanilla 166.90 & 15.50 & \textbf{10.75$\times$} \\
DecisionTree & 7 & 6 & \gibbonvanilla 299.70 & 110.90 & \textbf{2.70$\times$} & \gibbonvanilla 299.70 & 110.90 & \textbf{2.70$\times$} & \gibbonvanilla -- & -- & -- \\
DecisionTreeClassify & 7 & 6 & \gibbonvanilla 178.50 & 183.90 & 0.97$\times$ & \gibbonvanilla 178.50 & 183.90 & 0.97$\times$ & \gibbonvanilla -- & -- & -- \\
DomTree & 15 & 14 & \gibbonvanilla 358.20 & 147.80 & \textbf{2.42$\times$} & \gibbonvanilla 295.40 & 87.60 & \textbf{3.37$\times$} & \gibbonvanilla 62.80 & 60.20 & 1.04$\times$ \\
KDTree & 17 & 16 & \gibbonvanilla 265.90 & 167.40 & \textbf{1.59$\times$} & \gibbonvanilla 265.90 & 167.40 & \textbf{1.59$\times$} & \gibbonvanilla -- & -- & -- \\
LinearListReduction & 11 & 11 & \gibbonvanilla 1431.00 & 243.90 & \textbf{5.87$\times$} & \gibbonvanilla 418.60 & 17.10 & \textbf{24.55$\times$} & \gibbonvanilla 1012.00 & 226.90 & \textbf{4.46$\times$} \\
reduceNestedList & 3 & 3 & \gibbonvanilla 0.25 & 0.08 & \textbf{3.11$\times$} & \gibbonvanilla 0.25 & 0.08 & \textbf{3.11$\times$} & \gibbonvanilla -- & -- & -- \\
List & 2 & 2 & \gibbonvanilla -- & 326.10 & -- & \gibbonvanilla -- & 138.50 & -- & \gibbonvanilla -- & 187.50 & -- \\
MonoTree & 3 & 2 & \gibbonvanilla 556.70 & 222.40 & \textbf{2.50$\times$} & \gibbonvanilla 319.20 & 161.60 & \textbf{1.98$\times$} & \gibbonvanilla 237.50 & 60.90 & \textbf{3.90$\times$} \\
ObjectGraph & 5 & 4 & \gibbonvanilla 484.50 & 209.80 & \textbf{2.31$\times$} & \gibbonvanilla 303.40 & 145.80 & \textbf{2.08$\times$} & \gibbonvanilla 181.10 & 64.00 & \textbf{2.83$\times$} \\
TernaryTree & 5 & 3 & \gibbonvanilla 145.20 & 70.80 & \textbf{2.05$\times$} & \gibbonvanilla 49.40 & 30.60 & \textbf{1.62$\times$} & \gibbonvanilla 95.80 & 40.20 & \textbf{2.38$\times$} \\
Trie & 10 & 9 & \gibbonvanilla 306.60 & 89.90 & \textbf{3.41$\times$} & \gibbonvanilla 139.30 & 52.20 & \textbf{2.67$\times$} & \gibbonvanilla 167.30 & 37.70 & \textbf{4.44$\times$} \\
PiecewiseFunctions & 8 & 7 & \gibbonvanilla 580.80 & 174.80 & \textbf{3.32$\times$} & \gibbonvanilla 302.20 & 127.00 & \textbf{2.38$\times$} & \gibbonvanilla 278.60 & 47.80 & \textbf{5.83$\times$} \\
ColorOctree & 25 & 18 & \gibbonvanilla 647.00 & 356.50 & \textbf{1.81$\times$} & \gibbonvanilla 229.40 & 166.90 & \textbf{1.37$\times$} & \gibbonvanilla 417.70 & 189.60 & \textbf{2.20$\times$} \\
\bottomrule
\end{tabular}}
\end{table}
  \endgroup
  \begingroup
    \setlength{\textfloatsep}{4pt plus 1pt minus 1pt}%
    \setlength{\floatsep}{3pt plus 1pt minus 1pt}%
    \setlength{\intextsep}{3pt plus 1pt minus 1pt}%
    \setlength{\abovecaptionskip}{3pt}%
    \setlength{\belowcaptionskip}{2pt}%
    \begin{table}[htbp]
\centering
\caption{\emph{Clang 23.1.3.} Whole-program time (ms) with mutable and immutable cursors.}
\label{tab:cursor_comparison_times-clang}
\small
\gibbonfit{%
\begin{tabular}{p{5.5cm} r r r r}
\toprule
\textbf{Program} & $A_{rm}^{\scriptscriptstyle +av}$ & $A_{ri}^{\scriptscriptstyle +av}$ & $S_{rm}^{\scriptscriptstyle +av}$ & $S_{ri}$ \\
\midrule
Compiler & 7507.94 & 24462.02 & \textbf{6707.65} & -- \\
DBQuery & \textbf{7444.50} & 9372.91 & 8612.20 & 16357.20 \\
DecisionTree & 2850.06 & 6410.27 & \textbf{2478.64} & 11192.02 \\
DecisionTreeClassify & \textbf{4032.64} & 4111.57 & 4212.80 & 5432.92 \\
DomTree & 5346.49 & 8083.14 & \textbf{4450.28} & 12909.15 \\
KDTree & 4364.58 & 5749.98 & \textbf{3751.63} & 10177.83 \\
LinearListReduction & \textbf{18349.06} & 34455.93 & 20258.35 & -- \\
reduceNestedList & 135.54 & 489.78 & \textbf{135.30} & 486.75 \\
List & \textbf{9987.10} & -- & 11753.73 & -- \\
MonoTree & \textbf{7521.34} & 12190.24 & 8447.74 & 18178.81 \\
ObjectGraph & \textbf{6359.87} & 10854.50 & 7502.92 & 17107.75 \\
OctTree & \textbf{15588.60} & 16930.07 & 21382.53 & -- \\
PiecewiseFunctions\_norm2Estimate & \textbf{700.83} & 1338.26 & 710.83 & 3252.34 \\
PiecewiseFunctions\_truncateTolViolations & 662.33 & 1142.45 & \textbf{660.38} & 2906.26 \\
PiecewiseFunctions\_compressMass & 667.38 & 1110.21 & \textbf{632.65} & 2894.06 \\
PiecewiseFunctions\_autorefineMaxLevel & \textbf{822.36} & 1117.91 & 839.62 & 2865.24 \\
PiecewiseFunctions\_pmapCutHistogram & 695.42 & 1386.42 & \textbf{683.96} & 3242.61 \\
PiecewiseFunctions\_lbDeuxLoadProxy & \textbf{665.84} & 1179.03 & 678.64 & 3003.55 \\
PiecewiseFunctions\_addConstPW & \textbf{3112.46} & 3516.44 & 4846.64 & 6811.04 \\
PiecewiseFunctions\_diffPW & \textbf{3250.05} & 3694.44 & 5163.22 & 7666.17 \\
TernaryTree & \textbf{2480.51} & 3286.50 & 2743.56 & 6617.37 \\
Trie & \textbf{5307.39} & 7131.82 & 6927.67 & 14692.31 \\
ColorOctree & \textbf{14700.76} & 15961.19 & 23457.93 & -- \\
\bottomrule
\end{tabular}}
\end{table}
  \endgroup
  \begingroup
    \setlength{\textfloatsep}{4pt plus 1pt minus 1pt}%
    \setlength{\floatsep}{3pt plus 1pt minus 1pt}%
    \setlength{\intextsep}{3pt plus 1pt minus 1pt}%
    \setlength{\abovecaptionskip}{3pt}%
    \setlength{\belowcaptionskip}{2pt}%
    \begin{table}[htbp]
\centering
\caption{\emph{Clang 23.1.3.} Speedups between the cursor configurations of Table~\ref{tab:cursor_comparison_times-clang}.}
\label{tab:cursor_comparison_speedups-clang}
\small
\gibbonfit{%
\begin{tabular}{p{5.5cm} r r r r}
\toprule
\textbf{Program} & $A_{rm}^{\scriptscriptstyle +av}$/$S_{rm}^{\scriptscriptstyle +av}$ & $A_{ri}^{\scriptscriptstyle +av}$/$A_{rm}^{\scriptscriptstyle +av}$ & $A_{ri}^{\scriptscriptstyle +av}$/$S_{rm}^{\scriptscriptstyle +av}$ & $A_{ri}^{\scriptscriptstyle +av}$/$S_{ri}$ \\
\midrule
Compiler & \textbf{1.12$\times$} & \textbf{3.26$\times$} & \textbf{3.65$\times$} & -- \\
DBQuery & 0.86$\times$ & \textbf{1.26$\times$} & 1.09$\times$ & 0.57$\times$ \\
DecisionTree & \textbf{1.15$\times$} & \textbf{2.25$\times$} & \textbf{2.59$\times$} & 0.57$\times$ \\
DecisionTreeClassify & 0.96$\times$ & 1.02$\times$ & 0.98$\times$ & 0.76$\times$ \\
DomTree & \textbf{1.20$\times$} & \textbf{1.51$\times$} & \textbf{1.82$\times$} & 0.63$\times$ \\
KDTree & \textbf{1.16$\times$} & \textbf{1.32$\times$} & \textbf{1.53$\times$} & 0.56$\times$ \\
LinearListReduction & 0.91$\times$ & \textbf{1.88$\times$} & \textbf{1.70$\times$} & -- \\
reduceNestedList & 1.00$\times$ & \textbf{3.61$\times$} & \textbf{3.62$\times$} & 1.01$\times$ \\
List & 0.85$\times$ & -- & -- & -- \\
MonoTree & 0.89$\times$ & \textbf{1.62$\times$} & \textbf{1.44$\times$} & 0.67$\times$ \\
ObjectGraph & 0.85$\times$ & \textbf{1.71$\times$} & \textbf{1.45$\times$} & 0.63$\times$ \\
OctTree & 0.73$\times$ & 1.09$\times$ & 0.79$\times$ & -- \\
PiecewiseFunctions\_norm2Estimate & 0.99$\times$ & \textbf{1.91$\times$} & \textbf{1.88$\times$} & 0.41$\times$ \\
PiecewiseFunctions\_truncateTolViolations & 1.00$\times$ & \textbf{1.72$\times$} & \textbf{1.73$\times$} & 0.39$\times$ \\
PiecewiseFunctions\_compressMass & 1.05$\times$ & \textbf{1.66$\times$} & \textbf{1.75$\times$} & 0.38$\times$ \\
PiecewiseFunctions\_autorefineMaxLevel & 0.98$\times$ & \textbf{1.36$\times$} & \textbf{1.33$\times$} & 0.39$\times$ \\
PiecewiseFunctions\_pmapCutHistogram & 1.02$\times$ & \textbf{1.99$\times$} & \textbf{2.03$\times$} & 0.43$\times$ \\
PiecewiseFunctions\_lbDeuxLoadProxy & 0.98$\times$ & \textbf{1.77$\times$} & \textbf{1.74$\times$} & 0.39$\times$ \\
PiecewiseFunctions\_addConstPW & 0.64$\times$ & \textbf{1.13$\times$} & 0.73$\times$ & 0.52$\times$ \\
PiecewiseFunctions\_diffPW & 0.63$\times$ & \textbf{1.14$\times$} & 0.72$\times$ & 0.48$\times$ \\
TernaryTree & 0.90$\times$ & \textbf{1.32$\times$} & \textbf{1.20$\times$} & 0.50$\times$ \\
Trie & 0.77$\times$ & \textbf{1.34$\times$} & 1.03$\times$ & 0.49$\times$ \\
ColorOctree & 0.63$\times$ & 1.09$\times$ & 0.68$\times$ & -- \\
\bottomrule
\end{tabular}}
\end{table}
  \endgroup

\subsection{Comparison with GHC and MLton}\label{sec:appendix-clang-ghc-mlton}

Table~\ref{tab:comparison_ghc_mlton-clang} repeats
Table~\ref{tab:comparison_ghc_mlton} with the \colobus{} programs compiled by
Clang; GHC and MLton do not use a C compiler, so only the \colobus{} columns
change.

\begin{table}[htbp]
\centering
\caption{\emph{Clang 23.1.3.} Runtime comparison of \colobus{} (AoS/SoA), GHC, and MLton. Times are total median per iteration (ms). \textbf{Bold} marks the fastest time for each program. The Geomean row covers the 29 program(s) every column measured.}
\label{tab:comparison_ghc_mlton-clang}
\small
\gibbonfit{%
\begin{tabular}{l r r r r}
\toprule
\textbf{Program} & \textbf{\colobus{}-AoS} & \textbf{\colobus{}-SoA} & \textbf{GHC} & \textbf{MLton} \\
\midrule
Compiler & 310.30 & \textbf{273.50} & 1286.00 & 434.90 \\
DBQuery & \textbf{323.10} & 380.40 & 615.70 & 458.90 \\
DecisionTree & 119.90 & \textbf{107.30} & 411.00 & 336.40 \\
DecisionTreeClassify & 170.40 & 179.80 & 324.50 & \textbf{87.10} \\
DomTree & 214.80 & 175.30 & 427.60 & \textbf{145.70} \\
KDTree & 187.40 & \textbf{168.10} & 492.90 & 245.50 \\
LinearListReduction & 646.90 & 750.70 & 5629.00 & \textbf{553.10} \\
List & \textbf{383.90} & 454.00 & 13000.00 & 1547.00 \\
MonoTree & \textbf{80.80} & 91.00 & 591.00 & 169.40 \\
ObjectGraph & \textbf{271.60} & 331.20 & 672.10 & 366.00 \\
Oct:sumMass & 34.90 & \textbf{33.10} & 67.50 & 37.70 \\
Oct:sumEnergy & 40.70 & \textbf{37.30} & 171.70 & 51.30 \\
Oct:countActive & 34.80 & \textbf{29.60} & 79.70 & 40.20 \\
Oct:countParticles & 36.20 & \textbf{29.60} & 34.40 & 37.00 \\
Oct:barnesHutPotential & 41.60 & \textbf{37.50} & 169.30 & 52.50 \\
Oct:fmmPotential & \textbf{82.00} & 82.60 & 551.10 & 282.00 \\
Oct:scaleEnergy & \textbf{150.90} & 260.60 & 1186.00 & 170.90 \\
Oct:clearFlags & 145.20 & 261.90 & 288.40 & \textbf{83.10} \\
PiecewiseFunctions\_norm2Estimate & 24.60 & \textbf{21.60} & 61.00 & 33.30 \\
PiecewiseFunctions\_truncateTolViolations & 22.30 & \textbf{19.60} & 63.50 & 43.50 \\
PiecewiseFunctions\_compressMass & 22.40 & \textbf{18.60} & 56.30 & 39.80 \\
PiecewiseFunctions\_autorefineMaxLevel & 31.20 & \textbf{22.30} & 63.50 & 31.20 \\
PiecewiseFunctions\_pmapCutHistogram & 24.90 & \textbf{22.00} & 78.00 & 52.90 \\
PiecewiseFunctions\_lbDeuxLoadProxy & 22.70 & \textbf{22.10} & 60.10 & 30.10 \\
PiecewiseFunctions\_addConstPW & 121.50 & 200.90 & 182.60 & \textbf{88.00} \\
PiecewiseFunctions\_diffPW & 120.50 & 200.10 & 501.10 & \textbf{91.70} \\
TernaryTree & \textbf{107.90} & 118.70 & 313.00 & 187.90 \\
Trie & \textbf{219.20} & 294.80 & 507.10 & 292.00 \\
ColorOctree & 556.50 & 872.20 & 2927.00 & \textbf{474.10} \\
\midrule
\textbf{Geomean} & \textbf{95.30} & 101.30 & 310.20 & 123.00 \\
\bottomrule
\end{tabular}}
\end{table}

\subsection{Per-program results and hardware counters}\label{sec:appendix-clang-programs}

The supplementary material gives, for Clang, every program's fold and map
tables, the effect of each optimization pass by pass, and the hardware
counters, in the form of Appendix~\ref{sec:appendix-eval}.

}{}

\end{document}